\documentclass[aps,prc,twocolumn,showpacs,amssymb,amsmath,amsfonts,floatfix,nofootinbib]{revtex4-2}
\usepackage{graphicx}

\usepackage{amsmath}
\usepackage[usenames]{color}
\usepackage{epsfig}
\usepackage{epstopdf}
\usepackage{amssymb,amsmath}
\usepackage{amsmath}
\usepackage{epstopdf}
\usepackage{sidecap}
\usepackage{cases}
\usepackage{enumitem}
\usepackage{bm}
\usepackage{overpic}
\usepackage{upgreek}
\usepackage{morefloats}
\usepackage{hyperref}

\definecolor{grey}{rgb}{0.9,0.9,0.9}
\definecolor{black}{rgb}{0,0,0}

\newcommand{\be}{\begin{eqnarray}}
\newcommand{\ee}{\end{eqnarray}}
\newcommand{\bc}{\begin{center}}
\newcommand{\ec}{\end{center}}
\newcommand{\beq}{\begin{eqnarray}}
\newcommand{\eea}{\end{eqnarray}}

\newcommand{\Ocal}{\mathcal{O}}
\newcommand{\Ccal}{\mathcal{C}}

\newcommand{\HISKP}{Helmholtz-Institut f\"{u}r Strahlen- und Kernphysik, Universit\"{a}t Bonn, Germany}

\begin{document}


\title{Moravcsik's theorem on complete sets of polarization ob\-serva\-bles reexamined}
\author{Y.~Wunderlich}\email[Corresponding author: ]{wunderlich@hiskp.uni-bonn.de}\affiliation{\HISKP} 
\author{P.~Kroenert}
\affiliation{\HISKP} 
\author{F.~Afzal}
\affiliation{\HISKP} 
\author{A.~Thiel}
\affiliation{\HISKP} 

\date{\today}
\begin{abstract}
We revisit Moravcsik's theorem on the unique extraction of amplitudes from polarization ob\-serva\-bles, which has been originally published in 1985. The proof is (re-) written in a more formal and detailed way and the theorem is corrected for the special case of an odd number of amplitudes (this case was treated incorrectly in the original publication). Moravcsik's theorem, in the modified form, can be applied in principle to the extraction of an arbitrary number of $N$ helicity amplitudes. \\
The uniqueness theorem is then applied to hadronic reactions involving particles with spin. The most basic example is Pion-Nucleon scattering ($N=2$), the first non-trivial example is pseudoscalar meson photoproduction ($N=4$) and the most technically involved case treated here is given by pseudoscalar meson electroproduction ($N=6$). The application of Moravcsik's theorem to electroproduction yields new results, which for the first time provide insights into the structure and content of complete sets for this particular process. The uniqueness-statements for the various reactions are compared and an attempt is made to recognize general patterns, which emerge under the application of the theorem.
\end{abstract}

\maketitle

\section{Introduction} \label{sec:Intro}

In the field of hadron spectroscopy, reactions among particles with spin have long been used as tools to improve our understanding of QCD. In the case of baryon spectroscopy in particular~\cite{Klempt:2009pi, Ireland:2019uwn} most experimental activities have taken place at photon facilities all over the world like the CBELSA/TAPS experiment at Bonn~\cite{Sparks:2010vb, Thiel:2012yj, Gottschall:2013uha, Hartmann:2014mya}, A2 at MAMI (Mainz)~\cite{Schumann:2010js, Hornidge:2012ca, Sikora:2013vfa, Schumann:2015ypa, Adlarson:2015byy, Annand:2016ppc, Gardner:2016irh, Kashevarov:2017kqb, Dieterle:2017myg, Briscoe:2019cyo}, CLAS at JLab (Newport News)~\cite{Dugger:2013, Strauch:2015zob, Senderovich:2015lek, Mattione:2017fxc, Ho:2017kca, Collins:2017sgu, Kunkel:2017src} and LEPS at SPring-8 (Hyōgo Prefecture)~\cite{Kohri:2017kto}, in recent years. With these experiments, reactions containing one or multiple pseudoscalar mesons, as well as vector mesons, in the final state have been extracted. 
Recently, an increasing data set for the reaction of pseudoscalar meson electroproduction has become available~\cite{Zhao:2019fkm,Markov:2019fjy} and also the new CLAS12 experiment has started taking data using an electron beam~\cite{Burkert:2020akg}. Furthermore, it has to be mentioned that the GlueX Collaboration has published new data on $\pi^{0}$-, $\eta$- and $\eta'$-photoproduction recently~\cite{AlGhoul:2017nbp,Adhikari:2019gfa,Adhikari:2020cvz}.


A generic problem concerning reactions with spin is the extraction of $N$ so-called spin-amplitudes, which provide a model-independent parametrization of the $T$-matrix of the reaction, out of a set of $N^{2}$ polarization ob\-serva\-bles. In the context of such amplitude-extraction problems, it is natural to search for {\it complete experiments}~\cite{Barker:1975bp}, which denote minimal subsets of the full set of $N^{2}$ ob\-serva\-bles that allow for an unambiguous extraction of the $N$ amplitudes up to one overall phase. 

Following significant insights into discrete amplitude ambiguities established by Keaton and Workman~\cite{Keaton:1995pw, Keaton:1996pe}, Chiang and Tabakin~\cite{Chiang:1996em} found that in the case of pseudoscalar meson photoproduction ($N = 4$), a set of $8$ carefully selected ob\-serva\-bles can yield a complete experiment. While a complete proof has been lacking in the original publication~\cite{Chiang:1996em}, Nakayama~\cite{Nakayama:2018yzw} has recently given a rigorous algebraic treatment of all the relevant cases. His phase-fixing procedure utilizes the regularities in the definitions of the ob\-serva\-bles to a maximal extent. All the mathematical treatments of complete experiments mentioned up to this point assume the academic case of vanishing measurement uncertainty for the ob\-serva\-bles. As soon as ob\-serva\-bles have finite uncertainties, it is likely that more polarization ob\-serva\-bles are needed for a unique amplitude extraction. This fact has been substantiated in a number of recent works~\cite{Vrancx:2013pza, Vrancx:2014yja, Nys:2015kqa, Ireland:2010bi}.

For an arbitrary number of amplitudes $N$, the results of Keaton and Workman~\cite{Keaton:1995pw,Keaton:1996pe} and of Chiang and Tabakin~\cite{Chiang:1996em} generalize as follows (cf. in particular the footnote $1$ in the paper~\cite{Keaton:1996pe}): it is well-known that one can determine the $N$ amplitudes only up to one unknown overall phase. In order to achieve this, one needs at least $2 N - 1$ observables. However, with $2 N - 1$ observables, there generally still remain so-called {\it discrete ambiguities}~\cite{Keaton:1995pw,Chiang:1996em}, the resolution of which requires at least one additional observable. In this way, one arrives at a minimum number of $2N$ observables. Additionally, one has to bear in mind that the $2N$ observables are required to be known at each individual point in the kinematical phase space of the considered reaction. For a $2 \rightarrow 2$-process, this means at {\it each point in energy and angle}~\cite{Nakayama:2018yzw,MyDiploma,MyPhD}. Since the amplitude extraction also takes place at each point in phase space individually, the unknown overall phase can in principle have an arbitrary dependence on the full reaction kinematics.

While it is possible to find even more compelling heuristic arguments~\cite{GWUCommunication2016} that the number $2 N$ is indeed true for arbitrary $N$, a fully rigorous proof is, as far as we know at the moment, lacking. Actually, the theorem treated in this work can be understood as another nod in the direction that $2 N$ may indeed be the universally correct number. Nonetheless, this number has turned out to be true for all the specific reactions we found treated in the literature so far~\cite{Keaton:1995pw,Chiang:1996em,Nakayama:2018yzw,Tiator:2017cde,Anisovich:2013tij,Arenhoevel:2014dwa}.


In 1985, M.~J.~Moravcsik published a paper with a solution to the amplitude-extraction problem for an arbitrary number of amplitudes $N$~\cite{Moravcsik:1984uf}. However, instead of starting from the polarization ob\-serva\-bles mentioned above, he directly considered just the bilinear products $b_{i}^{\ast} b_{j}$ of amplitudes. Generally, polarization ob\-serva\-bles are invertible linear combinations of such bilinear products. Although the paper did not receive much attention at first, we feel that it deserves an explicit re-consideration, due to a number of features that make Moravcsik's theo\-rem attractive. Firstly, the theorem is formulated in the language of a 'geometrical analog'~\cite{Moravcsik:1984uf}, which leads to a lucid representation of complete sets in the form of graphs. Secondly, the theorem is valid for arbitrary $N$ and can thus be used as a master-approach for the (pre-) selection of complete sets of ob\-serva\-bles for in principle any reaction.  It is even possible to extract a standard procedure for this purpose, which can be automated on a computer. To be fair, Nakayamas phase-fixing procedure~\cite{Nakayama:2018yzw} can also be used for any number of amplitudes $N$. However, for larger $N$ (i.e. $N > 4$), it involves a rapidly growing number of different cases, which all need a single algebraic treatment. In contrast, the application of Moravcsik's theorem is in principle only limited by computation time. This makes it particularly useful for reactions with complicated spin-structures like~\cite{Tiator:2017cde,Roberts:2004mn,Arenhoevel:2014dwa,Pichowsky:1994gh}.

The paper is organized as follows: in section~\ref{sec:AlgebraicStartingPoint}, we describe the algebraic initial situation for an arbitrary number of amplitudes $N$. Moravcsik's theorem and its modified form are stated in section~\ref{sec:MoravcsikCompExp}. A proof of the modified theorem is included as well, in the appendix~\ref{sec:DetailedProof}. Section~\ref{sec:MoravcsikTheoremExamples} collects applications of the theorem to the simplest possible cases, which range from Pion-Nucleon scattering ($N=2$) to pseudoscalar meson photoproduction ($N = 4$). With the experience gained in the treatement of the most basic examples, it becomes clear how the theorem can be used in order to find complete sets of ob\-serva\-bles for the more complicated cases with higher $N$ in a fully automated procedure. This procedure is outlined in section~\ref{sec:UsefulnessHigherN}. Then, as a first example for such a more complicated case, the reaction of pseudoscalar meson electroproduction ($N=6$) is treated in section~\ref{sec:Electroproduction}. The application of the modified form of Moravcsik's theorem to electroproduction in particular yields interesting new findings concerning the structure of the corresponding complete sets. We summarize the applications to the cases of different $N$ and attempt to recognize general patterns in section~\ref{sec:ConclusionsAndOutlook}.


\section{Algebraic starting point} \label{sec:AlgebraicStartingPoint}

Usually, one considers subsets of polarization asymmetries $\Ocal^{\alpha}$ defined by
\begin{equation}
 \Ocal^{\alpha} = \bm{c}^{\alpha} \sum_{i,j = 1}^{N} b_{i}^{\ast} \tilde{\Gamma}^{\alpha}_{ij} b_{j}  , \text{ for } \alpha = 1,\ldots,N^{2}. \label{eq:GenericPolObservables}
\end{equation}
Here, $\bm{c}^{\alpha}$ are conventional pre-factors\footnote{For processes like Pion-Nucleon Scattering or pseudoscalar meson photoproduction, the factors $\bm{c}^{\alpha}$ are equal for all ob\-serva\-bles. However, for the example of pseudoscalar meson electroproduction, observable-dependent pre-factors need to be introduced in order to make the $\tilde{\Gamma}$-matrices satisfy the correct orthogonality relation (cf. section \ref{sec:MoravcsikTheoremExamples}).} and the $\tilde{\Gamma}^{\alpha}$ are a complete basis-system of orthogonal matrices: 
\begin{equation}
\mathrm{Tr} \left[ \tilde{\Gamma}^{\alpha} \tilde{\Gamma}^{\beta} \right] = \tilde{N} \delta_{\alpha \beta},  \label{eq:OrthogonalityRelation}
\end{equation}
with the usual Kronecker-symbol $\delta_{\alpha \beta}$. The normalization-factor $\tilde{N}$ can be equal to the number of amplitudes $N$, as is the case for Pion-Nucleon scattering (section~\ref{sec:PiN}) or for photoproduction (sec.~\ref{sec:PhotoprodExample}), as well as different from it (cf. electroproduction, section~\ref{sec:Electroproduction}). Without loss of generality, we can assume the $N$ complex amplitudes $b_{i}$ to be transversity amplitudes.

Moravcsik started at a point which is a bit different, i.e. he directly considered subsets of the bilinear products (called 'bicoms' in reference~\cite{Moravcsik:1984uf}) of amplitudes
\begin{equation}
 b_{j}^{\ast} b_{i}  , \text{ for } i,j=1,\ldots,N. \label{eq:GenericBicoms}
\end{equation}

We remark that due to the bilinear nature of the sets of quantities~\eqref{eq:GenericPolObservables} and~\eqref{eq:GenericBicoms}, the amplitudes generally can only be determined up to one unknown overall phase~\cite{Chiang:1996em, Nakayama:2018yzw}, which can depend on all kinematic variables of the problem. This means that the full information, which can be extracted, lies in the moduli and relative-phases of the $N$ amplitudes.

An important initial assumption by Moravcsik is that \textit{all} the $N$ moduli
\begin{equation}
  \left| b_{1} \right|,  \left| b_{2} \right|, \ldots,  \left| b_{N} \right|  , \label{eq:Moduli}
\end{equation}
have already been determined from a suitable subset composed of $N$ ob\-serva\-bles\footnote{For transversity amplitudes $b_{i}$, one can assume without loss of generality that such a subset can indeed always be found.}. This is a standard assumption for the algebraic analysis of complete experiments (cf. \cite{Chiang:1996em, Nakayama:2018yzw}) and therefore we shall also adopt it in this work.

A generic bilinear product is a complex number and thus can be decomposed into real- and imaginary parts: $b_{j}^{\ast} b_{i} = \mathrm{Re} \left[ b_{j}^{\ast} b_{i} \right] + i \hspace*{1pt} \mathrm{Im} \left[ b_{j}^{\ast} b_{i} \right] $. Upon introducing polar coordinates (i.e. modulus and phase) for each amplitude, the real parts of the bilinear products become
\begin{equation}
 \mathrm{Re} \left[ b_{j}^{\ast} b_{i} \right] = \left| b_{i} \right|\left| b_{j} \right|  \cos \phi_{ij}  . \label{eq:ReBicomEq}
\end{equation}
The real parts thus fix their corresponding relative-phase $\phi_{ij} := \phi_{i} - \phi_{j}$ up to the discrete ambiguity~\cite{Nakayama:2018yzw}
\begin{equation}
 \phi_{ij} \longrightarrow \phi_{ij}^{\pm} = \begin{cases}  + \alpha_{ij}, \\ - \alpha_{ij} ,   \end{cases} , \label{eq:CosTypeAmbiguity}
\end{equation}
with $\alpha_{ij}$ defined uniquely by the quantity $\mathrm{Re} \left[ b_{j}^{\ast} b_{i} \right]$, and on the interval $\alpha_{ij} \in \left[ 0, \pi \right]$. In the following, we refer to a discrete ambiguity of the form~\eqref{eq:CosTypeAmbiguity} as a 'cosine-type' ambiguity.

Similarly, the imaginary part is written as
\begin{equation}
 \mathrm{Im} \left[ b_{j}^{\ast} b_{i} \right] = \left| b_{i} \right|\left| b_{j} \right|  \sin \phi_{ij}  , \label{eq:ImBicomEq}
\end{equation}
and it fixes the corresponding relative-phase $\phi_{ij}$ up to the discrete phase-ambiguity~\cite{Nakayama:2018yzw}
\begin{equation}
 \phi_{ij} \longrightarrow \phi_{ij}^{\pm} = \begin{cases}  + \alpha_{ij}, \\ \pi - \alpha_{ij} ,   \end{cases} , \label{eq:SinTypeAmbiguity}
\end{equation}
where $\alpha_{ij}$ is defined uniquely by the quantity $\mathrm{Im} \left[ b_{j}^{\ast} b_{i} \right]$, and on the interval $\alpha_{ij} \in \left[ - \pi / 2, \pi / 2 \right]$. Accordingly, we refer to a discrete ambiguity of the form~\eqref{eq:SinTypeAmbiguity} as a 'sine-type' ambiguity.

Before elaborating on Moravcsik's result, we outline a simple technique on how to invert the definition~\eqref{eq:GenericPolObservables} for the bilinear products. Using the completeness relation for the $\tilde{\Gamma}$-matrices, one arrives at the following expression~\cite{Chiang:1996em,MyDiploma,MyPhD}:
\begin{equation}
 b_{i}^{\ast} b_{j} = \frac{1}{\tilde{N}} \sum_{\alpha = 1}^{N^{2}}  \left( \tilde{\Gamma}^{\alpha}_{ij} \right)^{\ast} \left( \frac{\Ocal^{\alpha}}{\bm{c}^{\alpha}} \right)  .  \label{eq:DirectInvNAmplitudes}
\end{equation}
Thus, the bilinear products have been extracted from a specific subset of ob\-serva\-bles, which follows from the algebra $\left\{ \tilde{\Gamma}^{\alpha} \right\}$. Since all bilinear products are now known, so are the moduli $\left| b_{i} \right| = \sqrt{b_{i}^{\ast} b_{i}}$. Furthermore, the relative-phases $\phi_{ij}$ are also known uniquely, since for each such phase, equation~\eqref{eq:DirectInvNAmplitudes} implies both $\cos \phi_{ij}$ {\it and} $\sin \phi_{ij}$\footnote{The complex exponential $\exp \phi_{jk}= \cos \phi_{jk} + i \sin \phi_{jk}$ is uniquely invertible on the interval $\phi_{jk} \in \left[ 0, 2 \pi \right)$. Compare this to the cosine- and sine-type ambiguities given in equations~\eqref{eq:CosTypeAmbiguity} and~\eqref{eq:SinTypeAmbiguity}.}.

Therefore, the amplitude-arrangement in the complex plane is fixed uniquely, whatever ob\-serva\-bles appear on the right-hand-side of~\eqref{eq:DirectInvNAmplitudes} for a specific (at best minimal) choice of bilinear products $b_{i}^{\ast} b_{j}$. Such a choice of bilinear products implies a combination of $n_{\text{ph.}}$ relative phases $\phi_{ij}$. For the unique extraction of the amplitudes using the direct-inversion technique~\eqref{eq:DirectInvNAmplitudes}, we thus need $2 n_{\text{ph.}}$ real trigonometric functions, i.e. both cosine and sine for each relative phase. 

Moravcsik's theorem, stated in the next section, is basically a technique on how to obtain unique solutions using minimally {\it only} $n_{\text{ph.}}$ instead of $2 n_{\text{ph.}}$ trigonometric functions. Halving of the real degrees of freedom required will then naturally also lead to a reduction of the number of necessary ob\-serva\-bles.

\section{Moravcsik's theorem} \label{sec:MoravcsikCompExp}

First, Moravcsik's theorem shall be cited in its original formulation, from page $2$ of reference~\cite{Moravcsik:1984uf}. Everything printed in italics in the following is a verbatim citation: \\

\textbf{\underline{Theorem 1 (Original theorem by Moravcsik):}} \\

\textit{The criterion and its proof will be described in the language of a geometrical analog, similar to the one used in a previous paper\footnote{Here, Moravcsik cites the reference~\cite{Goldstein:1973gn}.} discussing the determination of amplitudes. Let us denote each amplitude by a point, and each bilinear amplitude product by a line connecting the points that correspond to the amplitudes that appear in the product. The line is solid if we have $\mathrm{Re} \left[ b_{j}^{\ast} b_{i} \right]$, and broken (dashed) for $\mathrm{Im} \left[ b_{j}^{\ast} b_{i} \right]$. }

\textit{For even just a complete (and not fully complete) determination of the amplitudes, the set of lines in our diagram corresponding to a complete set of bilinear combinations of amplitudes ("bilinear products") must touch each amplitude point and must form a connected network. To be \textit{fully} complete}\footnote{Moravcsik distinguishes among \textit{fully complete} networks, which are those that yield a unique solution, and \textit{complete} ones, which still allow for discrete phase ambiguities, but do not allow for continuous ambiguities.}\textit{, the network must also satisfy the following two criteria:}
\begin{itemize}
 \item[(A)] \textit{Each amplitude must be included in a \textit{closed loop}}.
 \item[(B)] \textit{At least one closed loop belonging to each amplitude point must have an odd number of broken lines and an odd number of solid lines in it.}
\end{itemize}

This ends the verbatim citation of Moravcsik's theorem. In this work, we start with a slightly modified requirement, which represents a special case of the networks considered in Moravcsik's theorem, and then formulate and prove a modified version of the theorem: \\

\textbf{\underline{Theorem 2 (Modified version of Moravcsik's}} \\ \hspace*{69.5pt}  \textbf{\underline{result):}} \\

We consider the situation which Moravcsik calls the 'most economical'~\cite{Moravcsik:1984uf} version of a complete set in the geometrical analog, i.e. a large open chain which contains \textit{all} amplitude points, and consists of $N - 1$ lines for a problem with $N$ amplitudes. 

Then, we want to turn this open chain into a \textit{fully} complete set, by adding one additional connecting line which turns it into a closed loop of $N$ lines, and which has to contain all amplitude points exactly once. Furthermore, in such a closed loop every amplitude point is touched by exactly $2$ link-lines (or edges).

Such a closed loop corresponds to a unique solution, without discrete ambiguities, if it fulfills the following criterion, which is a bit different and seemingly simpler than in Moravcsik's case:
\begin{itemize}
 \item[(B')] The closed loop has to contain an odd number of dashed lines $n_{\text{d}} \geq 1$.
 
 %
 %
 %
 %
 
 In particular, contrary to criterium (B) of Moravcsik, the closed loop has to contain no solid lines at all, i.e. in case of an odd number of links $N$, the closed loop with $n_{\text{d}} = N$ is still a fully complete set.
 
 Furthermore, it is completely irrelevant which of the bilinear products are represented by the dashed lines, as long as the overall number $n_{\text{d}}$ is odd.
\end{itemize}

\begin{figure*}
 \begin{center}
\includegraphics[width = 0.31 \textwidth,trim={0 0 0 0.7cm},clip]{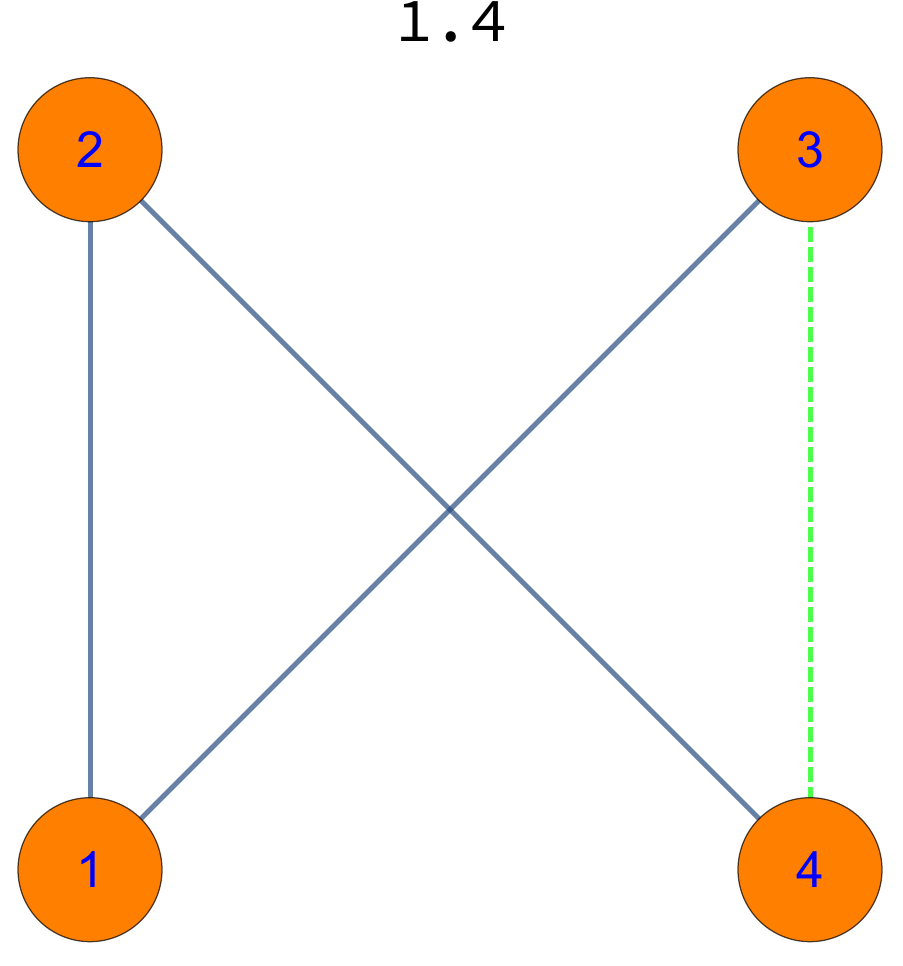} \hspace*{5pt}
\includegraphics[width = 0.31 \textwidth,trim={0 0 0 0.7cm},clip]{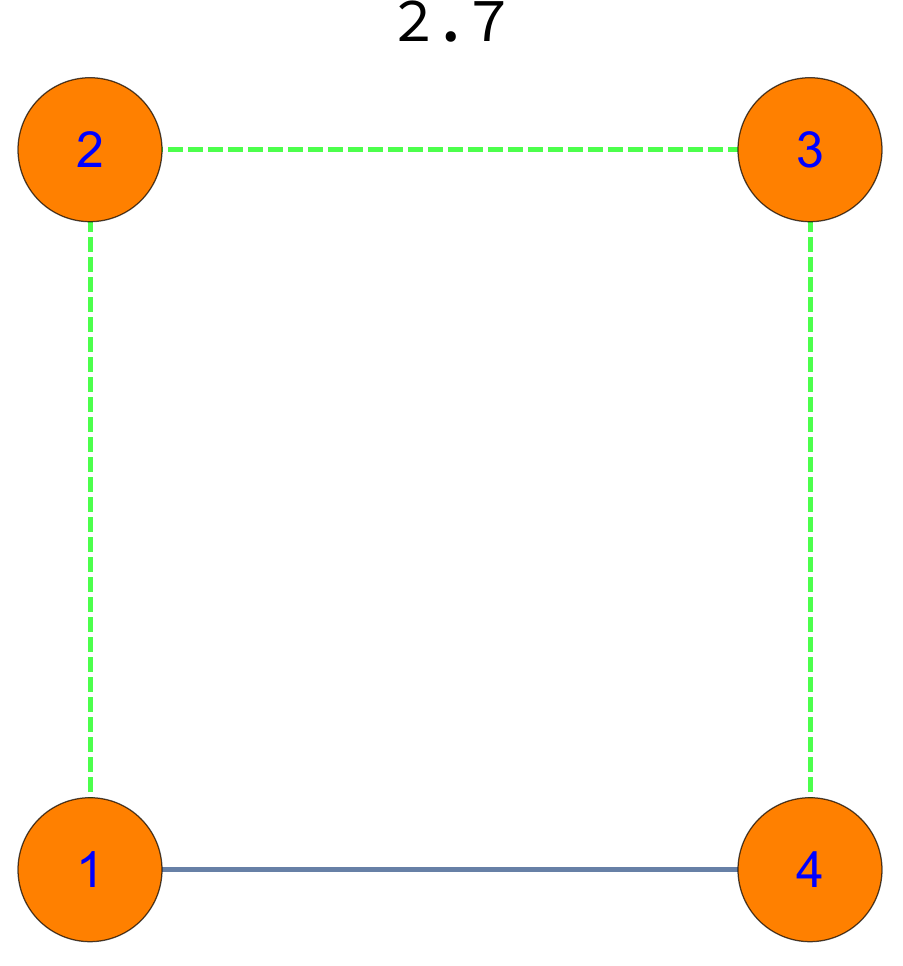} \hspace*{5pt}
\includegraphics[width = 0.31 \textwidth,trim={0 0 0 0.7cm},clip]{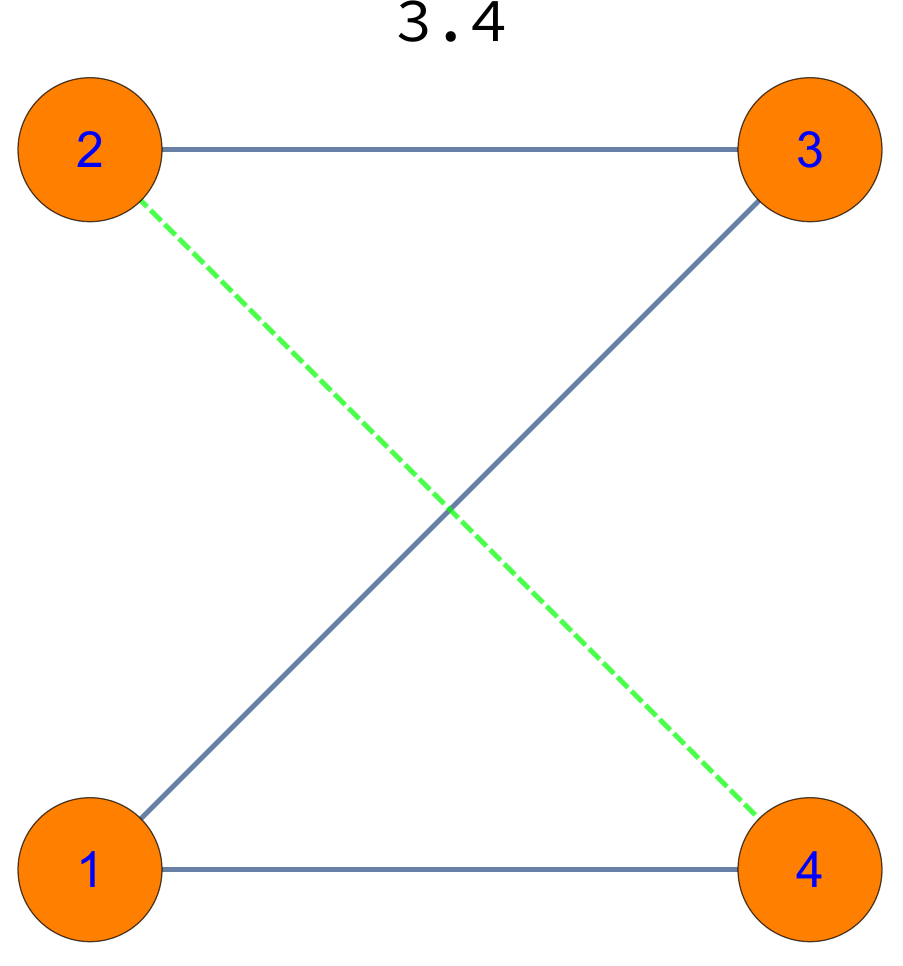}
\end{center}
\vspace*{-5pt}
\caption{The diagrams show three closed loops which satisfy the criteria posed in order to obtain a unique solution, according to Theorem 2. The exemplary case of pseudoscalar meson photoproduction, i.e. $N = 4$, was chosen. Green dashed lines represent the imaginary part of a bilinear product, while the real part is shown as a blue solid line. For more details, see section~\ref{sec:MoravcsikTheoremExamples}. (Color online)}
\label{fig:PhotoproductionExampleLoops}
\end{figure*}

In order to illustrate the somewhat abstract requirements formulated in the theorem, we show three examples for fully complete closed loops for the reaction of pseudoscalar meson photoproduction, i.e. $N = 4$ amplitudes, in Figure~\ref{fig:PhotoproductionExampleLoops}. A much more detailed discussion of this process, as well as others, can be found in section~\ref{sec:MoravcsikTheoremExamples}.

A detailed proof of Theorem 2 can be written, using the knowledge on the discrete ambiguity transformations~\eqref{eq:CosTypeAmbiguity} and~\eqref{eq:SinTypeAmbiguity} mentioned in section~\ref{sec:AlgebraicStartingPoint}. Due to its length, the proof has been relegated to appendix~\ref{sec:DetailedProof}. Within the proof, heavy use is made of consistency relations for the relative phases (cf. references~\cite{Nakayama:2018yzw,Chiang:1996em}) corresponding to the different bilinear products. The proof is in principle the same as in Moravcsik's paper~\cite{Moravcsik:1984uf}. However, we make an attempt at proving some crucial intermediate steps in a more formal way. Furthermore, the above-mentioned special case of $n_{\text{d}} = N$ (for $N$ odd) is discussed as well. At the end of the proof, we elaborate a bit more on singular special cases which can theoretically endanger the validity of the theorem. Such special cases occurr on sets of measure zero in the parameter-space comprised of the relative phases. Therefore, at least in a treatment of complete experiments that assumes the case of va\-nis\-hing measurement uncertainty, these special cases can pro\-bab\-ly be ignored (Moravcsik gives similar comments on page $2$ of his original work~\cite{Moravcsik:1984uf}). We remark that Nakayama finds very similar special conditions for the relative phases in his work on pseudoscalar meson photoproduction~\cite{Nakayama:2018yzw}. 

For the minimal possible case of a closed loop with all amplitudes appearing exactly once, we have exactly $N$ relative phases
\begin{equation}
  \underbrace{ \phi_{1i}, \phi_{ij}, \phi_{jr}, \ldots, \phi_{p q}, \phi_{q1}  .}_{\text{exactly } N \text{ links, or relative-phases}} \label{eq:CircularLoopRelPhases}
\end{equation}
Using binomial coefficients, the possible number of fully complete combinations according to Theorem 2 can be simply counted. In case the total number of amplitudes $N$ is \textit{odd}, the fully complete combinations amount to
\begin{equation}
  \mathcal{N}_{\text{comb.}} = \sum_{k = 0}^{(N - 1) / 2}  \binom{N}{2 k + 1} , \label{eq:NCircCombNodd}
\end{equation}
while for $N$ \textit{even}, the correct expression is
\begin{equation}
  \mathcal{N}_{\text{comb.}} = \sum_{k = 0}^{(N - 2) / 2}  \binom{N}{2 k + 1} . \label{eq:NCircCombNeven}
\end{equation}
These expressions are evaluated for the first few cases of $N = 2,\ldots,8$ in Table~\ref{tab:NCompleteCombinations}. 

It is seen that generally, the number of fully complete combinations in the minimal closed loop~\eqref{eq:CircularLoopRelPhases} scales with $N$ as:
\begin{equation}
  \mathcal{N}_{\text{comb.}} = 2^{(N - 1)} . \label{eq:NCircCombResult}
\end{equation}
Generally, Moravcsik's theorem in this slightly modified form always requires $2 N$ real quantities out of the bilinear products~\eqref{eq:GenericBicoms}, i.e. $N$ moduli~\eqref{eq:Moduli} plus $N$ real- or imaginary parts, in order to uniquely solve for the amplitudes. This already seems to imply that $2 N$ is also the absolute minimal number of required quantities from the ob\-serva\-bles-basis~\eqref{eq:GenericPolObservables}, which has been mentioned as an empirical fact in the introduction. The only question that remains is whether the quantities $b_{j}^{\ast} b_{i}$ map in a simple one-to-one way to the ob\-serva\-bles~$\Ocal^{\alpha}$. 

In order to address this last point, among others, we discuss examples for particular reactions in the next section. 

\begin{table}[ht]
 \begin{center}
 \begin{tabular}{l|rrrrrrr}
  $N$ & 2 & 3 & 4 & 5 & 6 & 7 & 8 \\
  \hline
  $\mathcal{N}_{\text{comb.}}$ & 2 &  4  & 8 &  16 &  32 &  64 &  128 
 \end{tabular}
 \end{center}
 \caption{The number of possible complete combinations for a minimal closed loop~\eqref{eq:CircularLoopRelPhases} is evaluated for $N = 2$ to $8$.}
 \label{tab:NCompleteCombinations}
\end{table}


\section{Basic Examples} \label{sec:MoravcsikTheoremExamples}

We continue with the consideration of the minimal closed loops with exactly $N$ links, which have been introduced in the previous section (see equation~\eqref{eq:CircularLoopRelPhases}). The examples of $N = 2$, $3$ and $4$ transversity amplitudes are treated in this section. 

For all these examples, we have to find the full number of possible topologies for the closed loops, with $N$ points connected via $N$ links and exactly $2$ links connected to each point. For $N = 2$, $3$ and $4$, this task can be completed by hand, while for all higher $N$ it becomes increasingly cumbersome (see the results for electroproduction, i.e. $N = 6$, in Figures~\ref{fig:ElectroproductionStartTopologies_I} to~\ref{fig:ElectroproductionStartTopologies_III} of section~\ref{sec:Electroproduction}).


Therefore, a Mathematica-code~\cite{Mathematica} has been developed for this work, which can complete this task automatically for in principle arbitrary numbers\footnote{Computing times of course set a limit on the numbers $N$ which can be treated numerically. Within acceptable times, we can obtain results for a maximal number of $N = 8$ amplitudes. Above that, computing times rise exponentially.} $N$. 

\subsection{$N = 2$ (Pion-Nucleon scattering)} \label{sec:PiN}

The reaction of Pion-Nucleon ($\pi N$) scattering is described model-independently by $2$ amplitudes, which are accompanied by $4$ ob\-serva\-bles~\cite{Hoehler84,Arndt:1994bu,Arndt:1995bj,Arndt:2006bf,Anisovich:2013tij}. Following the pioneering work by Hoehler and his group~\cite{Hoehler84}, highly significant contributions to the analysis of $\pi N$-scattering have been published by R.~A.~Arndt and collaborators~\cite{Arndt:1994bu,Arndt:1995bj,Arndt:2006bf}. The latter contributions have culminated in the GWU/SAID partial wave analysis, the results of which are publicly accessible~\cite{SAID}. 

\begin{table}[h]
 \begin{center}
 \begin{footnotesize}
 \begin{tabular}{lcr}
 \hline
 \hline
  Observable & Bil.-form & Shape-cl. \\
  \hline 
  $\Ocal^{1} =  \left| b_{1} \right|^{2} + \left| b_{2} \right|^{2} = \sigma_{0}$  &  $ \left< b \right| \hat{\sigma}^{1} \left|  b  \right>$  &    \\
  $\Ocal^{4} =  \left| b_{1} \right|^{2} - \left| b_{2} \right|^{2} = \check{P}$  &  $ \left< b \right| \hat{\sigma}^{4} \left|  b  \right>$  &  $\mathrm{D}$ \\
  \hline
   $\Ocal^{a}_{1} = - 2 \left| b_{1} \right| \left| b_{2} \right| \sin \phi_{12} = - 2 \mathrm{Im} \left[ b_{2}^{\ast} b_{1} \right] = - \check{R}$  & $ \left< b \right| \hat{\sigma}^{2} \left|  b  \right>$ &  \\
   $\Ocal^{a}_{2} = 2 \left| b_{1} \right| \left| b_{2} \right| \cos \phi_{12}  = 2 \mathrm{Re} \left[ b_{2}^{\ast} b_{1} \right] = \check{A}$ & $ \left< b \right| \hat{\sigma}^{3} \left|  b  \right>$ & $a = \mathrm{AD}$ \\
   \hline
   \hline
 \end{tabular}
 \end{footnotesize}
 \end{center}
 \caption{The definitions of the $4$ ob\-serva\-bles in $\pi N$ scattering are collected here (cf. \cite{Hoehler84,Anisovich:2013tij}). The definition of each observable as a bilinear form in terms of extended Pauli-matrices is indicated and the ob\-serva\-bles have been subdivided into two different classes, which correspond to the shapes of the defining matrices. The algebra of the matrices is given in appendix~\ref{sec:MatrixAlgebras}.}
 \label{tab:PiNObservables}
\end{table}

 All ob\-serva\-bles can be defined in terms of (extended) Pauli matrices $\hat{\sigma}^{\alpha}$, as written in Table~\ref{tab:PiNObservables}. The corresponding matrices are listed in appendix~\ref{sec:MatrixAlgebras}. The ob\-serva\-bles have the generic form~\eqref{eq:GenericPolObservables} with $\bm{c}^{\alpha} = 1$ for all $\alpha$ and the defining matrices satisfy the orthogonality relation~\eqref{eq:OrthogonalityRelation} with $\tilde{N} = 2$. For the sake of clarity, it is very useful to divide the matrices into different classes according to their shape (as it has been done in reference~\cite{Chiang:1996em} for the case of photoproduction). Here, we have two matrices in a diagonal shape-class 'D' and two matrices of anti-diagonal shape 'AD'.
 
\begin{figure}[h]
 \begin{center}
\includegraphics[width = 0.35 \textwidth]{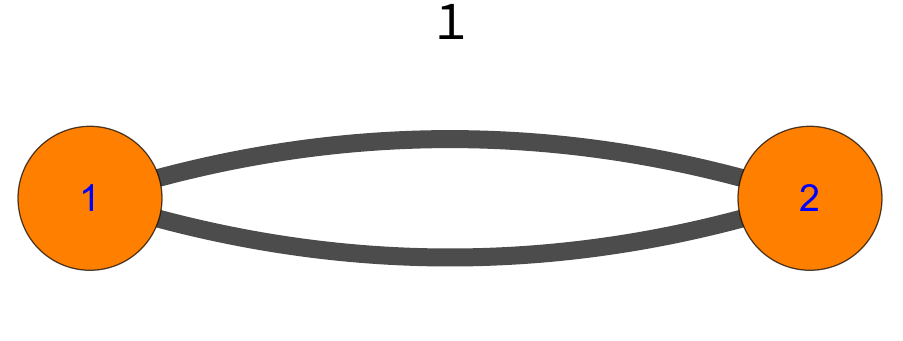}
 \end{center}
\vspace*{-5pt}
\caption{The diagram shows the only possible topology that exists for a closed loop formed out of $2$ points. As stated in section~\ref{sec:MoravcsikCompExp}, each point refers to one of the two amplitudes. A solid black line connecting two points represents in this case either a real- or imaginary part of a particular bilinear product. This topology can then be used to derive the fully complete sets of ob\-serva\-bles for Pion-Nucleon scattering, using Theorem~$2$ from section~\ref{sec:MoravcsikCompExp}. (Color online)}
\label{fig:PiNStartTopology}
\end{figure}

For the considered case of $N = 2$, there exists only one possible topology to form a minimal closed loop with this particular number of points, which is shown in Figure~\ref{fig:PiNStartTopology}. This single topology is the starting point in order to derive all the fully complete loops.

\begin{figure}[h]
 \begin{center}
\includegraphics[width = 0.235 \textwidth]{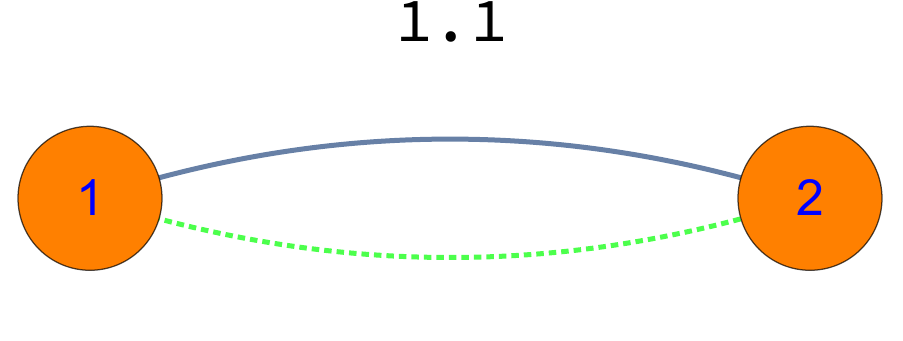}
\includegraphics[width = 0.235 \textwidth]{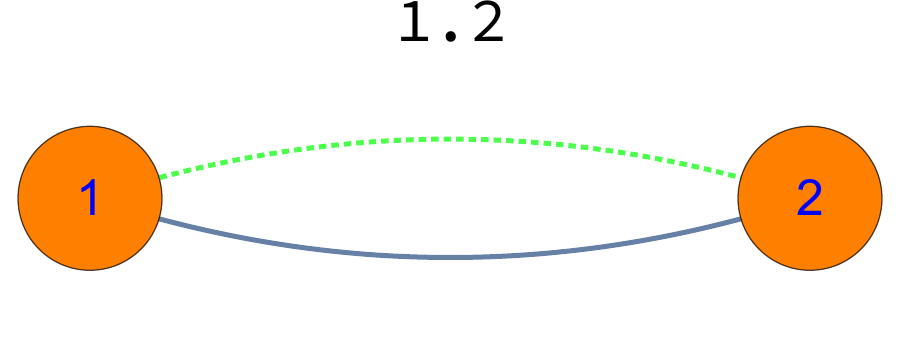}
 \end{center}
\vspace*{-5pt}
\caption{For $N=2$ transversity amplitudes ($\pi N$ scattering), the closed loops which yield unique solutions, i.e. the \textit{fully complete} loops are shown here. Each point refers to one of the two amplitudes. A solid blue line connecting two points represents the real part of a particular bilinear product and a dashed green line represents the imaginary part of the respective bilinear product. The direction of a line connecting two points is irrelevant. For instance, a dashed green line connecting points $1$ and $2$ can be tantamount to $\mathrm{Im} \left[ b_{1} b_{2}^{\ast} \right]$, i.e. $\sin \phi_{12}$, or to $\mathrm{Im} \left[ b_{2} b_{1}^{\ast} \right]$, i.e. $\sin \phi_{21} = - \sin \phi_{12}$, respectively. Thus, the sequence of indices in a bilinear product leads (in the case of dashed lines) only to a difference in sign, which is not important for the discussion of discrete ambiguities. (Color online)}
\label{fig:PiNCompleteBicoms}
\end{figure}

\begin{figure}[h]
 \begin{center}
 \vspace*{-10pt}
\includegraphics[width = 0.235 \textwidth]{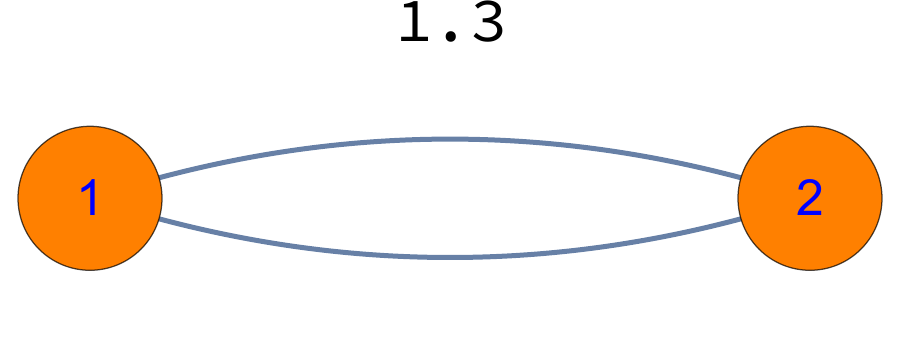}
\includegraphics[width = 0.235 \textwidth]{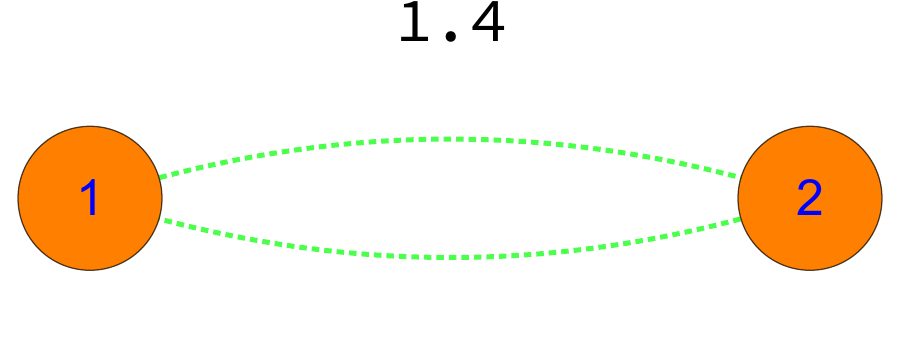}
\end{center}
\vspace*{-5pt}
\caption{All the closed loops which still leave discrete phase ambiguities unresolved are collected here for $N = 2$ transversity amplitudes ($\pi N$ scattering). (Color online)}
\label{fig:PiNInCompleteBicoms}
\end{figure}

All the closed loops which represent \textit{fully complete} sets, according to Theorem 2, are shown in Figure~\ref{fig:PiNCompleteBicoms}. All the remaining closed loops, which correspond to combinations that still leave some discrete phase ambiguities unresolved, are collected in Figure~\ref{fig:PiNInCompleteBicoms}. 

Looking at the possibilities of fully complete combinations in Figure~\ref{fig:PiNCompleteBicoms}, it is seen that, assuming the moduli $\left| b_{1} \right|$ were known, these combinations would be equivalent to
\begin{equation}
  \left( \cos \phi_{12}, \sin \phi_{21} \right), \text{ or } \left( \sin \phi_{12}, \cos \phi_{21} \right). \label{eq:PiNCompleteBicoms}
\end{equation}
Comparing with Table~\ref{tab:PiNObservables}, we see indeed that the theorem reproduces the well-known statement which says that for Pion-Nucleon scattering, one needs all $4$ ob\-serva\-bles in order to obtain a complete set (cf. the introduction of~\cite{Anisovich:2013tij}).

It is clear that for a single relative-phase $\phi_{12}$, both sine and cosine are needed to uniquely fix $\phi_{12} \in \left[ 0, 2 \pi \right)$. For illustrative purposes, we look at the $\pi N$-problem again, but in the language of the modified form of Moravcsik's theorem. The following derivation is the simplest possible special case of the general theorem proven in appendix~\ref{sec:DetailedProof}. The consistency relation for $\pi N$ scattering looks simple. It reads:
\begin{equation}
 \phi_{12} + \phi_{21} = 0 . \label{eq:PiNRelPhasesConstraint}
\end{equation}
%


Now, take for instance the fully complete combination from example 1.2 in Figure~\ref{fig:PiNCompleteBicoms}, i.e. $\left( \sin \phi_{12}, \cos \phi_{21} \right)$. We write down all the possible cases for the consistency relation~\eqref{eq:PiNRelPhasesConstraint} for this set, i.e. (cf. equations~\eqref{eq:CosTypeAmbiguity} and~\eqref{eq:SinTypeAmbiguity}):
\begin{align}
 \alpha_{12} + \alpha_{21} &= 0  , \label{eq:PiNComplSetCaseI} \\
 \alpha_{12} - \alpha_{21} &= 0  , \label{eq:PiNComplSetCaseII} \\
 \pi - \alpha_{12} + \alpha_{21} &= 0  , \label{eq:PiNComplSetCaseIII} \\
 \pi - \alpha_{12} - \alpha_{21} &= 0  . \label{eq:PiNComplSetCaseIV}
\end{align}
It is seen that indeed all of these equations are linearly independent (cf. appendix~\ref{sec:DetailedProof}) and thus all discrete phase ambiguities are resolved. We look next at the example number 1.4 from Figure~\ref{fig:PiNInCompleteBicoms}, which is not fully complete, i.e. $\left( \sin \phi_{12}, \sin \phi_{21} \right)$. The cases for the consistency relation~\eqref{eq:PiNRelPhasesConstraint} read
\begin{align}
 \alpha_{12} + \alpha_{21} &= 0  , \label{eq:PiNInComplSetCaseI} \\
 \alpha_{12} + \pi - \alpha_{21} &= 0  , \label{eq:PiNInComplSetCaseII} \\
 \pi - \alpha_{12} + \alpha_{21} &= 0  , \label{eq:PiNInComplSetCaseIII} \\
 \pi - \alpha_{12} + \pi - \alpha_{21} = - \alpha_{12} - \alpha_{21} &= 0  . \label{eq:PiNInComplSetCaseIV}
\end{align}
In this case, the equations~\eqref{eq:PiNInComplSetCaseI} and~\eqref{eq:PiNInComplSetCaseIV}, as well as equations~\eqref{eq:PiNInComplSetCaseII} and~\eqref{eq:PiNInComplSetCaseIII}, form linearly dependent pairs. Therefore, there still remain unresolved discrete phase ambiguities.

In summary, all $2 N = 4$ ob\-serva\-bles are required for completeness in Pion-Nucleon scattering, which thus exhaust all the available shape-classes. 

\subsection{$N = 3$ (mathematical example, constructed using the Gell-Mann matrices)} \label{sec:3Amplitudes}

We consider an example with $N=3$ transversity amplitudes, which does not seem to have a physical analogue\footnote{In the literature consulted for this work, we did not find a physical example for a process described by $N = 3$ amplitudes. The simplest example for an odd number of amplitudes, at least as far as we know right now, is Nucleon-Nucleon ($NN$) elastic scattering with $N = 5$ (see reference~\cite{Bystricky:1976jr}). \\ The group of F.~Lehar at Saclay collected 10 observables for $NN$
elastic scattering, i.e. for the processes of $pp$-~\cite{Bystricky:1998rh} and $np$-scattering~\cite{Ball:1998jj}. The direct reconstruction analysis yielded several solutions (up to 4) and $\chi^{2}$-criteria did not allow to distinguish a unique
solution. More details are given in reference~\cite{Workman:2016ysf}.}, but can still be studied for purely academic purposes. In order to obtain a mathematical example with $3$ amplitudes, it makes sense to define the ob\-serva\-bles via the Gell-Mann matrices $\tilde{\lambda}^{\alpha}$, extended with the identity (for a listing of the Gell-Mann matrices, see for instance~\cite{Peskin:1995ev}). The definitions of the full set of $9$ resulting ob\-serva\-bles are written in Table~\ref{tab:ToyModelObservables}. The defining matrices are collected in appendix~\ref{sec:MatrixAlgebras}.

All ob\-serva\-bles have the generic form~\eqref{eq:GenericPolObservables} with $\bm{c}^{\alpha} = 1 / 2$ and the extended Gell-Mann matrices, in the normalization chosen here, satisfy equation~\eqref{eq:OrthogonalityRelation} with $\tilde{N} = 2$. The shape-classes consist of one class with $3$ diagonal ob\-serva\-bles 'D', as well as three non-diagonal shape-classes with $2$ ob\-serva\-bles each. The non-diagonal ob\-serva\-bles are divided into one class of anti-diagonal shape ('AD') and two classes of matrices with parallelogram-shape ('P1' and 'P2').


\begin{table*}
 \begin{center}
 \begin{tabular}{lcr}
 \hline
 \hline
  Observable & Bilinear form & Shape-class \\
  \hline 
  $\Ocal^{1} = \frac{1}{2} \left( \sqrt{\frac{2}{3}} \left| b_{1} \right|^{2} + \sqrt{\frac{2}{3}} \left| b_{2} \right|^{2} + \sqrt{\frac{2}{3}} \left| b_{3} \right|^{2} \right)$  &  $\frac{1}{2} \left< b \right| \tilde{\lambda}^{1} \left|  b  \right>$  &    \\
  $\Ocal^{4} = \frac{1}{2} \left( \left| b_{1} \right|^{2} - \left| b_{2} \right|^{2}  \right)$  &  $\frac{1}{2} \left< b \right| \tilde{\lambda}^{4} \left|  b  \right>$  &  $\mathrm{D}$ \\
  $ \Ocal^{9} = \frac{1}{2} \left( \frac{ \left| b_{1} \right|^{2}}{\sqrt{3}} + \frac{\left| b_{2} \right|^{2}}{\sqrt{3}} - \frac{2 \left| b_{3} \right|^{2}}{\sqrt{3}} \right)$  &  $\frac{1}{2} \left< b \right| \tilde{\lambda}^{9} \left|  b  \right>$  &   \\
  \hline
   $\Ocal^{a}_{1} = \left| b_{1} \right| \left| b_{3} \right| \sin \phi_{13} = \mathrm{Im} \left[ b_{3}^{\ast} b_{1} \right]$  & $\frac{1}{2} \left< b \right| \tilde{\lambda}^{6} \left|  b  \right>$ &  \\
   $\Ocal^{a}_{2} = \left| b_{1} \right| \left| b_{3} \right| \cos \phi_{13}  = \mathrm{Re} \left[ b_{3}^{\ast} b_{1} \right]$ & $\frac{1}{2} \left< b \right| \tilde{\lambda}^{5} \left|  b  \right>$ & $a = \mathrm{AD}$ \\
   \hline
   $\Ocal^{b}_{1} = \left| b_{2} \right| \left| b_{3} \right| \sin \phi_{23}  = \mathrm{Im} \left[ b_{3}^{\ast} b_{2} \right]$  & $\frac{1}{2} \left< b \right| \tilde{\lambda}^{8} \left|  b  \right>$ &  \\
   $\Ocal^{b}_{2} = \left| b_{2} \right| \left| b_{3} \right| \cos \phi_{23}   = \mathrm{Re} \left[ b_{3}^{\ast} b_{2} \right]$ & $\frac{1}{2} \left< b \right| \tilde{\lambda}^{7} \left|  b  \right>$  & $b = \mathrm{P1}$ \\
   \hline
   $\Ocal^{c}_{1} = \left| b_{1} \right| \left| b_{2} \right| \sin \phi_{12} = \mathrm{Im} \left[ b_{2}^{\ast} b_{1} \right]$  &  $\frac{1}{2} \left< b \right| \tilde{\lambda}^{3} \left|  b  \right>$  &  \\
   $\Ocal^{c}_{2} = \left| b_{1} \right| \left| b_{2} \right| \cos \phi_{12}  = \mathrm{Re} \left[ b_{2}^{\ast} b_{1} \right]$ & $\frac{1}{2} \left< b \right| \tilde{\lambda}^{2} \left|  b  \right>$ & $c = \mathrm{P2}$ \\
   \hline
   \hline
 \end{tabular}
 \end{center}
 \caption{The definitions of ob\-serva\-bles in the example with $N = 3$ amplitudes are collected here. The matrices defining the bilinear forms are given in appendix~\ref{sec:MatrixAlgebras}.}
 \label{tab:ToyModelObservables}
\end{table*}

\begin{figure}
 \begin{center}
\includegraphics[width = 0.35 \textwidth]{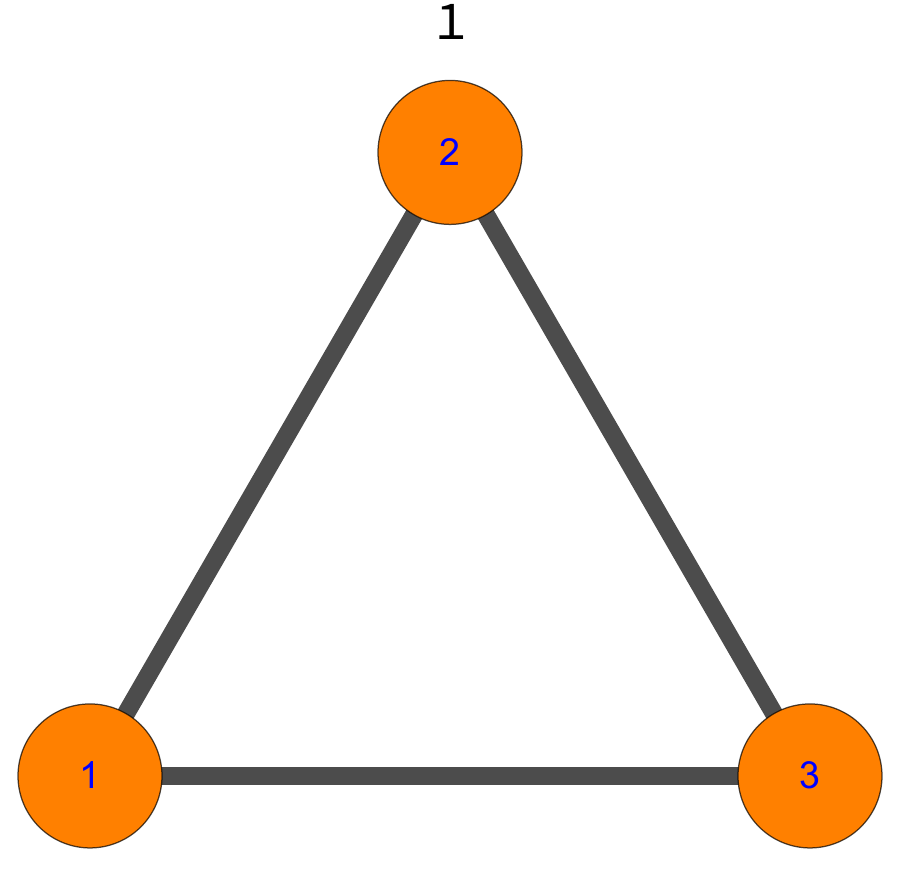}
\end{center}
\vspace*{-5pt}
\caption{The only possibility that exists for a closed loop formed out of $3$ points is shown here. This topology can then be used to derive the fully complete sets of ob\-serva\-bles for the example with $N = 3$ amplitudes, using Theorem~$2$ from section~\ref{sec:MoravcsikCompExp}.}
\label{fig:3AmplitudesStartTopology}
\end{figure}
\begin{figure}
 \begin{center}
 \vspace*{-2.5pt}
\includegraphics[width = 0.233 \textwidth]{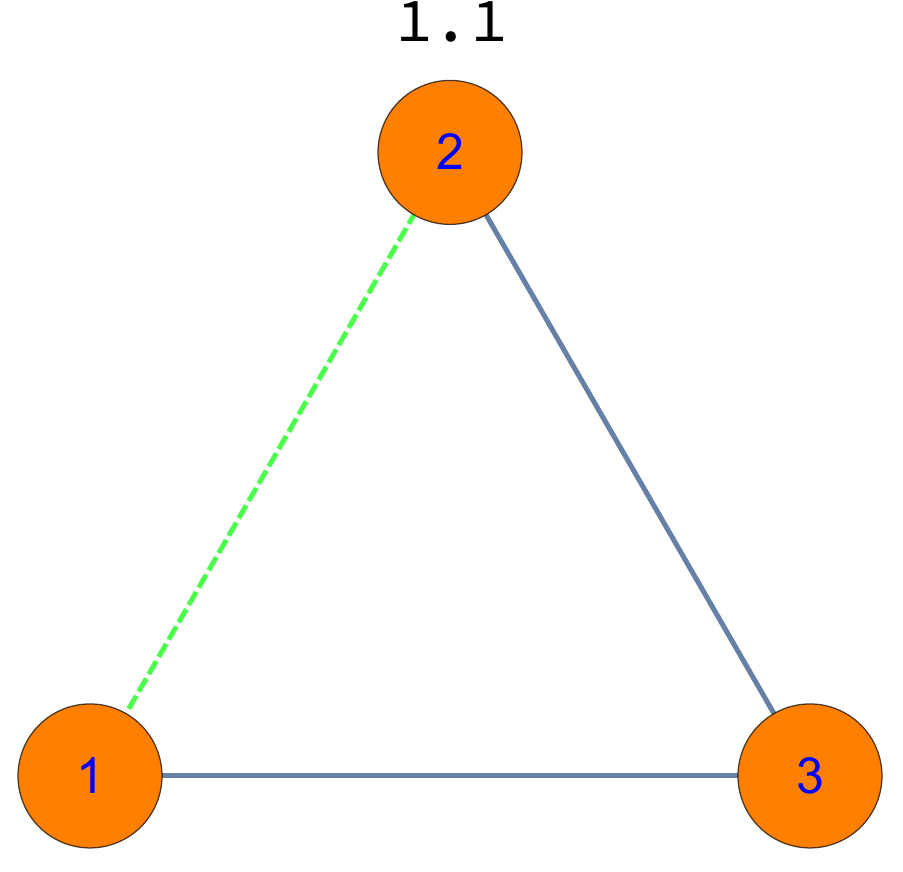} 
\includegraphics[width = 0.233 \textwidth]{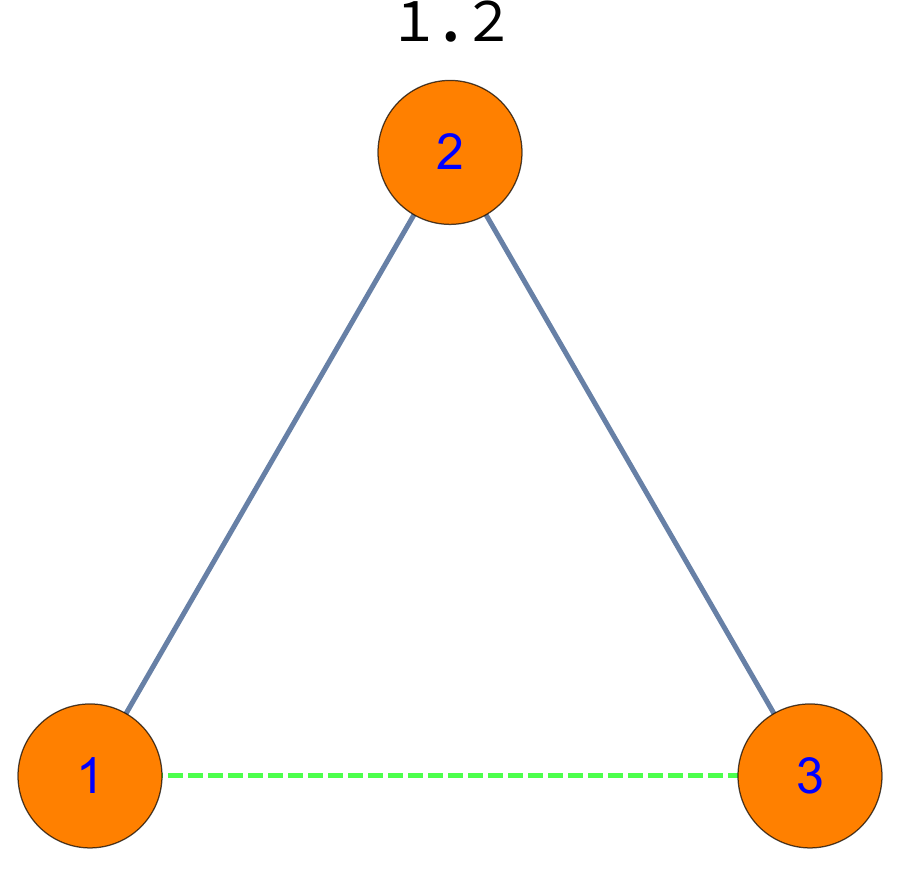} \\
\includegraphics[width = 0.233 \textwidth]{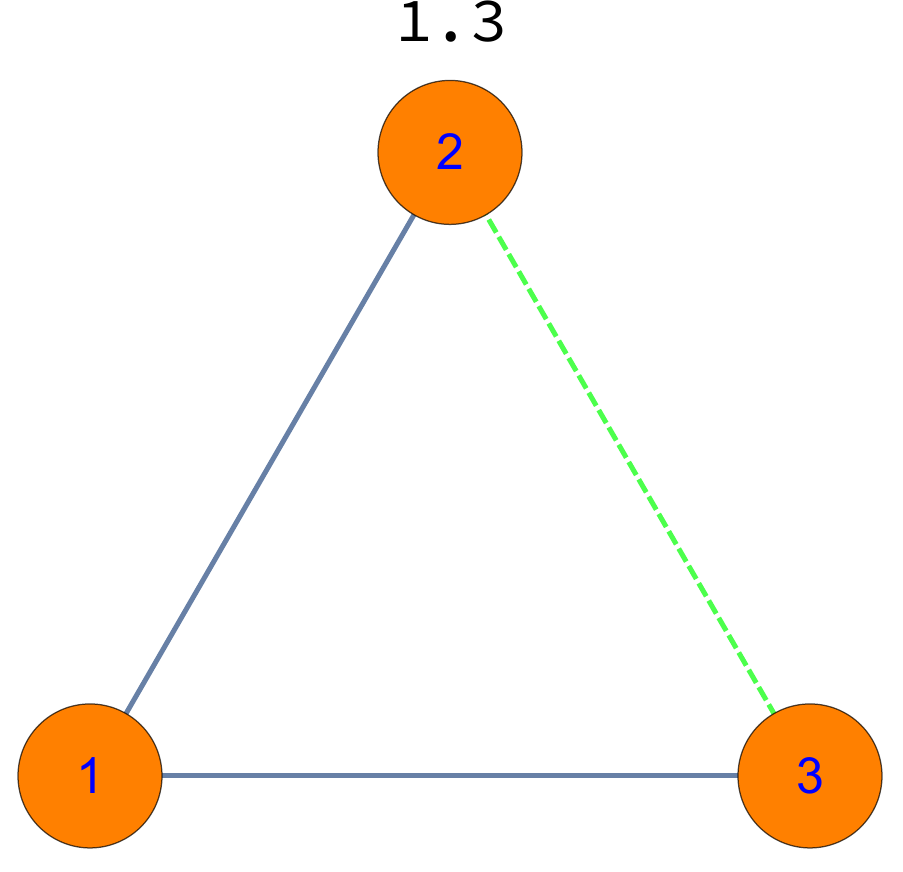} 
\includegraphics[width = 0.233 \textwidth]{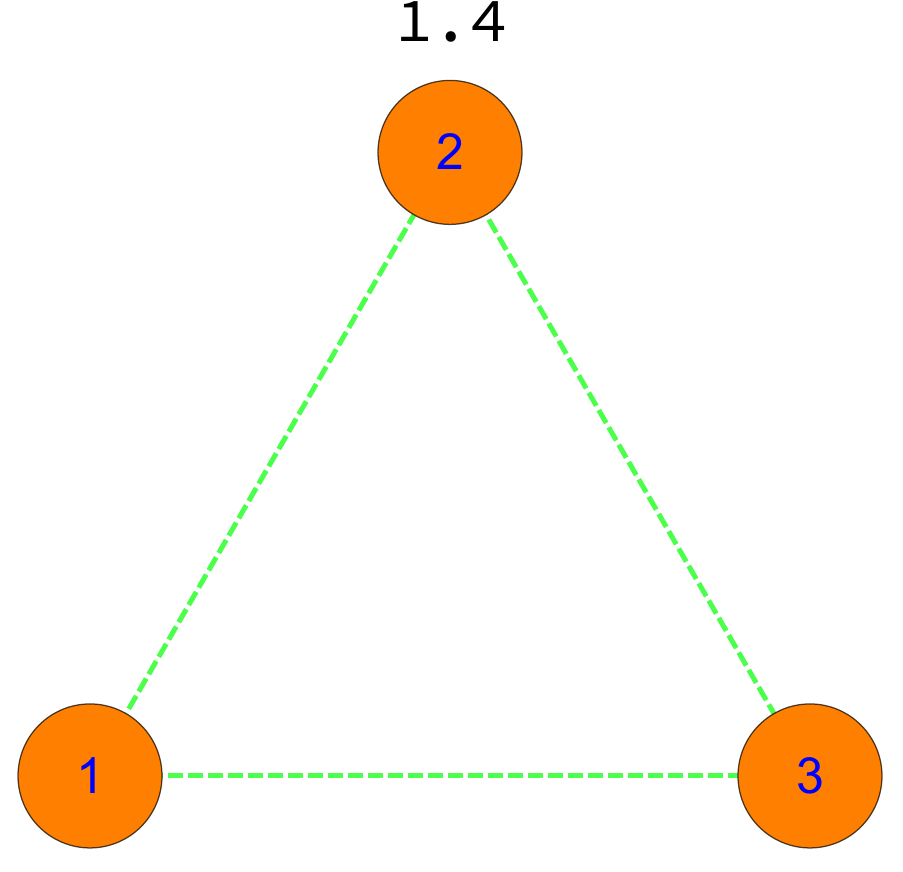} 
\end{center}
\vspace*{-5pt}
\caption{For the example with $N=3$ transversity amplitudes, the closed loops which yield unique solutions, i.e. the \textit{fully complete} loops, are shown here. (Color online)}
\label{fig:3AmplitudesCompleteBicoms}
\end{figure}

For $N = 3$, as well as for the case of $N = 2$, there exists only one possible start-topology usable to form minimal closed loops, which is shown in Figure~\ref{fig:3AmplitudesStartTopology}. The closed circular loops corresponding, according to Theorem 2, to fully complete sets are shown in Figure~\ref{fig:3AmplitudesCompleteBicoms}. Since $N$ is odd in this case, we also have a fully complete combination which is composed of dashed lines exclusively (cf. the special case remarked in the proof in appendix~\ref{sec:DetailedProof}). The loops which are not yet fully complete, i.e. which have an even number of dashed lines and thus leave unresolved discrete ambiguities, are not shown explicitly.

%


We remark that the standard assumption has been made that the moduli $\left| b_{i} \right|$ are already known, or equivalently, that the three ob\-serva\-bles $\left( \Ocal^{1}, \Ocal^{4}, \Ocal^{9} \right)$ have already been measured. Then, comparing to the definitions in Table~\ref{tab:ToyModelObservables}, it is seen that the fully complete sets correspond in each case to the three quantities from shape-class 'D' plus one of the following four possible sets of ob\-serva\-bles
\begin{align}
 &\left( \Ocal^{a}_{2}, \Ocal^{b}_{2}, \Ocal^{c}_{1} \right),  \hspace*{1.5pt} \left( \Ocal^{a}_{2}, \Ocal^{b}_{1}, \Ocal^{c}_{2} \right),  \hspace*{1.5pt} \left( \Ocal^{a}_{1}, \Ocal^{b}_{2}, \Ocal^{c}_{2} \right), \nonumber \\
 &\text{ and } \left( \Ocal^{a}_{1}, \Ocal^{b}_{1}, \Ocal^{c}_{1} \right). \label{eq:3AmplitudesCompExps}
\end{align}
The completeness of these sets follows from the fundamental consistency relation (cf. the discussion in section~\ref{sec:PiN} and the general proof in appendix~\ref{sec:DetailedProof})
\begin{equation}
 \phi_{12} + \phi_{23} + \phi_{31} = 0 . \label{eq:NEquals3RelPhasesConstraint}
\end{equation}

We see that $2 N = 6$ ob\-serva\-bles can yield a unique solution for the amplitudes. Therefore, a reduction from the full set of $9$ ob\-serva\-bles has occurred. Again, as was the case for $N = 2$, the complete sets have to be selected from all $4$ available shape-classes. The completeness of the sets~\eqref{eq:3AmplitudesCompExps} has been verified numerically.

%



\subsection{$N = 4$ (pseudoscalar meson photoproduction)} \label{sec:PhotoprodExample}

In case of pseudoscalar meson photoprodcution, one has $4$ amplitudes and $16$ polarization ob\-serva\-bles~\cite{Chiang:1996em, Nakayama:2018yzw}. The latter are defined using the $4 \times 4$ Dirac-matrices $\tilde{\Gamma}^{\alpha}$, as written in Table~\ref{tab:PhotoObservables}. The Dirac-matrices are listed in appendix~\ref{sec:MatrixAlgebras} (reference~\cite{Chiang:1996em} originally pointed out the importance of these matrices in this particular context).


\begin{table*}
 \begin{center}
 \begin{tabular}{lcr}
 \hline
 \hline
  Observable & Bilinear form & Shape-class \\
  \hline 
  $\sigma_{0} = \frac{1}{2} \left( \left| b_{1} \right|^{2} + \left| b_{2} \right|^{2} + \left| b_{3} \right|^{2} + \left| b_{4} \right|^{2} \right)$  &  $\frac{1}{2} \left< b \right| \tilde{\Gamma}^{1} \left|  b  \right>$  &    \\
  $- \check{\Sigma} = \frac{1}{2} \left( \left| b_{1} \right|^{2} + \left| b_{2} \right|^{2} - \left| b_{3} \right|^{2} - \left| b_{4} \right|^{2} \right)$  &  $\frac{1}{2} \left< b \right| \tilde{\Gamma}^{4} \left|  b  \right>$  &  $\mathcal{S} = \mathrm{D}$ \\
  $- \check{T} = \frac{1}{2} \left( - \left| b_{1} \right|^{2} + \left| b_{2} \right|^{2} + \left| b_{3} \right|^{2} - \left| b_{4} \right|^{2} \right)$  &  $\frac{1}{2} \left< b \right| \tilde{\Gamma}^{10} \left|  b  \right>$  &    \\
  $\check{P} = \frac{1}{2} \left( - \left| b_{1} \right|^{2} + \left| b_{2} \right|^{2} - \left| b_{3} \right|^{2} + \left| b_{4} \right|^{2} \right)$  &  $\frac{1}{2} \left< b \right| \tilde{\Gamma}^{12} \left|  b  \right>$  &    \\
  \hline
   $\Ocal^{a}_{1+} = \left| b_{1} \right| \left| b_{3} \right| \sin \phi_{13} + \left| b_{2} \right| \left| b_{4} \right| \sin \phi_{24} = \mathrm{Im} \left[ b_{3}^{\ast} b_{1} + b_{4}^{\ast} b_{2} \right] = - \check{G}$  & $\frac{1}{2} \left< b \right| \tilde{\Gamma}^{3} \left|  b  \right>$ &  \\
   $\Ocal^{a}_{1-} = \left| b_{1} \right| \left| b_{3} \right| \sin \phi_{13} - \left| b_{2} \right| \left| b_{4} \right| \sin \phi_{24}  = \mathrm{Im} \left[ b_{3}^{\ast} b_{1} - b_{4}^{\ast} b_{2} \right] = \check{F}$ & $\frac{1}{2} \left< b \right| \tilde{\Gamma}^{11} \left|  b  \right>$ & $a = \mathcal{BT} = \mathrm{PR}$ \\
   $\Ocal^{a}_{2+} = \left| b_{1} \right| \left| b_{3} \right| \cos \phi_{13} + \left| b_{2} \right| \left| b_{4} \right| \cos \phi_{24}  = \mathrm{Re} \left[ b_{3}^{\ast} b_{1} + b_{4}^{\ast} b_{2} \right] = - \check{E}$  & $\frac{1}{2} \left< b \right| \tilde{\Gamma}^{9} \left|  b  \right>$ &  \\
   $\Ocal^{a}_{2-} = \left| b_{1} \right| \left| b_{3} \right| \cos \phi_{13} - \left| b_{2} \right| \left| b_{4} \right| \cos \phi_{24} = \mathrm{Re} \left[ b_{3}^{\ast} b_{1} - b_{4}^{\ast} b_{2} \right] =  \check{H}$  & $\frac{1}{2} \left< b \right| \tilde{\Gamma}^{5} \left|  b  \right>$ &  \\
   \hline
   $\Ocal^{b}_{1+} = \left| b_{1} \right| \left| b_{4} \right| \sin \phi_{14} + \left| b_{2} \right| \left| b_{3} \right| \sin \phi_{23} = \mathrm{Im} \left[ b_{4}^{\ast} b_{1} + b_{3}^{\ast} b_{2} \right] = \check{O}_{z'}$  & $\frac{1}{2} \left< b \right| \tilde{\Gamma}^{7} \left|  b  \right>$ &  \\
   $\Ocal^{b}_{1-} = \left| b_{1} \right| \left| b_{4} \right| \sin \phi_{14} - \left| b_{2} \right| \left| b_{3} \right| \sin \phi_{23}  = \mathrm{Im} \left[ b_{4}^{\ast} b_{1} - b_{3}^{\ast} b_{2} \right] = - \check{C}_{x'}$ & $\frac{1}{2} \left< b \right| \tilde{\Gamma}^{16} \left|  b  \right>$  & $b = \mathcal{BR} = \mathrm{AD}$ \\
   $\Ocal^{b}_{2+} = \left| b_{1} \right| \left| b_{4} \right| \cos \phi_{14} + \left| b_{2} \right| \left| b_{3} \right| \cos \phi_{23}  = \mathrm{Re} \left[ b_{4}^{\ast} b_{1} + b_{3}^{\ast} b_{2} \right] = - \check{C}_{z'}$  & $\frac{1}{2} \left< b \right| \tilde{\Gamma}^{2} \left|  b  \right>$ &  \\
   $\Ocal^{b}_{2-} = \left| b_{1} \right| \left| b_{4} \right| \cos \phi_{14} - \left| b_{2} \right| \left| b_{3} \right| \cos \phi_{23} = \mathrm{Re} \left[ b_{4}^{\ast} b_{1} - b_{3}^{\ast} b_{2} \right] = - \check{O}_{x'}$  & $\frac{1}{2} \left< b \right| \tilde{\Gamma}^{14} \left|  b  \right>$ &  \\
   \hline
   $\Ocal^{c}_{1+} = \left| b_{1} \right| \left| b_{2} \right| \sin \phi_{12} + \left| b_{3} \right| \left| b_{4} \right| \sin \phi_{34} = \mathrm{Im} \left[ b_{2}^{\ast} b_{1} + b_{4}^{\ast} b_{3} \right] = - \check{L}_{x'}$  &  $\frac{1}{2} \left< b \right| \tilde{\Gamma}^{8} \left|  b  \right>$  &  \\
   $\Ocal^{c}_{1-} = \left| b_{1} \right| \left| b_{2} \right| \sin \phi_{12} - \left| b_{3} \right| \left| b_{4} \right| \sin \phi_{34}  = \mathrm{Im} \left[ b_{2}^{\ast} b_{1} - b_{4}^{\ast} b_{3} \right] = - \check{T}_{z'}$ & $\frac{1}{2} \left< b \right| \tilde{\Gamma}^{13} \left|  b  \right>$ & $c = \mathcal{TR} = \mathrm{PL}$ \\
   $\Ocal^{c}_{2+} = \left| b_{1} \right| \left| b_{2} \right| \cos \phi_{12} + \left| b_{3} \right| \left| b_{4} \right| \cos \phi_{34}  = \mathrm{Re} \left[ b_{2}^{\ast} b_{1} + b_{4}^{\ast} b_{3} \right] = - \check{L}_{z'}$  &  $\frac{1}{2} \left< b \right| \tilde{\Gamma}^{15} \left|  b  \right>$ &  \\
   $\Ocal^{c}_{2-} = \left| b_{1} \right| \left| b_{2} \right| \cos \phi_{12} - \left| b_{3} \right| \left| b_{4} \right| \cos \phi_{34} = \mathrm{Re} \left[ b_{2}^{\ast} b_{1} - b_{4}^{\ast} b_{3} \right] = \check{T}_{x'}$  &  $\frac{1}{2} \left< b \right| \tilde{\Gamma}^{6} \left|  b  \right>$  &  \\
   \hline
   \hline
 \end{tabular}
 \end{center}
 \caption{The definitions of the $16$ polarization ob\-serva\-bles in pseudoscalar meson photoproduction (cf. ref.~\cite{Chiang:1996em}) are collected here. The definitions and sign-conventions for the ob\-serva\-bles are consistent with the PhD-thesis~\cite{MyPhD}. The matrices defining the bilinear forms have been collected in appendix~\ref{sec:MatrixAlgebras}.}
 \label{tab:PhotoObservables}
\end{table*}


The ob\-serva\-bles have the generic form~\eqref{eq:GenericPolObservables} with $\bm{c}^{\alpha} = 1 / 2$ and the Dirac-matrices satisfy orthogonality~\eqref{eq:OrthogonalityRelation} with $\tilde{N} = 4$. There exist $4$ shape-classes of diagonal ('D'), right-parallelogram ('PR'), anti-diagonal ('AD') and left-parallelogram ('PL') type (cf. ref.~\cite{Chiang:1996em}). Every shape-class contains $4$ ob\-serva\-bles. The diagonal shape-class 'D' contains the unpolarized differential cross section and the $3$ single-spin ob\-serva\-bles. Each of the $3$ non-diagonal shape-classes matches exactly to one of the three groups of Beam-Target ($\mathcal{BT}$), Beam-Recoil ($\mathcal{BR}$),  and Target-Recoil ($\mathcal{TR}$) experiments, as indicated in Table~\ref{tab:PhotoObservables}. For the ob\-serva\-bles in the non-diagonal shape-classes, we use the intuitive systematic notation introduced by Nakayama~\cite{Nakayama:2018yzw}.

\begin{figure*}
 \begin{center}
\includegraphics[width = 0.31 \textwidth]{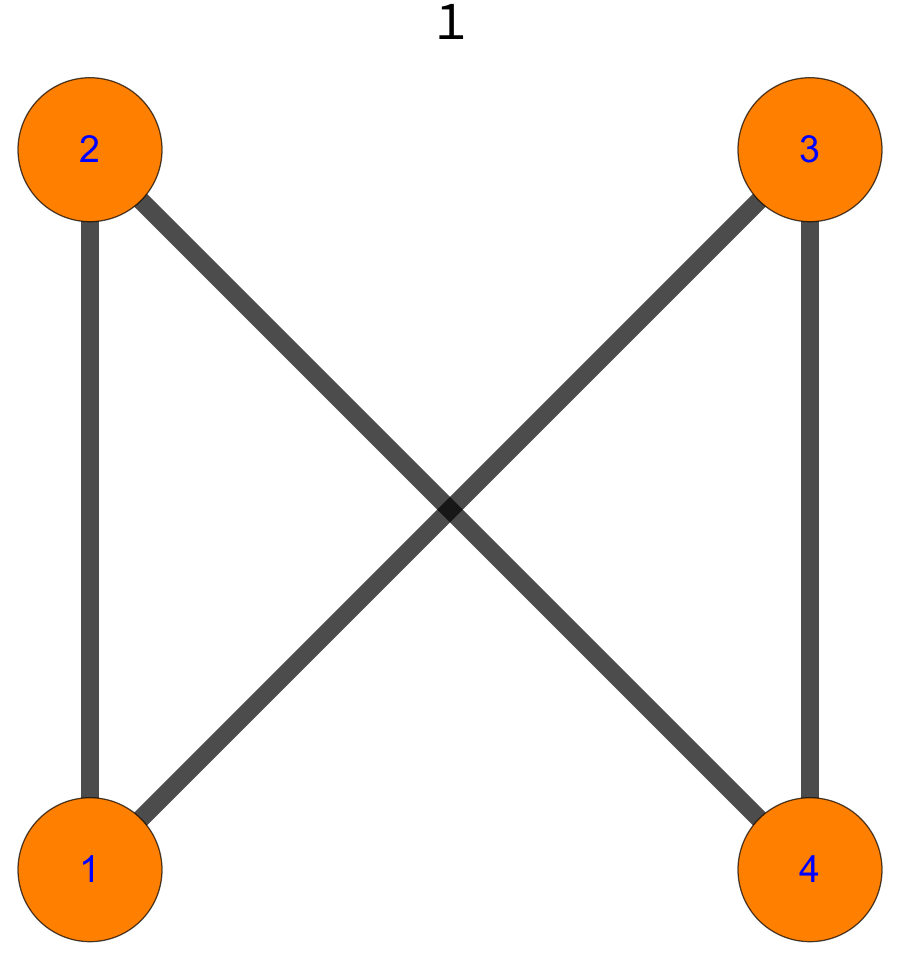} \hspace*{5pt}
\includegraphics[width = 0.31 \textwidth]{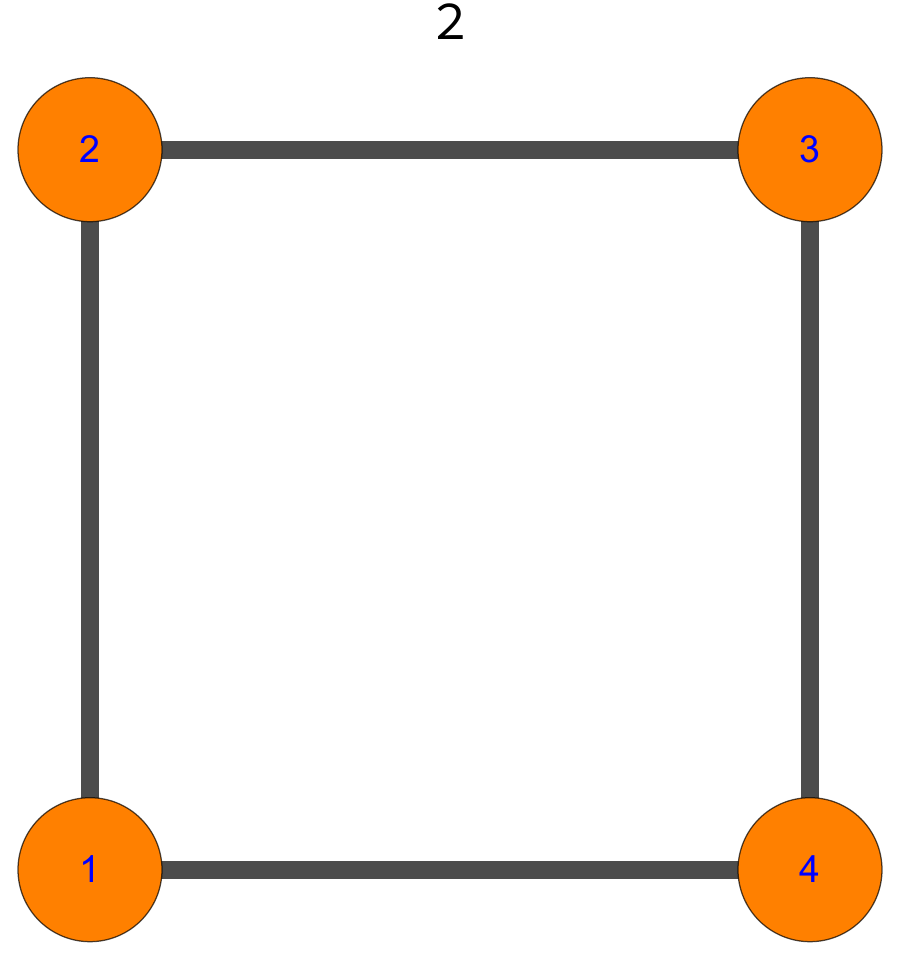} \hspace*{5pt}
\includegraphics[width = 0.31 \textwidth]{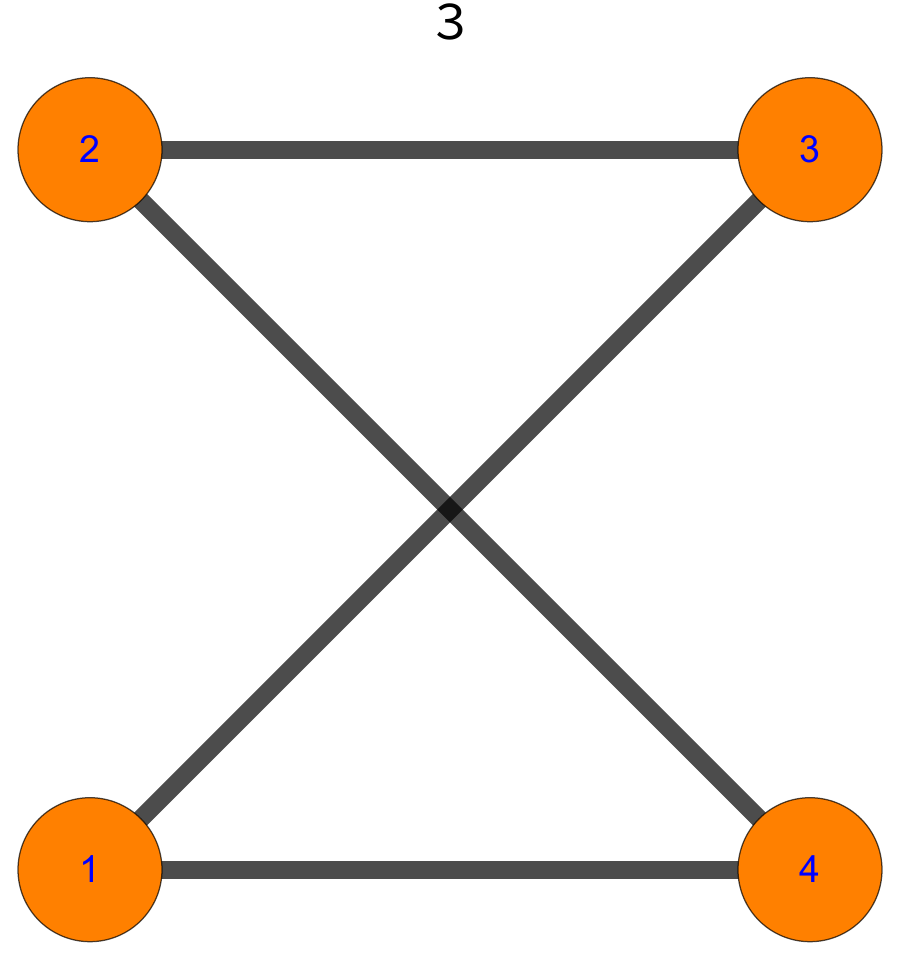}
\end{center}
\vspace*{-5pt}
\caption{The diagrams show the $3$ disctinct possible topologies that exist for closed loops formed using $N = 4$ points. Each topology can then be used as a starting point to derive fully complete sets for pseudoscalar meson photoproduction, according to Theorem~$2$ from section~\ref{sec:MoravcsikCompExp}.}
\label{fig:PhotoproductionStartTopologies}
\end{figure*}

For $N = 4$, there exists a novelty compared to both cases treated previously. Now, there exists $3$ possible topologies for the minimal closed loops formed out of $N = 4$ points, which are all shown in Figure~\ref{fig:PhotoproductionStartTopologies}. Each of these topologies can now be used as a starting point to derive fully complete closed loops.

\begin{figure*}
 \begin{center}
\includegraphics[width = 0.2425 \textwidth]{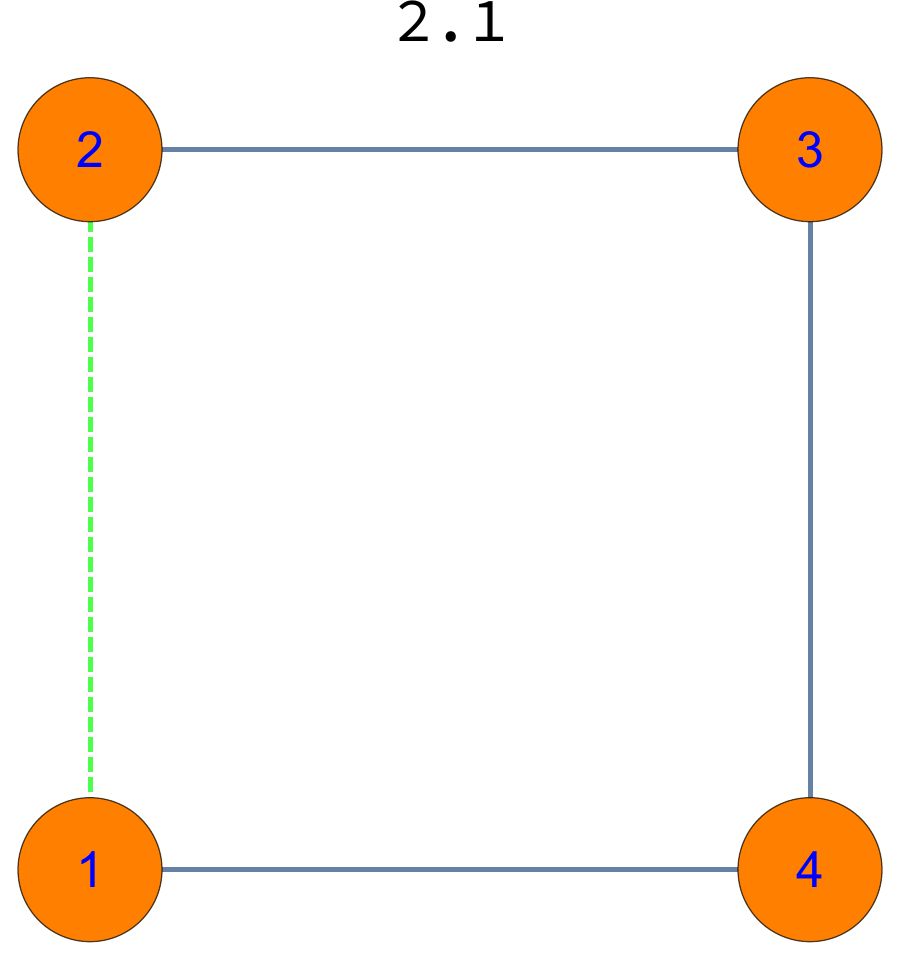} 
\includegraphics[width = 0.2425 \textwidth]{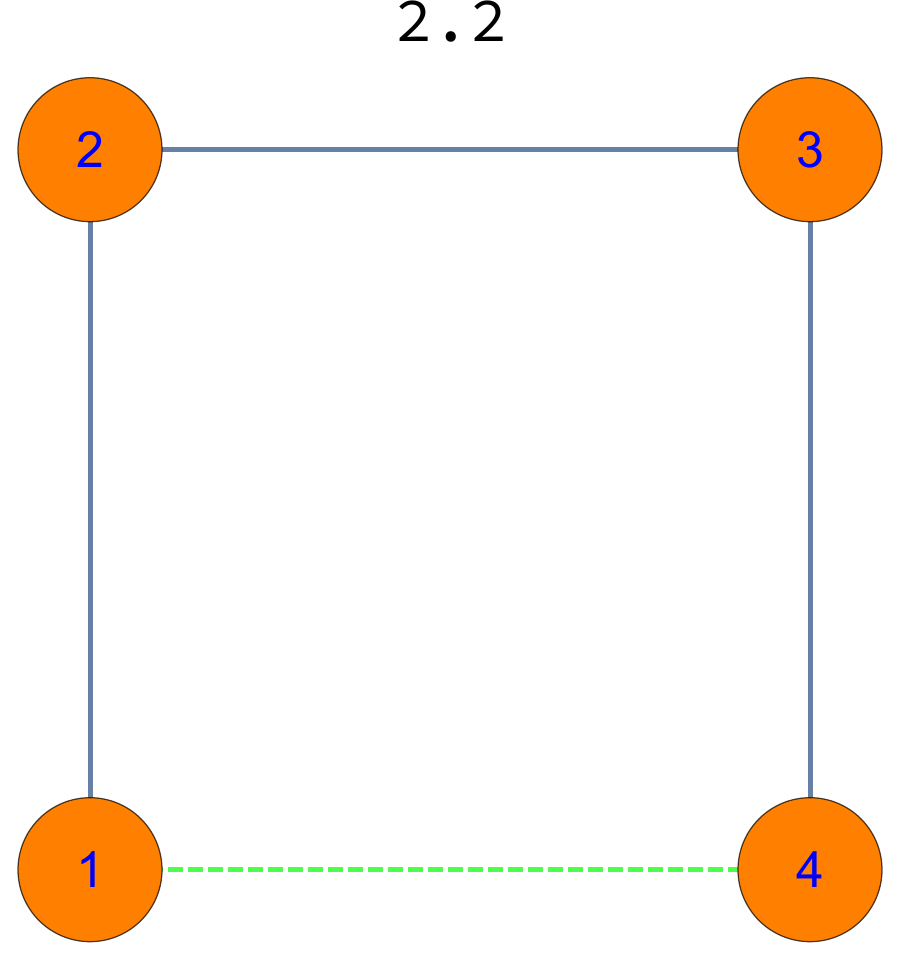}
\includegraphics[width = 0.2425 \textwidth]{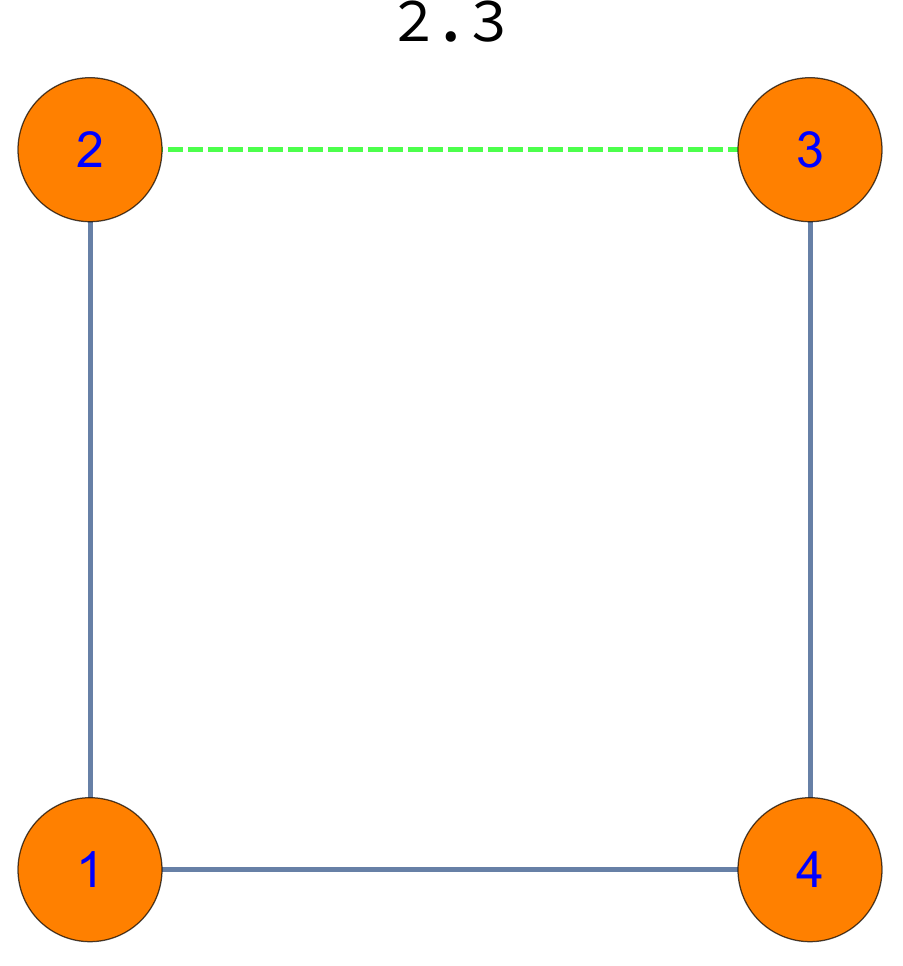} 
\includegraphics[width = 0.2425 \textwidth]{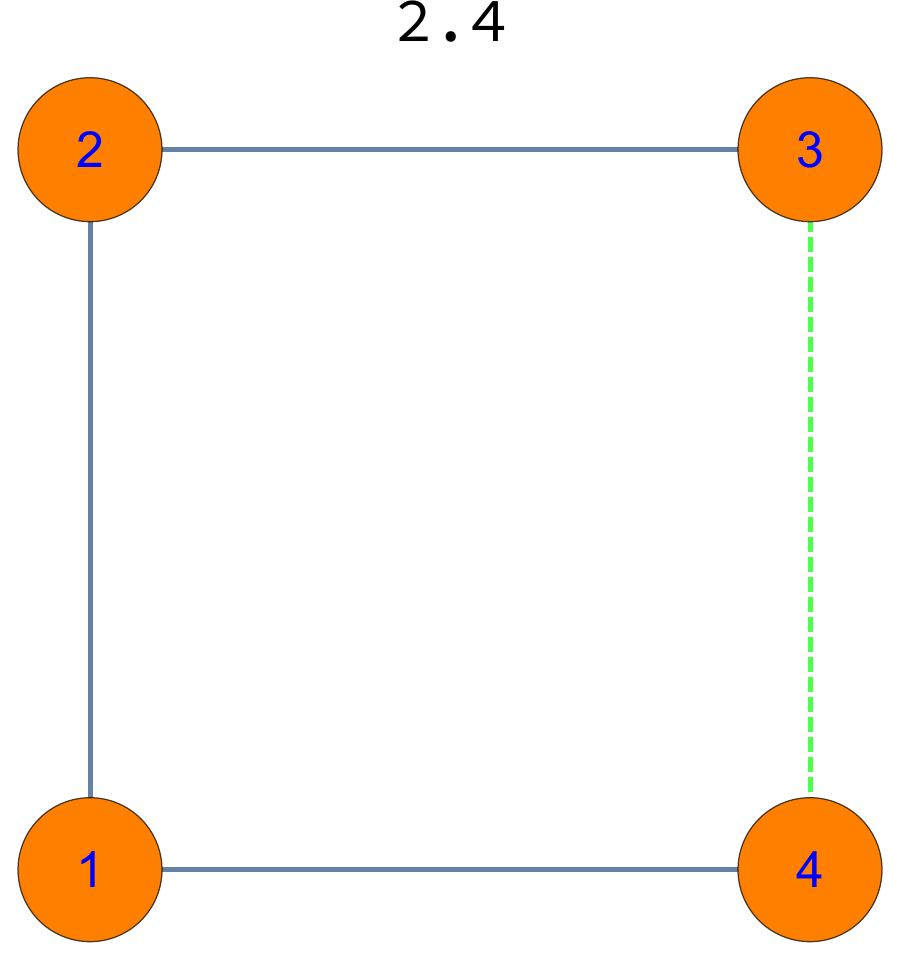} \\
\vspace*{2.5pt}
\includegraphics[width = 0.2425 \textwidth]{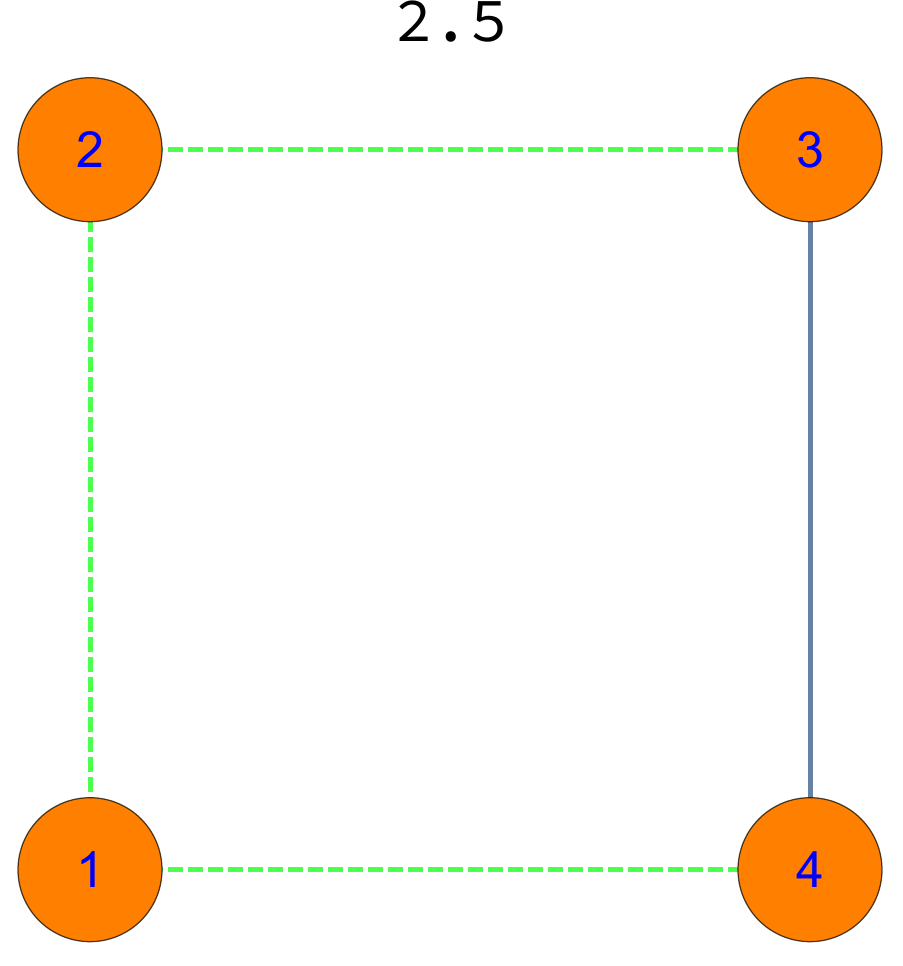} 
\includegraphics[width = 0.2425 \textwidth]{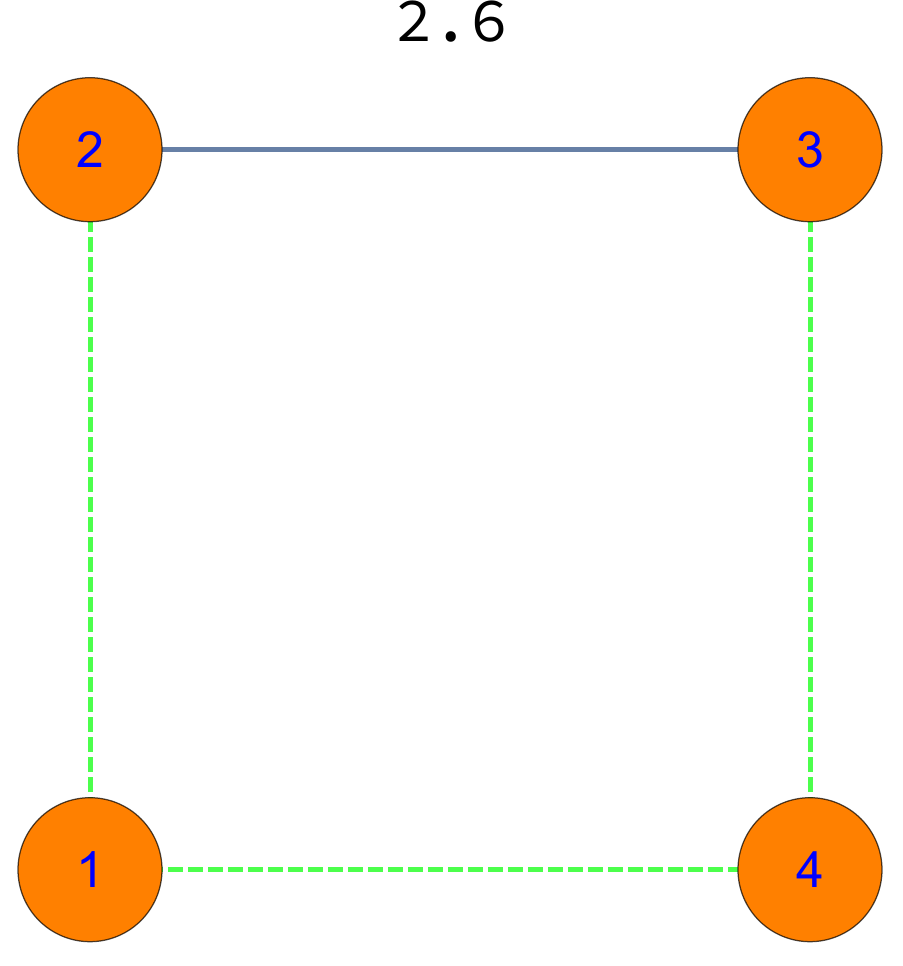} 
\includegraphics[width = 0.2425 \textwidth]{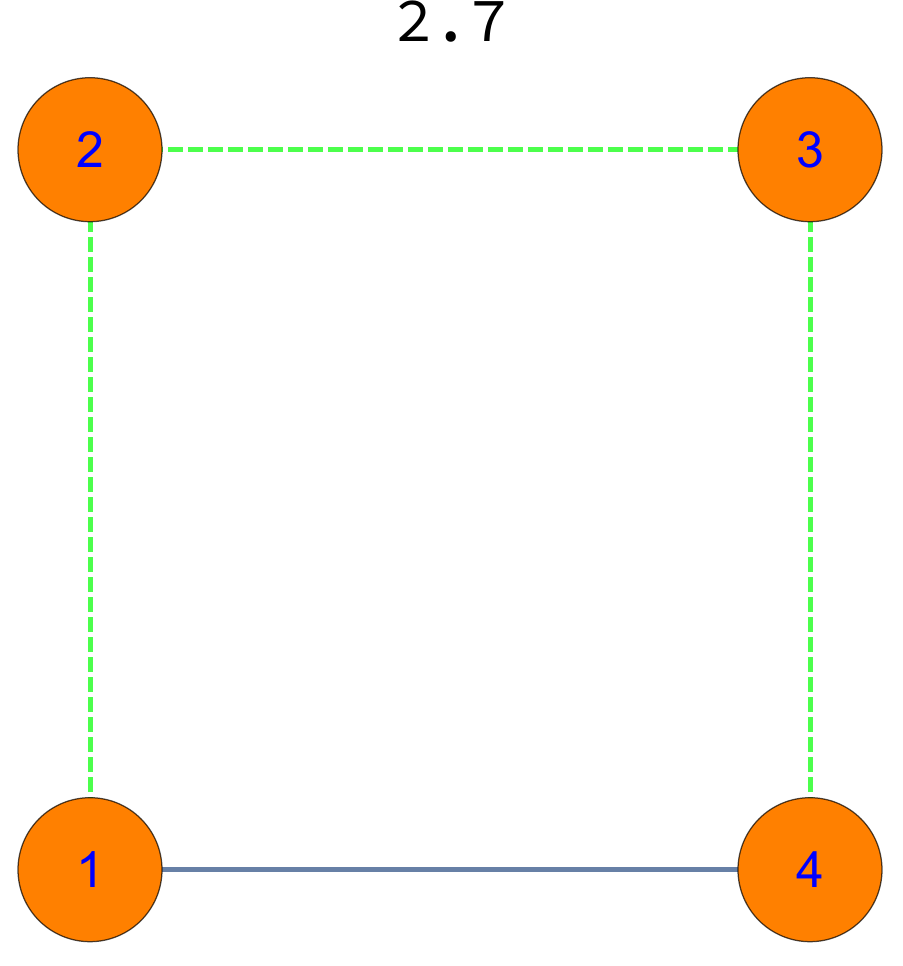} 
\includegraphics[width = 0.2425 \textwidth]{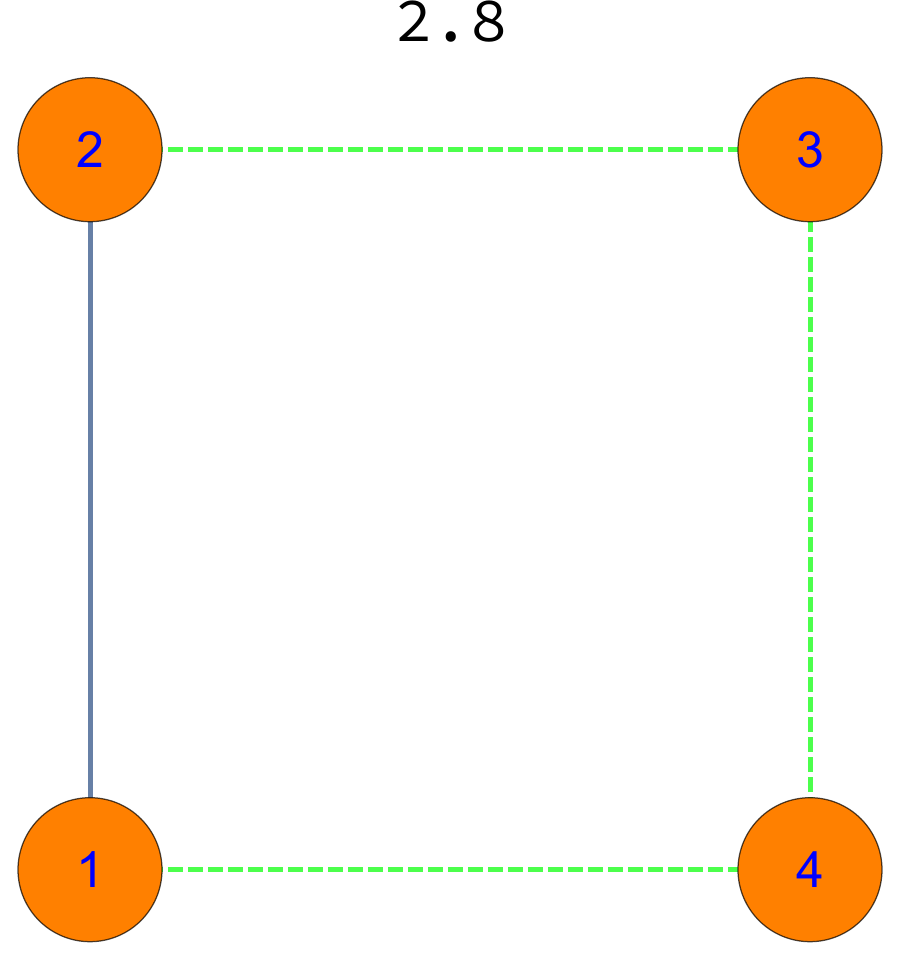} 
\end{center}
\vspace*{-5pt}
\caption{For the example with $N=4$ transversity amplitudes (photoproduction), the closed circular loops which yield unique solutions, i.e. those coming from topology~$2$ in Figure~\ref{fig:PhotoproductionStartTopologies}, are shown here. These are all the closed loops with an \textit{odd} number of dashed green lines (i.e. imaginary parts). (Color online)}
\label{fig:PhotoproductionCompleteBicoms}
\end{figure*}

All the fully complete combinations of real- and imaginary parts of bilinear products can again be found using Theorem 2. For instance, for the closed loops stemming from topology $2$ of Figure~\ref{fig:PhotoproductionStartTopologies}, all the \textit{fully complete} combinations are shown in Figure~\ref{fig:PhotoproductionCompleteBicoms}. We again refrain from showing all the remaining closed loops, which still leave unresolved discrete phase ambiguities, explicitly. Examples for fully complete closed loops that do \textit{not} derive from the circular type topology $2$ can be seen in Figure~\ref{fig:PhotoproductionExampleLoops} of section~\ref{sec:MoravcsikCompExp}.

It is worth spending more time and effort on the elaboration of the differences between the case of photoproduction, i.e. $N = 4$, and the cases of $N = 2$ and $3$ which have been discussed above.

%

%
%

When considering the definitions collected in Table~\ref{tab:PhotoObservables}, it becomes clear that in the case of $N = 4$, there does not exist a direct connection between ob\-serva\-bles and real- and imaginary parts of bilinear products any more. Rather, any non-diagonal observable $\Ocal^{n}_{\nu \pm}$ (for $n = a,b,c$) mixes two such real- and/or imaginary parts. 

The physical reason for this behavior is not fully clear right now. Mathematically it is simply true due to the defining properties of the Dirac $\tilde{\Gamma}$-matrices, as opposed to Pauli- or Gell-Mann matrices in the cases treated before.

Thus, the actual ob\-serva\-bles $\Ocal$ in photoproduction do \textit{not} isolate real- and imaginary parts of bilinear products any more. To accomplish this task, one has to define \textit{modified ob\-serva\-bles} $\tilde{\Ocal}$, according to\footnote{This definition does not include the diagonal ob\-serva\-bles.}:
\begin{align}
  \tilde{\Ocal}^{n}_{1 \pm} &:= \frac{1}{2} \left( \Ocal^{n}_{1+} \pm \Ocal^{n}_{1-}  \right) \text{, } n = a,b,c, \label{eq:PhotoprodModObsI} \\
  \tilde{\Ocal}^{n}_{2 \pm} &:= \frac{1}{2} \left( \Ocal^{n}_{2+} \pm \Ocal^{n}_{2-}  \right) \text{, } n = a,b,c. \label{eq:PhotoprodModObsII} 
\end{align}
Comparing to Table~\ref{tab:PhotoObservables}, it is seen quickly that indeed each $\tilde{\Ocal}$ is exactly equal to the real- or imaginary part of a particular bilinear product. However, the necessity to change ob\-serva\-bles as defined in equations~\eqref{eq:PhotoprodModObsI} and~\eqref{eq:PhotoprodModObsII} leads to the fact that the complete sets according to Moravcsik's theorem do \textit{not} map to the complete sets according to Chiang/Tabakin~\cite{Nakayama:2018yzw,Chiang:1996em} in a simple way any more. 

As an example, we consider the fully complete loop from case 2.1 shown in Figure~\ref{fig:PhotoproductionCompleteBicoms}. According to the rules established before, this case corresponds to the following set of real- and imaginary parts
\begin{equation}
 \mathrm{Im} \left[ b_{1} b_{2}^{\ast} \right]  , \mathrm{Re} \left[ b_{2} b_{3}^{\ast} \right]  , \mathrm{Re} \left[ b_{3} b_{4}^{\ast} \right]  , \mathrm{Re} \left[ b_{4} b_{1}^{\ast} \right]  . \label{eq:PhotoprodMoravcsikExample}
\end{equation}
This combination is equivalent to the following set of modified ob\-serva\-bles $\tilde{\Ocal}$:
\begin{align}
 \left| b_{1} \right| \left| b_{2} \right| \sin \phi_{12} &\equiv \tilde{\Ocal}^{c}_{1+} = \frac{1}{2} \left( \Ocal^{c}_{1+} + \Ocal^{c}_{1-} \right)  , \label{eq:PhotoprodExampleObsI} \\
 \left| b_{2} \right| \left| b_{3} \right| \cos \phi_{23} &\equiv \tilde{\Ocal}^{b}_{2-} = \frac{1}{2} \left( \Ocal^{b}_{2+} - \Ocal^{b}_{2-} \right)  , \label{eq:PhotoprodExampleObsII} \\
 \left| b_{3} \right| \left| b_{4} \right| \cos \phi_{34} &\equiv \tilde{\Ocal}^{c}_{2-} = \frac{1}{2} \left( \Ocal^{c}_{2+} - \Ocal^{c}_{2-} \right)  , \label{eq:PhotoprodExampleObsIII} \\
 \left| b_{1} \right| \left| b_{4} \right| \cos \phi_{14} &\equiv \tilde{\Ocal}^{b}_{2+} = \frac{1}{2} \left( \Ocal^{b}_{2+} + \Ocal^{b}_{2-} \right)  , \label{eq:PhotoprodExampleObsIV} 
\end{align}
where for comparison, we also write down the corresponding definitions in terms of actual ob\-serva\-bles $\Ocal$. Finally, since we assume the moduli $\left| b_{i} \right|$ as already known, our example for a fully complete closed loop is equivalent to the following set of sines and cosines:
\begin{equation}
 \sin \phi_{12}, \cos \phi_{23}, \cos \phi_{34} , \cos \phi_{14}. \label{eq:PhotoprodMoravcsikExampleRelPhases}
\end{equation}
One can check quickly that this set is complete according to Theorem 2, by writing down all the possible discrete phase ambiguities and enumerating all the cases for the fundamental consistency relation
\begin{equation}
 \phi_{12} + \phi_{23} + \phi_{34} + \phi_{41}  = 0  . \label{eq:PhotoFundamentalConsistencyRel}
\end{equation}
In this way, a set of linearly independent relations will emerge, as shown in the proof of Theorem 2 (cf. appendix~\ref{sec:DetailedProof}).

However, the mismatch to the results of Chiang/Tabakin becomes apparent, since in order to evaluate the $4$ modified ob\-serva\-bles 
\begin{equation}
 \left\{ \tilde{\Ocal}^{b}_{2+}, \tilde{\Ocal}^{b}_{2-}, \tilde{\Ocal}^{c}_{1+}, \tilde{\Ocal}^{c}_{2-}   \right\}  , \label{eq:PhotoprodMoravcsikExampleModObs}
\end{equation}
one needs the following set of $6$ actual ob\-serva\-bles
\begin{equation}
 \left\{ \Ocal^{b}_{2+}, \Ocal^{b}_{2-}, \Ocal^{c}_{1+}, \Ocal^{c}_{1-}, \Ocal^{c}_{2+},   \Ocal^{c}_{2-}   \right\}  . \label{eq:PhotoprodMoravcsikExampleActualObs}
\end{equation}

Combined with the $4$ group $\mathcal{S}$ ob\-serva\-bles, which have to be used in order to fix the $4$ moduli, it is seen that the set~\eqref{eq:PhotoprodMoravcsikExampleActualObs} coming from the modified version of Moravcsik's theorem actually amounts to a total of $10$ ob\-serva\-bles, and not $8$ as in the case of Chiang/Tabakin~\cite{Nakayama:2018yzw,Chiang:1996em}. This seems to be generally the case for photoproduction: the minimal fully complete closed loops according to Moravcsik lead to sets of polarization ob\-serva\-bles which are slightly over-complete (by $2$ ob\-serva\-bles, to be exact). Due to this mismatch of results, we will in the following distinguish 'Moravcsik-complete' sets, which may contain more than $2N$ ob\-serva\-bles, and '(absolutely) minimal complete sets' with exactly $2N$ ob\-serva\-bles.

Every possible topology from Figure~\ref{fig:PhotoproductionStartTopologies} leads to $8$ possible fully complete closed loops and each of these loops can again, using Table~\ref{tab:PhotoObservables}, be seen to imply a set of ob\-serva\-bles, just as for the loop 2.1 discussed above. In this way, we can derive $24$ Moravcsik-complete sets of ob\-serva\-bles, from which however only $12$ turn out to be non-redundant. For these $12$ complete sets, the non-diagonal ob\-serva\-bles contained in them are listed in Table~\ref{tab:MinimalMoravcsikSetsPhoto}.


It is only reasonable to assume that the complete sets of ob\-serva\-bles derived from the fully complete loops according to Moravcsik contain the complete experiments according to Chiang/Tabakin as subsets. In all cases we considered so far, this was indeed the case.

In case of the six ob\-serva\-bles~\eqref{eq:PhotoprodMoravcsikExampleActualObs} from our example, i.e. loop 2.1 of Figure~\ref{fig:PhotoproductionCompleteBicoms}, a comparison with the results in references~\cite{Nakayama:2018yzw,Chiang:1996em} shows that they contain the following absolutely minimally complete subsets:
\begin{align}
 &\left\{ \Ocal^{b}_{2+}, \Ocal^{b}_{2-}, \Ocal^{c}_{1+}, \Ocal^{c}_{2+}   \right\} \text{, or } \label{eq:PhotoprodMoravcsikExampleActualObsMinCompI} \\
 &\left\{  \Ocal^{b}_{2+}, \Ocal^{b}_{2-}, \Ocal^{c}_{1-},   \Ocal^{c}_{2-}   \right\} . \label{eq:PhotoprodMoravcsikExampleActualObsMinCompII} 
\end{align}
In case either~\eqref{eq:PhotoprodMoravcsikExampleActualObsMinCompI} or~\eqref{eq:PhotoprodMoravcsikExampleActualObsMinCompII} were considered, one would always have a set of two redundant (or superfluous) ob\-serva\-bles within the example~\eqref{eq:PhotoprodMoravcsikExampleActualObs}, i.e. either
\begin{equation}
 \left( \Ocal^{c}_{1-}, \Ocal^{c}_{2-} \right)  \text{ or } \left( \Ocal^{c}_{1+}, \Ocal^{c}_{2+} \right)   . \label{eq:PhotoprodRedundantObsCases}
\end{equation}

\begin{table}[h]
\begin{tabular}{r|ccc||r|ccc}
Set-Nr. & \multicolumn{3}{c}{Observables} & Set-Nr. & \multicolumn{3}{c}{Observables} \\
\hline   
 1 & $\Ocal^{a}_{2 \pm}$ & $\Ocal^{c}_{1 \pm}$ & $\Ocal^{c}_{2 \pm}$ & 7 & $\Ocal^{b}_{1 \pm}$ & $\Ocal^{c}_{1 \pm}$ & $\Ocal^{c}_{2 \pm}$  \\
 2 & $\Ocal^{a}_{1 \pm}$ & $\Ocal^{a}_{2 \pm}$ & $\Ocal^{c}_{2 \pm}$ & 8 & $\Ocal^{b}_{1 \pm}$ & $\Ocal^{b}_{2 \pm}$ & $\Ocal^{c}_{1 \pm}$  \\
 3 & $\Ocal^{a}_{1 \pm}$ & $\Ocal^{c}_{1 \pm}$ & $\Ocal^{c}_{2 \pm}$ & 9 & $\Ocal^{a}_{1 \pm}$ & $\Ocal^{a}_{2 \pm}$ & $\Ocal^{b}_{2 \pm}$  \\
 4 & $\Ocal^{a}_{1 \pm}$ & $\Ocal^{a}_{2 \pm}$ & $\Ocal^{c}_{1 \pm}$ & 10 & $\Ocal^{a}_{2 \pm}$ & $\Ocal^{b}_{1 \pm}$ & $\Ocal^{b}_{2 \pm}$ \\
 5 & $\Ocal^{b}_{2 \pm}$ & $\Ocal^{c}_{1 \pm}$ & $\Ocal^{c}_{2 \pm}$ & 11 & $\Ocal^{a}_{1 \pm}$ & $\Ocal^{a}_{2 \pm}$ & $\Ocal^{b}_{1 \pm}$ \\
 6 & $\Ocal^{b}_{1 \pm}$ & $\Ocal^{b}_{2 \pm}$ & $\Ocal^{c}_{2 \pm}$ & 12 & $\Ocal^{a}_{1 \pm}$ & $\Ocal^{b}_{1 \pm}$ & $\Ocal^{b}_{2 \pm}$ \\
\end{tabular}
\caption{The $12$ distinct possibilities to form complete sets according to Moravcsik for photoproduction are shown here. In each case, $6$ ob\-serva\-bles are listed which have to be picked in addition to the $4$ single-spin ob\-serva\-bles $\left\{ \sigma_{0}, \Sigma, T, P \right\}$ (cf. Table~\ref{tab:PhotoObservables}).}
\label{tab:MinimalMoravcsikSetsPhoto}
\end{table}

Then, it should always be possible to use the $4$ ob\-serva\-bles in the minimal complete set, i.e. sets like~\eqref{eq:PhotoprodMoravcsikExampleActualObsMinCompI} or~\eqref{eq:PhotoprodMoravcsikExampleActualObsMinCompII}, in order to determine the redundant ones.

As a proof of concept, we want to demonstrate this procedure here for the example-set~\eqref{eq:PhotoprodMoravcsikExampleActualObsMinCompI}. This means, we determine the superfluous ob\-serva\-bles $\left( \Ocal^{c}_{1-}, \Ocal^{c}_{2-} \right)$ from the minimal complete set~\eqref{eq:PhotoprodMoravcsikExampleActualObsMinCompI}. In the following, we only sketch the derivation. All further details can be found in appendix~\ref{sec:MoravSetReductionPhoto}.

We begin by employing constraints among the $4$ ob\-serva\-bles within the group $c$ (similar to the Fierz-identities listed in reference~\cite{Chiang:1996em}). Considering the definitions in Table~\ref{tab:PhotoObservables}, as well as the definitions of the modified ob\-serva\-bles~\eqref{eq:PhotoprodModObsI} and~\eqref{eq:PhotoprodModObsII}, it becomes apparent that the following constraints hold:
\begin{align}
 \left( \tilde{\Ocal}^{c}_{1+} \right)^{2} + \left(  \tilde{\Ocal}^{c}_{2+} \right)^{2}  &= \left| b_{1} \right|^{2}  \left| b_{2} \right|^{2} , \label{eq:GroupCInternalConstraintI} \\
 \left( \tilde{\Ocal}^{c}_{1-} \right)^{2} + \left(  \tilde{\Ocal}^{c}_{2-} \right)^{2}  &= \left| b_{3} \right|^{2}  \left| b_{4} \right|^{2}  . \label{eq:GroupCInternalConstraintII}
\end{align}
Using these two quadratic equations, it is possible to derive expressions for the ob\-serva\-bles $\left( \Ocal^{c}_{1-}, \Ocal^{c}_{2-} \right)$ in terms of the two quantities $\left( \Ocal^{c}_{1+}, \Ocal^{c}_{2+} \right)$. However, there remains a four-fold discrete ambiguity in the result. In appendix~\ref{sec:MoravSetReductionPhoto}, we denote the corresponding solutions as $\left( \Ocal^{c}_{2-} \right)_{\text{I,II}}$ and $\left(\Ocal^{c}_{1-}\right)_{\left(\pm,\left\{\text{I,II}\right\}\right)}$, with sub-scripts 'I', 'II', '+' and '-' that label the $4$ ambiguities.

These remaining ambiguities cannot be resolved using ob\-serva\-bles from the group $c$ alone. Instead, one has to find some way to transfer information from the group $b$ ob\-serva\-bles, which are also contained in the complete set~\eqref{eq:PhotoprodMoravcsikExampleActualObsMinCompI}, to the group $c$ ob\-serva\-bles. This task is accomplished using the following consistency relation, which represents a unique relation connecting the relative phases from both groups:
\begin{equation}
 \underbrace{\phi_{12} + \phi_{34}}_{c} = \underbrace{\phi_{14} - \phi_{23}}_{b}  . \label{eq:ConsistencyRelPhotoprodExampleMainText}
\end{equation}
Taking the cosine and sine of both sides of this relation, the following constraints can be derived among the ob\-serva\-bles of group $c$ and the relative phases belonging to group $b$
\begin{align}
  \tilde{\Ocal}^{c}_{2+}  \tilde{\Ocal}^{c}_{2-} -  \tilde{\Ocal}^{c}_{1+} \tilde{\Ocal}^{c}_{1-} &= \left| b_{1} \right|  \left| b_{2} \right| \left| b_{3} \right|  \left| b_{4} \right| \cos  \left(  \phi_{14} - \phi_{23} \right)  , \label{eq:CosAddTheoReformMainText} \\
  \tilde{\Ocal}^{c}_{1+}  \tilde{\Ocal}^{c}_{2-} +  \tilde{\Ocal}^{c}_{2+} \tilde{\Ocal}^{c}_{1-} &= \left| b_{1} \right|  \left| b_{2} \right| \left| b_{3} \right|  \left| b_{4} \right| \sin  \left(  \phi_{14} - \phi_{23} \right)  . \label{eq:SinAddTheoReformMainText}
\end{align}
As can be seen from the definitions in Table~\ref{tab:PhotoObservables}, the two ob\-serva\-bles from class $b$ in the set~\eqref{eq:PhotoprodMoravcsikExampleActualObsMinCompI} are fully equivalent to the cosines of the respective relative-phases. Therefore, we obtain the following $4$-fold discrete phase-ambiguity for the phases of the terms on the right-hand-sides of~\eqref{eq:CosAddTheoReformMainText} and~\eqref{eq:SinAddTheoReformMainText}:
\begin{equation}
 \phi^{\pm}_{14} - \phi^{\pm}_{23} . \label{eq:GroupBRelPhasesDiscreteAmbsMainText}
\end{equation}
Appendix~\ref{sec:MoravSetReductionPhoto} contains more details on how the two additional constraints~\eqref{eq:CosAddTheoReformMainText} and~\eqref{eq:SinAddTheoReformMainText}, together with the phase-information contained in~\eqref{eq:GroupBRelPhasesDiscreteAmbsMainText}, can be used in order to resolve the four-fold discrete ambiguity and thus complete the unique determination of the two quantities $\left( \Ocal^{c}_{1-}, \Ocal^{c}_{2-} \right)$. We remark that all the derivations and mathematical statements made here and in appendix~\ref{sec:MoravSetReductionPhoto} have been checked using Mathematica~\cite{Mathematica}. Furthermore, the approach is very similar in spirit to reductions performed in the work by Arenh\"{o}vel and Fix~\cite{Arenhoevel:2014dwa}. However, the details of both procedures are slightly different.

In summary, the modified form of Moravcsik's theorem implies complete sets of $10$ ob\-serva\-bles for pseudoscalar meson photoproduction (see Table~\ref{tab:MinimalMoravcsikSetsPhoto}), which represents a reduction from the full set of $16$ ob\-serva\-bles. We see that these ob\-serva\-bles have to be picked from $3$ of the $4$ available shape-classes. Thus, also with respect to shape-classes, a reduction has occurred, other than in the cases of $N = 2$ and $3$. Starting from a Moravcsik-complete set of $10$ ob\-serva\-bles, it is possible to reduce further down to an absolutely minimal subset containing $8$ ob\-serva\-bles.

\section{Usefulness of Moravcsik's theorem for cases of higher $N$} \label{sec:UsefulnessHigherN}



The real strength of Moravcsik's theorem lies in the fact that it is formulated for an arbitrary number of amplitudes $N$. Thus, it may become really useful as a simple criterion for the (pre-) selection of slightly over-complete experiments for processes which feature even more then $4$ amplitudes! Examples of current experimental interest are here certainly the electroproduction of one pseudoscalar meson ($N = 6$)~\cite{Tiator:2017cde}, the photoproduction of two pseudoscalar mesons ($N = 8$)~\cite{Roberts:2004mn}, or the photoproduction of vector mesons ($N = 12$)~\cite{Pichowsky:1994gh}.

Furthermore, this pre-selection of complete sets according to the modified version of Moravcsik's theorem can be completely automated on a computer, as has been done in the Mathematica-code~\cite{Mathematica} written for this work. From the results obtained in section~\ref{sec:MoravcsikTheoremExamples}, we can extract the following set of steps for a generic problem with $N$ amplitudes:
\begin{itemize}
 \item[(i)] Find all possible topologies for a closed loop with $N$ points (or vertices) and $N$ links (or edges). Each point has to attach to exactly $2$ link-lines. The number of possible topologies is equal to $1$ for $N = 2$ and $N = 3$, and equal to $\frac{(N - 1)!}{2}$ for all $N \geq 4$.
 \item[(ii)] Use each topology obtained in step (i) as a starting-point to derive $\mathcal{N}_{\text{comb.}}$ complete sets of real- and imaginary parts of bilinear products, or equivalently of cosines and sines of relative phases, according to Theorem 2 (cf. Table~\ref{tab:NCompleteCombinations}).
 \item[(iii)] Implement an association which assigns to each real- or imaginary part of a particular bilinear product a certain set of polarization ob\-serva\-bles. The particular association depends on each case of $N$ amplitudes under consideration and can be extracted from the Tables which collect the definitions of the ob\-serva\-bles (cf. Tables~\ref{tab:PiNObservables},~\ref{tab:ToyModelObservables} and~\ref{tab:PhotoObservables} from section~\ref{sec:MoravcsikTheoremExamples}.). Apply this association to the results of step (ii). In this way, the complete sets according to the modified form of Moravcsik's theorem are obtained. The number of such sets amounts to exactly:
 \begin{equation}
  \hspace*{10pt} \left[ \text{no. of topologies from step (i) } \right] \times \mathcal{N}_{\text{comb.}}  . \label{eq:NumberCompSetMoravcsik}
 \end{equation}
 \item[(iv)] Investigate the complete sets obtained in step (iii) and remove possibly redundant combinations, in case they are present. This yields the final result, i.e. a unique collection of complete sets of ob\-serva\-bles according to the modified version of Moravcsik's theorem, with every combination appearing exactly once.  
\end{itemize}

It is possible to write a code which just needs the number $N$, as well as the association needed in step (iii), as input. Then, the complete sets according to the modified form of Moravcsik's theorem are obtained automatically as a list.

For all $N \geq 4$, step (iv) will probably yield complete sets which contain more ob\-serva\-bles than the absolute minimal number $2 N$ required for a complete experiment. In order to show that a particular set of $2N$ ob\-serva\-bles is complete, one then has to apply algebraic reductions to a suitable Moravcsik-complete set obtained from step (iv). An example for such a reduction has been discussed at the end of section~\ref{sec:PhotoprodExample} for $N = 4$ amplitudes. It is very likely that similar tactics work for cases of higher $N$, but the complexity of the required calculations promises to increase rapidly.

In order to illustrate the above-given procedure, the case of pseudoscalar meson electroproduction is treated as an example in the next section.

\section{Pseudoscalar meson electroproduction ($N = 6$)} \label{sec:Electroproduction}

The reaction of pseudoscalar meson electroproduction is described by $N = 6$ amplitudes, which are accompanied by $36$ polarization ob\-serva\-bles~\cite{Tiator:2017cde}. The expressions for the ob\-serva\-bles are collected in Table~\ref{tab:ElectroObservablesI} and the $6 \times 6$ Dirac-matrices $\tilde{\Gamma}^{\alpha}$, which define the ob\-serva\-bles as bilinear forms, are shown in appendix~\ref{sec:MatrixAlgebras}. For the ob\-serva\-bles in non-diagonal shape-classes, we again use the systematic notation introduced by Nakayama~\cite{Nakayama:2018yzw}. However, in Table~\ref{tab:ElectroObservablesI}, we also give the observables in the usual physical notation, which is taken from the paper by Tiator and collaborators~\cite{Tiator:2017cde} and which proceeds as follows: each observable corresponds to a so-called 'response-function' $R^{\beta \alpha}_{i}$. The super-script $\alpha$ indicates the target-polarization, $\beta$ describes the recoil-polarization and the sub-script $i$ represents the polarization of the virtual photon. The latter can take the following configurations: $i \in \left\{ L, T, LT, TT, LT', TT' \right\}$, which correspond to purely longitudinal, purely transverse or 'mixed' (for instance longitudinal-transverse) interference contributions to the differential cross section. An additional super-script '$s$' or '$c$' on the left of the response-function indicates a possible sine- or cosine-dependence of the respective contribution, which depends on the azimuthal angle of the produced meson.

\begin{table*}
 \begin{center}
 \begin{tabular}{lcr}
  \hline
  \hline
  Observable & Bilinear form & Shape-class \\
  \hline 
  $R^{00}_{T} = \frac{1}{2} \left( \left| b_{1} \right|^{2} + \left| b_{2} \right|^{2} + \left| b_{3} \right|^{2} + \left| b_{4} \right|^{2} \right)$  &  $\frac{1}{2} \left< b \right| \tilde{\Gamma}^{1} \left|  b  \right>$  &    \\
  $- \hspace*{1pt}^{c} R^{00}_{TT} = \frac{1}{2} \left( \left| b_{1} \right|^{2} + \left| b_{2} \right|^{2} - \left| b_{3} \right|^{2} - \left| b_{4} \right|^{2} \right)$  &  $\frac{1}{2} \left< b \right| \tilde{\Gamma}^{4} \left|  b  \right>$  &  $ \mathrm{D1}$ \\
  $- R_{T}^{0y} = \frac{1}{2} \left( - \left| b_{1} \right|^{2} + \left| b_{2} \right|^{2} + \left| b_{3} \right|^{2} - \left| b_{4} \right|^{2} \right)$  &  $\frac{1}{2} \left< b \right| \tilde{\Gamma}^{10} \left|  b  \right>$  &    \\
  $ - R^{y' 0}_{T} = \frac{1}{2} \left( - \left| b_{1} \right|^{2} + \left| b_{2} \right|^{2} - \left| b_{3} \right|^{2} + \left| b_{4} \right|^{2} \right)$  &  $\frac{1}{2} \left< b \right| \tilde{\Gamma}^{12} \left|  b  \right>$  &    \\
  \hline
   $\Ocal^{a}_{1+} = \left| b_{1} \right| \left| b_{3} \right| \sin \phi_{13} + \left| b_{2} \right| \left| b_{4} \right| \sin \phi_{24} = \mathrm{Im} \left[ b_{3}^{\ast} b_{1} + b_{4}^{\ast} b_{2} \right] = - \hspace*{1pt}^{s} R_{TT}^{0z}$  & $\frac{1}{2} \left< b \right| \tilde{\Gamma}^{3} \left|  b  \right>$ &  \\
   $\Ocal^{a}_{1-} = \left| b_{1} \right| \left| b_{3} \right| \sin \phi_{13} - \left| b_{2} \right| \left| b_{4} \right| \sin \phi_{24}  = \mathrm{Im} \left[ b_{3}^{\ast} b_{1} - b_{4}^{\ast} b_{2} \right] = R_{TT'}^{0x}$ & $\frac{1}{2} \left< b \right| \tilde{\Gamma}^{11} \left|  b  \right>$ & $a = \mathrm{PR1}$ \\
   $\Ocal^{a}_{2+} = \left| b_{1} \right| \left| b_{3} \right| \cos \phi_{13} + \left| b_{2} \right| \left| b_{4} \right| \cos \phi_{24}  = \mathrm{Re} \left[ b_{3}^{\ast} b_{1} + b_{4}^{\ast} b_{2} \right] = R_{TT'}^{0z}$  & $\frac{1}{2} \left< b \right| \tilde{\Gamma}^{9} \left|  b  \right>$ &  \\
   $\Ocal^{a}_{2-} = \left| b_{1} \right| \left| b_{3} \right| \cos \phi_{13} - \left| b_{2} \right| \left| b_{4} \right| \cos \phi_{24} = \mathrm{Re} \left[ b_{3}^{\ast} b_{1} - b_{4}^{\ast} b_{2} \right] =  \hspace*{1pt}^{s} R^{0x}_{TT}$  & $\frac{1}{2} \left< b \right| \tilde{\Gamma}^{5} \left|  b  \right>$ &  \\
   \hline
   $\Ocal^{b}_{1+} = \left| b_{1} \right| \left| b_{4} \right| \sin \phi_{14} + \left| b_{2} \right| \left| b_{3} \right| \sin \phi_{23} = \mathrm{Im} \left[ b_{4}^{\ast} b_{1} + b_{3}^{\ast} b_{2} \right] = - \hspace*{1pt}^{s} R^{z'0}_{TT}$  & $\frac{1}{2} \left< b \right| \tilde{\Gamma}^{7} \left|  b  \right>$ &  \\
   $\Ocal^{b}_{1-} = \left| b_{1} \right| \left| b_{4} \right| \sin \phi_{14} - \left| b_{2} \right| \left| b_{3} \right| \sin \phi_{23}  = \mathrm{Im} \left[ b_{4}^{\ast} b_{1} - b_{3}^{\ast} b_{2} \right] = - R^{x'0}_{TT'}$ & $\frac{1}{2} \left< b \right| \tilde{\Gamma}^{16} \left|  b  \right>$  & $b = \mathrm{AD1}$ \\
   $\Ocal^{b}_{2+} = \left| b_{1} \right| \left| b_{4} \right| \cos \phi_{14} + \left| b_{2} \right| \left| b_{3} \right| \cos \phi_{23}  = \mathrm{Re} \left[ b_{4}^{\ast} b_{1} + b_{3}^{\ast} b_{2} \right] = R^{z'0}_{TT'}$  & $\frac{1}{2} \left< b \right| \tilde{\Gamma}^{2} \left|  b  \right>$ &  \\
   $\Ocal^{b}_{2-} = \left| b_{1} \right| \left| b_{4} \right| \cos \phi_{14} - \left| b_{2} \right| \left| b_{3} \right| \cos \phi_{23} = \mathrm{Re} \left[ b_{4}^{\ast} b_{1} - b_{3}^{\ast} b_{2} \right] = - \hspace*{1pt}^{s} R^{x'0}_{TT}$  & $\frac{1}{2} \left< b \right| \tilde{\Gamma}^{14} \left|  b  \right>$ &  \\
   \hline
   $\Ocal^{c}_{1+} = \left| b_{1} \right| \left| b_{2} \right| \sin \phi_{12} + \left| b_{3} \right| \left| b_{4} \right| \sin \phi_{34} = \mathrm{Im} \left[ b_{2}^{\ast} b_{1} + b_{4}^{\ast} b_{3} \right] = - R^{x'z}_{T}$  &  $\frac{1}{2} \left< b \right| \tilde{\Gamma}^{8} \left|  b  \right>$  &  \\
   $\Ocal^{c}_{1-} = \left| b_{1} \right| \left| b_{2} \right| \sin \phi_{12} - \left| b_{3} \right| \left| b_{4} \right| \sin \phi_{34}  = \mathrm{Im} \left[ b_{2}^{\ast} b_{1} - b_{4}^{\ast} b_{3} \right] = R^{z'x}_{T}$ & $\frac{1}{2} \left< b \right| \tilde{\Gamma}^{13} \left|  b  \right>$ & $c = \mathrm{PL1}$ \\
   $\Ocal^{c}_{2+} = \left| b_{1} \right| \left| b_{2} \right| \cos \phi_{12} + \left| b_{3} \right| \left| b_{4} \right| \cos \phi_{34}  = \mathrm{Re} \left[ b_{2}^{\ast} b_{1} + b_{4}^{\ast} b_{3} \right] = R_{T}^{z' z}$  &  $\frac{1}{2} \left< b \right| \tilde{\Gamma}^{15} \left|  b  \right>$ &  \\
   $\Ocal^{c}_{2-} = \left| b_{1} \right| \left| b_{2} \right| \cos \phi_{12} - \left| b_{3} \right| \left| b_{4} \right| \cos \phi_{34} = \mathrm{Re} \left[ b_{2}^{\ast} b_{1} - b_{4}^{\ast} b_{3} \right] = R_{T}^{x' x}$  &  $\frac{1}{2} \left< b \right| \tilde{\Gamma}^{6} \left|  b  \right>$  &  \\
   \hline
  $ R_{L}^{00} =\left| b_{5} \right|^{2} + \left| b_{6} \right|^{2}$  &  $\frac{1}{\sqrt{2}} \left< b \right| \tilde{\Gamma}^{17} \left|  b  \right>$  & $\mathrm{D2}$ \\
   $ R_{L}^{0y} = \left| b_{5} \right|^{2} - \left| b_{6} \right|^{2}$ & $\frac{1}{\sqrt{2}} \left< b \right| \tilde{\Gamma}^{18} \left|  b  \right>$ &   \\
   \hline 
  $ \Ocal^{d}_{1} = 2 \left| b_{5} \right| \left| b_{6} \right| \sin \phi_{56} = 2 \mathrm{Im} \left[ b_{6}^{\ast} b_{5} \right] =  R_{L}^{z'x}$  &  $\frac{1}{\sqrt{2}} \left< b \right| \tilde{\Gamma}^{20} \left|  b  \right>$  & $d = \mathrm{AD2}$ \\
   $ \Ocal^{d}_{2} = 2 \left| b_{5} \right| \left| b_{6} \right| \cos \phi_{56}  = 2 \mathrm{Re} \left[ b_{6}^{\ast} b_{5} \right] = - R_{L}^{x'x}$ & $\frac{1}{\sqrt{2}} \left< b \right| \tilde{\Gamma}^{19} \left|  b  \right>$ &   \\
  \hline 
   $\Ocal^{e}_{1+} = \left| b_{3} \right| \left| b_{6} \right| \sin \phi_{36} + \left| b_{4} \right| \left| b_{5} \right| \sin \phi_{45} = \mathrm{Im} \left[ b_{6}^{\ast} b_{3} + b_{5}^{\ast} b_{4} \right] = - \hspace*{1pt}^{s} R^{00}_{LT'}$  &  $\frac{1}{2} \left< b \right| \tilde{\Gamma}^{31} \left|  b  \right>$  &  \\
   $\Ocal^{e}_{1-} = \left| b_{3} \right| \left| b_{6} \right| \sin \phi_{36} - \left| b_{4} \right| \left| b_{5} \right| \sin \phi_{45}  = \mathrm{Im} \left[ b_{6}^{\ast} b_{3} - b_{5}^{\ast} b_{4} \right] = \hspace*{1pt}^{s} R^{0y}_{LT'}$ & $\frac{1}{2} \left< b \right| \tilde{\Gamma}^{29} \left|  b  \right>$ & $e = \mathrm{AD3}$ \\
   $\Ocal^{e}_{2+} = \left| b_{3} \right| \left| b_{6} \right| \cos \phi_{36} + \left| b_{4} \right| \left| b_{5} \right| \cos \phi_{45}  = \mathrm{Re} \left[ b_{6}^{\ast} b_{3} + b_{5}^{\ast} b_{4} \right] = \hspace*{1pt}^{c} R^{00}_{LT}$  &  $\frac{1}{2} \left< b \right| \tilde{\Gamma}^{21} \left|  b  \right>$ &  \\
   $\Ocal^{e}_{2-} = \left| b_{3} \right| \left| b_{6} \right| \cos \phi_{36} - \left| b_{4} \right| \left| b_{5} \right| \cos \phi_{45} = \mathrm{Re} \left[ b_{6}^{\ast} b_{3} - b_{5}^{\ast} b_{4} \right] = - \hspace*{1pt}^{c} R^{0y}_{LT}$  &  $\frac{1}{2} \left< b \right| \tilde{\Gamma}^{23} \left|  b  \right>$  &  \\
   \hline 
   $\Ocal^{f}_{1+} = \left| b_{1} \right| \left| b_{6} \right| \sin \phi_{16} + \left| b_{2} \right| \left| b_{5} \right| \sin \phi_{25} = \mathrm{Im} \left[ b_{6}^{\ast} b_{1} + b_{5}^{\ast} b_{2} \right] = - \hspace*{1pt}^{s} R^{0z}_{LT}$  &  $\frac{1}{2} \left< b \right| \tilde{\Gamma}^{30} \left|  b  \right>$  &  \\
   $\Ocal^{f}_{1-} = \left| b_{1} \right| \left| b_{6} \right| \sin \phi_{16} - \left| b_{2} \right| \left| b_{5} \right| \sin \phi_{25}  = \mathrm{Im} \left[ b_{6}^{\ast} b_{1} - b_{5}^{\ast} b_{2} \right] = \hspace*{1pt}^{c} R^{0x}_{LT'}$ & $\frac{1}{2} \left< b \right| \tilde{\Gamma}^{24} \left|  b  \right>$ & $f = \mathrm{AD4}$ \\
   $\Ocal^{f}_{2+} = \left| b_{1} \right| \left| b_{6} \right| \cos \phi_{16} + \left| b_{2} \right| \left| b_{5} \right| \cos \phi_{25}  = \mathrm{Re} \left[ b_{6}^{\ast} b_{1} + b_{5}^{\ast} b_{2} \right] = \hspace*{1pt}^{c} R^{0z}_{LT'}$  &  $\frac{1}{2} \left< b \right| \tilde{\Gamma}^{32} \left|  b  \right>$ &  \\
   $\Ocal^{f}_{2-} = \left| b_{1} \right| \left| b_{6} \right| \cos \phi_{16} - \left| b_{2} \right| \left| b_{5} \right| \cos \phi_{25} = \mathrm{Re} \left[ b_{6}^{\ast} b_{1} - b_{5}^{\ast} b_{2} \right] =  \hspace*{1pt}^{s} R^{0x}_{LT}$  &  $\frac{1}{2} \left< b \right| \tilde{\Gamma}^{22} \left|  b  \right>$  &  \\
   \hline 
   $\Ocal^{g}_{1+} = \left| b_{1} \right| \left| b_{5} \right| \sin \phi_{15} + \left| b_{2} \right| \left| b_{6} \right| \sin \phi_{26} = \mathrm{Im} \left[ b_{5}^{\ast} b_{1} + b_{6}^{\ast} b_{2} \right] = -  \hspace*{1pt}^{s} R^{z'0}_{LT}$  &  $\frac{1}{2} \left< b \right| \tilde{\Gamma}^{33} \left|  b  \right>$  &  \\
   $\Ocal^{g}_{1-} = \left| b_{1} \right| \left| b_{5} \right| \sin \phi_{15} - \left| b_{2} \right| \left| b_{6} \right| \sin \phi_{26}  = \mathrm{Im} \left[ b_{5}^{\ast} b_{1} - b_{6}^{\ast} b_{2} \right] = - \hspace*{1pt}^{c} R^{x'0}_{LT'}$ & $\frac{1}{2} \left< b \right| \tilde{\Gamma}^{26} \left|  b  \right>$ & $g = \mathrm{PR2}$ \\
   $\Ocal^{g}_{2+} = \left| b_{1} \right| \left| b_{5} \right| \cos \phi_{15} + \left| b_{2} \right| \left| b_{6} \right| \cos \phi_{26}  = \mathrm{Re} \left[ b_{5}^{\ast} b_{1} + b_{6}^{\ast} b_{2} \right] = \hspace*{1pt}^{c} R^{z'0}_{LT'} $  &  $\frac{1}{2} \left< b \right| \tilde{\Gamma}^{34} \left|  b  \right>$ &  \\
   $\Ocal^{g}_{2-} = \left| b_{1} \right| \left| b_{5} \right| \cos \phi_{15} - \left| b_{2} \right| \left| b_{6} \right| \cos \phi_{26} = \mathrm{Re} \left[ b_{5}^{\ast} b_{1} - b_{6}^{\ast} b_{2} \right] = - \hspace*{1pt}^{s} R^{x'0}_{LT}$  &  $\frac{1}{2} \left< b \right| \tilde{\Gamma}^{25} \left|  b  \right>$  &  \\
   \hline 
   $\Ocal^{h}_{1+} = \left| b_{3} \right| \left| b_{5} \right| \sin \phi_{35} + \left| b_{4} \right| \left| b_{6} \right| \sin \phi_{46} = \mathrm{Im} \left[ b_{5}^{\ast} b_{3} + b_{6}^{\ast} b_{4} \right] = \hspace*{1pt}^{s} R^{x'x}_{LT'}$  &  $\frac{1}{2} \left< b \right| \tilde{\Gamma}^{35} \left|  b  \right>$  &  \\
   $\Ocal^{h}_{1-} = \left| b_{3} \right| \left| b_{5} \right| \sin \phi_{35} - \left| b_{4} \right| \left| b_{6} \right| \sin \phi_{46}  = \mathrm{Im} \left[ b_{5}^{\ast} b_{3} - b_{6}^{\ast} b_{4} \right] = - \hspace*{1pt}^{c} R^{z'x}_{LT}$ & $\frac{1}{2} \left< b \right| \tilde{\Gamma}^{28} \left|  b  \right>$ & $h = \mathrm{PR3}$ \\
   $\Ocal^{h}_{2+} = \left| b_{3} \right| \left| b_{5} \right| \cos \phi_{35} + \left| b_{4} \right| \left| b_{6} \right| \cos \phi_{46}  = \mathrm{Re} \left[ b_{5}^{\ast} b_{3} + b_{6}^{\ast} b_{4} \right] = - \hspace*{1pt}^{c} R^{x'x}_{LT}$  &  $\frac{1}{2} \left< b \right| \tilde{\Gamma}^{36} \left|  b  \right>$ &  \\
   $\Ocal^{h}_{2-} = \left| b_{3} \right| \left| b_{5} \right| \cos \phi_{35} - \left| b_{4} \right| \left| b_{6} \right| \cos \phi_{46} = \mathrm{Re} \left[ b_{5}^{\ast} b_{3} - b_{6}^{\ast} b_{4} \right] = - \hspace*{1pt}^{s} R^{z'x}_{LT'}$  &  $\frac{1}{2} \left< b \right| \tilde{\Gamma}^{27} \left|  b  \right>$  &  \\
   \hline
   \hline
 \end{tabular}
 \end{center}
 \caption{The definitions of electroproduction ob\-serva\-bles are collected here for the diagonal ob\-serva\-bles of types $\mathrm{D1}$ and $\mathrm{D2}$, as well as for the non-diagonal shape-classes $\left\{ a,b,c,d,e,f,g,h \right\}$. The definitions and sign-conventions for the ob\-serva\-bles have been taken over from reference~\cite{Tiator:2017cde}.  The matrices $ \tilde{\Gamma}^{\alpha}$ have been collected in appendix~\ref{sec:MatrixAlgebras}. \\ For each ob\-serva\-ble from a non-diagonal shape-class, we give the systematic notation introduced by Nakayama~\cite{Nakayama:2018yzw}. In addition, we also give the usual physical notation for the observables, which is as follows~\cite{Tiator:2017cde}: every observable corresponds to a 'response-function' $R^{\beta \alpha}_{i}$. The superscript $\alpha$ indicates the target-polarization, $\beta$ describes the recoil-polarization and the sub-script $i$ represents the polarization of the virtual photon, which can take the following configurations: $i \in \left\{ T, L, TL, TT, TL', TT' \right\}$ (corresponding to purely longitudinal, purely transverse or 'mixed' interference contributions to the differential cross section). An additional super-script '$s$' or '$c$' on the left of the response-function indicates a possible sine- or cosine-dependence of the respective contribution to the differential cross section (with sine or cosine depending on the azimuthal angle of the produced meson).}
 \label{tab:ElectroObservablesI}
\end{table*}

In the normalization-convention chosen here, the Dirac matrices for electroproduction satisfy the orthogonality relation~\eqref{eq:OrthogonalityRelation} with $\tilde{N} = 4$. The matrices (and thus also the ob\-serva\-bles) can be grouped into $10$ overall shape-classes. Two shape-classes contain diagonal ob\-serva\-bles. One of them, called 'D1', contains $4$ ob\-serva\-bles which correspond to matrices with non-vanishing entries in the first $4$ diagonal elements. The second diagonal shape-class, called 'D2', contains $2$ matrices with non-vanishing entries in the fifth and sixth diagonal element.


The remaining $30$ ob\-serva\-bles are divided into $8$ non-diagonal shape-classes, which comprise four shape-classes of anti-diagonal structure ('AD1'$,\ldots,$'AD4'), three shape-classes of right-parallelogram type ('PR1'$,\ldots,$'PR3') and one class of left-parallelogram structure ('PL1'). All non-diagonal shape-classes contain $4$ ob\-serva\-bles each, apart from class 'AD2' which contains just $2$ quantities. The normalization-factor $\bm{c}^{\alpha}$ is equal to $1/2$ for all shape-classes except for 'D2' and 'AD2'. For the latter two classes, we have $\bm{c}^{\alpha} = 1/\sqrt{2}$.

\begin{figure*}
 \begin{center}
\includegraphics[width = 0.98 \textwidth]{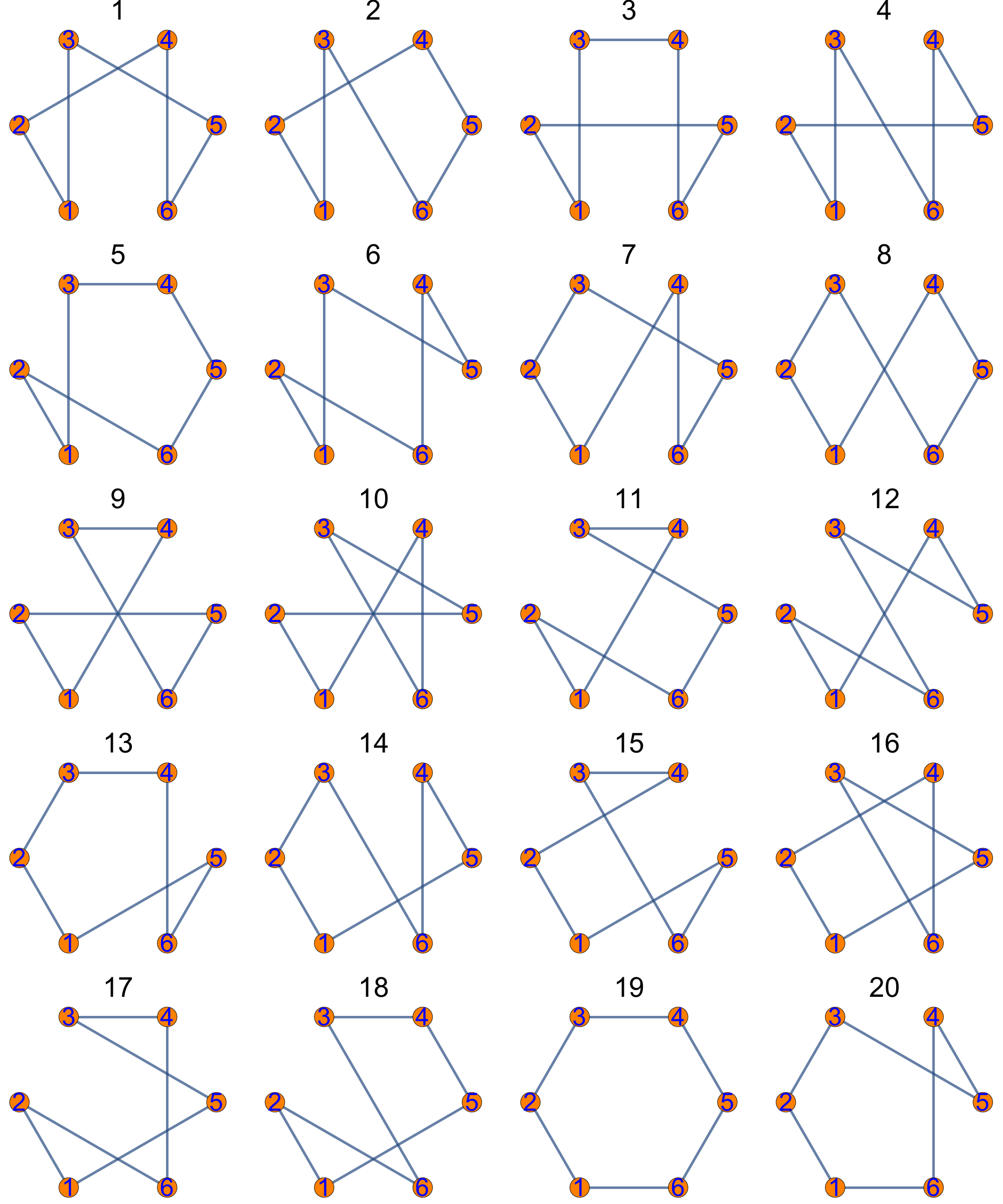}
\end{center}
\vspace*{-5pt}
\caption{The first set of $20$ possibilities out of an overall number of $60$ possible topologies are shown that exist for a minimal closed loop formed out of $N = 6$ points. Each topology can be used to derive the fully complete sets of ob\-serva\-bles for electroproduction, according to Theorem~$2$.}
\label{fig:ElectroproductionStartTopologies_I}
\end{figure*}
\begin{figure*}
 \begin{center}
\includegraphics[width = 0.98 \textwidth]{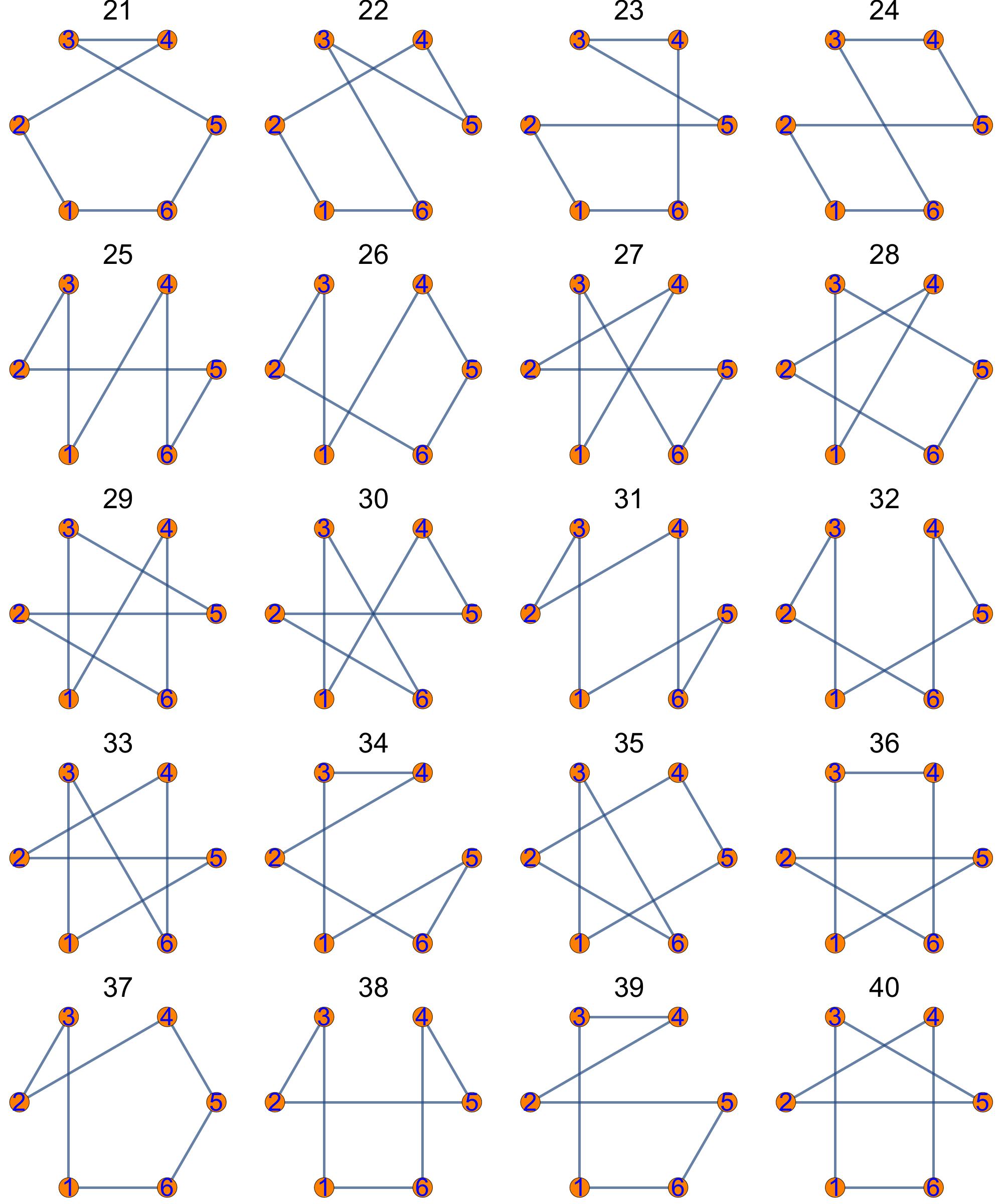}
\end{center}
\vspace*{-5pt}
\caption{Figure~\ref{fig:ElectroproductionStartTopologies_I} is continued here. The second set of $20$ out of an overall list of $60$ possible topologies is shown, which exists for electroproduction.}
\label{fig:ElectroproductionStartTopologies_II}
\end{figure*}
\begin{figure*}
 \begin{center}
\includegraphics[width = 0.98 \textwidth]{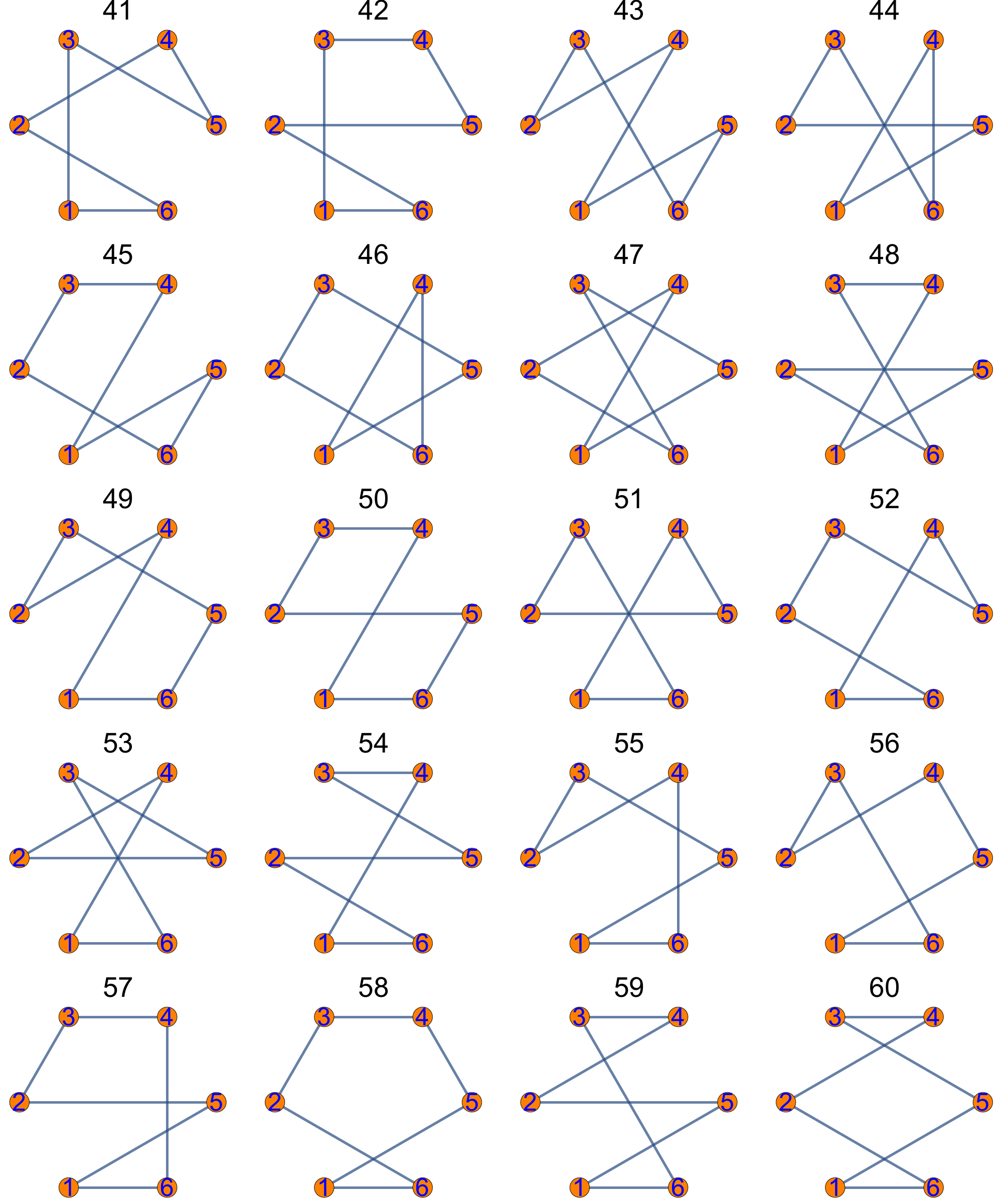}
\end{center}
\vspace*{-5pt}
\caption{Figure~\ref{fig:ElectroproductionStartTopologies_II} is continued here. The third set of $20$ out of an overall list of $60$ possible topologies is shown, which exists for electroproduction.}
\label{fig:ElectroproductionStartTopologies_III}
\end{figure*}

For the electroproduction problem with $N = 6$, the number of possible topologies that exist for minimal closed loops made of $6$ points has increased rapidly compared to the case of photoproduction, i.e. $N = 4$ (cf. Figure~\ref{fig:PhotoproductionStartTopologies}). Here, we obtain $60$ topologies from our Mathematica code, which are all collected in Figures~\ref{fig:ElectroproductionStartTopologies_I} to~\ref{fig:ElectroproductionStartTopologies_III}.

\begin{figure}
 \begin{center}
\includegraphics[width = 0.27 \textwidth]{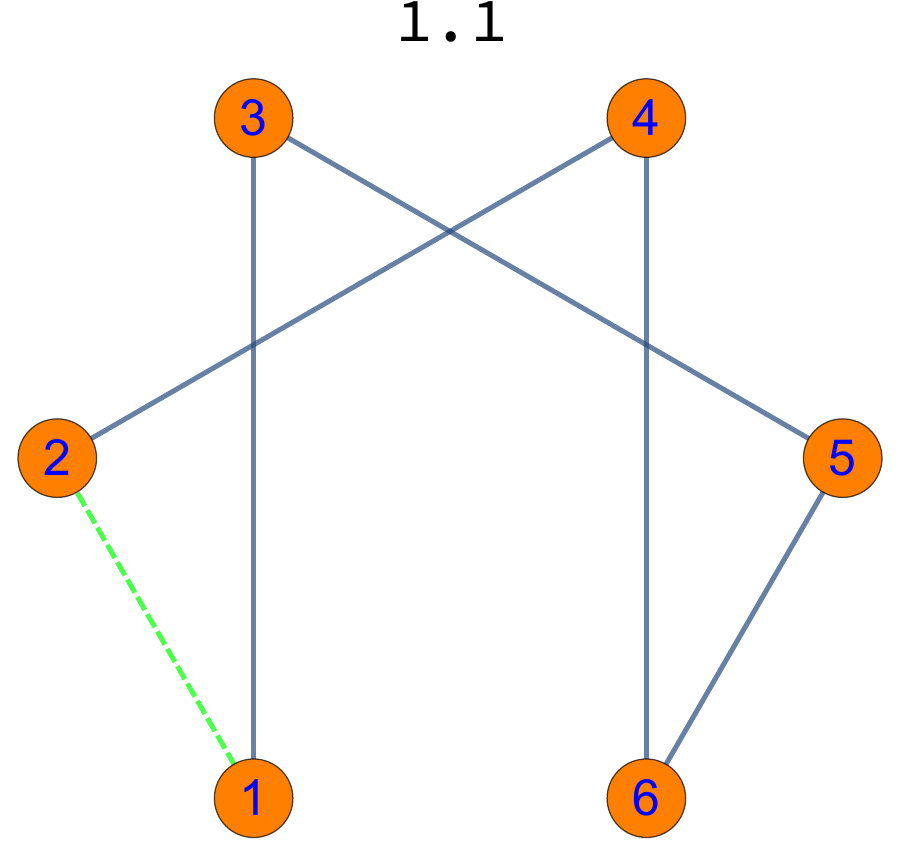} \\
\includegraphics[width = 0.27 \textwidth]{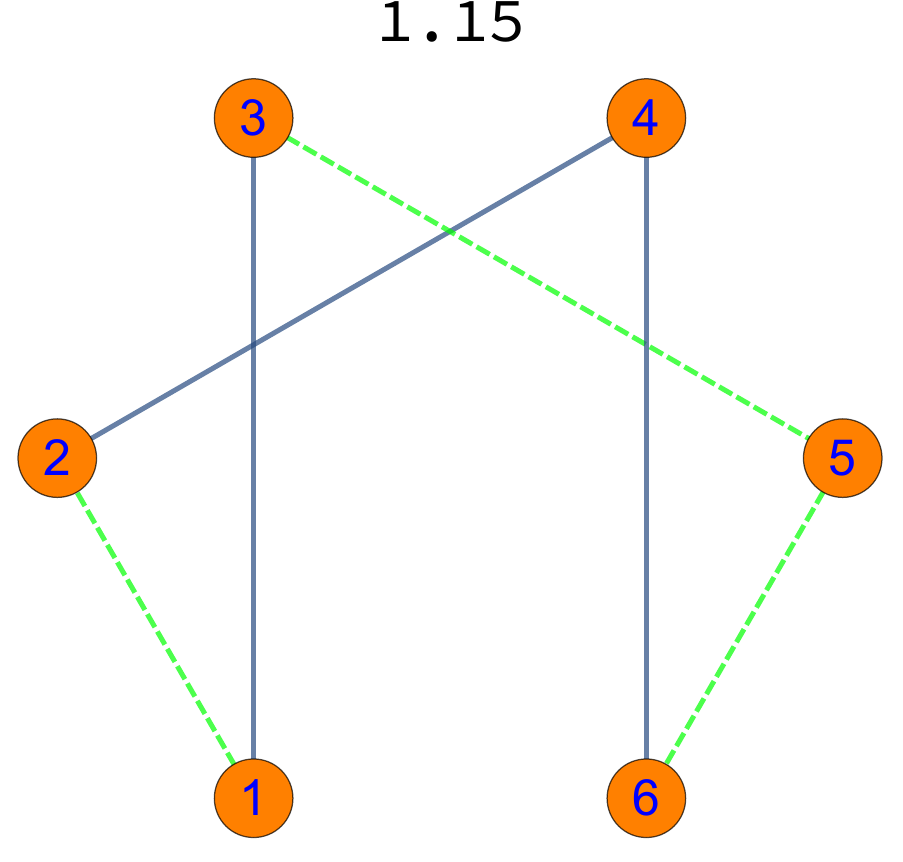} \\
\includegraphics[width = 0.27 \textwidth]{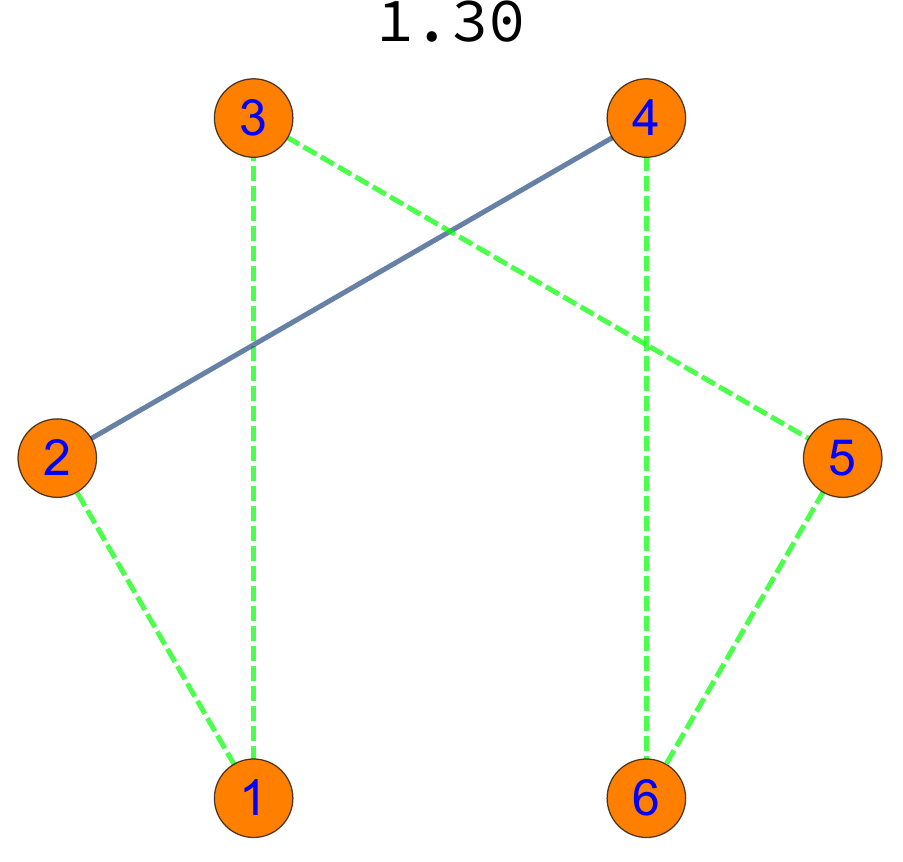} \\
\end{center}
\vspace*{-5pt}
\caption{For pseudoscalar meson electroproduction ($N = 6$), three examples out of $32$ possible closed loops which yield unique solutions are shown here, constructed by starting from topology~$1$ in Figure~\ref{fig:ElectroproductionStartTopologies_I}. (Color online)}
\label{fig:ElectroproductionCompleteLoops_Examples}
\end{figure}

Each of these start topologies can be used in order to derive $32$ possible fully complete loops according to Theorem~$2$ from section~\ref{sec:MoravcsikCompExp}. In Figure~\ref{fig:ElectroproductionCompleteLoops_Examples}, we show some illustrative examples for fully complete loops deduced from topology~$1$ of Figure~\ref{fig:ElectroproductionStartTopologies_I}. Each of the start topologies contained in Figures~\ref{fig:ElectroproductionStartTopologies_I} to~\ref{fig:ElectroproductionStartTopologies_III} also implies $32$ loops that still leave discrete ambiguities (those with an even number of dashed lines). These possibilities are not illustrated here explicitly.

As an example for the deduction of a Moravcsik-complete set, we pick the possibility 1.1 shown in Figure~\ref{fig:ElectroproductionCompleteLoops_Examples}. This loop implies the following combination of cosines and sines (or equivalently, real- and imaginary parts):
\begin{equation}
 \sin \phi_{12}, \cos \phi_{24}, \cos \phi_{46} , \cos \phi_{56}, \cos \phi_{35}, \cos \phi_{13}. \label{eq:ElectroprodMoravcsikExampleRelPhases}
\end{equation}

Looking at Table~\ref{tab:ElectroObservablesI}, we again observe the problem that the polarization ob\-serva\-bles do not isolate the real- and imaginary part of the bilinear products. Instead, we can again define modified ob\-serva\-bles according to the equations~\eqref{eq:PhotoprodModObsI} and~\eqref{eq:PhotoprodModObsII} used for photoproduction. These two equations can be used for all non-diagonal shape-classes except for the class 'AD2', where such a separation is however also not necessary (cf. Table~\ref{tab:ElectroObservablesI}).

In this way, we obtain the following set of non-diagonal ob\-serva\-bles which corresponds to the loop 1.1 from Figure~\ref{fig:ElectroproductionCompleteLoops_Examples} (this set is listed with Nr. (A.i.1) in Table~\ref{tab:MinimalMoravcsikSetsElectroI} of appendix~\ref{sec:MoravComplSetsElectro} and in Table~\ref{tab:MinimalMoravcsikSetsElectroExampleSets13} further below):
\begin{equation}
 \left\{ \Ocal^{a}_{2+}, \Ocal^{a}_{2-}, \Ocal^{c}_{1+}, \Ocal^{c}_{1-}, \Ocal^{d}_{2} , \Ocal^{h}_{2+},   \Ocal^{h}_{2-}   \right\}  . \label{eq:ElectroprodMoravcsikExampleActualObs}
\end{equation}
We see that these $7$ quantities, in combination with the $6$ diagonal ob\-serva\-bles which are always assumed to be measured, form a Moravcsik-complete set composed of $13$ polarization ob\-serva\-bles. The number $13$ is the minimal number of ob\-serva\-bles contained in any Moravcsik-complete set we found for electroproduction, using the topologies from Figures~\ref{fig:ElectroproductionStartTopologies_I} to~\ref{fig:ElectroproductionStartTopologies_III}. Interestingly, this is only one observable above the $2 N = 12$ quantities which constitute an absolutely minimally complete set.

Applying the procedure described in section~\ref{sec:UsefulnessHigherN} using Mathematica~\cite{Mathematica}, we find overall $776$ non-redundant complete sets according to Moravcsik in the considered case of $N = 6$. These sets contain $64$ complete sets of $13$ ob\-serva\-bles and furthermore $96$ complete sets composed of $14$ ob\-serva\-bles, which have been collected in Tables~\ref{tab:MinimalMoravcsikSetsElectroI} to~\ref{tab:MinimalMoravcsikSetsElectroII2OneTwo} of appendix~\ref{sec:MoravComplSetsElectro}. We refrain from showing the remaining $616$ Moravcsik-complete sets, which all contain more than $14$ ob\-serva\-bles. We note however that the largest number of ob\-serva\-bles in a Moravcsik-complete set found in this study is $18$.

We find a mismatch between the number of ob\-serva\-bles contained in the Moravcsik-complete sets and the absolutely minimal complete sets, similarly to the case of photoproduction (sec.~\ref{sec:PhotoprodExample}). As was argued for photoproduction, it is reasonable to assume that the absolutely minimal complete sets of $12$ can be found as subsets of the Moravcsik-complete sets. Then, one should in each case be able to do an algebraic reduction just as demonstrated in section~\ref{sec:PhotoprodExample} for photoproduction.

We sketch this reduction for one particular case for electroproduction and it turns out that the Moravcsik-complete sets of $14$ ob\-serva\-bles are particularly well-suited for this procedure.


%
%


%

\clearpage

%


%

\newpage

As an example, we consider the first set listed in Table~\ref{tab:MinimalMoravcsikSetsElectroII}, i.e. the set (B.i.1) (see also Table~\ref{tab:MinimalMoravcsikSetsElectroExampleSets14} further below):
\begin{equation}
 \left\{ \Ocal^{c}_{1+}, \Ocal^{c}_{1-}, \Ocal^{c}_{2+}, \Ocal^{c}_{2-}, \Ocal^{g}_{2+}, \Ocal^{g}_{2-} , \Ocal^{h}_{2+},   \Ocal^{h}_{2-}   \right\}  . \label{eq:ElectroprodMoravcsikExampleWith14Obs}
\end{equation}
Using the method of phase-fixing worked out by Nakayama\footnote{To be a bit more precise: we extended the case of \textbf{(2 + 2)} ob\-serva\-bles, which Nakayama describes in section~III of his work on photoproduction~\cite{Nakayama:2018yzw}, to the corresponding case \textbf{(2 + 2 + 2)} for electroproduction. This means that all cases are considered where $2$ ob\-serva\-bles are selected from $3$ different shape-classes. The generalization is relatively straightforward and yields the set~\eqref{eq:ElectroprodMoravcsikCompleteExampleWith12Obs}. However, in order to extend Nakayamas full discussion of all cases from photoproduction to electroproduction, one has to discuss a number of different cases which grows very rapidly.}~\cite{Nakayama:2018yzw}, as well as a complementary numerical check, we found for example the following absolutely minimal complete set as a subset of~\eqref{eq:ElectroprodMoravcsikExampleWith14Obs}:
\begin{equation}
 \left\{ \Ocal^{c}_{1+}, \Ocal^{c}_{2-}, \Ocal^{g}_{2+}, \Ocal^{g}_{2-} , \Ocal^{h}_{2+},   \Ocal^{h}_{2-}   \right\}  . \label{eq:ElectroprodMoravcsikCompleteExampleWith12Obs}
\end{equation}
For the reduction from~\eqref{eq:ElectroprodMoravcsikExampleWith14Obs} down to~\eqref{eq:ElectroprodMoravcsikCompleteExampleWith12Obs}, we can use very similar steps to those performed in section~\ref{sec:PhotoprodExample} and appendix~\ref{sec:MoravSetReductionPhoto}. First, we employ the internal constraints of the shape-class $c$, which are formally exactly the same as the constraints expressed in equations~\eqref{eq:GroupCInternalConstraintI} and~\eqref{eq:GroupCInternalConstraintII} for photoproduction (sec.~\ref{sec:PhotoprodExample}).

Then, we obtain expressions for $\left( \Ocal^{c}_{1-}, \Ocal^{c}_{2+} \right)$ in terms of $\left( \Ocal^{c}_{1+}, \Ocal^{c}_{2-} \right)$ but again, as in the case of photoproduction, a four-fold discrete ambiguity remains.

In order to resolve this discrete ambiguity, we need information from the remaining $4$ ob\-serva\-bles in set~\eqref{eq:ElectroprodMoravcsikCompleteExampleWith12Obs}, which stem from the shape-classes $g$ and $h$. A consistency relation among the relative-phases of the shape-classes $c$, $g$ and $h$ exists and it reads as follows
\begin{equation}
 \underbrace{\phi_{12} - \phi_{34}}_{c} = \underbrace{\phi_{15} - \phi_{26}}_{g} - \underbrace{\phi_{35} + \phi_{46}}_{h}  . \label{eq:GroupCGHConsistencyRelation}
\end{equation}

We follow a similar tactic as in section~\ref{sec:PhotoprodExample} (as well as appendix~\ref{sec:MoravSetReductionPhoto}) and take the sine and cosine of both sides of this phase-constraint. Then, using addition theorems, we arrive at the following set of constraints
\begin{align}
  \tilde{\Ocal}^{c}_{2+}  \tilde{\Ocal}^{c}_{2-} +  \tilde{\Ocal}^{c}_{1+} \tilde{\Ocal}^{c}_{1-} &= \left| b_{1} \right|  \left| b_{2} \right| \left| b_{3} \right|  \left| b_{4} \right| \nonumber \\ 
  & \times \cos  \left(  \phi_{15} - \phi_{26} - \phi_{35} + \phi_{46} \right)  , \label{eq:CosAddTheoReform_ElPro} \\
  \tilde{\Ocal}^{c}_{1+}  \tilde{\Ocal}^{c}_{2-} -  \tilde{\Ocal}^{c}_{2+} \tilde{\Ocal}^{c}_{1-} &= \left| b_{1} \right|  \left| b_{2} \right| \left| b_{3} \right|  \left| b_{4} \right| \nonumber \\
  & \times \sin  \left(  \phi_{15} - \phi_{26} - \phi_{35} + \phi_{46} \right)  . \label{eq:SinAddTheoReform_ElPro}
\end{align}
Since the ob\-serva\-bles from classes $g$ and $h$ in the set~\eqref{eq:ElectroprodMoravcsikCompleteExampleWith12Obs} are fully equivalent to the cosines of the appearing relative-phases (cf. Table~\ref{tab:ElectroObservablesI}), we obtain the following $16$-fold discrete phase-ambiguity for the terms on the right-hand-sides of~\eqref{eq:CosAddTheoReform_ElPro} and~\eqref{eq:SinAddTheoReform_ElPro}:
\begin{equation}
 \phi_{15}^{\pm} - \phi_{26}^{\pm} - \phi_{35}^{\pm} + \phi_{46}^{\pm} . \label{eq:ElProReductionPhaseAmbiguitiesGHGroups}
\end{equation}
These $16$ ambiguities stem entirely from discrete ambiguities of the cosine-type~\eqref{eq:CosTypeAmbiguity}. Now, inserting the solutions corresponding to the $4$-fold discrete ambiguity for $\left( \Ocal^{c}_{1-}, \Ocal^{c}_{2+} \right)$ on the left-hand-sides and the $16$ possible phase-ambiguities~\eqref{eq:ElProReductionPhaseAmbiguitiesGHGroups} on the right-hand-sides of equations~\eqref{eq:CosAddTheoReform_ElPro} and~\eqref{eq:SinAddTheoReform_ElPro}, both equations will be able to distinguish which is the correct solution for the ob\-serva\-bles $\left( \Ocal^{c}_{1-}, \Ocal^{c}_{2+} \right)$.

Having finished the mathematical discussion, we now extract the physical content from the results stemming from our application of the modified form of Moravcsik's theorem to electroproduction. In order to do this, we focus on the Moravcsik-complete sets composed of $13$ and $14$ observables, which are collected in Tables~\ref{tab:MinimalMoravcsikSetsElectroI} to~\ref{tab:MinimalMoravcsikSetsElectroII2OneTwo} of appendix~\ref{sec:MoravComplSetsElectro}. In order to illustrate our discussion, we have extracted illustrative example-sets from the Tables in appendix~\ref{sec:MoravComplSetsElectro}, which are shown in Tables~\ref{tab:MinimalMoravcsikSetsElectroExampleSets13} and~\ref{tab:MinimalMoravcsikSetsElectroExampleSets14}.

\begin{table}
\begin{tabular}{l|ccccccc}
Set-Nr. & \multicolumn{7}{c}{Observables} \\
\hline   
A.i.1  &  $R_{TT'}^{0z}$ & $\hspace*{1pt}^{s} R^{0x}_{TT}$ & $R^{x'z}_{T}$ & $R^{z'x}_{T}$ & $R_{L}^{x'x}$ & $\hspace*{1pt}^{c} R^{x'x}_{LT}$ & $\hspace*{1pt}^{s} R^{z'x}_{LT'}$   \\ 
& $\Ocal^{a}_{2+}$ & $\Ocal^{a}_{2-}$ & $\Ocal^{c}_{1+}$ & $\Ocal^{c}_{1-}$ & $\Ocal^{d}_{2}$ & $\Ocal^{h}_{2+}$ & $\Ocal^{h}_{2-}$  \\  
A.ii.4  &   $R_{TT'}^{0z}$  &  $\hspace*{1pt}^{s} R^{0x}_{TT}$ & $R^{x'z}_{T}$  &  $R^{z'x}_{T}$ & $R_{L}^{x'x}$  &  $\hspace*{1pt}^{s} R^{00}_{LT'}$  &  $\hspace*{1pt}^{s} R^{0y}_{LT'}$  \\ 
& $\Ocal^{a}_{2+}$ & $\Ocal^{a}_{2-}$ & $\Ocal^{c}_{1+}$  &  $\Ocal^{c}_{1-}$ & $\Ocal^{d}_{2}$ & $\Ocal^{e}_{1+}$ & $\Ocal^{e}_{1-}$  \\  
 A.iii.8  &  $\hspace*{1pt}^{s} R^{z'0}_{TT}$ & $R^{x'0}_{TT'}$  &  $R_{T}^{z' z}$ & $R_{T}^{x' x}$ & $R_{L}^{z'x}$  &  $\hspace*{1pt}^{s} R^{x'x}_{LT'}$  &  $\hspace*{1pt}^{c} R^{z'x}_{LT}$   \\ 
   & $\Ocal^{b}_{1+}$ & $\Ocal^{b}_{1-}$ & $\Ocal^{c}_{2+}$  &  $\Ocal^{c}_{2-}$ & $\Ocal^{d}_{1}$ & $\Ocal^{h}_{1+}$ & $\Ocal^{h}_{1-}$  \\  
 A.iv.6  &  $R^{z'0}_{TT'}$  &  $\hspace*{1pt}^{s} R^{x'0}_{TT}$ & $R_{T}^{z' z}$  &  $R_{T}^{x' x}$ & $R_{L}^{z'x}$  &  $\hspace*{1pt}^{s} R^{00}_{LT'}$  &  $\hspace*{1pt}^{s} R^{0y}_{LT'}$  \\  
  & $\Ocal^{b}_{2+}$ & $\Ocal^{b}_{2-}$ & $\Ocal^{c}_{2+}$  &  $\Ocal^{c}_{2-}$ & $\Ocal^{d}_{1}$ & $\Ocal^{e}_{1+}$ & $\Ocal^{e}_{1-}$  \\  
  A.v.2   &  $R_{TT'}^{0z}$  &  $\hspace*{1pt}^{s} R^{0x}_{TT}$ & $R_{T}^{z' z}$  &  $R_{T}^{x' x}$ & $R_{L}^{z'x}$  &  $\hspace*{1pt}^{c} R^{z'0}_{LT'}$  &  $\hspace*{1pt}^{s} R^{x'0}_{LT}$  \\ 
& $\Ocal^{a}_{2+}$ & $\Ocal^{a}_{2-}$ & $\Ocal^{c}_{2+}$  &  $\Ocal^{c}_{2-}$ & $\Ocal^{d}_{1}$ & $\Ocal^{g}_{2+}$ & $\Ocal^{g}_{2-}$  \\  
 A.vi.4  &  $\hspace*{1pt}^{s} R_{TT}^{0z}$ & $R_{TT'}^{0x}$  &  $R_{T}^{z' z}$ & $R_{T}^{x' x}$ & $R_{L}^{z'x}$  &  $\hspace*{1pt}^{c} R^{0z}_{LT'}$  &  $\hspace*{1pt}^{s} R^{0x}_{LT}$  \\  
 & $\Ocal^{a}_{1+}$ & $\Ocal^{a}_{1-}$ & $\Ocal^{c}_{2+}$  &  $\Ocal^{c}_{2-}$ & $\Ocal^{d}_{1}$ & $\Ocal^{f}_{2+}$ & $\Ocal^{f}_{2-}$   \\ 
 A.vii.2  &  $R^{z'0}_{TT'}$  &  $\hspace*{1pt}^{s} R^{x'0}_{TT}$ & $R_{T}^{z' z}$  &  $R_{T}^{x' x}$ & $R_{L}^{z'x}$  &  $\hspace*{1pt}^{c} R^{z'0}_{LT'}$  &  $\hspace*{1pt}^{s} R^{x'0}_{LT}$  \\  
 & $\Ocal^{b}_{2+}$ & $\Ocal^{b}_{2-}$ & $\Ocal^{c}_{2+}$  &  $\Ocal^{c}_{2-}$ & $\Ocal^{d}_{1}$ & $\Ocal^{g}_{2+}$ & $\Ocal^{g}_{2-}$ \\  
  A.viii.1  &  $R^{z'0}_{TT'}$  &  $\hspace*{1pt}^{s} R^{x'0}_{TT}$ & $R^{x'z}_{T}$  &  $R^{z'x}_{T}$ & $R_{L}^{x'x}$  &  $\hspace*{1pt}^{c} R^{0z}_{LT'}$  &  $\hspace*{1pt}^{s} R^{0x}_{LT}$   \\
 & $\Ocal^{b}_{2+}$ & $\Ocal^{b}_{2-}$ & $\Ocal^{c}_{1+}$  &  $\Ocal^{c}_{1-}$ & $\Ocal^{d}_{2}$ & $\Ocal^{f}_{2+}$ & $\Ocal^{f}_{2-}$ 
\end{tabular}
\caption{Here we collect $8$ selected examples for the Moravcsik-complete sets composed of $13$ observables. For each of the $8$ possible combinations of shape-classes in these complete sets, one example has been chosen. In every case, the $7$ ob\-serva\-bles given here have to be picked in addition to the $6$ diagonal ob\-serva\-bles $\left\{ R^{00}_{T},  R^{00}_{TT}, R_{T}^{0y}, R^{y' 0}_{T}, R_{L}^{00} , R_{L}^{0y}  \right\}$. \\ Every example is given in the physical notation $R^{\beta \alpha}_{i}$ and also (directly below) in Nakayama's systematic notation. Furthermore, the labelling-scheme for the sets is chosen as follows: the letter 'A' denotes the Moravcsik-complete sets of $13$ observables, the roman numerals i, $\ldots$, viii indicate the different possible combinations of shape-classes and the regular numbers ($1$, $2$, $3$, $\ldots$) at the end count the number of the set from the respective group of shape-class combinations (cf. Tables~\ref{tab:MinimalMoravcsikSetsElectroI} to~\ref{tab:MinimalMoravcsikSetsElectroI2} in appendix~\ref{sec:MoravComplSetsElectro}).}
\label{tab:MinimalMoravcsikSetsElectroExampleSets13}
\end{table}

\newpage

\begin{table}
\begin{scriptsize}
\begin{tabular}{l|cccccccc}
Set-Nr. & \multicolumn{8}{c}{Observables} \\
\hline   
B.i.1  &  $R^{x'z}_{T}$ & $R^{z'x}_{T}$ & $R_{T}^{z' z}$ & $R_{T}^{x' x}$ & $\hspace*{1pt}^{c} R^{z'0}_{LT'}$  &  $\hspace*{1pt}^{s} R^{x'0}_{LT}$ & $\hspace*{1pt}^{c} R^{x'x}_{LT}$  &  $\hspace*{1pt}^{s} R^{z'x}_{LT'}$  \\ 
& $\Ocal^{c}_{1+}$ & $\Ocal^{c}_{1-}$ & $\Ocal^{c}_{2+}$  &  $\Ocal^{c}_{2-}$ & $\Ocal^{g}_{2+}$ & $\Ocal^{g}_{2-}$ & $\Ocal^{h}_{2+}$ & $\Ocal^{h}_{2-}$  \\  
B.ii.3  &  $R_{T}^{z' z}$  &  $R_{T}^{x' x}$ & $\hspace*{1pt}^{s} R^{00}_{LT'}$  &  $\hspace*{1pt}^{s} R^{0y}_{LT'}$  &  $\hspace*{1pt}^{c} R^{00}_{LT}$  &  $\hspace*{1pt}^{c} R^{0y}_{LT}$  &  $\hspace*{1pt}^{c} R^{z'0}_{LT'}$  &  $\hspace*{1pt}^{s} R^{x'0}_{LT}$  \\ 
  & $\Ocal^{c}_{2+}$ & $\Ocal^{c}_{2-}$ & $\Ocal^{e}_{1+}$  &  $\Ocal^{e}_{1-}$ & $\Ocal^{e}_{2+}$ & $\Ocal^{e}_{2-}$ & $\Ocal^{g}_{2+}$ & $\Ocal^{g}_{2-}$   \\  
  B.iii.10  &  $R^{x'z}_{T}$  &  $R^{z'x}_{T}$ & $\hspace*{1pt}^{s} R^{0z}_{LT}$  &  $\hspace*{1pt}^{c} R^{0x}_{LT'}$  &  $\hspace*{1pt}^{s} R^{x'x}_{LT'}$  &  $\hspace*{1pt}^{c} R^{z'x}_{LT}$  &  $\hspace*{1pt}^{c} R^{x'x}_{LT}$  &  $\hspace*{1pt}^{s} R^{z'x}_{LT'}$    \\ 
  & $\Ocal^{c}_{1+}$ & $\Ocal^{c}_{1-}$ & $\Ocal^{f}_{1+}$  &  $\Ocal^{f}_{1-}$ & $\Ocal^{h}_{1+}$ & $\Ocal^{h}_{1-}$ & $\Ocal^{h}_{2+}$ & $\Ocal^{h}_{2-}$  \\  
   B.iv.5  &  $R^{x'z}_{T}$  &  $R^{z'x}_{T}$ & $\hspace*{1pt}^{c} R^{00}_{LT}$  &  $\hspace*{1pt}^{c} R^{0y}_{LT}$  &  $\hspace*{1pt}^{s} R^{0z}_{LT}$  &  $\hspace*{1pt}^{c} R^{0x}_{LT'}$  &  $\hspace*{1pt}^{c} R^{0z}_{LT'}$  &  $\hspace*{1pt}^{s} R^{0x}_{LT}$    \\  
  & $\Ocal^{c}_{1+}$ & $\Ocal^{c}_{1-}$ & $\Ocal^{e}_{2+}$  &  $\Ocal^{e}_{2-}$ & $\Ocal^{f}_{1+}$ & $\Ocal^{f}_{1-}$ & $\Ocal^{f}_{2+}$ & $\Ocal^{f}_{2-}$    \\  
  B.v.10  &  $\hspace*{1pt}^{s} R_{TT}^{0z}$ & $R_{TT'}^{0x}$  &  $\hspace*{1pt}^{s} R^{00}_{LT'}$  &  $\hspace*{1pt}^{s} R^{0y}_{LT'}$  &  $\hspace*{1pt}^{c} R^{00}_{LT}$  &  $\hspace*{1pt}^{c} R^{0y}_{LT}$  &  $\hspace*{1pt}^{s} R^{z'0}_{LT}$  &  $\hspace*{1pt}^{c} R^{x'0}_{LT'}$   \\ 
  & $\Ocal^{a}_{1+}$ & $\Ocal^{a}_{1-}$ & $\Ocal^{e}_{1+}$  &  $\Ocal^{e}_{1-}$ & $\Ocal^{e}_{2+}$ & $\Ocal^{e}_{2-}$ & $\Ocal^{g}_{1+}$ & $\Ocal^{g}_{1-}$    \\  
   B.vi.3  &  $R_{TT'}^{0z}$  &  $\hspace*{1pt}^{s} R^{0x}_{TT}$  &  $\hspace*{1pt}^{c} R^{0z}_{LT'}$  &  $\hspace*{1pt}^{s} R^{0x}_{LT}$  &  $\hspace*{1pt}^{s} R^{x'x}_{LT'}$  &  $\hspace*{1pt}^{c} R^{z'x}_{LT}$  &  $\hspace*{1pt}^{c} R^{x'x}_{LT}$  &  $\hspace*{1pt}^{s} R^{z'x}_{LT'}$  \\  
  & $\Ocal^{a}_{2+}$ & $\Ocal^{a}_{2-}$ & $\Ocal^{f}_{2+}$  &  $\Ocal^{f}_{2-}$ & $\Ocal^{h}_{1+}$ & $\Ocal^{h}_{1-}$ & $\Ocal^{h}_{2+}$ & $\Ocal^{h}_{2-}$ \\ 
  B.vii.2   &  $R^{z'0}_{TT'}$  &  $\hspace*{1pt}^{s} R^{x'0}_{TT}$  &  $\hspace*{1pt}^{s} R^{z'0}_{LT}$  &  $\hspace*{1pt}^{c} R^{x'0}_{LT'}$  &  $\hspace*{1pt}^{c} R^{z'0}_{LT'}$  &  $\hspace*{1pt}^{s} R^{x'0}_{LT}$  &  $\hspace*{1pt}^{c} R^{x'x}_{LT}$  &  $\hspace*{1pt}^{s} R^{z'x}_{LT'}$    \\  
  & $\Ocal^{b}_{2+}$ & $\Ocal^{b}_{2-}$ & $\Ocal^{g}_{1+}$  &  $\Ocal^{g}_{1-}$ & $\Ocal^{g}_{2+}$ & $\Ocal^{g}_{2-}$ & $\Ocal^{h}_{2+}$ & $\Ocal^{h}_{2-}$     \\  
  B.viii.4   &  $\hspace*{1pt}^{s} R^{z'0}_{TT}$ & $R^{x'0}_{TT'}$  &  $\hspace*{1pt}^{c} R^{00}_{LT}$  &  $\hspace*{1pt}^{c} R^{0y}_{LT}$  &  $\hspace*{1pt}^{s} R^{0z}_{LT}$  &  $\hspace*{1pt}^{c} R^{0x}_{LT'}$  &  $\hspace*{1pt}^{c} R^{0z}_{LT'}$  &  $\hspace*{1pt}^{s} R^{0x}_{LT}$   \\
   & $\Ocal^{b}_{1+}$ & $\Ocal^{b}_{1-}$ & $\Ocal^{e}_{2+}$  &  $\Ocal^{e}_{2-}$ & $\Ocal^{f}_{1+}$ & $\Ocal^{f}_{1-}$ & $\Ocal^{f}_{2+}$ & $\Ocal^{f}_{2-}$ 
\end{tabular}
\end{scriptsize}
\caption{Here, we collect $8$ selected examples for the Moravcsik-complete sets composed of $14$ observables. The layout and notation used is very similar to Table~\ref{tab:MinimalMoravcsikSetsElectroExampleSets13}. The only exception is given by the fact that in the labelling-scheme, the letter 'B' denotes the complete sets of $14$ observables (cf. Tables~\ref{tab:MinimalMoravcsikSetsElectroII} to~\ref{tab:MinimalMoravcsikSetsElectroII2OneTwo} in appendix~\ref{sec:MoravComplSetsElectro}).}
\label{tab:MinimalMoravcsikSetsElectroExampleSets14}
\end{table}

As already mentioned above, we found $64$ Moravcsik-complete sets composed of $13$ observables in this work and $96$ sets composed of $14$ observables. Both the complete sets of $13$ and $14$ observables have in common that they each divide into $8$ different groups, sorted by the combination of different shape-classes that occur in the respective complete sets. The notation chosen for the set-numbers in the Tables~\ref{tab:MinimalMoravcsikSetsElectroExampleSets13},~\ref{tab:MinimalMoravcsikSetsElectroExampleSets14} and~\ref{tab:MinimalMoravcsikSetsElectroI} to~\ref{tab:MinimalMoravcsikSetsElectroII2OneTwo} takes this into account. For the complete sets of $13$ observables, each group contains $8$ different sets and for the complete sets of $14$, each group contains $12$ sets.

All the different combinations of shape-classes have in common, for both the Moravcsik-complete sets of $13$ and of $14$ observables, that at least one shape-class occurrs which contains recoil-polarisation observables. The non-diagonal shape-classes without recoil-polarization are $a$, $e$ and $f$ (cf. Table~\ref{tab:ElectroObservablesI}). The remaining classes of $b$, $c$, $d$, $g$ and $h$ all contain recoil-polarisation observables and indeed at least one of these latter classes occurrs in each Moravcsik-complete set listed in the Tables~\ref{tab:MinimalMoravcsikSetsElectroI} to~\ref{tab:MinimalMoravcsikSetsElectroII2OneTwo}. In this respect, our results for the Moravcsik-complete sets in electroproduction are very similar to the well-known results for absolutely minimal complete sets in photoproduction~\cite{Keaton:1995pw,Chiang:1996em,Nakayama:2018yzw}, in that the recoil-polarization observabes cannot be completely avoided for a unique amplitude extraction. 

It is interesting to further investigate the structure of the obtained Moravcsik-complete sets in electroproduction. The complete sets of $13$ observables each contain $4$ quantities from the purely transverse shape-classes $a$, $b$ and $c$ (cf. Table~\ref{tab:ElectroObservablesI}), one quantity of the purely longitudinal shape-class $d$ and two observables taken from one of the 'mixed' longitudinal-transverse shape-classes $e$, $\ldots$, $h$. Therefore, the amount of measurements with longitudinal polarization of the virtual photon, which is new to electroproduction as compared to photoproduction, is relatively small with just $3$ such observables in each case. However, this has the disadvantage of requiring more observables with recoil-polarization. The minimal amount of observables with recoil-polarization in a complete set of $13$ (apart from the diagonal observable $R_{T}^{y' 0}$ which always has to be measured) is $3$. The set (A.ii.4) shown in Table~\ref{tab:MinimalMoravcsikSetsElectroExampleSets13} as one particular example. There exist Moravcsik-complete sets of $13$ with up to $7$ recoil-polarization observables, such as the set (A.iii.8) in Table~\ref{tab:MinimalMoravcsikSetsElectroExampleSets13}.

For the complete sets of $14$ observables (cf. Table~\ref{tab:MinimalMoravcsikSetsElectroExampleSets14}) the situation is different. In this case, there exist some sets with just $2$ recoil-polarization observables, such as the set (B.v.10) shown in Table~\ref{tab:MinimalMoravcsikSetsElectroExampleSets14}. The maximal amount of recoil-polarization is again given in cases where the full set consists of such observables (e.g. the set (B.vii.2) in Table~\ref{tab:MinimalMoravcsikSetsElectroExampleSets14}). Furthermore, every Moravcsik-complete set with $14$ observables contains either $2$ or $4$ quantities from the purely transverse shape-classes ($a$, $b$ and $c$) and either $6$ or $4$ observables from the mixed longitudinal-transverse shape-classes ($e$, $\ldots$, $h$). Thus, the amount of measurements which use the longitudinal polarization has increased substantially compared to the complete sets of $13$ observables.

Finally, we remark that our results are consistent with the statements made in the work on complete sets in electroproduction published by Tiator and collaborators~\cite{Tiator:2017cde}, in the sense that the complete sets constructed in section~IV of their paper also cannot fully avoid recoil polarization. However, the main method of construction used in reference~\cite{Tiator:2017cde} is different from ours. There, the authors first select complete sets of $8$ fully transverse observables (i.e. complete photoproduction-sets) and then select in addition a full shape-class of $4$ longitudinal-transverse observables, in order to arrive at $12$ quantities. Tiator and collaborators also mention the approach of first starting with the $6$ diagonal observables $\left\{ R^{00}_{T},  R^{00}_{TT}, R_{T}^{0y}, R^{y' 0}_{T}, R_{L}^{00} , R_{L}^{0y}  \right\}$ and the choosing of non-diagonal observables in addition, which corresponds in principle to the approach used in our work. In reference~\cite{Tiator:2017cde}, the authors state that within their work, it has not been obvious how to choose the complete sets, using this latter approach. At this point, the application of the modified form of Moravcsik's theorem performed in this work can yield a significant contribution, although the number of observables contained in a complete set is at least $13$ and not $12$.

In summary, we have found Moravcsik-complete sets with a minimal number of $13$ ob\-serva\-bles for electroproduction. This already represents a considerable reduction from the full set of $36$ ob\-serva\-bles. Furthermore, these complete sets of $13$ have to be selected from $6$ different shape-classes, namely the $2$ diagonal classes plus $4$ non-diagonal ones (see Tables~\ref{tab:MinimalMoravcsikSetsElectroI} to~\ref{tab:MinimalMoravcsikSetsElectroI2}). However, we also found complete sets composed of $14$ ob\-serva\-bles, which have to be selected from only $5$ different shape-classes (Tables~\ref{tab:MinimalMoravcsikSetsElectroII} to~\ref{tab:MinimalMoravcsikSetsElectroII2OneTwo}). Thus, the reduction from the full set of $10$ shape-classes down to $5$ is even more substantial than in the case of photoproduction (sec.~\ref{sec:PhotoprodExample}).


\section{Conclusions and Outlook} \label{sec:ConclusionsAndOutlook}

\begin{table*}
 \begin{center}
\begin{tabular}{l|cr|lr|r|r}
Process & $N$ & $n_{\text{obs.}} = N^{2}$ & $n_{\text{obs.,min.}}^{\text{Moravcsik}}$ & $N^{2} - n_{\text{obs.,min.}}^{\text{Moravcsik}}$ & $n_{\text{total}}^{\text{shape-cl.}}$ & $n_{\text{min.}}^{\text{shape-cl.}}$ \\
\hline
$\pi N$-scattering & $2$ & $4$ & $4 = 2 N$ & $0$ & $2$ & $2$ \\
Math. example & $3$ & $9$ & $6 = 2 N$ & $3$ & $4$ & $4$ \\
Photoproduction & $4$ & $16$ & $10 = 2 N + 2$ & $6$ & $4$ & $3$ \\
Electroproduction & $6$ & $36$ & $13 = 2 N + 1$ & $23$ & $10$ & $5$ \\
\end{tabular}
\end{center}
\caption{The results obtained in this work are summarized for the amplitude-extraction problems with different numbers of amplitudes $N$. For each problem, one has $n_{\text{obs.}} = N^{2}$ ob\-serva\-bles overall. The number of ob\-serva\-bles $n_{\text{obs.,min.}}^{\text{Moravcsik}}$ contained in a minimal complete set according to the modified form of Moravcsik's theorem is shown and compared to the absolute minimal number $2 N$. The minimal number of different shape-classes, from which the ob\-serva\-bles have to be selected, is also shown and compared to the total number of shape-classes in the problem. The minimal number of classes is always composed of the shape-class(es) of the diagonal ob\-serva\-bles (which are always assumed to be measured), plus the minimal number of non-diagonal shape-classes implied by the modified version of Moravcsik's theorem.}
\label{tab:SummaryTableRisingN}
\end{table*}

This work treated the theorem by Moravcsik~\cite{Moravcsik:1984uf} on the unique extraction of amplitudes, as well as the implications drawn from it for applications to physical reactions. The theorem has been re-stated in a slightly modified form. For this form, the proof has been worked out in detail, trying to make it as formally complete and accessible as possible. 

While this theorem is valid for an arbitrary number of amplitudes $N$, we have applied it to a number of specific reactions with a rising number of amplitudes, i.e. to Pion-Nucleon scattering ($N = 2$), a purely mathematical example with $N = 3$ amplitudes, pseudoscalar meson photoproduction ($N = 4$) and electroproduction ($N = 6$). For the electroproduction reaction in particular, the above-mentioned theorem has been applied for the first time and it has yielded interesting insights into the structure of the corresponding complete sets. The results are summarized in Table~\ref{tab:SummaryTableRisingN} for all the considered reactions.

It is interesting to try to extract patterns from the results Moravcsik's theorem yields for examples with different $N$ and to compare them to the known treatments of complete sets of ob\-serva\-bles with an absolutely minimal number of $2 N$~\cite{Chiang:1996em,Tiator:2017cde,Nakayama:2018yzw}. We observe the following:



\begin{itemize}
 \item For $N = 2$ and $3$, the minimally complete sets according to the modified form of Moravcsik's theorem are equal in number and content to the absolutely minimal complete sets of $2N$. This is not the case for the higher $N \geq 4$.
 \item Compared to the full number of available ob\-serva\-bles $N^{2}$, the modified version of Moravcsik's theorem implies a reduction for the number of necessary ob\-serva\-bles to obtain a unique solution, for all cases but the simplest one, i.e. $N = 2$. For ascending numbers $N$, the degree of this reduction increases in the sense that the number of ob\-serva\-bles from the full set which are not needed for a minimally complete Moravcsik-set rises (cf. numbers given in the fifth column of Table~\ref{tab:SummaryTableRisingN}).
 \item Another feature found in other treatments of complete experiments~\cite{Chiang:1996em,Tiator:2017cde} re-emerges: the minimal number of different shape-classes, from which the ob\-serva\-bles in the complete sets have to be picked, does {\it not} correspond to the smallest possible one for $N > 2$. For instance, from earlier treatments on photoproduction~\cite{Chiang:1996em,Nakayama:2018yzw}, it is known that one has to combine ob\-serva\-bles from at least $3$ different shape-classes, i.e. the diagonal ob\-serva\-bles plus $2$ different non-diagonal shape-classes. In case one uses the diagonal ob\-serva\-bles, plus four additional quantities from one single non-diagonal shape-class, continuous ambiguities remain~\cite{Nakayama:2018yzw}. This behaviour is reproduced by the modified Moravcsik's theorem (see Table~\ref{tab:SummaryTableRisingN}).
 
 In the case of electroproduction, the modified Moravcsik's theorem implies that complete sets have to be selected from at least $5$ different shape-classes, i.e. $2$ diagonal classes plus $3$ non-diagonal ones. While one could also theoretically select the corresponding number of ob\-serva\-bles from just $4$ different shape-classes (i.e. $2$ diagonal plus $2$ non-diagonal ones), this possibility is ruled out.
 
 Furthermore, as a general feature, the difference of the total number of shape-classes to the minimal number of classes required for a complete set also rises with ascending numbers for $N$ (see Table~\ref{tab:SummaryTableRisingN}).
\end{itemize}

The applications of the modified form of Moravcsik's theorem have already yielded interesting results. Furthermore, the possibility to extract the complete sets in a fully automated procedure is attractive. However, the mismatch between the size of the complete sets according to Moravcsik and the absolute minimum number of $2 N$ ob\-serva\-bles, which is present for $N \geq 4$, remains a problem. How to obtain all the minimal sets of $2 N$ ob\-serva\-bles using (maybe) a modified version of the approach presented here seems not at all obvious. Therefore, this mismatch deserves further investigation and it is highly probable that it can be traced back to general mathematical properties of the matrix-algebras $\left\{ \tilde{\Gamma}^{\alpha} \right\}$ used in the formulation of the ob\-serva\-bles as bilinear forms. Still, since a master-approach on how to obtain minimal complete sets of $2 N$ ob\-serva\-bles for problems with arbitrary $N$ is the ultimate goal, this avenue of exploration is important.

Another possible direction of future research lies in the application of the modified approach according to Moravcsik as presented here to more complicated reactions, which are of current practical interest. In this vein, a detailed treatment of the photoproduction of two pseudoscalar mesons is prepared at the moment~\cite{PhilippEtAl:2MesonPaper}.



\appendix

\section{Proof of Theorem 2}   \label{sec:DetailedProof}


Choose $b_{1}$ as an anchor amplitude for the whole procedure and demand it to be real: $b_{1} \equiv \left| b_{1}  \right| >0$. In this way, the unknown overall phase $\phi^{b} (W, \theta)$ is fixed. Note that the choice of $b_{1}$ as anchor amplitude is just a convention.

Now, connect amplitude-points in order to form an un-branched open chain of bilinear products:
\begin{equation}
 b_{1} b_{i}^{\ast}, b_{i} b_{j}^{\ast}  , \ldots, b_{p} b_{q}^{\ast}, \label{eq:OpenChainBicoms}
\end{equation}
where the indices
\begin{equation}
 1,i,j,\ldots,p,q \in \left\{ 1, \ldots, N \right\} , \label{eq:IndicesOpenChain}
\end{equation}
are demanded to be \textit{all different}. Thus, the open chain has exactly $N - 1$ links.

Furthermore, as stated above, we can assume the moduli~\eqref{eq:Moduli} to be already determined. Therefore, one can consider the following set of relative-phases corresponding to the open chain:
\begin{equation}
 \phi_{1i}, \phi_{ij}, \ldots , \phi_{pq} . \label{eq:RelPhasesOpenChain}
\end{equation}
First, let us assume that all links in the chain are provided by \textit{solid} lines, i.e. that only the real parts
\begin{equation}
 \mathrm{Re} \left[ b_{1} b_{i}^{\ast} \right], \mathrm{Re} \left[ b_{i} b_{j}^{\ast} \right] , \ldots, \mathrm{Re} \left[ b_{p} b_{q}^{\ast} \right], \label{eq:OpenChainBicomsRealParts}
\end{equation}
are considered. Then, we know from the discussion in the beginning that all these real parts leave a cosine-type ambiguity (eq.~\eqref{eq:CosTypeAmbiguity}) for their corresponding relative phase, i.e. one has the collection of $2^{N-1}$ discrete phase ambiguities
\begin{equation}
 \phi_{1i}^{\pm} = \begin{cases} + \alpha_{1i} \\ - \alpha_{1i} \end{cases} , \phi_{ij}^{\pm} = \begin{cases} + \alpha_{ij} \\ - \alpha_{ij} \end{cases} , \ldots , \phi_{pq}^{\pm} = \begin{cases} + \alpha_{pq} \\ - \alpha_{pq} \end{cases} . \label{eq:2toNdiscreteambiguitiesRealParts}
\end{equation}

As a means to remove (a large part of) these discrete ambiguities, we connect the amplitudes (or, points) $q$ and $1$ by a solid line, in order to close the chain and thus form a closed loop with $N$ solid link-lines. The relative-phase $\phi_{q1}$ therefore also has the cosine-type ambiguity~\eqref{eq:CosTypeAmbiguity}.

Then, we know that whatever the correct solution is, it has to satisfy the following \textit{consistency relation} for the relative phases, which has to be valid for any arrangement of $N$ amplitudes in the complex plane\footnote{This consistency relation is also quickly verified by plugging in the definition $\phi_{ij} = \phi_{i} - \phi_{j}$ of the relative phases.}
\begin{equation}
 \phi_{1i} + \phi_{ij} + \ldots + \phi_{pq} + \phi_{q1} = 0  . \label{eq:GenericConsistencyRel}
\end{equation}
When written down for all the possible discrete ambiguities~\eqref{eq:2toNdiscreteambiguitiesRealParts}, the consistency relation reads
\begin{equation}
 \phi_{1i}^{\pm} + \phi_{ij}^{\pm} + \ldots + \phi_{pq}^{\pm} + \phi_{q1}^{\pm} = 0  , \label{eq:ConsistencyRelPmNotation}
\end{equation}
or, when written in an alternative notation to collect the different cases (the one which we use in the remainder of the proof, and which Moravcsik also uses in his paper~\cite{Moravcsik:1984uf}), it becomes
\begin{equation}
  \left\{ \begin{array}{c} + \alpha_{1i} \\ - \alpha_{1i} \end{array} \right\} + \left\{ \begin{array}{c} + \alpha_{ij} \\ - \alpha_{ij} \end{array} \right\} + \ldots + \left\{ \begin{array}{c} + \alpha_{pq} \\ - \alpha_{pq} \end{array} \right\} + \left\{ \begin{array}{c} + \alpha_{q1} \\ - \alpha_{q1} \end{array} \right\} = 0  . \label{eq:ConsistencyRelCasesNotation}
\end{equation}
The basic logic is now the following: one of the above-given $2^{N}$ discrete ambiguities is the correct, i.e. 'true', solution to the problem. For it to be truly unique, its corresponding consistency constraint, which is one of the cases~\eqref{eq:ConsistencyRelCasesNotation}, has to be \textit{linearly independent}~\cite{Nakayama:2018yzw} from the consistency constraints of all the other cases. In case linearly dependent constraints remain, the solution is \textit{not} unique.

To illustrate this point, suppose that the true solution to our problem would be given by the sign-combination
\begin{equation}
 \alpha_{1i} - \alpha_{ij} + \ldots + \alpha_{pq} - \alpha_{q1} = 0  . \label{eq:ReExampleCorrectSol}
\end{equation}
When we multiply both sides of this equation by $(-1)$, it becomes
\begin{equation}
 - \alpha_{1i} + \alpha_{ij} - \ldots - \alpha_{pq} + \alpha_{q1} = 0  , \label{eq:ReExampleDegenerateSol}
\end{equation}
and we see that this constraint is also one of the possible cases~\eqref{eq:ConsistencyRelCasesNotation}. Therefore, the consistency relations~\eqref{eq:ReExampleCorrectSol} and~\eqref{eq:ReExampleDegenerateSol} are linearly dependent and thus a $2$-fold discrete ambiguity remains. We call the corresponding solutions \textit{degenerate} (cf. reference~\cite{Nakayama:2018yzw}).

One can convince oneself that the same statement remains to be true, no matter which sign-combination other than~\eqref{eq:ReExampleCorrectSol} we would have assumed to be the true solution. Therefore, for the case with only solid lines, or real parts of bilinear products, in the loop, the general consistency relation~\eqref{eq:GenericConsistencyRel} can reduce the $2^{N}$-fold discrete ambiguity~\eqref{eq:2toNdiscreteambiguitiesRealParts} to a $2$-fold discrete phase ambiguity, and the combination with only real parts is therefore not fully complete.

We note that in general situations, one can find linearly dependent pairs of consistency relations via the following two manipulations performed on the whole equation

\begin{itemize}
 \item[$\ast)$] Multiplication by $(-1)$,
 \item[$\ast)$] Addition or subtraction of multiples of $2 \pi$ on both sides of the equation. (We can do this manipulation since relations among phases are always valid modulo addition of multiples of $2 \pi$.)
\end{itemize}
Now, assume that the link between amplitudes $q$ and $1$ were changed from a solid to a dashed line, i.e. we would exchange only the real part $\mathrm{Re} \left[ b_{q} b_{1}^{\ast} \right]$ for the imaginary part $\mathrm{Im} \left[ b_{q} b_{1}^{\ast} \right]$. Then, the discrete ambiguity for the relative phase $\phi_{q1}$ changes to a sine-type ambiguity~\eqref{eq:SinTypeAmbiguity} and therefore the relevant cases for the consistency relation become
\begin{align}
  &\left\{ \begin{array}{c} + \alpha_{1i} \\ - \alpha_{1i} \end{array} \right\} + \left\{ \begin{array}{c} + \alpha_{ij} \\ - \alpha_{ij} \end{array} \right\} + \ldots + \left\{ \begin{array}{c} + \alpha_{pq} \\ - \alpha_{pq} \end{array} \right\} + \left\{ \begin{array}{c} + \alpha_{q1} \\ \pi - \alpha_{q1} \end{array} \right\} \nonumber \\
  &= 0  . \label{eq:ConsistRelCasesOneDashed}
\end{align}
Again, assume some (in principle arbitrary) combination to be the true solution of the problem, for instance
\begin{equation}
 - \alpha_{1i} + \alpha_{ij} + \ldots - \alpha_{pq} + \alpha_{q1} = 0  . \label{eq:OneDashedCorrectSol}
\end{equation}
Multiply the whole equation by $(-1)$ in order to get
\begin{equation}
 + \alpha_{1i} - \alpha_{ij} - \ldots + \alpha_{pq} - \alpha_{q1} = 0  . \label{eq:OneDashedCandidateDegSol}
\end{equation}
Now, no further transformation can lead from this equation to any of the cases contained in equation~\eqref{eq:ConsistRelCasesOneDashed} for which $\alpha_{q1}$ has a minus-sign, since for any of these cases, there has to be one additional single summand of $\pi$ on the left-hand-side. This means, the candidate for a possibly degenerate constraint would then read
\begin{equation}
 + \alpha_{1i} - \alpha_{ij} - \ldots + \alpha_{pq} + \pi - \alpha_{q1} = 0  , \label{eq:OneDashedSummandPiSol}
\end{equation}
but it simply cannot be obtained from equation~\eqref{eq:OneDashedCandidateDegSol} by use of the allowed transformations. The same is true in case any other starting-combination other than~\eqref{eq:OneDashedCorrectSol} were assumed to be the true solution. This means that \textit{no} degenerate solutions exist any more and the summand of $\pi$ in equation~\eqref{eq:ConsistRelCasesOneDashed} has fully lifted the degeneracy. This means that the considered closed loop is fully complete!

The same argument as above holds in case any link other than the connection between $q$ and $1$ (i.e. $1 \leftrightarrow i$, $i \leftrightarrow j$, $\ldots$) were assumed to be the single dashed line present in a closed loop of otherwise only solid lines. Therefore, we see that in the case of only a single dashed line, the closed loop is always fully complete.

Therefore, we have learned that the sine-type ambiguities~\eqref{eq:SinTypeAmbiguity} are very important to lift degeneracies, due to the appearance of additional single 'summands of $\pi$'.

As a next step, consider the case of multiple dashed lines (i.e., imaginary parts) present in the closed loop. For instance, assume the last $n_{\text{d}}$ links in the loop to be dashed, i.e. we look at the following possible cases for the consistency relation:
\begin{widetext}
\begin{equation}
  \left\{ \begin{array}{c} + \alpha_{1i} \\ - \alpha_{1i} \end{array} \right\} + \ldots + \left\{ \begin{array}{c} + \alpha_{\ell r} \\ - \alpha_{\ell r} \end{array} \right\} + \underbrace{ \left\{ \begin{array}{c} + \alpha_{rs} \\ \pi - \alpha_{rs} \end{array} \right\} + \left\{ \begin{array}{c} + \alpha_{st} \\ \pi - \alpha_{st} \end{array} \right\} + \ldots + \left\{ \begin{array}{c} + \alpha_{q1} \\ \pi - \alpha_{q1} \end{array} \right\}}_{n_{\text{d}} \text{ terms}} = 0  . \label{eq:ConsistRelCasesMultipleDashed}
\end{equation}
\end{widetext}
In order to see in which general cases the degeneracies are resolved, we need some additional Lemmas\footnote{To avoid an unnecessarily cluttered notation in the proofs, we change the notation for the formulation of the Lemmas.}. The first Lemma is concerned with the fact that, due to relations among phases always being valid up to addition of multiples of $2 \pi$, at most a single summand of $\pi$ can remain in all the cases collected in equation~\eqref{eq:ConsistRelCasesMultipleDashed}.

\vspace*{10pt}

\begin{itemize}
\item[$\bullet$] \textbf{\underline{Lemma 1 (Sum of phase-variables under}} \\ \hspace*{51pt} \textbf{\underline{sine-type ambiguities):}} \\

Consider the sum of $n$ phase-variables $\varphi_{i}$ ($n \geq 2$), i.e. real variables taking values on the interval $\left[- \pi / 2, \pi / 2 \right]$ and which are arguments of $\sin$-functions:
\begin{equation}
 \varphi_{1} + \ldots + \varphi_{n} . \label{eq:SumOfPhaseVariables}
\end{equation}
Each of the phase-variables is subject to a $2$-fold discrete ambiguity of the 'sine-type', i.e.
\begin{equation}
 \varphi_{i}  \longrightarrow \begin{cases} \varphi_{i} ,  \\ \pi - \varphi_{i}  , \end{cases} \text{for } i = 1,\ldots,n. \label{eq:SinTypeAmbiguityVarPhi}
\end{equation}
Then, there exist $2^{n}$ cases for the form of the sum~\eqref{eq:SumOfPhaseVariables}, and these cases can be summarized as follows:
\begin{align}
 &\left\{ \begin{array}{c}  \varphi_{1} \\ \pi - \varphi_{1}  \end{array}  \right\} + \left\{ \begin{array}{c}  \varphi_{2} \\ \pi - \varphi_{2}  \end{array}  \right\} + \ldots + \left\{ \begin{array}{c}  \varphi_{n} \\ \pi - \varphi_{n}  \end{array}  \right\} \nonumber \\
 &= \left\{ \begin{array}{l}  \sum_{k = 1}^{n} \Ccal^{(j'')}_{k} \varphi_{k} \\ \pi + \sum_{k = 1}^{n} \Ccal^{(j')}_{k} \varphi_{k}  \end{array} \right\}  . \label{eq:IdentityPhaseVariablesPhiAmbiguity}
\end{align}
On the right-hand-side of this equation, all the $\Ccal$-coefficients take values of either $(+1)$ or $(-1)$. The sums over double-primed coefficients $\Ccal^{(j'')}_{k}$ denote all possible cases of linear combinations of phases with an \textit{even} number of $(-1)$-signs, while the sums over single-primed coefficients $\Ccal^{(j')}_{k}$ denote all possible cases with an \textit{odd} number of $(-1)$-signs. Moravcsik has stated this result in his paper~\cite{Moravcsik:1984uf}. A proof of the statement can be found towards the end of this appendix.
\end{itemize}

The second Lemma is concerned with the transformation-behaviour of the $\Ccal$-coefficients under the multiplication by $(-1)$, since this is the most important transformation in the search for degenerate consistency relations.


\vspace*{5pt}

\begin{itemize}
\item[$\bullet$] \textbf{\underline{Lemma 2 (Multiplication by $(-1)$):}} \\

Assume that Lemma 1 and equation~\eqref{eq:IdentityPhaseVariablesPhiAmbiguity} are valid for a set of $n$ phase-variables $\varphi_{i}$, which are all subject to the 'sine-type' ambiguity~\eqref{eq:SinTypeAmbiguityVarPhi}.

Suppose that $n$ were an \textit{even} number. Then the following relations among $\Ccal$-coefficients are valid:
\begin{align}
 (-1) \times \Ccal^{(j')}_{k} &= \Ccal^{(\tilde{j}')}_{k} , \label{eq:RelationAmongCoeffEven} \\
 (-1) \times \Ccal^{(j'')}_{k} &= \Ccal^{(\tilde{j}'')}_{k} , \label{eq:RelationAmongCoeffEvenII}
\end{align}
for suitable indices $\tilde{j}'$, $\tilde{j}''$. This means that for an even number of variables, the single-primed coefficients transform into each other under multiplication by $(-1)$ and the double-primed coefficients transform into each other under the same transformation. No mixing of single- and double-primed coefficients occurs!

If however, the number $n$ is \textit{odd}, one has
\begin{equation}
 (-1) \times \Ccal^{(j')}_{k} = \Ccal^{(\tilde{j}'')}_{k} , \label{eq:RelationAmongCoeffOdd}
\end{equation}
for suitable indices $\tilde{j}''$. Thus, for an odd number of variables, single-primed and double-primed coefficients are transformed into each other under multiplication by $(-1)$. A proof of this Lemma is given at the end of this appendix.
\end{itemize}
We return now to the cases for the consistency constraints collected in equation~\eqref{eq:ConsistRelCasesMultipleDashed} and re-write this equation using Lemma 1:
\begin{widetext}
\begin{align}
  &\left\{ \begin{array}{c} + \alpha_{1i} \\ - \alpha_{1i} \end{array} \right\} + \ldots + \left\{ \begin{array}{c} + \alpha_{\ell r} \\ - \alpha_{\ell r} \end{array} \right\} + \left\{ \begin{array}{c} + \alpha_{rs} \\ \pi - \alpha_{rs} \end{array} \right\} + \ldots + \left\{ \begin{array}{c} + \alpha_{q1} \\ \pi - \alpha_{q1} \end{array} \right\} \nonumber \\
  &= \left\{ \begin{array}{c} + \alpha_{1i} \\ - \alpha_{1i} \end{array} \right\} + \ldots + \left\{ \begin{array}{c} + \alpha_{\ell r} \\ - \alpha_{\ell r} \end{array} \right\} + \left\{ \begin{array}{l}  \sum_{\mu, \nu} \Ccal^{(j'')}_{\mu \nu} \alpha_{\mu \nu} \\ \pi + \sum_{\mu, \nu} \Ccal^{(j')}_{\mu \nu} \alpha_{\mu \nu}  \end{array} \right\}  = 0  .
  \label{eq:ConsistRelLemmaAppliedMultipleDashed}
\end{align}
\end{widetext}

\vspace*{0.5pt}

The notation with a single summation-index '$k$', which has been used in the Lemmas above, has been generalized to multiple summation-indices $\mu, \nu$, but this is straightforward.

Now, assume that $n_{\text{d}}$ is \textit{even}. Suppose further that the true solution corresponds to a combination of signs without a summand of $\pi$, i.e.
\begin{equation}
 \alpha_{1i} - \ldots + \alpha_{\ell r} + \sum_{\mu, \nu}   \Ccal^{(j'')}_{\mu \nu} \alpha_{\mu \nu} = 0 . \label{eq:NevenAssumedTrueSolution}
\end{equation}
Multiplying this equation by $(-1)$, we get
\begin{align}
 & - \alpha_{1i} + \ldots - \alpha_{\ell r} + \sum_{\mu, \nu}  \left[ - \Ccal^{(j'')}_{\mu \nu} \right] \alpha_{\mu \nu} \nonumber \\
 &= - \alpha_{1i} + \ldots - \alpha_{\ell r} + \sum_{\mu, \nu}   \Ccal^{(\tilde{j}'')}_{\mu \nu} \alpha_{\mu \nu} = 0 , \label{eq:NevenAssumedDegenerateSolution}
\end{align}
where Lemma 2 has been used in order to obtain the final form of the result. Since according to Lemma 2, the coefficients $\Ccal^{(j'')}_{\mu \nu}$ remain double-primed under the multiplication by $(-1)$, it is clear that a degenerate solution can always be found. The same is true if any combination other than equation~\eqref{eq:NevenAssumedTrueSolution}, which could also include a summand of $\pi$, would have been assumed to be the true solution. Therefore, for $n_{\text{d}}$ even, there always remains a $2$-fold discrete phase-ambiguity.

Assume now that $n_{\text{d}}$ is \textit{odd}. Furthermore, suppose that some combination with a summand of $\pi$ were the true solution, i.e.
\begin{equation}
 - \alpha_{1i} + \ldots + \alpha_{\ell r} + \pi + \sum_{\mu, \nu}   \Ccal^{(j')}_{\mu \nu} \alpha_{\mu \nu} = 0 . \label{eq:NoddAssumedTrueSolution}
\end{equation}
We multiply this equation by $(-1)$ and use Lemma 2 in order to obtain 
\begin{equation}
  + \alpha_{1i} - \ldots - \alpha_{\ell r} - \pi + \sum_{\mu, \nu}   \Ccal^{(\tilde{j}'')}_{\mu \nu} \alpha_{\mu \nu} = 0 . \label{eq:NoddAssumedDegenerateSolution}
\end{equation}
We add $2 \pi$ on the left-hand-side of this equation and get:
\begin{equation}
  + \alpha_{1i} - \ldots - \alpha_{\ell r} + \pi + \sum_{\mu, \nu}   \Ccal^{(\tilde{j}'')}_{\mu \nu} \alpha_{\mu \nu} = 0 . \label{eq:NoddAssumedDegenerateSolutionII}
\end{equation}
Since, according to Lemma 2, the coefficients $\Ccal^{(j')}_{\mu \nu}$ have turned into double-primed coefficients $\Ccal^{(\tilde{j}'')}_{\mu \nu}$, the sign-combination in equation~\eqref{eq:NoddAssumedDegenerateSolutionII} corresponds to one of the cases \textit{without} $\pi$ in equation~\eqref{eq:ConsistRelLemmaAppliedMultipleDashed}. However, the summand of $\pi$ in equation~\eqref{eq:NoddAssumedDegenerateSolutionII} cannot be removed by any of the allowed transformations. Therefore, no degeneracy of solutions exists! The same is true is we assume any combination other than the one in equation~\eqref{eq:NoddAssumedTrueSolution} (also one without summand of $\pi$) as the true solution. Therefore, for $n_{\text{d}}$ odd, the closed loop always represents a fully complete set.

Note that all statements made above remain true in case one does not assume the last $n_{d}$ lines in the chain to be dashed, as was done in equation~\eqref{eq:ConsistRelCasesMultipleDashed}, but indeed any other combination of $n_{\text{d}}$ link-lines in the closed loop.

The special case of a closed loop which includes only (i.e. exclusively) dashed lines remains to be discussed. This is also the case which Moravcsik~\cite{Moravcsik:1984uf} is (seemingly) not treating correctly.

We again collect all the consistency relations for all possible cases of discrete ambiguities and employ Lemma 1 in order to obtain: 
\begin{align}
 &\left\{ \begin{array}{c} + \alpha_{1i} \\ \pi - \alpha_{1i} \end{array} \right\} + \ldots + \left\{ \begin{array}{c} + \alpha_{p q} \\ \pi - \alpha_{p q} \end{array} \right\} + \left\{ \begin{array}{c} + \alpha_{q1} \\ \pi - \alpha_{q1} \end{array} \right\}  \nonumber \\
   &= \left\{ \begin{array}{l}  \sum_{\mu, \nu} \Ccal^{(j'')}_{\mu \nu} \alpha_{\mu \nu} \\ \pi + \sum_{\mu, \nu} \Ccal^{(j')}_{\mu \nu} \alpha_{\mu \nu}  \end{array} \right\}  = 0   .  \label{eq:ConsistRelAllLinksInLoopBroken}
\end{align}
Here, the same arguments as above, under repeated use of Lemma 2, lead to the result that also in this case the number of dashed lines has to be \textit{odd} in order to lift all degeneracies and thus to obtain a fully complete set.

What remains to be discussed are the special, singular numerical configurations for which the theorem proven here loses its validity. Consideration of the arguments given above shows that, also in case $n_{\text{d}}$ is odd, the sine-type ambiguities would lose their power to lift the degeneracies in case only one linear combination of $\alpha$'s corresponding to all the $n_{\text{d}}$ dashed lines of the true solution would \textit{vanish}, for instance (cf. equations~\eqref{eq:ConsistRelCasesMultipleDashed} and~\eqref{eq:ConsistRelLemmaAppliedMultipleDashed})
\begin{equation}
 \sum_{\mu, \nu} \Ccal_{\mu \nu}^{(j'')} \alpha_{\mu \nu} = 0 , \label{eq:SingularCaseAlphasI}
\end{equation}
for some particular index $j''$.

This could be due to one of the following cases:
\begin{itemize}
 \item[$\ast)$] All $\alpha$'s vanish individually, i.e.: $\alpha_{\mu \nu} = 0$ for all the terms included in the sum~\eqref{eq:SingularCaseAlphasI}.
 \item[$\ast)$] None of the $\alpha$'s vanish individually, but there exists a complicated singular sub-manifold in the space of relative-phases, which is defined by the validity of equation~\eqref{eq:SingularCaseAlphasI} and upon which the considered closed loop is not a complete set any more.
 \item[$\ast)$] A mixture of the two cases before, i.e.: some $\alpha$'s vanish individually $\alpha_{\tilde{\mu} \tilde{\nu}} = 0$, for a certain collection of indices $\tilde{\mu}$, $\tilde{\nu}$. Furthermore, singular sub-manifolds exist in the parameter space of the remaining relative phases (i.e. those with $\alpha_{\mu \nu} \neq 0$), defined by the validity of the constraint
\begin{equation}
 \sum_{\mu \neq \tilde{\mu}, \nu \neq \tilde{\nu}} \Ccal_{\mu \nu}^{(j'')} \alpha_{\mu \nu} = 0 . \label{eq:SingularCaseAlphasIPrime}
\end{equation}
\end{itemize}

We suspect that the probability of such singular parameter-configurations to occur grows with a larger number $N$ of transversity amplitudes. However, these cases probably only become relevant as soon as one introduces measurement errors into the problem. Even for simulations with pseudo-data of finite numerical precision, such cases can become relevant for higher $N$. However, in the academic case of an \textit{exactly} solvable complete experiment, such singular cases can probably be ignored.

\textcolor{white}{blank} \hfill \textbf{QED} \\

This concludes our derivation of the core result of this work. For the sake of completeness, in the following we provide proofs for the more technical Lemmas 1 and 2 introduced above: \\

{\bf \underline{Proof of Lemma 1:}}

We construct a proof via complete induction and start with the lowest non-trivial case of $n=2$ phase-variables.

\begin{itemize}
 \item[$\ast)$] \underline{$n = 2$}:
 
 Consider the possible cases
 \begin{equation}
   \left\{ \begin{array}{c}  \varphi_{1} \\ \pi - \varphi_{1}  \end{array}  \right\} + \left\{ \begin{array}{c}  \varphi_{2} \\ \pi - \varphi_{2}  \end{array}  \right\} . \label{eq:NEqualsTwoPossibleCases}
 \end{equation}
 Remembering that relations among phases are always valid up to addition of multiples of $2 \pi$, we see that the total set of possible cases becomes
 \begin{align}
   &\left\{ \begin{array}{c}  \varphi_{1} \\ \pi - \varphi_{1}  \end{array}  \right\} + \left\{ \begin{array}{c}  \varphi_{2} \\ \pi - \varphi_{2}  \end{array}  \right\} \nonumber \\
   &= \begin{cases} + \varphi_{1} + \varphi_{2}, \\ \pi - \varphi_{1} + \varphi_{2}, \\ \varphi_{1} + \pi - \varphi_{2} , \\ \pi - \varphi_{1} + \pi - \varphi_{2} . \end{cases} = \begin{cases} + \varphi_{1} + \varphi_{2}, \\ \pi - \varphi_{1} + \varphi_{2}, \\ \pi + \varphi_{1} - \varphi_{2} , \\ - \varphi_{1} - \varphi_{2} . \end{cases} . \label{eq:NEqualsTwoPossibleCasesEval}
 \end{align}
 \item[$\ast)$] \underline{$n \longrightarrow n + 1$}:
 
 Assume now that equation~\eqref{eq:IdentityPhaseVariablesPhiAmbiguity} is valid for a set of $n$ phase-variables $\varphi_{i}$. Add a further variable $\varphi_{n+1}$, which is also subject to the 'sine-type' ambiguity~\eqref{eq:SinTypeAmbiguityVarPhi}. Then, we have:
 \begin{align}
  &\left\{ \begin{array}{c}  \varphi_{1} \\ \pi - \varphi_{1}  \end{array}  \right\} + \ldots + \left\{ \begin{array}{c}  \varphi_{n} \\ \pi - \varphi_{n}  \end{array}  \right\} + \left\{ \begin{array}{c}  \varphi_{n+1} \\ \pi - \varphi_{n+1}  \end{array}  \right\} \nonumber \\
  &= \left\{ \begin{array}{l}  \sum_{k = 1}^{n} \Ccal^{(j'')}_{k} \varphi_{k} \\ \pi + \sum_{k = 1}^{n} \Ccal^{(j')}_{k} \varphi_{k}  \end{array} \right\} + \left\{ \begin{array}{c}  \varphi_{n+1} \\ \pi - \varphi_{n+1}  \end{array}  \right\} \nonumber \\
  &= \left\{ \begin{array}{l}  \sum_{k = 1}^{n} \Ccal^{(j'')}_{k} \varphi_{k} + \varphi_{n+1} \\ \sum_{k = 1}^{n} \Ccal^{(j'')}_{k} \varphi_{k} + \pi -  \varphi_{n+1} \\ \pi + \sum_{k = 1}^{n} \Ccal^{(j')}_{k} \varphi_{k} + \varphi_{n+1}  \\ \pi + \sum_{k = 1}^{n} \Ccal^{(j')}_{k} \varphi_{k} + \pi - \varphi_{n+1}   \end{array} \right\} \nonumber \\
  &= \begin{cases} \sum_{k = 1}^{n} \Ccal^{(j'')}_{k} \varphi_{k} + \varphi_{n+1} , \\ \pi + \sum_{k = 1}^{n} \Ccal^{(j'')}_{k} \varphi_{k} -  \varphi_{n+1} , \\ \pi + \sum_{k = 1}^{n} \Ccal^{(j')}_{k} \varphi_{k} + \varphi_{n+1} , \\ \sum_{k = 1}^{n} \Ccal^{(j')}_{k} \varphi_{k} - \varphi_{n+1}. \end{cases}  \label{eq:PhaseRelationInductiveProof}
 \end{align}
 One can see that the number of $(-1)$-signs in the first and fourth combination in the end result of~\eqref{eq:PhaseRelationInductiveProof} is \textit{even} and that the number of $(-1)$-signs in the second and third combination is \textit{odd}. \\
 \textcolor{white}{Hallo!} \hfill \textbf{QED}
\end{itemize}

{\bf \underline{Proof of Lemma 2:}}

First, we treat the even numbers $n$. We prove again via induction:
\begin{itemize}
 \item[$\ast)$] \underline{$n=2$:}
 
 We have (cf. the above)
 \begin{align}
   & \left\{ \begin{array}{c}  \varphi_{1} \\ \pi - \varphi_{1}  \end{array}  \right\} + \left\{ \begin{array}{c}  \varphi_{2} \\ \pi - \varphi_{2}  \end{array}  \right\} = \begin{cases} + \varphi_{1} + \varphi_{2}, \\ \pi - \varphi_{1} + \varphi_{2}, \\ \varphi_{1} + \pi - \varphi_{2} , \\ - \varphi_{1} - \varphi_{2} . \end{cases} \nonumber \\
   & =: \begin{cases} \Ccal^{(1'')}_{1} \varphi_{1} + \Ccal^{(1'')}_{2} \varphi_{2}, \\ \pi + \Ccal^{(1')}_{1} \varphi_{1} + \Ccal^{(1')}_{2} \varphi_{2}, \\ \pi + \Ccal^{(2')}_{1} \varphi_{1} + \Ccal^{(2')}_{2} \varphi_{2} , \\ \Ccal^{(2'')}_{1} \varphi_{1} + \Ccal^{(2'')}_{2} \varphi_{2} . \end{cases} \label{eq:NEqualsTwoPossibleCasesEvalII}
 \end{align}
 We can read off the following relations
 \begin{align}
  (-1) \times \Ccal^{(1')}_{1} &= \Ccal^{(2')}_{1} , \label{eq:ReadOffRelI} \\
  (-1) \times \Ccal^{(1')}_{2} &= \Ccal^{(2')}_{2} , \label{eq:ReadOffRelII} \\
  (-1) \times \Ccal^{(1'')}_{1} &= \Ccal^{(2'')}_{1} , \label{eq:ReadOffRelIII} \\
  (-1) \times \Ccal^{(1'')}_{2} &= \Ccal^{(2'')}_{2} , \label{eq:ReadOffRelIV}
 \end{align}
 and see a special case of Lemma 2 fulfilled.
 \item[$\ast)$] \underline{$n \longrightarrow n + 2$:}
 
 We use Lemma 1 in order to decompose as follows:
 \begin{widetext}
 %
\begin{align}
 & \hspace*{12.5pt} \left\{ \begin{array}{c}  \varphi_{1} \\ \pi - \varphi_{1}  \end{array}  \right\} +  \ldots + \left\{ \begin{array}{c}  \varphi_{n} \\ \pi - \varphi_{n}  \end{array}  \right\} + \left\{ \begin{array}{c}  \varphi_{n+1} \\ \pi - \varphi_{n+1}  \end{array}  \right\} + \left\{ \begin{array}{c}  \varphi_{n+2} \\ \pi - \varphi_{n+2}  \end{array}  \right\} \nonumber \\
 &= \left\{ \begin{array}{l}  \sum_{k = 1}^{n} \Ccal^{(j'')}_{k} \varphi_{k} \\ \pi + \sum_{k = 1}^{n} \Ccal^{(j')}_{k} \varphi_{k}  \end{array} \right\} + \left\{ \begin{array}{c}  \varphi_{n+1} \\ \pi - \varphi_{n+1}  \end{array}  \right\} + \left\{ \begin{array}{c}  \varphi_{n+2} \\ \pi - \varphi_{n+2}  \end{array}  \right\} \nonumber \\
 &= \left\{ \begin{array}{l}  \sum_{k = 1}^{n} \Ccal^{(j'')}_{k} \varphi_{k} + \varphi_{n+1} + \varphi_{n+2} \\ \sum_{k = 1}^{n} \Ccal^{(j'')}_{k} \varphi_{k} + \pi - \varphi_{n+1} + \varphi_{n+2} \\ \sum_{k = 1}^{n} \Ccal^{(j'')}_{k} \varphi_{k} + \varphi_{n+1} + \pi - \varphi_{n+2} \\ \sum_{k = 1}^{n} \Ccal^{(j'')}_{k} \varphi_{k} - \varphi_{n+1} - \varphi_{n+2} \\ \pi + \sum_{k = 1}^{n} \Ccal^{(j')}_{k} \varphi_{k} + \varphi_{n+1} + \varphi_{n+2} \\ \sum_{k = 1}^{n} \Ccal^{(j')}_{k} \varphi_{k} - \varphi_{n+1} + \varphi_{n+2} \\  \sum_{k = 1}^{n} \Ccal^{(j')}_{k} \varphi_{k} + \varphi_{n+1} - \varphi_{n+2} \\ \pi + \sum_{k = 1}^{n} \Ccal^{(j')}_{k} \varphi_{k} - \varphi_{n+1} - \varphi_{n+2}  \end{array} \right\} . \label{eq:EvenCaseGeneralN}
\end{align}
%
\end{widetext}
We see that the relations~\eqref{eq:RelationAmongCoeffEven} and~\eqref{eq:RelationAmongCoeffEvenII} remain intact, also for the case of $n+2$, since they already have been assumed to be valid among all coefficients with $k \leq n$.

For instance, multiply the last case in the result~\eqref{eq:EvenCaseGeneralN} (minus $\pi$) by $(-1)$ to obtain:
{\allowdisplaybreaks
\begin{align}
 & (-1) \left(  \sum_{k = 1}^{n} \Ccal^{(j')}_{k} \varphi_{k} - \varphi_{n+1} - \varphi_{n+2} \right) \nonumber \\ &=  \sum_{k = 1}^{n} \left[ - \Ccal^{(j')}_{k} \right] \varphi_{k} + \varphi_{n+1} + \varphi_{n+2} \nonumber \\
 &=  \sum_{k = 1}^{n} \Ccal^{(\tilde{j}')}_{k} \varphi_{k} + \varphi_{n+1} + \varphi_{n+2} . \label{eq:ExampleCalcEven}
\end{align}
}
The last expression (with $\pi$ added again) resembles the one which is written in the fifth case of equation~\eqref{eq:EvenCaseGeneralN} and which also is of single-primed type, since a summand of $\pi$ appears in it.
\end{itemize}
Next we treat the case of an \textit{odd} number $n$. Again, we proceed via induction:
\begin{itemize}
 \item[$\ast)$] \underline{$n=3$:}
 
 We have
 \begin{widetext}
 \begin{align}
    \left\{ \begin{array}{c}  \varphi_{1} \\ \pi - \varphi_{1}  \end{array}  \right\} + \left\{ \begin{array}{c}  \varphi_{2} \\ \pi - \varphi_{2}  \end{array}  \right\} + \left\{ \begin{array}{c}  \varphi_{3} \\ \pi - \varphi_{3}  \end{array}  \right\} &= \begin{cases} + \varphi_{1} + \varphi_{2} + \varphi_{3}, \\ \pi - \varphi_{1} + \varphi_{2} + \varphi_{3}, \\ + \varphi_{1} + \pi - \varphi_{2} + \varphi_{3}, \\ + \varphi_{1} + \varphi_{2} + \pi - \varphi_{3}, \\ \pi - \varphi_{1} + \pi - \varphi_{2} + \varphi_{3}, \\ \pi - \varphi_{1} + \varphi_{2} + \pi - \varphi_{3}, \\ + \varphi_{1} + \pi - \varphi_{2} + \pi - \varphi_{3}, \\ \pi - \varphi_{1} + \pi - \varphi_{2} + \pi - \varphi_{3}. \end{cases} = \begin{cases} + \varphi_{1} + \varphi_{2} + \varphi_{3}, \\ \pi - \varphi_{1} + \varphi_{2} + \varphi_{3}, \\ \pi + \varphi_{1} - \varphi_{2} + \varphi_{3}, \\ \pi + \varphi_{1} + \varphi_{2} - \varphi_{3}, \\ - \varphi_{1} - \varphi_{2} + \varphi_{3}, \\ - \varphi_{1} + \varphi_{2}  - \varphi_{3}, \\ + \varphi_{1} - \varphi_{2} - \varphi_{3}, \\ \pi - \varphi_{1}  - \varphi_{2} - \varphi_{3}. \end{cases} \nonumber \\
   &=: \begin{cases} \Ccal^{(1'')}_{1} \varphi_{1} + \Ccal^{(1'')}_{2} \varphi_{2} + \Ccal^{(1'')}_{3} \varphi_{3}, \\  \pi + \Ccal^{(1')}_{1} \varphi_{1} + \Ccal^{(1')}_{2} \varphi_{2} + \Ccal^{(1')}_{3} \varphi_{3}, \\  \pi + \Ccal^{(2')}_{1} \varphi_{1} + \Ccal^{(2')}_{2} \varphi_{2} + \Ccal^{(2')}_{3} \varphi_{3}, \\  \pi + \Ccal^{(3')}_{1} \varphi_{1} + \Ccal^{(3')}_{2} \varphi_{2} + \Ccal^{(3')}_{3} \varphi_{3}, \\\Ccal^{(2'')}_{1} \varphi_{1} + \Ccal^{(2'')}_{2} \varphi_{2} + \Ccal^{(2'')}_{3} \varphi_{3}, \\\Ccal^{(3'')}_{1} \varphi_{1} + \Ccal^{(3'')}_{2} \varphi_{2} + \Ccal^{(3'')}_{3} \varphi_{3}, \\\Ccal^{(4'')}_{1} \varphi_{1} + \Ccal^{(4'')}_{2} \varphi_{2} + \Ccal^{(4'')}_{3} \varphi_{3}, \\ \pi + \Ccal^{(4')}_{1} \varphi_{1} + \Ccal^{(4')}_{2} \varphi_{2} + \Ccal^{(4')}_{3} \varphi_{3} . \end{cases} \label{eq:NEqualsThreePossibleCasesEvalIII}
 \end{align}
 \end{widetext}
 We read off the following relations
 \begin{align}
  (-1) \times \Ccal^{(4')}_{k} &= \Ccal^{(1'')}_{k} , \text{ for all } k = 1,\ldots,3,  \label{eq:OddReadOffRelI} \\
  (-1) \times \Ccal^{(3')}_{k} &= \Ccal^{(2'')}_{k} , 
  \text{ for all } k = 1,\ldots,3,\label{eq:OddReadOffRelII} \\
  (-1) \times \Ccal^{(2')}_{k} &= \Ccal^{(3'')}_{k} , \text{ for all } k = 1,\ldots,3, \label{eq:OddReadOffRelIII} \\
  (-1) \times \Ccal^{(1')}_{k} &= \Ccal^{(4'')}_{k} , \text{ for all } k = 1,\ldots,3, \label{eq:OddReadOffRelIV}
 \end{align}
 and see that primed and unprimed coefficients transform into each other under a multiplication by $(-1)$.
 \item[$\ast)$] \underline{$n \longrightarrow n + 2$:}
 
 Also for the case of $n$ odd, we use Lemma 1 in order to decompose as follows:
 \begin{widetext}
\begin{align}
 & \hspace*{12.5pt} \left\{ \begin{array}{c}  \varphi_{1} \\ \pi - \varphi_{1}  \end{array}  \right\} +  \ldots + \left\{ \begin{array}{c}  \varphi_{n} \\ \pi - \varphi_{n}  \end{array}  \right\} + \left\{ \begin{array}{c}  \varphi_{n+1} \\ \pi - \varphi_{n+1}  \end{array}  \right\} + \left\{ \begin{array}{c}  \varphi_{n+2} \\ \pi - \varphi_{n+2}  \end{array}  \right\} \nonumber \\
 &= \left\{ \begin{array}{l}  \sum_{k = 1}^{n} \Ccal^{(j'')}_{k} \varphi_{k} \\ \pi + \sum_{k = 1}^{n} \Ccal^{(j')}_{k} \varphi_{k}  \end{array} \right\} + \left\{ \begin{array}{c}  \varphi_{n+1} \\ \pi - \varphi_{n+1}  \end{array}  \right\} + \left\{ \begin{array}{c}  \varphi_{n+2} \\ \pi - \varphi_{n+2}  \end{array}  \right\} \nonumber \\ 
 &= \left\{ \begin{array}{l}  \sum_{k = 1}^{n} \Ccal^{(j'')}_{k} \varphi_{k} + \varphi_{n+1} + \varphi_{n+2} \\ \sum_{k = 1}^{n} \Ccal^{(j'')}_{k} \varphi_{k} + \pi - \varphi_{n+1} + \varphi_{n+2} \\ \sum_{k = 1}^{n} \Ccal^{(j'')}_{k} \varphi_{k} + \varphi_{n+1} + \pi - \varphi_{n+2} \\ \sum_{k = 1}^{n} \Ccal^{(j'')}_{k} \varphi_{k} - \varphi_{n+1} - \varphi_{n+2} \\ \pi + \sum_{k = 1}^{n} \Ccal^{(j')}_{k} \varphi_{k} + \varphi_{n+1} + \varphi_{n+2} \\ \sum_{k = 1}^{n} \Ccal^{(j')}_{k} \varphi_{k} - \varphi_{n+1} + \varphi_{n+2} \\  \sum_{k = 1}^{n} \Ccal^{(j')}_{k} \varphi_{k} + \varphi_{n+1} - \varphi_{n+2} \\ \pi + \sum_{k = 1}^{n} \Ccal^{(j')}_{k} \varphi_{k} - \varphi_{n+1} - \varphi_{n+2}  \end{array} \right\} . \label{eq:OddCaseGeneralN}
\end{align}
\end{widetext}
We see that the relation~\eqref{eq:RelationAmongCoeffOdd} remains intact, also for the case of $n+2$.

For instance, multiply the last case in the result~\eqref{eq:OddCaseGeneralN} (minus $\pi$) by $(-1)$ to obtain:
{\allowdisplaybreaks
\begin{align}
 & (-1) \left(  \sum_{k = 1}^{n} \Ccal^{(j')}_{k} \varphi_{k} - \varphi_{n+1} - \varphi_{n+2} \right) \nonumber \\
 &=  \sum_{k = 1}^{n} \left[ - \Ccal^{(j')}_{k} \right] \varphi_{k} + \varphi_{n+1} + \varphi_{n+2} \nonumber \\
 &=  \sum_{k = 1}^{n} \Ccal^{(\tilde{j}'')}_{k} \varphi_{k} + \varphi_{n+1} + \varphi_{n+2} . \label{eq:ExampleCalcOdd}
\end{align}
}
The last expression resembles the one which is written in the first case of equation~\eqref{eq:OddCaseGeneralN}, which is however of double-primed type, since no summand of $\pi$ appears in it. \textcolor{white}{Hello!} \hfill \textbf{QED}
\end{itemize}

\section{Matrix-algebras for different spin-reactions} \label{sec:MatrixAlgebras}

In this appendix, we list the complete and orthogonal basis-systems of matrices for all the processes discussed in sections~\ref{sec:MoravcsikTheoremExamples} and~\ref{sec:Electroproduction}. \\

\begin{itemize}
 \item \underline{Pion-Nucleon scattering ($N=2$; cf. section~\ref{sec:PiN}):}

\begin{align}
\hat{\sigma}_{\mathrm{D}} = 
\left[ \begin{array}{cc} 
\bm{\hat{a}} & 0  \\
0 & \bm{\hat{b}}  \end{array} \right] 
\quad \mathrm{,}& \quad 
\begin{array}{lccccc}
\mathrm{  } & \mathrm{ } & \bm{\hat{a}} & \bm{\hat{b}} \\
\hline
\sigma_{0} & \hat{\sigma}^{1} & +1 & +1 \\
\check{P} & \hat{\sigma}^{4} & +1 & -1 \\
\end{array} \label{eq:PauliMatD}
\\
\hat{\sigma}_{\mathrm{AD}} = 
\left[ \begin{array}{cc} 
0 & \bm{\hat{a}} \\
\bm{\hat{b}} & 0  \end{array} \right] 
\quad \mathrm{,}& \quad 
\begin{array}{lccccc}
 \mathrm{  } & \mathrm{ } & \bm{\hat{a}} & \bm{\hat{b}} \\
\hline
 \Ocal_{1}^{a} = - \check{R} & \hat{\sigma}^{2} & -i & i \\
 \Ocal_{2}^{a} = \check{A} & \hat{\sigma}^{3} & +1 & +1 
\end{array} \label{eq:PauliMatAD}
\end{align}
\end{itemize}

\begin{itemize}

\item \underline{Mathematical example for $N=3$ (cf. section~\ref{sec:3Amplitudes}):}

\begin{align}
\tilde \lambda_{\mathrm{D}} = 
\left[ \begin{array}{ccc} 
\bm{\hat{a}} & 0 & 0 \\
0 & \bm{\hat{b}} & 0 \\
0 & 0 & \bm{\hat{c}} \end{array} \right] 
\hspace*{5pt} \mathrm{,}& \hspace*{5pt} 
\begin{array}{lcccc}
\mathrm{  } & \mathrm{ } & \bm{\hat{a}} & \bm{\hat{b}} & \bm{\hat{c}} \\
\hline
\Ocal^{1} & \tilde \lambda^{1} & +\sqrt{2/3} & +\sqrt{2/3} & +\sqrt{2/3} \\
\Ocal^{4} & \tilde \lambda^{4} & +1 & -1 & 0 \\
\Ocal^{9} & \tilde \lambda^{9} & +1/\sqrt{3} & +1/\sqrt{3} & -2/\sqrt{3} 
\end{array} , \label{eq:LambdaTildeD}
\\
\tilde \lambda_{\mathrm{AD}} =
\left[ \begin{array}{ccc} 
0 & 0 & \bm{\hat{a}} \\
0 & 0 & 0 \\
\bm{\hat{b}} & 0 & 0 \end{array} \right] 
\hspace*{5pt} \mathrm{,}& \hspace*{5pt} 
\begin{array}{lccc}
 \mathrm{  } & \mathrm{ } & \bm{\hat{a}} & \bm{\hat{b}}  \\
\hline
 \Ocal_{1}^{a} & \tilde \lambda^{6} & +i & -i  \\
 \Ocal_{2}^{a} & \tilde \lambda^{5} & +1 & +1  \\
\end{array} , \label{eq:LambdaTildePR}
\\
\tilde \lambda_{\mathrm{P1}} = 
\left[ \begin{array}{ccc} 
0 & 0 & 0 \\
0 & 0 & \bm{\hat{a}}  \\
0 & \bm{\hat{b}} & 0  \end{array} \right] 
\hspace*{5pt} \mathrm{,}& \hspace*{5pt} 
\begin{array}{lccc}
 \mathrm{  } & \mathrm{ } & \bm{\hat{a}} & \bm{\hat{b}}  \\
\hline
 \Ocal_{1}^{b} & \tilde \lambda^{8} & +i & -i  \\
 \Ocal_{2}^{b} & \tilde \lambda^{7} & +1 & +1  \\
\end{array} , \label{eq:LambdaTildeAD}
\end{align}

\begin{align}
\tilde \lambda_{\mathrm{P2}} = 
\left[ \begin{array}{ccc} 
0 & \bm{\hat{a}} & 0 \\
\bm{\hat{b}} & 0 & 0 \\
0 & 0 & 0 \end{array} \right] 
\hspace*{5pt} \mathrm{,}& \hspace*{5pt} 
\begin{array}{lccc}
 \mathrm{  } & \mathrm{ } & \bm{\hat{a}} & \bm{\hat{b}}  \\
\hline
 \Ocal_{1}^{c} & \tilde \lambda^{3} & +i & -i  \\
 \Ocal_{2}^{c} & \tilde \lambda^{2} & +1 & +1  \\
\end{array} . \label{eq:LambdaTildePL}
\end{align}
\end{itemize}


\vspace*{10pt}

\begin{itemize}
\item \underline{Photoproduction ($N=4$; cf. section~\ref{sec:PhotoprodExample}):}

\begin{align}
\tilde \Gamma_{\mathrm{D}} = 
\left[ \begin{array}{cccc} 
\bm{\hat{a}} & 0 & 0 & 0 \\
0 & \bm{\hat{b}} & 0 & 0 \\
0 & 0 & \bm{\hat{c}} & 0 \\
0 & 0 & 0 & \bm{\hat{d}} \end{array} \right] 
\hspace*{2.5pt} \mathrm{,}& \hspace*{2.5pt} 
\begin{array}{lccccc}
\mathrm{  } & \mathrm{ } & \bm{\hat{a}} & \bm{\hat{b}} & \bm{\hat{c}} & \bm{\hat{d}} \\
\hline
\sigma_{0} & \tilde \Gamma^{1} & +1 & +1 & +1 & +1 \\
 - \check{\Sigma} & \tilde \Gamma^{4} & +1 & +1 & -1 & -1 \\
 - \check{T} & \tilde \Gamma^{10} & -1 & +1 & +1 & -1 \\
\check{P} & \tilde \Gamma^{12} & -1 & +1 & -1 & +1 \\
\end{array} \label{eq:GammaTildeD}
\\
\tilde \Gamma_{\mathrm{PR}} = 
\left[ \begin{array}{cccc} 
0 & 0 & \bm{\hat{a}} & 0 \\
0 & 0 & 0 & \bm{\hat{b}} \\
\bm{\hat{c}} & 0 & 0 & 0 \\
0 & \bm{\hat{d}} & 0 & 0 \end{array} \right] 
\hspace*{2.5pt} \mathrm{,}& \hspace*{2.5pt} 
\begin{array}{lccccc}
 \mathrm{  } & \mathrm{ } & \bm{\hat{a}} & \bm{\hat{b}} & \bm{\hat{c}} & \bm{\hat{d}} \\
\hline
 \Ocal_{1+}^{a}  & \tilde \Gamma^{3} & +i & +i & - i & -i \\
 \Ocal_{2-}^{a}  & \tilde \Gamma^{5} & +1 & -1 & +1 & -1 \\
 \Ocal_{2+}^{a}  & \tilde \Gamma^{9} & +1 & +1 & +1 & +1 \\
 \Ocal_{1-}^{a}  & \tilde \Gamma^{11} & +i & -i & -i & +i \\
\end{array} \label{eq:GammaTildePR}
\\
\tilde \Gamma_{\mathrm{AD}} =  
\left[ \begin{array}{cccc} 
0 & 0 & 0 & \bm{\hat{a}} \\
0 & 0 & \bm{\hat{b}} & 0 \\
0 & \bm{\hat{c}} & 0 & 0 \\
\bm{\hat{d}} & 0 & 0 & 0 \end{array} \right] 
\hspace*{2.5pt} \mathrm{,}& \hspace*{2.5pt} 
\begin{array}{lccccc}
 \mathrm{  }  & \mathrm{ } & \bm{\hat{a}} & \bm{\hat{b}} & \bm{\hat{c}} & \bm{\hat{d}} \\
\hline
 \Ocal_{2-}^{b}  & \tilde \Gamma^{14} & +1 & -1 & -1 & +1 \\
 \Ocal_{1+}^{b}  & \tilde \Gamma^{7} & +i & +i & -i & -i \\
 \Ocal_{1-}^{b}  & \tilde \Gamma^{16} & +i & -i & +i & -i \\
 \Ocal_{2+}^{b}  & \tilde \Gamma^{2} & +1 & +1 & +1 & +1 \\
\end{array} \label{eq:GammaTildeAD}
\\
\tilde \Gamma_{\mathrm{PL}} = 
\left[ \begin{array}{cccc} 
0 & \bm{\hat{a}} & 0 & 0 \\
\bm{\hat{b}} & 0 & 0 & 0 \\
0 & 0 & 0 & \bm{\hat{c}} \\
0 & 0 & \bm{\hat{d}} & 0 \end{array} \right] 
\hspace*{2.5pt} \mathrm{,}& \hspace*{2.5pt} 
\begin{array}{lccccc}
 \mathrm{  } & \mathrm{ } & \bm{\hat{a}} & \bm{\hat{b}} & \bm{\hat{c}} & \bm{\hat{d}} \\
\hline
 \Ocal_{2-}^{c}  & \tilde \Gamma^{6} & +1 & +1 & -1 & -1 \\
 \Ocal_{1-}^{c}  & \tilde \Gamma^{13} & +i & -i & -i & +i \\
 \Ocal_{1+}^{c}  & \tilde \Gamma^{8} & +i & -i & +i & -i \\
 \Ocal_{2+}^{c}  & \tilde \Gamma^{15} & +1 & +1 & +1 & +1 \\
\end{array} \label{eq:GammaTildePL}
\end{align}

\item \underline{Electroproduction ($N=6$; cf. section~\ref{sec:Electroproduction}):}

\begin{align}
&\tilde \Gamma_{\mathrm{D1}} = 
\left[ \begin{array}{cccccc} 
\bm{\hat{a}} & 0 & 0 & 0 & 0 & 0 \\
0 & \bm{\hat{b}} & 0 & 0 & 0 & 0 \\
0 & 0 & \bm{\hat{c}} & 0 & 0 & 0 \\
0 & 0 & 0 & \bm{\hat{d}} & 0 & 0 \\
0 & 0 & 0 & 0 & 0 & 0 \\
0 & 0 & 0 & 0 & 0 & 0\end{array} \right] 
\hspace*{0.5pt} \mathrm{,} \hspace*{1.5pt} 
\begin{array}{lccccc}
\mathrm{  } & \mathrm{ } & \bm{\hat{a}} & \bm{\hat{b}} & \bm{\hat{c}} & \bm{\hat{d}} \\
\hline
R_{T}^{00} & \tilde \Gamma^{1} & +1 & +1 & +1 & +1 \\
- \hspace*{1pt}^{c}R^{00}_{TT} & \tilde \Gamma^{4} & +1 & +1 & -1 & -1 \\
- R^{0y}_{T} & \tilde \Gamma^{10} & -1 & +1 & +1 & -1 \\
- R^{y'0}_{T} & \tilde \Gamma^{12} & -1 & +1 & -1 & +1 \\
\end{array} \label{eq:ElProdGammaTildeD} \\
&\tilde \Gamma_{\mathrm{PR1}} = 
\left[ \begin{array}{cccccc} 
0 & 0 & \bm{\hat{a}} & 0 & 0 & 0 \\
0 & 0 & 0 & \bm{\hat{b}} & 0 & 0 \\
\bm{\hat{c}} & 0 & 0 & 0 & 0 & 0 \\
0 & \bm{\hat{d}} & 0 & 0 & 0 & 0 \\
0 & 0 & 0 & 0 & 0 & 0 \\
0 & 0 & 0 & 0 & 0 & 0 
\end{array} \right] 
\hspace*{0.5pt} \mathrm{,} \hspace*{1.5pt} 
\begin{array}{lccccc}
 \mathrm{  } & \mathrm{ } & \bm{\hat{a}} & \bm{\hat{b}} & \bm{\hat{c}} & \bm{\hat{d}} \\
\hline
 \Ocal_{1+}^{a}  & \tilde \Gamma^{3} & +i & +i & - i & - i \\
 \Ocal_{2-}^{a}  & \tilde \Gamma^{5} & +1 & -1 & +1 & -1 \\
 \Ocal_{2+}^{a}  & \tilde \Gamma^{9} & +1 & +1 & +1 & +1 \\
 \Ocal_{1-}^{a}  & \tilde \Gamma^{11} & +i & -i & -i & +i \\
\end{array} \label{eq:ElProdGammaTildePR}
\end{align}

\begin{align}
\tilde \Gamma_{\mathrm{AD1}} =
\left[ \begin{array}{cccccc} 
0 & 0 & 0 & \bm{\hat{a}} & 0 & 0 \\
0 & 0 & \bm{\hat{b}} & 0 & 0 & 0 \\
0 & \bm{\hat{c}} & 0 & 0 & 0 & 0 \\
\bm{\hat{d}} & 0 & 0 & 0 & 0 & 0 \\
0 & 0 & 0 & 0 & 0 & 0 \\
0 & 0 & 0 & 0 & 0 & 0 \end{array} \right] 
\hspace*{3pt} \mathrm{,}& \hspace*{3pt} 
\begin{array}{lccccc}
 \mathrm{  }  & \mathrm{ } & \bm{\hat{a}} & \bm{\hat{b}} & \bm{\hat{c}} & \bm{\hat{d}} \\
\hline
 \Ocal_{2-}^{b}  & \tilde \Gamma^{14} & +1 & -1 & -1 & +1 \\
 \Ocal_{1+}^{b}  & \tilde \Gamma^{7} & +i & +i & -i & -i \\
 \Ocal_{1-}^{b}  & \tilde \Gamma^{16} & +i & -i & +i & -i \\
 \Ocal_{2+}^{b}  & \tilde \Gamma^{2} & +1 & +1 & +1 & +1 \\
\end{array} \label{eq:ElProdGammaTildeAD}
\\
\tilde \Gamma_{\mathrm{PL1}} = 
\left[ \begin{array}{cccccc} 
0 & \bm{\hat{a}} & 0 & 0 & 0 & 0 \\
\bm{\hat{b}} & 0 & 0 & 0 & 0 & 0 \\
0 & 0 & 0 & \bm{\hat{c}} & 0 & 0 \\
0 & 0 & \bm{\hat{d}} & 0 & 0 & 0 \\
0 & 0 & 0 & 0 & 0 & 0 \\
0 & 0 & 0 & 0 & 0 & 0 \end{array} \right] 
\hspace*{3pt} \mathrm{,}& \hspace*{3pt} 
\begin{array}{lccccc}
 \mathrm{  } & \mathrm{ } & \bm{\hat{a}} & \bm{\hat{b}} & \bm{\hat{c}} & \bm{\hat{d}} \\
\hline
 \Ocal_{2-}^{c}  & \tilde \Gamma^{6} & +1 & +1 & -1 & -1 \\
 \Ocal_{1-}^{c}  & \tilde \Gamma^{13} & +i & -i & -i & +i \\
 \Ocal_{1+}^{c}  & \tilde \Gamma^{8} & +i & -i & +i & -i \\
 \Ocal_{2+}^{c}  & \tilde \Gamma^{15} & +1 & +1 & +1 & +1 \\
\end{array} \label{eq:ElProdGammaTildePL} \\
\tilde \Gamma_{\mathrm{D2}} = 
\left[ \begin{array}{cccccc} 
0 & 0 & 0 & 0 & 0 & 0 \\
0 & 0 & 0 & 0 & 0 & 0 \\
0 & 0 & 0 & 0 & 0 & 0 \\
0 & 0 & 0 & 0 & 0 & 0 \\
0 & 0 & 0 & 0 & \bm{\hat{a}} & 0 \\
0 & 0 & 0 & 0 & 0 & \bm{\hat{b}}\end{array} \right] 
\hspace*{3pt} \mathrm{,}& \hspace*{3pt} 
\begin{array}{lccccc}
\mathrm{  } & \mathrm{ } & \bm{\hat{a}} & \bm{\hat{b}} \\
\hline
R^{00}_{L} & \tilde \Gamma^{17} & +\sqrt{2} & +\sqrt{2} \\
R^{0y}_{L}& \tilde \Gamma^{18} & +\sqrt{2} & - \sqrt{2}
\end{array} \label{eq:ElProdGammaTildeDII}
\\
\tilde \Gamma_{\mathrm{AD2}} = 
\left[ \begin{array}{cccccc} 
0 & 0 & 0 & 0 & 0 & 0 \\
0 & 0 & 0 & 0 & 0 & 0 \\
0 & 0 & 0 & 0 & 0 & 0 \\
0 & 0 & 0 & 0 & 0 & 0 \\
0 & 0 & 0 & 0 & 0 & \bm{\hat{a}} \\
0 & 0 & 0 & 0 & \bm{\hat{b}} & 0\end{array} \right] 
\hspace*{3pt} \mathrm{,}& \hspace*{3pt} 
\begin{array}{lccccc}
\mathrm{  } & \mathrm{ } & \bm{\hat{a}} & \bm{\hat{b}} \\
\hline
\Ocal_{2}^{d}  & \tilde \Gamma^{19} & \sqrt{2} & \sqrt{2} \\
\Ocal_{1}^{d}  & \tilde \Gamma^{20} & + i \sqrt{2} & - i \sqrt{2}
\end{array} \label{eq:ElProdGammaTildeADSmall} \\
\tilde \Gamma_{\mathrm{AD3}} = 
\left[ \begin{array}{cccccc} 
0 & 0 & 0 & 0 & 0 & 0 \\
0 & 0 & 0 & 0 & 0 & 0 \\
0 & 0 & 0 & 0 & 0 & \bm{\hat{a}} \\
0 & 0 & 0 & 0 & \bm{\hat{b}} & 0 \\
0 & 0 & 0 & \bm{\hat{c}} & 0 & 0 \\
0 & 0 & \bm{\hat{d}} & 0 & 0 & 0 
\end{array} \right] 
\hspace*{3pt} \mathrm{,}& \hspace*{3pt} 
\begin{array}{lccccc}
 \mathrm{  } & \mathrm{ } & \bm{\hat{a}} & \bm{\hat{b}} & \bm{\hat{c}} & \bm{\hat{d}} \\
\hline
 \Ocal_{2+}^{e}  & \tilde \Gamma^{21} & +1 & +1 & +1 & +1 \\
 \Ocal_{2-}^{e}  & \tilde \Gamma^{23} & +1 & -1 & -1 & +1 \\
 \Ocal_{1-}^{e}  & \tilde \Gamma^{29} & +i & -i & +i & -i \\
 \Ocal_{1+}^{e}  & \tilde \Gamma^{31} & +i & +i & -i & -i \\
\end{array} \label{eq:ElProdGammaTildeADIII} \\
\tilde \Gamma_{\mathrm{AD4}} = 
\left[ \begin{array}{cccccc} 
0 & 0 & 0 & 0 & 0 & \bm{\hat{a}} \\
0 & 0 & 0 & 0 & \bm{\hat{b}} & 0 \\
0 & 0 & 0 & 0 & 0 & 0 \\
0 & 0 & 0 & 0 & 0 & 0 \\
0 & \bm{\hat{c}} & 0 & 0 & 0 & 0 \\
\bm{\hat{d}} & 0 & 0 & 0 & 0 & 0 
\end{array} \right] 
\hspace*{3pt} \mathrm{,}& \hspace*{3pt} 
\begin{array}{lccccc}
 \mathrm{  } & \mathrm{ } & \bm{\hat{a}} & \bm{\hat{b}} & \bm{\hat{c}} & \bm{\hat{d}} \\
\hline
 \Ocal_{2-}^{f}  & \tilde \Gamma^{22} & +1 & -1 & -1 & +1 \\
 \Ocal_{1-}^{f}  & \tilde \Gamma^{24} & +i & -i & +i & -i \\
 \Ocal_{1+}^{f}  & \tilde \Gamma^{30} & +i & +i & -i & -i \\
 \Ocal_{2+}^{f}  & \tilde \Gamma^{32} & +1 & +1 & +1 & +1 \\
\end{array} \label{eq:ElProdGammaTildeADIV}
\end{align}

\begin{align}
\tilde \Gamma_{\mathrm{PR2}} = 
\left[ \begin{array}{cccccc} 
0 & 0 & 0 & 0 & \bm{\hat{a}} & 0 \\
0 & 0 & 0 & 0 & 0 & \bm{\hat{b}} \\
0 & 0 & 0 & 0 & 0 & 0 \\
0 & 0 & 0 & 0 & 0 & 0 \\
\bm{\hat{c}} & 0 & 0 & 0 & 0 & 0 \\
0 & \bm{\hat{d}} & 0 & 0 & 0 & 0 
\end{array} \right] 
\hspace*{3pt} \mathrm{,}& \hspace*{3pt} 
\begin{array}{lccccc}
 \mathrm{  } & \mathrm{ } & \bm{\hat{a}} & \bm{\hat{b}} & \bm{\hat{c}} & \bm{\hat{d}} \\
\hline
 \Ocal_{2-}^{g} & \tilde \Gamma^{25} & +1 & -1 & +1 & -1 \\
 \Ocal_{1-}^{g} & \tilde \Gamma^{26} & +i & -i & -i & +i \\
 \Ocal_{1+}^{g} & \tilde \Gamma^{33} & +i & +i & -i & -i \\
 \Ocal_{2+}^{g} & \tilde \Gamma^{34} & +1 & +1 & +1 & +1 \\
\end{array} \label{eq:ElProdGammaTildePRII} \\
\tilde \Gamma_{\mathrm{PR3}} = 
\left[ \begin{array}{cccccc} 
0 & 0 & 0 & 0 & 0 & 0 \\
0 & 0 & 0 & 0 & 0 & 0 \\
0 & 0 & 0 & 0 & \bm{\hat{a}} & 0 \\
0 & 0 & 0 & 0 & 0 & \bm{\hat{b}} \\
0 & 0 & \bm{\hat{c}} & 0 & 0 & 0 \\
0 & 0 & 0 & \bm{\hat{d}} & 0 & 0 
\end{array} \right] 
\hspace*{3pt} \mathrm{,}& \hspace*{3pt} 
\begin{array}{lccccc}
 \mathrm{  } & \mathrm{ } & \bm{\hat{a}} & \bm{\hat{b}} & \bm{\hat{c}} & \bm{\hat{d}} \\
\hline
 \Ocal_{2-}^{h}  & \tilde \Gamma^{27} & +1 & -1 & +1 & -1 \\
 \Ocal_{1-}^{h}  & \tilde \Gamma^{28} & +i & -i & -i & +i \\
 \Ocal_{1+}^{h}  & \tilde \Gamma^{35} & +i & +i & -i & -i \\
 \Ocal_{2+}^{h}  & \tilde \Gamma^{36} & +1 & +1 & +1 & +1 \\
\end{array} \label{eq:ElProdGammaTildePRIII} 
\end{align}

\end{itemize}

\section{Determination of the superfluous ob\-serva\-bles $\left( \Ocal^{c}_{1-}, \Ocal^{c}_{2-} \right)$ from the minimal complete set~\eqref{eq:PhotoprodMoravcsikExampleActualObsMinCompI}} \label{sec:MoravSetReductionPhoto}

For the following derivations, we can treat the right-hand-sides of equations~\eqref{eq:GroupCInternalConstraintI} and~\eqref{eq:GroupCInternalConstraintII} as constants, since it is assumed that the moduli $\left| b_{i} \right|$ have already been determined. Inserting the definitions of the modified ob\-serva\-bles~\eqref{eq:PhotoprodModObsI} and~\eqref{eq:PhotoprodModObsII}, the quadratic constraints become
\begin{align}
  &\left( \Ocal^{c}_{1+}  +  \Ocal^{c}_{1-} \right)^{2} + \left(  \Ocal^{c}_{2+}  +  \Ocal^{c}_{2-} \right)^{2} \nonumber \\
  &=  \left( \Ocal^{c}_{1+} \right)^{2} + 2 \Ocal^{c}_{1+} \Ocal^{c}_{1-} + \left( \Ocal^{c}_{1-} \right)^{2} + \left( \Ocal^{c}_{2+} \right)^{2} \nonumber \\
  & \hspace*{11.5pt} + 2 \Ocal^{c}_{2+} \Ocal^{c}_{2-} + \left( \Ocal^{c}_{2-} \right)^{2}  = 4 \left| b_{1} \right|^{2}  \left| b_{2} \right|^{2} , \label{eq:GroupCInternalConstraintPrimeI} \\
  &\left( \Ocal^{c}_{1+}  -  \Ocal^{c}_{1-} \right)^{2} + \left(  \Ocal^{c}_{2+}  -  \Ocal^{c}_{2-} \right)^{2} \nonumber \\
  &=  \left( \Ocal^{c}_{1+} \right)^{2} - 2 \Ocal^{c}_{1+} \Ocal^{c}_{1-} + \left( \Ocal^{c}_{1-} \right)^{2} + \left( \Ocal^{c}_{2+} \right)^{2} \nonumber \\
  & \hspace*{11.5pt} - 2 \Ocal^{c}_{2+} \Ocal^{c}_{2-} + \left( \Ocal^{c}_{2-} \right)^{2}  = 4 \left| b_{3} \right|^{2}  \left| b_{4} \right|^{2} . \label{eq:GroupCInternalConstraintPrimeII}
\end{align}
We can add both of these equations in order to isolate the purely quadratic terms, or subtract both equations in order to isolate the cross-terms. Doing both, we get the following equivalent set of equations
\begin{align}
 2 \kappa_{1} &:= 2 \left( \left| b_{1} \right|^{2}  \left| b_{2} \right|^{2} + \left| b_{3} \right|^{2}  \left| b_{4} \right|^{2} \right) \nonumber \\
 &= \left\{  \left( \Ocal^{c}_{1+} \right)^{2} + \left( \Ocal^{c}_{1-} \right)^{2} + \left( \Ocal^{c}_{2+} \right)^{2} + \left( \Ocal^{c}_{2-} \right)^{2}  \right\} , \label{eq:GroupCInternalConstraintIFinal} \\
 \kappa_{2} &:= \left| b_{1} \right|^{2}  \left| b_{2} \right|^{2} - \left| b_{3} \right|^{2}  \left| b_{4} \right|^{2} \nonumber \\
 &= \left\{  \Ocal^{c}_{1+}  \Ocal^{c}_{1-} + \Ocal^{c}_{2+} \Ocal^{c}_{2-} \right\} , \label{eq:GroupCInternalConstraintIIFinal} 
\end{align}
where two new constants, $\kappa_{1}$ and $\kappa_{2}$, have been introduced as well.

The purely quadratic equation~\eqref{eq:GroupCInternalConstraintIFinal} can be quickly solved for either of the two redundant ob\-serva\-bles. Solving for $\Ocal^{c}_{1-}$, we get
\begin{equation}
 \Ocal^{c}_{1-} = \eta \left| \Ocal^{c}_{1-} \right| = \eta \sqrt{2 \kappa_{1} -  \left( \Ocal^{c}_{1+} \right)^{2} - \left( \Ocal^{c}_{2+} \right)^{2} - \left( \Ocal^{c}_{2-} \right)^{2} }   , \label{eq:OcalC1minusResolved}
\end{equation}
where the variable $\eta$ can take both values $\eta = \pm 1$ and it keeps track of the fact that the sign of $\Ocal^{c}_{1-}$ is as of yet undetermined. Next, we re-write the constraint among the cross-terms, equation~\eqref{eq:GroupCInternalConstraintIIFinal}, as
\begin{equation}
  \Ocal^{c}_{1+}  \Ocal^{c}_{1-} = \kappa_{2} - \Ocal^{c}_{2+}  \Ocal^{c}_{2-}  . \label{eq:PhotoprodSecondConstraintResolved}
\end{equation}
The strategy is now to introduce the result~\eqref{eq:OcalC1minusResolved} for $\Ocal^{c}_{1-}$ into this equation and then solve for $\Ocal^{c}_{2-}$. However, in equation~\eqref{eq:OcalC1minusResolved}, the observable $\Ocal^{c}_{2-}$ appears under the square-root. Therefore, it is advisable to introduce the result~\eqref{eq:OcalC1minusResolved} into~\eqref{eq:PhotoprodSecondConstraintResolved} and then square the whole equation. This results in:
\begin{align}
 & \left( \Ocal^{c}_{1+} \right)^{2} \left\{  2 \kappa_{1} -  \left( \Ocal^{c}_{1+} \right)^{2} - \left( \Ocal^{c}_{2+} \right)^{2} - \left( \Ocal^{c}_{2-} \right)^{2}  \right\} \nonumber \\
 &= \kappa_{2}^{2} - 2 \kappa_{2} \Ocal^{c}_{2+}  \Ocal^{c}_{2-} + \left( \Ocal^{c}_{2+}  \right)^{2}  \left( \Ocal^{c}_{2-} \right)^{2}  . \label{eq:PhotoprodSecondConstraintSquared}
\end{align}
Dividing this equation by $\left( \Ocal^{c}_{1+} \right)^{2}$ and doing some more algebra, we arrive at the following quadratic equation for $\Ocal^{c}_{2-}$
\begin{align}
 &\left( \Ocal^{c}_{2-} \right)^{2} - \frac{2 \kappa_{2} \Ocal^{c}_{2+}}{\left( \Ocal^{c}_{1+} \right)^{2} + \left( \Ocal^{c}_{2+} \right)^{2}} \Ocal^{c}_{2-} + \Bigg[ \left(  \Ocal^{c}_{1+} \right)^{2} \nonumber \\
 &+ \frac{\kappa_{2}^{2}}{\left( \Ocal^{c}_{1+} \right)^{2} + \left( \Ocal^{c}_{2+} \right)^{2}} - 2 \frac{\kappa_{1} \left( \Ocal^{c}_{1+} \right)^{2}}{\left( \Ocal^{c}_{1+} \right)^{2} + \left( \Ocal^{c}_{2+} \right)^{2}} \Bigg] = 0 . \label{eq:QuadraticEqForOc2Minus}
\end{align}
This equation generally has two solutions, which read
\begin{widetext}
\begin{align}
 \left( \Ocal^{c}_{2-} \right)_{\text{I,II}}  &=  \frac{ \kappa_{2} \Ocal^{c}_{2+}}{\left( \Ocal^{c}_{1+} \right)^{2} + \left( \Ocal^{c}_{2+} \right)^{2}} \pm \sqrt{\frac{ \kappa_{2}^{2} \left(\Ocal^{c}_{2+}\right)^{2}}{\left[\left( \Ocal^{c}_{1+} \right)^{2} + \left( \Ocal^{c}_{2+} \right)^{2}\right]^{2}} - \left(  \Ocal^{c}_{1+} \right)^{2} - \frac{\kappa_{2}^{2}}{\left( \Ocal^{c}_{1+} \right)^{2} + \left( \Ocal^{c}_{2+} \right)^{2}} + 2 \frac{\kappa_{1} \left( \Ocal^{c}_{1+} \right)^{2}}{\left( \Ocal^{c}_{1+} \right)^{2} + \left( \Ocal^{c}_{2+} \right)^{2}} } \nonumber \\
& =  \frac{\kappa_{2} \Ocal^{c}_{2+} \pm \sqrt{ 2 \kappa_{1} \left( \Ocal^{c}_{1+} \right)^{2} \left( \left( \Ocal^{c}_{1+} \right)^{2} + \left( \Ocal^{c}_{2+} \right)^{2} \right) - \kappa_{2}^{2} \left( \Ocal^{c}_{1+} \right)^{2} - \left( \Ocal^{c}_{1+} \right)^{2} \left[ \left( \Ocal^{c}_{1+} \right)^{2} + \left( \Ocal^{c}_{2+} \right)^{2} \right]^{2}}}{\left( \Ocal^{c}_{1+} \right)^{2} + \left( \Ocal^{c}_{2+} \right)^{2}}  . \label{eq:QuadraticEqForOc2MinusSolutions}
\end{align}
\end{widetext}
Thus, the observable $\Ocal^{c}_{2-}$ is determined up to a two-fold discrete ambiguity. Inserting the result~\eqref{eq:QuadraticEqForOc2MinusSolutions} into the equation~\eqref{eq:OcalC1minusResolved} obtained above, we get the following possible solutions for $\Ocal^{c}_{1-}$:
\begin{align}
 &\left(\Ocal^{c}_{1-}\right)_{\left(\pm,\left\{\text{I,II}\right\}\right)}  = \nonumber \\
 & \hspace*{5pt} \pm \sqrt{2 \kappa_{1} -  \left( \Ocal^{c}_{1+} \right)^{2} - \left( \Ocal^{c}_{2+} \right)^{2} - \left[\left( \Ocal^{c}_{2-} \right)_{\text{I,II}}\right]^{2} }   . \label{eq:OcalC1minusFourSolutions}
\end{align}
We see that $\Ocal^{c}_{1-}$ is determined up to a four-fold discrete ambiguity, which also makes the overall ambiguity for the determination of the two superfluous ob\-serva\-bles $\left( \Ocal^{c}_{1-}, \Ocal^{c}_{2-} \right)$ from just the group $c$ alone, a four-fold one.

In order to resolve the ambiguities contained in the results~\eqref{eq:QuadraticEqForOc2MinusSolutions} and~\eqref{eq:OcalC1minusFourSolutions}, one has to include the additional information provided by the two ob\-serva\-bles from group $b$, which are contained in the minimal complete set~\eqref{eq:PhotoprodMoravcsikExampleActualObsMinCompI}. As a preparatory step, we note that the fundamental consistency relation~\eqref{eq:PhotoFundamentalConsistencyRel} can be re-written as follows, in order to connect ob\-serva\-bles from the groups $c$ and $b$:
\begin{equation}
 \underbrace{\phi_{12} + \phi_{34}}_{c} = \underbrace{\phi_{14} - \phi_{23}}_{b}  . \label{eq:ConsistencyRelPhotoprodExample}
\end{equation}
Furthermore, we note the following two important addition-theorems for the cosine and sine, evaluated on the relative-phases corresponding to the group $c$
\begin{align}
 \cos \left( \phi_{12} + \phi_{34} \right) &=  \cos \phi_{12} \cos \phi_{34} - \sin \phi_{12} \sin \phi_{34}  , \label{eq:CosAddTheo}  \\
 \sin \left( \phi_{12} + \phi_{34} \right) &= \sin \phi_{12} \cos \phi_{34} + \cos \phi_{12} \sin \phi_{34}  . \label{eq:SinAddTheo}
\end{align}
Combining the consistency relation~\eqref{eq:ConsistencyRelPhotoprodExample} with the cosine-theorem~\eqref{eq:CosAddTheo} and the definitions of the modified ob\-serva\-bles $\tilde{\Ocal}$, we get the following result
\begin{align}
  &\cos \left( \phi_{14} - \phi_{23} \right) \equiv  \cos \left( \phi_{12} + \phi_{34} \right) \nonumber \\
  &= \frac{\tilde{\Ocal}^{c}_{2+}}{\left| b_{1} \right|  \left| b_{2} \right|} \frac{\tilde{\Ocal}^{c}_{2-}}{\left| b_{3} \right|  \left| b_{4} \right|} - \frac{\tilde{\Ocal}^{c}_{1+}}{\left| b_{1} \right|  \left| b_{2} \right|}  \frac{\tilde{\Ocal}^{c}_{1-}}{\left| b_{3} \right|  \left| b_{4} \right|}   , \label{eq:CosAddTheoIntermediateStep}
\end{align}
which is equivalent to
\begin{equation}
  \tilde{\Ocal}^{c}_{2+}  \tilde{\Ocal}^{c}_{2-} -  \tilde{\Ocal}^{c}_{1+} \tilde{\Ocal}^{c}_{1-} = \left| b_{1} \right|  \left| b_{2} \right| \left| b_{3} \right|  \left| b_{4} \right| \cos  \left(  \phi_{14} - \phi_{23} \right)  . \label{eq:CosAddTheoReform}
\end{equation}
Similarly, when starting from the sine-theorem~\eqref{eq:SinAddTheo}, one obtains
\begin{equation}
  \tilde{\Ocal}^{c}_{1+}  \tilde{\Ocal}^{c}_{2-} +  \tilde{\Ocal}^{c}_{2+} \tilde{\Ocal}^{c}_{1-} = \left| b_{1} \right|  \left| b_{2} \right| \left| b_{3} \right|  \left| b_{4} \right| \sin  \left(  \phi_{14} - \phi_{23} \right)  . \label{eq:SinAddTheoReform}
\end{equation}
Thus, we managed to establish a connection between cross-terms of group-$c$ ob\-serva\-bles on the left-hand-side, and relative-phases from group $b$ on the right-hand-side. 

Next, we have to determine what the two group-$b$ ob\-serva\-bles $\left( \Ocal^{b}_{2+}, \Ocal^{b}_{2-} \right)$, which are both contained in the minimal complete set~\eqref{eq:PhotoprodMoravcsikExampleActualObsMinCompI}, can tell us about said relative-phases. In this case, the actual ob\-serva\-bles $\left( \Ocal^{b}_{2+}, \Ocal^{b}_{2-} \right)$ are fully equivalent to the modified ob\-serva\-bles $\left( \tilde{\Ocal}^{b}_{2+}, \tilde{\Ocal}^{b}_{2-} \right)$, and therefore also to the cosines $\left(\cos \phi_{14} , \cos \phi_{23} \right)$. Therefore, both relative-phases are known up to the discrete cosine-type ambiguities
\begin{equation}
 \phi_{14}^{\pm} = \begin{cases} + \alpha_{14}, \\ - \alpha_{14},  \end{cases}   \text{ and }  \phi_{23}^{\pm} = \begin{cases} + \alpha_{23}, \\ - \alpha_{23}.  \end{cases} \label{eq:CosTypeAmbiguitiesGroupB}
\end{equation}
These ambiguities furthermore imply the following possible set of discrete values for the right-hand-side of the consistency relation~\eqref{eq:ConsistencyRelPhotoprodExample}: 
\begin{align}
 \phi_{14}^{+} - \phi_{23}^{+} &= \alpha_{14} - \alpha_{23} , \label{eq:ConsistRelGroupBCasesI} \\
 \phi_{14}^{+} - \phi_{23}^{-} &= \alpha_{14} + \alpha_{23} , \label{eq:ConsistRelGroupBCasesII} \\
 \phi_{14}^{-} - \phi_{23}^{+} &= - \alpha_{14} - \alpha_{23} , \label{eq:ConsistRelGroupBCasesIII} \\
 \phi_{14}^{-} - \phi_{23}^{-} &= - \alpha_{14} + \alpha_{23} . \label{eq:ConsistRelGroupBCasesIV}
\end{align}
It is now time to turn to our cosine-constraint~\eqref{eq:CosAddTheoReform}. However, due to the symmetry of the cosine-function [$\cos (x) = \cos (-x)$], the right-hand-side of this constraint can only take two possible different values under the four possible cases given in equations~\eqref{eq:ConsistRelGroupBCasesI} to~\eqref{eq:ConsistRelGroupBCasesIV}. We denote these two possibilities by introducing two new variables, $\gamma_{1}$ and $\gamma_{2}$:
\begin{equation}
  \left| b_{1} \right|  \left| b_{2} \right| \left| b_{3} \right|  \left| b_{4} \right|  \cos \left( \pm \alpha_{14} \mp \alpha_{23} \right) =: \frac{1}{4} \gamma_{1,2}  . \label{eq:DegGammaCos}
\end{equation}
Inserting the definitions of the modified ob\-serva\-bles $\tilde{\Ocal}$, the constraint~\eqref{eq:CosAddTheoReform} therefore becomes
\begin{align}
 &\left( \Ocal^{c}_{2+} + \Ocal^{c}_{2-} \right)  \left( \Ocal^{c}_{2+} - \Ocal^{c}_{2-}  \right)  \nonumber \\
 &-  \left( \Ocal^{c}_{1+} + \Ocal^{c}_{1-}  \right)  \left( \Ocal^{c}_{1+} - \Ocal^{c}_{1-}  \right) = \gamma_{1,2}  . \label{eq:CosAddTheoRelationActualObservables}
\end{align}
Multiplying out all the brackets, we see that only the purely quadratic terms remain
\begin{equation}
 \left( \Ocal^{c}_{2+} \right)^{2} -  \left( \Ocal^{c}_{2-}  \right)^{2} -  \left( \Ocal^{c}_{1+}  \right)^{2}   + \left( \Ocal^{c}_{1-}  \right)^{2} = \gamma_{1,2}  . \label{eq:CosAddTheoRelationActualObservablesII}
\end{equation}
Inserting the solutions~\eqref{eq:QuadraticEqForOc2MinusSolutions} and~\eqref{eq:OcalC1minusFourSolutions} into this equation, we obtain
\begin{equation}
 - 2 \left[ \left( \Ocal^{c}_{2-} \right)_{\text{I,II}} \right]^{2}  - 2 \left( \Ocal^{c}_{1+} \right)^{2}  +  2 \kappa_{1} = \gamma_{1,2} . \label{eq:CosAddTheoFinalFormConstraint}
\end{equation}
We suppose that this equation can be used, at least numerically, to decide which of the two solutions $\left(\Ocal^{c}_{2-} \right)_{\text{I,II}}$, as well as which of the two possible $\gamma$'s, is the correct one. This probably works up to highly singular numerical special cases, where multiple possibilities survive the check with equation~\eqref{eq:CosAddTheoFinalFormConstraint}.

Suppose equation~\eqref{eq:CosAddTheoFinalFormConstraint} can resolve all of the ambiguities as mentioned above. Then, since we also know which $\gamma$ is the correct one, the four cases given in equations~\eqref{eq:ConsistRelGroupBCasesI} to~\eqref{eq:ConsistRelGroupBCasesIV} reduce to two possibilities, which are only distinct by an overall sign. Denote the modulus of both those possibilities as $\tilde{\alpha}$. Then, the right-hand-side of the sine-constraint~\eqref{eq:SinAddTheoReform} can also only take two possible values: 
\begin{equation}
  \left| b_{1} \right|  \left| b_{2} \right| \left| b_{3} \right|  \left| b_{4} \right|  \sin \left( \pm \tilde{\alpha} \right) =: \frac{1}{4} \xi_{1,2}  . \label{eq:DegXiSin}
\end{equation}
Here, two new variables $\xi_{1}$ and $\xi_{2}$ have been defined. Introducing the definitions of the modified ob\-serva\-bles $\tilde{\Ocal}$, the sine-constraint becomes
\begin{align}
 &\left( \Ocal^{c}_{1+} + \Ocal^{c}_{1-} \right)  \left( \Ocal^{c}_{2+} - \Ocal^{c}_{2-}  \right) \nonumber \\
 &+  \left( \Ocal^{c}_{2+} + \Ocal^{c}_{2-}  \right)  \left( \Ocal^{c}_{1+} - \Ocal^{c}_{1-}  \right) = \xi_{1,2}  . \label{eq:SinAddTheoRelationActualObservables}
\end{align}
Multiplying out all the brackets, only the following cross-terms remain:
\begin{equation}
  \Ocal^{c}_{1+}  \Ocal^{c}_{2+} - \Ocal^{c}_{1-} \Ocal^{c}_{2-} = \frac{1}{2} \xi_{1,2}  . \label{eq:SinAddTheoRelationActualObservablesII}
\end{equation}
While the ambiguity for the observable $\Ocal^{c}_{2-}$ has been resolved, there remains the overall sign-ambiguity for $\Ocal^{c}_{1-}$. Thus, we have to consider the cases
\begin{equation}
  \Ocal^{c}_{1+}  \Ocal^{c}_{2+} \mp \left| \Ocal^{c}_{1-}  \right|  \Ocal^{c}_{2-} = \frac{1}{2} \xi_{1,2}  . \label{eq:SinAddTheoRelationActualObservablesFinalForm}
\end{equation}
Equation~\eqref{eq:SinAddTheoRelationActualObservablesFinalForm} is capable of determining the correct sign of $\Ocal^{c}_{1-}$, as well as which of the $\xi$'s is the correct one. This is probably possible up to some singular numerical cases.

Therefore, we managed to determine the redundant ob\-serva\-bles $\left( \Ocal^{c}_{1-}, \Ocal^{c}_{2-} \right)$ in terms of all the ob\-serva\-bles contained in the minimal complete set~\eqref{eq:PhotoprodMoravcsikExampleActualObsMinCompI}.

\section{Moravcsik-complete sets for pseudoscalar meson electroproduction} \label{sec:MoravComplSetsElectro}

The Moravcsik-complete sets implied for the case of electroproduction are listed here (see section~\ref{sec:Electroproduction}). We list only the cases with the minimal number of $7$ non-diagonal ob\-serva\-bles in Tables~\ref{tab:MinimalMoravcsikSetsElectroI} to~\ref{tab:MinimalMoravcsikSetsElectroI2}, as well as the next-to-minimal cases with a number of $8$ non-diagonal ob\-serva\-bles in Tables~\ref{tab:MinimalMoravcsikSetsElectroII} to~\ref{tab:MinimalMoravcsikSetsElectroII2OneTwo}.

\begin{table}
\begin{tabular}{l|ccccccc}
Set-Nr. & \multicolumn{7}{c}{Observables}  \\
\hline   
  A.i.1 & $R_{TT'}^{0z}$ & $\hspace*{1pt}^{s} R^{0x}_{TT}$ & $R^{x'z}_{T}$ & $R^{z'x}_{T}$ & $R_{L}^{x'x}$ & $\hspace*{1pt}^{c} R^{x'x}_{LT}$ & $\hspace*{1pt}^{s} R^{z'x}_{LT'}$      \\ 
   &  $\Ocal^{a}_{2+}$  &  $\Ocal^{a}_{2-}$  &  $\Ocal^{c}_{1+}$  &  $\Ocal^{c}_{1-}$  &  $\Ocal^{d}_{2}$  & $\Ocal^{h}_{2+}$ &  $\Ocal^{h}_{2-}$     \\  
   A.i.2 & $R_{TT'}^{0z}$ & $\hspace*{1pt}^{s} R^{0x}_{TT}$ & $R_{T}^{z' z}$ & $R_{T}^{x' x}$ & $R_{L}^{z'x}$ & $\hspace*{1pt}^{c} R^{x'x}_{LT}$ & $\hspace*{1pt}^{s} R^{z'x}_{LT'}$    \\ 
   & $\Ocal^{a}_{2+}$ & $\Ocal^{a}_{2-}$ & $\Ocal^{c}_{2+}$ & $\Ocal^{c}_{2-}$ & $\Ocal^{d}_{1}$ & $\Ocal^{h}_{2+}$ & $\Ocal^{h}_{2-}$    \\  
   A.i.3 & $\hspace*{1pt}^{s} R_{TT}^{0z}$ & $R_{TT'}^{0x}$ & $R^{x'z}_{T}$ & $R^{z'x}_{T}$ & $R_{L}^{x'x}$ & $\hspace*{1pt}^{c} R^{x'x}_{LT}$ & $\hspace*{1pt}^{s} R^{z'x}_{LT'}$    \\    
    & $\Ocal^{a}_{1+}$ & $\Ocal^{a}_{1-}$ & $\Ocal^{c}_{1+}$ & $\Ocal^{c}_{1-}$ & $\Ocal^{d}_{2}$ & $\Ocal^{h}_{2+}$ & $\Ocal^{h}_{2-}$     \\ 
    A.i.4 & $R_{TT'}^{0z}$ & $\hspace*{1pt}^{s} R^{0x}_{TT}$ & $R^{x'z}_{T}$ & $R^{z'x}_{T}$ & $R_{L}^{x'x}$ & $\hspace*{1pt}^{s} R^{x'x}_{LT'}$ & $\hspace*{1pt}^{c} R^{z'x}_{LT}$    \\  
    & $\Ocal^{a}_{2+}$ & $\Ocal^{a}_{2-}$ & $\Ocal^{c}_{1+}$ & $\Ocal^{c}_{1-}$ & $\Ocal^{d}_{2}$ & $\Ocal^{h}_{1+}$ & $\Ocal^{h}_{1-}$     \\ 
    A.i.5 & $\hspace*{1pt}^{s} R_{TT}^{0z}$ & $R_{TT'}^{0x}$ & $R_{T}^{z' z}$ & $R_{T}^{x' x}$ & $R_{L}^{z'x}$ & $\hspace*{1pt}^{c} R^{x'x}_{LT}$ & $\hspace*{1pt}^{s} R^{z'x}_{LT'}$    \\ 
    & $\Ocal^{a}_{1+}$ & $\Ocal^{a}_{1-}$ & $\Ocal^{c}_{2+}$ & $\Ocal^{c}_{2-}$ & $\Ocal^{d}_{1}$ & $\Ocal^{h}_{2+}$ & $\Ocal^{h}_{2-}$    \\ 
    A.i.6 & $R_{TT'}^{0z}$ & $\hspace*{1pt}^{s} R^{0x}_{TT}$ & $R_{T}^{z' z}$ & $R_{T}^{x' x}$ & $R_{L}^{z'x}$ & $\hspace*{1pt}^{s} R^{x'x}_{LT'}$ & $\hspace*{1pt}^{c} R^{z'x}_{LT}$   \\  
    & $\Ocal^{a}_{2+}$ & $\Ocal^{a}_{2-}$ & $\Ocal^{c}_{2+}$ & $\Ocal^{c}_{2-}$ & $\Ocal^{d}_{1}$ & $\Ocal^{h}_{1+}$ & $\Ocal^{h}_{1-}$    \\ 
    A.i.7 & $\hspace*{1pt}^{s} R_{TT}^{0z}$ & $R_{TT'}^{0x}$ & $R^{x'z}_{T}$ & $R^{z'x}_{T}$ & $R_{L}^{x'x}$ & $\hspace*{1pt}^{s} R^{x'x}_{LT'}$ & $\hspace*{1pt}^{c} R^{z'x}_{LT}$   \\     
    & $\Ocal^{a}_{1+}$ & $\Ocal^{a}_{1-}$ & $\Ocal^{c}_{1+}$ & $\Ocal^{c}_{1-}$ & $\Ocal^{d}_{2}$ & $\Ocal^{h}_{1+}$ & $\Ocal^{h}_{1-}$    \\ 
    A.i.8  &  $\hspace*{1pt}^{s} R_{TT}^{0z}$ & $R_{TT'}^{0x}$  &  $R_{T}^{z' z}$ & $R_{T}^{x' x}$ & $R_{L}^{z'x}$  &  $\hspace*{1pt}^{s} R^{x'x}_{LT'}$  &  $\hspace*{1pt}^{c} R^{z'x}_{LT}$   \\  
    & $\Ocal^{a}_{1+}$ & $\Ocal^{a}_{1-}$ & $\Ocal^{c}_{2+}$  &  $\Ocal^{c}_{2-}$ & $\Ocal^{d}_{1}$ & $\Ocal^{h}_{1+}$ & $\Ocal^{h}_{1-}$    \\
   \hline 
    A.ii.1 & $R_{TT'}^{0z}$ & $\hspace*{1pt}^{s} R^{0x}_{TT}$ & $R^{x'z}_{T}$  &  $R^{z'x}_{T}$ & $R_{L}^{x'x}$  &  $\hspace*{1pt}^{c} R^{00}_{LT}$  &  $\hspace*{1pt}^{c} R^{0y}_{LT}$   \\  
    & $\Ocal^{a}_{2+}$ & $\Ocal^{a}_{2-}$ & $\Ocal^{c}_{1+}$  &  $\Ocal^{c}_{1-}$ & $\Ocal^{d}_{2}$ & $\Ocal^{e}_{2+}$ & $\Ocal^{e}_{2-}$   \\ 
    A.ii.2  &  $R_{TT'}^{0z}$  &  $\hspace*{1pt}^{s} R^{0x}_{TT}$ & $R_{T}^{z' z}$  &  $R_{T}^{x' x}$ & $R_{L}^{z'x}$  &  $\hspace*{1pt}^{c} R^{00}_{LT}$  &  $\hspace*{1pt}^{c} R^{0y}_{LT}$    \\ 
    & $\Ocal^{a}_{2+}$ & $\Ocal^{a}_{2-}$ & $\Ocal^{c}_{2+}$  &  $\Ocal^{c}_{2-}$ & $\Ocal^{d}_{1}$ & $\Ocal^{e}_{2+}$ & $\Ocal^{e}_{2-}$   \\ 
    A.ii.3  &  $\hspace*{1pt}^{s} R_{TT}^{0z}$ & $R_{TT'}^{0x}$  &  $R^{x'z}_{T}$ & $R^{z'x}_{T}$ & $R_{L}^{x'x}$  &  $\hspace*{1pt}^{c} R^{00}_{LT}$  &  $\hspace*{1pt}^{c} R^{0y}_{LT}$   \\   
    & $\Ocal^{a}_{1+}$ & $\Ocal^{a}_{1-}$ & $\Ocal^{c}_{1+}$  &  $\Ocal^{c}_{1-}$ & $\Ocal^{d}_{2}$ & $\Ocal^{e}_{2+}$ & $\Ocal^{e}_{2-}$   \\
    A.ii.4  &  $R_{TT'}^{0z}$  &  $\hspace*{1pt}^{s} R^{0x}_{TT}$ & $R^{x'z}_{T}$  &  $R^{z'x}_{T}$ & $R_{L}^{x'x}$  &  $\hspace*{1pt}^{s} R^{00}_{LT'}$  &  $\hspace*{1pt}^{s} R^{0y}_{LT'}$   \\  
    & $\Ocal^{a}_{2+}$ & $\Ocal^{a}_{2-}$ & $\Ocal^{c}_{1+}$  &  $\Ocal^{c}_{1-}$ & $\Ocal^{d}_{2}$ & $\Ocal^{e}_{1+}$ & $\Ocal^{e}_{1-}$    \\
    A.ii.5  &  $\hspace*{1pt}^{s} R_{TT}^{0z}$ & $R_{TT'}^{0x}$  &  $R_{T}^{z' z}$ & $R_{T}^{x' x}$ & $R_{L}^{z'x}$  &  $\hspace*{1pt}^{c} R^{00}_{LT}$  &  $\hspace*{1pt}^{c} R^{0y}_{LT}$   \\  
    & $\Ocal^{a}_{1+}$ & $\Ocal^{a}_{1-}$ & $\Ocal^{c}_{2+}$  &  $\Ocal^{c}_{2-}$ & $\Ocal^{d}_{1}$ & $\Ocal^{e}_{2+}$ & $\Ocal^{e}_{2-}$    \\  
    A.ii.6  &  $R_{TT'}^{0z}$  &  $\hspace*{1pt}^{s} R^{0x}_{TT}$ & $R_{T}^{z' z}$  &  $R_{T}^{x' x}$ & $R_{L}^{z'x}$  &  $\hspace*{1pt}^{s} R^{00}_{LT'}$  &  $\hspace*{1pt}^{s} R^{0y}_{LT'}$   \\   
    & $\Ocal^{a}_{2+}$ & $\Ocal^{a}_{2-}$ & $\Ocal^{c}_{2+}$  &  $\Ocal^{c}_{2-}$ & $\Ocal^{d}_{1}$ & $\Ocal^{e}_{1+}$ & $\Ocal^{e}_{1-}$   \\  
    A.ii.7  &  $\hspace*{1pt}^{s} R_{TT}^{0z}$ & $R_{TT'}^{0x}$  &  $R^{x'z}_{T}$ & $R^{z'x}_{T}$ & $R_{L}^{x'x}$  &  $\hspace*{1pt}^{s} R^{00}_{LT'}$  &  $\hspace*{1pt}^{s} R^{0y}_{LT'}$  \\     
    & $\Ocal^{a}_{1+}$ & $\Ocal^{a}_{1-}$ & $\Ocal^{c}_{1+}$  &  $\Ocal^{c}_{1-}$ & $\Ocal^{d}_{2}$ & $\Ocal^{e}_{1+}$ & $\Ocal^{e}_{1-}$   \\  
    A.ii.8  &  $\hspace*{1pt}^{s} R_{TT}^{0z}$ & $R_{TT'}^{0x}$  &  $R_{T}^{z' z}$ & $R_{T}^{x' x}$ & $R_{L}^{z'x}$  &  $\hspace*{1pt}^{s} R^{00}_{LT'}$  &  $\hspace*{1pt}^{s} R^{0y}_{LT'}$  \\  
   & $\Ocal^{a}_{1+}$ & $\Ocal^{a}_{1-}$ & $\Ocal^{c}_{2+}$  &  $\Ocal^{c}_{2-}$ & $\Ocal^{d}_{1}$ & $\Ocal^{e}_{1+}$ & $\Ocal^{e}_{1-}$  
\end{tabular}
\caption{The first subset of $16$ from a total of $64$ distinct possibilities to form Moravcsik-complete sets for electroproduction with a minimal number of ob\-serva\-bles is listed here (cf. section~\ref{sec:Electroproduction}).  In each case, $7$ ob\-serva\-bles are listed which have to be picked in addition to the $6$ diagonal ob\-serva\-bles $\left\{ R^{00}_{T},  R^{00}_{TT}, R_{T}^{0y}, R^{y' 0}_{T}, R_{L}^{00} , R_{L}^{0y} \right\}$ for electroproduction (cf. Table~\ref{tab:ElectroObservablesI}). This implies $13$ ob\-serva\-bles in total for each case. Observe that each case contains exactly one observable from the shape-class $d$. \\ The labelling-scheme for the sets is chosen as follows: the letter 'A' denotes the Moravcsik-complete sets of $13$ observables, the roman numerals i, $\ldots$, viii indicate the different possible combinations of shape-classes and the regular number ($1$, $2$, $3$, $\ldots$) at the end counts the number of the set from the respective group of shape-class combinations. Furthermore, we list each set in the physical notation $R^{\beta \alpha}_{i}$~\cite{Tiator:2017cde} (cf. Table~\ref{tab:ElectroObservablesI}) and also, directly below, in Nakayama's systematic mathematical notation $\Ocal^{n}_{\nu \pm}$.}
\label{tab:MinimalMoravcsikSetsElectroI}
\end{table}

\begin{table}
\begin{tabular}{l|ccccccc}
Set-Nr. & \multicolumn{7}{c}{Observables}  \\
\hline   
    A.iii.1  &  $R^{z'0}_{TT'}$  &  $\hspace*{1pt}^{s} R^{x'0}_{TT}$ & $R^{x'z}_{T}$  &  $R^{z'x}_{T}$ & $R_{L}^{x'x}$  &  $\hspace*{1pt}^{c} R^{x'x}_{LT}$  &  $\hspace*{1pt}^{s} R^{z'x}_{LT'}$ \\ 
        & $\Ocal^{b}_{2+}$ & $\Ocal^{b}_{2-}$ & $\Ocal^{c}_{1+}$  &  $\Ocal^{c}_{1-}$ & $\Ocal^{d}_{2}$ & $\Ocal^{h}_{2+}$ & $\Ocal^{h}_{2-}$  \\
    A.iii.2  &  $R^{z'0}_{TT'}$  &  $\hspace*{1pt}^{s} R^{x'0}_{TT}$ & $R_{T}^{z' z}$  &  $R_{T}^{x' x}$ & $R_{L}^{z'x}$  &  $\hspace*{1pt}^{c} R^{x'x}_{LT}$  &  $\hspace*{1pt}^{s} R^{z'x}_{LT'}$   \\
      & $\Ocal^{b}_{2+}$ & $\Ocal^{b}_{2-}$ & $\Ocal^{c}_{2+}$  &  $\Ocal^{c}_{2-}$ & $\Ocal^{d}_{1}$ & $\Ocal^{h}_{2+}$ & $\Ocal^{h}_{2-}$  \\
    A.iii.3  &  $\hspace*{1pt}^{s} R^{z'0}_{TT}$ & $R^{x'0}_{TT'}$  &  $R^{x'z}_{T}$ & $R^{z'x}_{T}$ & $R_{L}^{x'x}$  &  $\hspace*{1pt}^{c} R^{x'x}_{LT}$  &  $\hspace*{1pt}^{s} R^{z'x}_{LT'}$  \\
      &  $\Ocal^{b}_{1+}$ & $\Ocal^{b}_{1-}$ & $\Ocal^{c}_{1+}$  &  $\Ocal^{c}_{1-}$ & $\Ocal^{d}_{2}$ & $\Ocal^{h}_{2+}$ & $\Ocal^{h}_{2-}$  \\
    A.iii.4  &  $R^{z'0}_{TT'}$  &  $\hspace*{1pt}^{s} R^{x'0}_{TT}$ & $R^{x'z}_{T}$  &  $R^{z'x}_{T}$ & $R_{L}^{x'x}$  &  $\hspace*{1pt}^{s} R^{x'x}_{LT'}$  &  $\hspace*{1pt}^{c} R^{z'x}_{LT}$   \\
      & $\Ocal^{b}_{2+}$ & $\Ocal^{b}_{2-}$ & $\Ocal^{c}_{1+}$  &  $\Ocal^{c}_{1-}$ & $\Ocal^{d}_{2}$ & $\Ocal^{h}_{1+}$ & $\Ocal^{h}_{1-}$  \\
     A.iii.5  &  $\hspace*{1pt}^{s} R^{z'0}_{TT}$ & $R^{x'0}_{TT'}$  &  $R_{T}^{z' z}$ & $R_{T}^{x' x}$ & $R_{L}^{z'x}$  &  $\hspace*{1pt}^{c} R^{x'x}_{LT}$  &  $\hspace*{1pt}^{s} R^{z'x}_{LT'}$   \\
      & $\Ocal^{b}_{1+}$ & $\Ocal^{b}_{1-}$ & $\Ocal^{c}_{2+}$  &  $\Ocal^{c}_{2-}$ & $\Ocal^{d}_{1}$ & $\Ocal^{h}_{2+}$ & $\Ocal^{h}_{2-}$  \\
     A.iii.6  &  $R^{z'0}_{TT'}$  &  $\hspace*{1pt}^{s} R^{x'0}_{TT}$ & $R_{T}^{z' z}$  &  $R_{T}^{x' x}$ & $R_{L}^{z'x}$  &  $\hspace*{1pt}^{s} R^{x'x}_{LT'}$  &  $\hspace*{1pt}^{c} R^{z'x}_{LT}$   \\
      & $\Ocal^{b}_{2+}$ & $\Ocal^{b}_{2-}$ & $\Ocal^{c}_{2+}$  &  $\Ocal^{c}_{2-}$ & $\Ocal^{d}_{1}$ & $\Ocal^{h}_{1+}$ & $\Ocal^{h}_{1-}$  \\
    A.iii.7  &  $\hspace*{1pt}^{s} R^{z'0}_{TT}$ & $R^{x'0}_{TT'}$  &  $R^{x'z}_{T}$ & $R^{z'x}_{T}$ & $R_{L}^{x'x}$  &  $\hspace*{1pt}^{s} R^{x'x}_{LT'}$  &  $\hspace*{1pt}^{c} R^{z'x}_{LT}$  \\
       & $\Ocal^{b}_{1+}$ & $\Ocal^{b}_{1-}$ & $\Ocal^{c}_{1+}$  &  $\Ocal^{c}_{1-}$ & $\Ocal^{d}_{2}$ & $\Ocal^{h}_{1+}$ & $\Ocal^{h}_{1-}$  \\
      A.iii.8  &  $\hspace*{1pt}^{s} R^{z'0}_{TT}$ & $R^{x'0}_{TT'}$  &  $R_{T}^{z' z}$ & $R_{T}^{x' x}$ & $R_{L}^{z'x}$  &  $\hspace*{1pt}^{s} R^{x'x}_{LT'}$  &  $\hspace*{1pt}^{c} R^{z'x}_{LT}$  \\
      & $\Ocal^{b}_{1+}$ & $\Ocal^{b}_{1-}$ & $\Ocal^{c}_{2+}$  &  $\Ocal^{c}_{2-}$ & $\Ocal^{d}_{1}$ & $\Ocal^{h}_{1+}$ & $\Ocal^{h}_{1-}$ \\
    \hline
    A.iv.1  &  $R^{z'0}_{TT'}$  &  $\hspace*{1pt}^{s} R^{x'0}_{TT}$ & $R^{x'z}_{T}$  &  $R^{z'x}_{T}$ & $R_{L}^{x'x}$  &  $\hspace*{1pt}^{c} R^{00}_{LT}$  &  $\hspace*{1pt}^{c} R^{0y}_{LT}$   \\
       & $\Ocal^{b}_{2+}$ & $\Ocal^{b}_{2-}$ & $\Ocal^{c}_{1+}$  &  $\Ocal^{c}_{1-}$ & $\Ocal^{d}_{2}$ & $\Ocal^{e}_{2+}$ & $\Ocal^{e}_{2-}$    \\
     A.iv.2   &  $R^{z'0}_{TT'}$  &  $\hspace*{1pt}^{s} R^{x'0}_{TT}$ & $R_{T}^{z' z}$  &  $R_{T}^{x' x}$ & $R_{L}^{z'x}$  &  $\hspace*{1pt}^{c} R^{00}_{LT}$  &  $\hspace*{1pt}^{c} R^{0y}_{LT}$  \\
    & $\Ocal^{b}_{2+}$ & $\Ocal^{b}_{2-}$ & $\Ocal^{c}_{2+}$  &  $\Ocal^{c}_{2-}$ & $\Ocal^{d}_{1}$ & $\Ocal^{e}_{2+}$ & $\Ocal^{e}_{2-}$   \\
     A.iv.3   &  $\hspace*{1pt}^{s} R^{z'0}_{TT}$ & $R^{x'0}_{TT'}$  &  $R^{x'z}_{T}$ & $R^{z'x}_{T}$ & $R_{L}^{x'x}$  &  $\hspace*{1pt}^{c} R^{00}_{LT}$  &  $\hspace*{1pt}^{c} R^{0y}_{LT}$   \\
     & $\Ocal^{b}_{1+}$ & $\Ocal^{b}_{1-}$ & $\Ocal^{c}_{1+}$  &  $\Ocal^{c}_{1-}$ & $\Ocal^{d}_{2}$ & $\Ocal^{e}_{2+}$ & $\Ocal^{e}_{2-}$  \\ 
     A.iv.4   &  $R^{z'0}_{TT'}$  &  $\hspace*{1pt}^{s} R^{x'0}_{TT}$ & $R^{x'z}_{T}$  &  $R^{z'x}_{T}$ & $R_{L}^{x'x}$  &  $\hspace*{1pt}^{s} R^{00}_{LT'}$  &  $\hspace*{1pt}^{s} R^{0y}_{LT'}$ \\
      & $\Ocal^{b}_{2+}$ & $\Ocal^{b}_{2-}$ & $\Ocal^{c}_{1+}$  &  $\Ocal^{c}_{1-}$ & $\Ocal^{d}_{2}$ & $\Ocal^{e}_{1+}$ & $\Ocal^{e}_{1-}$ \\
    A.iv.5   &  $\hspace*{1pt}^{s} R^{z'0}_{TT}$ & $R^{x'0}_{TT'}$  &  $R_{T}^{z' z}$ & $R_{T}^{x' x}$ & $R_{L}^{z'x}$  &  $\hspace*{1pt}^{c} R^{00}_{LT}$  &  $\hspace*{1pt}^{c} R^{0y}_{LT}$  \\
   & $\Ocal^{b}_{1+}$ & $\Ocal^{b}_{1-}$ & $\Ocal^{c}_{2+}$  &  $\Ocal^{c}_{2-}$ & $\Ocal^{d}_{1}$ & $\Ocal^{e}_{2+}$ & $\Ocal^{e}_{2-}$  \\
    A.iv.6  &  $R^{z'0}_{TT'}$  &  $\hspace*{1pt}^{s} R^{x'0}_{TT}$ & $R_{T}^{z' z}$  &  $R_{T}^{x' x}$ & $R_{L}^{z'x}$  &  $\hspace*{1pt}^{s} R^{00}_{LT'}$  &  $\hspace*{1pt}^{s} R^{0y}_{LT'}$ \\
    & $\Ocal^{b}_{2+}$ & $\Ocal^{b}_{2-}$ & $\Ocal^{c}_{2+}$  &  $\Ocal^{c}_{2-}$ & $\Ocal^{d}_{1}$ & $\Ocal^{e}_{1+}$ & $\Ocal^{e}_{1-}$ \\
     A.iv.7   &  $\hspace*{1pt}^{s} R^{z'0}_{TT}$ & $R^{x'0}_{TT'}$  &  $R^{x'z}_{T}$ & $R^{z'x}_{T}$ & $R_{L}^{x'x}$  &  $\hspace*{1pt}^{s} R^{00}_{LT'}$  &  $\hspace*{1pt}^{s} R^{0y}_{LT'}$ \\
    & $\Ocal^{b}_{1+}$ & $\Ocal^{b}_{1-}$ & $\Ocal^{c}_{1+}$  &  $\Ocal^{c}_{1-}$ & $\Ocal^{d}_{2}$ & $\Ocal^{e}_{1+}$ & $\Ocal^{e}_{1-}$  \\
    A.iv.8   &  $\hspace*{1pt}^{s} R^{z'0}_{TT}$ & $R^{x'0}_{TT'}$  &  $R_{T}^{z' z}$ & $R_{T}^{x' x}$ & $R_{L}^{z'x}$  &  $\hspace*{1pt}^{s} R^{00}_{LT'}$  &  $\hspace*{1pt}^{s} R^{0y}_{LT'}$ \\ 
    & $\Ocal^{b}_{1+}$ & $\Ocal^{b}_{1-}$ & $\Ocal^{c}_{2+}$  &  $\Ocal^{c}_{2-}$ & $\Ocal^{d}_{1}$ & $\Ocal^{e}_{1+}$ & $\Ocal^{e}_{1-}$  \\
    \hline
      A.v.1   &  $R_{TT'}^{0z}$  &  $\hspace*{1pt}^{s} R^{0x}_{TT}$ & $R^{x'z}_{T}$  &  $R^{z'x}_{T}$ & $R_{L}^{x'x}$  &  $\hspace*{1pt}^{c} R^{z'0}_{LT'}$  &  $\hspace*{1pt}^{s} R^{x'0}_{LT}$  \\ 
    & $\Ocal^{a}_{2+}$ & $\Ocal^{a}_{2-}$ & $\Ocal^{c}_{1+}$  &  $\Ocal^{c}_{1-}$ & $\Ocal^{d}_{2}$ & $\Ocal^{g}_{2+}$ & $\Ocal^{g}_{2-}$   \\ 
   A.v.2   &  $R_{TT'}^{0z}$  &  $\hspace*{1pt}^{s} R^{0x}_{TT}$ & $R_{T}^{z' z}$  &  $R_{T}^{x' x}$ & $R_{L}^{z'x}$  &  $\hspace*{1pt}^{c} R^{z'0}_{LT'}$  &  $\hspace*{1pt}^{s} R^{x'0}_{LT}$   \\ 
   & $\Ocal^{a}_{2+}$ & $\Ocal^{a}_{2-}$ & $\Ocal^{c}_{2+}$  &  $\Ocal^{c}_{2-}$ & $\Ocal^{d}_{1}$ & $\Ocal^{g}_{2+}$ & $\Ocal^{g}_{2-}$   \\
   A.v.3   &  $\hspace*{1pt}^{s} R_{TT}^{0z}$ & $R_{TT'}^{0x}$  &  $R^{x'z}_{T}$ & $R^{z'x}_{T}$ & $R_{L}^{x'x}$  &  $\hspace*{1pt}^{c} R^{z'0}_{LT'}$  &  $\hspace*{1pt}^{s} R^{x'0}_{LT}$   \\ 
    & $\Ocal^{a}_{1+}$ & $\Ocal^{a}_{1-}$ & $\Ocal^{c}_{1+}$  &  $\Ocal^{c}_{1-}$ & $\Ocal^{d}_{2}$ & $\Ocal^{g}_{2+}$ & $\Ocal^{g}_{2-}$  \\ 
   A.v.4  &  $\hspace*{1pt}^{s} R_{TT}^{0z}$ & $R_{TT'}^{0x}$  &  $R_{T}^{z' z}$ & $R_{T}^{x' x}$ & $R_{L}^{z'x}$  &  $\hspace*{1pt}^{c} R^{z'0}_{LT'}$  &  $\hspace*{1pt}^{s} R^{x'0}_{LT}$  \\ 
    & $\Ocal^{a}_{1+}$ & $\Ocal^{a}_{1-}$ & $\Ocal^{c}_{2+}$  &  $\Ocal^{c}_{2-}$ & $\Ocal^{d}_{1}$ & $\Ocal^{g}_{2+}$ & $\Ocal^{g}_{2-}$  \\
   A.v.5   &  $R_{TT'}^{0z}$  &  $\hspace*{1pt}^{s} R^{0x}_{TT}$ & $R^{x'z}_{T}$  &  $R^{z'x}_{T}$ & $R_{L}^{x'x}$  &  $\hspace*{1pt}^{s} R^{z'0}_{LT}$  &  $\hspace*{1pt}^{c} R^{x'0}_{LT'}$   \\ 
    & $\Ocal^{a}_{2+}$ & $\Ocal^{a}_{2-}$ & $\Ocal^{c}_{1+}$  &  $\Ocal^{c}_{1-}$ & $\Ocal^{d}_{2}$ & $\Ocal^{g}_{1+}$ & $\Ocal^{g}_{1-}$   \\ 
   A.v.6   &  $R_{TT'}^{0z}$  &  $\hspace*{1pt}^{s} R^{0x}_{TT}$ & $R_{T}^{z' z}$  &  $R_{T}^{x' x}$ & $R_{L}^{z'x}$  &  $\hspace*{1pt}^{s} R^{z'0}_{LT}$  &  $\hspace*{1pt}^{c} R^{x'0}_{LT'}$  \\
    & $\Ocal^{a}_{2+}$ & $\Ocal^{a}_{2-}$ & $\Ocal^{c}_{2+}$  &  $\Ocal^{c}_{2-}$ & $\Ocal^{d}_{1}$ & $\Ocal^{g}_{1+}$ & $\Ocal^{g}_{1-}$  \\ 
   A.v.7  &  $\hspace*{1pt}^{s} R_{TT}^{0z}$ & $R_{TT'}^{0x}$  &  $R^{x'z}_{T}$ & $R^{z'x}_{T}$ & $R_{L}^{x'x}$  &  $\hspace*{1pt}^{s} R^{z'0}_{LT}$  &  $\hspace*{1pt}^{c} R^{x'0}_{LT'}$  \\
   & $\Ocal^{a}_{1+}$ & $\Ocal^{a}_{1-}$ & $\Ocal^{c}_{1+}$  &  $\Ocal^{c}_{1-}$ & $\Ocal^{d}_{2}$ & $\Ocal^{g}_{1+}$ & $\Ocal^{g}_{1-}$   \\ 
   A.v.8  &  $\hspace*{1pt}^{s} R_{TT}^{0z}$ & $R_{TT'}^{0x}$  &  $R_{T}^{z' z}$ & $R_{T}^{x' x}$ & $R_{L}^{z'x}$  &  $\hspace*{1pt}^{s} R^{z'0}_{LT}$  &  $\hspace*{1pt}^{c} R^{x'0}_{LT'}$  \\
    & $\Ocal^{a}_{1+}$ & $\Ocal^{a}_{1-}$ & $\Ocal^{c}_{2+}$  &  $\Ocal^{c}_{2-}$ & $\Ocal^{d}_{1}$ & $\Ocal^{g}_{1+}$ & $\Ocal^{g}_{1-}$  
\end{tabular}
\caption{Table~\ref{tab:MinimalMoravcsikSetsElectroI} is continued here. The second subset of $24$ from a total of $64$ distinct possibilities is shown. The labelling-scheme for the sets is explained in Table~\ref{tab:MinimalMoravcsikSetsElectroI}.}
\label{tab:MinimalMoravcsikSetsElectroIOneTwo}
\end{table}

\begin{table}
\begin{tabular}{l|ccccccc}
Set-Nr. & \multicolumn{7}{c}{Observables}  \\ 
   \hline
   A.vi.1   &  $R_{TT'}^{0z}$  &  $\hspace*{1pt}^{s} R^{0x}_{TT}$ & $R^{x'z}_{T}$  &  $R^{z'x}_{T}$ & $R_{L}^{x'x}$  &  $\hspace*{1pt}^{c} R^{0z}_{LT'}$  &  $\hspace*{1pt}^{s} R^{0x}_{LT}$   \\
 &  $\Ocal^{a}_{2+}$ & $\Ocal^{a}_{2-}$ & $\Ocal^{c}_{1+}$  &  $\Ocal^{c}_{1-}$ & $\Ocal^{d}_{2}$ & $\Ocal^{f}_{2+}$ & $\Ocal^{f}_{2-}$  \\
 A.vi.2  &  $R_{TT'}^{0z}$  &  $\hspace*{1pt}^{s} R^{0x}_{TT}$ & $R_{T}^{z' z}$  &  $R_{T}^{x' x}$ & $R_{L}^{z'x}$  &  $\hspace*{1pt}^{c} R^{0z}_{LT'}$  &  $\hspace*{1pt}^{s} R^{0x}_{LT}$  \\
  & $\Ocal^{a}_{2+}$ & $\Ocal^{a}_{2-}$ & $\Ocal^{c}_{2+}$  &  $\Ocal^{c}_{2-}$ & $\Ocal^{d}_{1}$ & $\Ocal^{f}_{2+}$ & $\Ocal^{f}_{2-}$  \\ 
 A.vi.3  &  $\hspace*{1pt}^{s} R_{TT}^{0z}$ & $R_{TT'}^{0x}$  &  $R^{x'z}_{T}$ & $R^{z'x}_{T}$ & $R_{L}^{x'x}$  &  $\hspace*{1pt}^{c} R^{0z}_{LT'}$  &  $\hspace*{1pt}^{s} R^{0x}_{LT}$  \\
  & $\Ocal^{a}_{1+}$ & $\Ocal^{a}_{1-}$ & $\Ocal^{c}_{1+}$  &  $\Ocal^{c}_{1-}$ & $\Ocal^{d}_{2}$ & $\Ocal^{f}_{2+}$ & $\Ocal^{f}_{2-}$    \\
 A.vi.4  &  $\hspace*{1pt}^{s} R_{TT}^{0z}$ & $R_{TT'}^{0x}$  &  $R_{T}^{z' z}$ & $R_{T}^{x' x}$ & $R_{L}^{z'x}$  &  $\hspace*{1pt}^{c} R^{0z}_{LT'}$  &  $\hspace*{1pt}^{s} R^{0x}_{LT}$  \\
 & $\Ocal^{a}_{1+}$ & $\Ocal^{a}_{1-}$ & $\Ocal^{c}_{2+}$  &  $\Ocal^{c}_{2-}$ & $\Ocal^{d}_{1}$ & $\Ocal^{f}_{2+}$ & $\Ocal^{f}_{2-}$    \\
A.vi.5  &  $R_{TT'}^{0z}$  &  $\hspace*{1pt}^{s} R^{0x}_{TT}$ & $R^{x'z}_{T}$  &  $R^{z'x}_{T}$ & $R_{L}^{x'x}$  &  $\hspace*{1pt}^{s} R^{0z}_{LT}$  &  $\hspace*{1pt}^{c} R^{0x}_{LT'}$  \\ 
& $\Ocal^{a}_{2+}$ & $\Ocal^{a}_{2-}$ & $\Ocal^{c}_{1+}$  &  $\Ocal^{c}_{1-}$ & $\Ocal^{d}_{2}$ & $\Ocal^{f}_{1+}$ & $\Ocal^{f}_{1-}$   \\  
A.vi.6  &  $R_{TT'}^{0z}$  &  $\hspace*{1pt}^{s} R^{0x}_{TT}$ & $R_{T}^{z' z}$  &  $R_{T}^{x' x}$ & $R_{L}^{z'x}$  &  $\hspace*{1pt}^{s} R^{0z}_{LT}$  &  $\hspace*{1pt}^{c} R^{0x}_{LT'}$  \\ 
& $\Ocal^{a}_{2+}$ & $\Ocal^{a}_{2-}$ & $\Ocal^{c}_{2+}$  &  $\Ocal^{c}_{2-}$ & $\Ocal^{d}_{1}$ & $\Ocal^{f}_{1+}$ & $\Ocal^{f}_{1-}$  \\  
A.vi.7  &  $\hspace*{1pt}^{s} R_{TT}^{0z}$ & $R_{TT'}^{0x}$  &  $R^{x'z}_{T}$ & $R^{z'x}_{T}$ & $R_{L}^{x'x}$  &  $\hspace*{1pt}^{s} R^{0z}_{LT}$  &  $\hspace*{1pt}^{c} R^{0x}_{LT'}$  \\ 
 & $\Ocal^{a}_{1+}$ & $\Ocal^{a}_{1-}$ & $\Ocal^{c}_{1+}$  &  $\Ocal^{c}_{1-}$ & $\Ocal^{d}_{2}$ & $\Ocal^{f}_{1+}$ & $\Ocal^{f}_{1-}$    \\  
 A.vi.8  &  $\hspace*{1pt}^{s} R_{TT}^{0z}$ & $R_{TT'}^{0x}$  &  $R_{T}^{z' z}$ & $R_{T}^{x' x}$ & $R_{L}^{z'x}$  &  $\hspace*{1pt}^{s} R^{0z}_{LT}$  &  $\hspace*{1pt}^{c} R^{0x}_{LT'}$  \\  
 & $\Ocal^{a}_{1+}$ & $\Ocal^{a}_{1-}$ & $\Ocal^{c}_{2+}$  &  $\Ocal^{c}_{2-}$ & $\Ocal^{d}_{1}$ & $\Ocal^{f}_{1+}$ & $\Ocal^{f}_{1-}$  \\
 \hline
    A.vii.1  &  $R^{z'0}_{TT'}$  &  $\hspace*{1pt}^{s} R^{x'0}_{TT}$ & $R^{x'z}_{T}$  &  $R^{z'x}_{T}$ & $R_{L}^{x'x}$  &  $\hspace*{1pt}^{c} R^{z'0}_{LT'}$  &  $\hspace*{1pt}^{s} R^{x'0}_{LT}$  \\ 
      & $\Ocal^{b}_{2+}$ & $\Ocal^{b}_{2-}$ & $\Ocal^{c}_{1+}$  &  $\Ocal^{c}_{1-}$ & $\Ocal^{d}_{2}$ & $\Ocal^{g}_{2+}$ & $\Ocal^{g}_{2-}$  \\ 
    A.vii.2  &  $R^{z'0}_{TT'}$  &  $\hspace*{1pt}^{s} R^{x'0}_{TT}$ & $R_{T}^{z' z}$  &  $R_{T}^{x' x}$ & $R_{L}^{z'x}$  &  $\hspace*{1pt}^{c} R^{z'0}_{LT'}$  &  $\hspace*{1pt}^{s} R^{x'0}_{LT}$  \\ 
    & $\Ocal^{b}_{2+}$ & $\Ocal^{b}_{2-}$ & $\Ocal^{c}_{2+}$  &  $\Ocal^{c}_{2-}$ & $\Ocal^{d}_{1}$ & $\Ocal^{g}_{2+}$ & $\Ocal^{g}_{2-}$ \\
    A.vii.3  &  $\hspace*{1pt}^{s} R^{z'0}_{TT}$ & $R^{x'0}_{TT'}$  &  $R^{x'z}_{T}$ & $R^{z'x}_{T}$ & $R_{L}^{x'x}$  &  $\hspace*{1pt}^{c} R^{z'0}_{LT'}$  &  $\hspace*{1pt}^{s} R^{x'0}_{LT}$  \\ 
     & $\Ocal^{b}_{1+}$ & $\Ocal^{b}_{1-}$ & $\Ocal^{c}_{1+}$  &  $\Ocal^{c}_{1-}$ & $\Ocal^{d}_{2}$ & $\Ocal^{g}_{2+}$ & $\Ocal^{g}_{2-}$  \\ 
    A.vii.4  &  $\hspace*{1pt}^{s} R^{z'0}_{TT}$ & $R^{x'0}_{TT'}$  &  $R_{T}^{z' z}$ & $R_{T}^{x' x}$ & $R_{L}^{z'x}$  &  $\hspace*{1pt}^{c} R^{z'0}_{LT'}$  &  $\hspace*{1pt}^{s} R^{x'0}_{LT}$  \\ 
     & $\Ocal^{b}_{1+}$ & $\Ocal^{b}_{1-}$ & $\Ocal^{c}_{2+}$  &  $\Ocal^{c}_{2-}$ & $\Ocal^{d}_{1}$ & $\Ocal^{g}_{2+}$ & $\Ocal^{g}_{2-}$  \\
     A.vii.5  &  $R^{z'0}_{TT'}$  &  $\hspace*{1pt}^{s} R^{x'0}_{TT}$ & $R^{x'z}_{T}$  &  $R^{z'x}_{T}$ & $R_{L}^{x'x}$  &  $\hspace*{1pt}^{s} R^{z'0}_{LT}$  &  $\hspace*{1pt}^{c} R^{x'0}_{LT'}$  \\ 
     & $\Ocal^{b}_{2+}$ & $\Ocal^{b}_{2-}$ & $\Ocal^{c}_{1+}$  &  $\Ocal^{c}_{1-}$ & $\Ocal^{d}_{2}$ & $\Ocal^{g}_{1+}$ & $\Ocal^{g}_{1-}$  \\ 
     A.vii.6  &  $R^{z'0}_{TT'}$  &  $\hspace*{1pt}^{s} R^{x'0}_{TT}$ & $R_{T}^{z' z}$  &  $R_{T}^{x' x}$ & $R_{L}^{z'x}$  &  $\hspace*{1pt}^{s} R^{z'0}_{LT}$  &  $\hspace*{1pt}^{c} R^{x'0}_{LT'}$  \\
      & $\Ocal^{b}_{2+}$ & $\Ocal^{b}_{2-}$ & $\Ocal^{c}_{2+}$  &  $\Ocal^{c}_{2-}$ & $\Ocal^{d}_{1}$ & $\Ocal^{g}_{1+}$ & $\Ocal^{g}_{1-}$  \\ 
    A.vii.7  &  $\hspace*{1pt}^{s} R^{z'0}_{TT}$ & $R^{x'0}_{TT'}$  &  $R^{x'z}_{T}$ & $R^{z'x}_{T}$ & $R_{L}^{x'x}$  &  $\hspace*{1pt}^{s} R^{z'0}_{LT}$  &  $\hspace*{1pt}^{c} R^{x'0}_{LT'}$  \\
    & $\Ocal^{b}_{1+}$ & $\Ocal^{b}_{1-}$ & $\Ocal^{c}_{1+}$  &  $\Ocal^{c}_{1-}$ & $\Ocal^{d}_{2}$ & $\Ocal^{g}_{1+}$ & $\Ocal^{g}_{1-}$  \\ 
    A.vii.8  &  $\hspace*{1pt}^{s} R^{z'0}_{TT}$ & $R^{x'0}_{TT'}$  &  $R_{T}^{z' z}$ & $R_{T}^{x' x}$ & $R_{L}^{z'x}$  &  $\hspace*{1pt}^{s} R^{z'0}_{LT}$  &  $\hspace*{1pt}^{c} R^{x'0}_{LT'}$  \\
     & $\Ocal^{b}_{1+}$ & $\Ocal^{b}_{1-}$ & $\Ocal^{c}_{2+}$  &  $\Ocal^{c}_{2-}$ & $\Ocal^{d}_{1}$ & $\Ocal^{g}_{1+}$ & $\Ocal^{g}_{1-}$  \\ 
   \hline
    A.viii.1  &  $R^{z'0}_{TT'}$  &  $\hspace*{1pt}^{s} R^{x'0}_{TT}$ & $R^{x'z}_{T}$  &  $R^{z'x}_{T}$ & $R_{L}^{x'x}$  &  $\hspace*{1pt}^{c} R^{0z}_{LT'}$  &  $\hspace*{1pt}^{s} R^{0x}_{LT}$   \\
  & $\Ocal^{b}_{2+}$ & $\Ocal^{b}_{2-}$ & $\Ocal^{c}_{1+}$  &  $\Ocal^{c}_{1-}$ & $\Ocal^{d}_{2}$ & $\Ocal^{f}_{2+}$ & $\Ocal^{f}_{2-}$  \\
  A.viii.2  &  $R^{z'0}_{TT'}$  &  $\hspace*{1pt}^{s} R^{x'0}_{TT}$ & $R_{T}^{z' z}$  &  $R_{T}^{x' x}$ & $R_{L}^{z'x}$  &  $\hspace*{1pt}^{c} R^{0z}_{LT'}$  &  $\hspace*{1pt}^{s} R^{0x}_{LT}$  \\
  & $\Ocal^{b}_{2+}$ & $\Ocal^{b}_{2-}$ & $\Ocal^{c}_{2+}$  &  $\Ocal^{c}_{2-}$ & $\Ocal^{d}_{1}$ & $\Ocal^{f}_{2+}$ & $\Ocal^{f}_{2-}$  \\ 
   A.viii.3  &  $\hspace*{1pt}^{s} R^{z'0}_{TT}$ & $R^{x'0}_{TT'}$  &  $R^{x'z}_{T}$ & $R^{z'x}_{T}$ & $R_{L}^{x'x}$  &  $\hspace*{1pt}^{c} R^{0z}_{LT'}$  &  $\hspace*{1pt}^{s} R^{0x}_{LT}$   \\
   & $\Ocal^{b}_{1+}$ & $\Ocal^{b}_{1-}$ & $\Ocal^{c}_{1+}$  &  $\Ocal^{c}_{1-}$ & $\Ocal^{d}_{2}$ & $\Ocal^{f}_{2+}$ & $\Ocal^{f}_{2-}$    \\
  A.viii.4  &  $\hspace*{1pt}^{s} R^{z'0}_{TT}$ & $R^{x'0}_{TT'}$  &  $R_{T}^{z' z}$ & $R_{T}^{x' x}$ & $R_{L}^{z'x}$  &  $\hspace*{1pt}^{c} R^{0z}_{LT'}$  &  $\hspace*{1pt}^{s} R^{0x}_{LT}$  \\
  & $\Ocal^{b}_{1+}$ & $\Ocal^{b}_{1-}$ & $\Ocal^{c}_{2+}$  &  $\Ocal^{c}_{2-}$ & $\Ocal^{d}_{1}$ & $\Ocal^{f}_{2+}$ & $\Ocal^{f}_{2-}$   \\
 A.viii.5  &  $R^{z'0}_{TT'}$  &  $\hspace*{1pt}^{s} R^{x'0}_{TT}$ & $R^{x'z}_{T}$  &  $R^{z'x}_{T}$ & $R_{L}^{x'x}$  &  $\hspace*{1pt}^{s} R^{0z}_{LT}$  &  $\hspace*{1pt}^{c} R^{0x}_{LT'}$  \\ 
 & $\Ocal^{b}_{2+}$ & $\Ocal^{b}_{2-}$ & $\Ocal^{c}_{1+}$  &  $\Ocal^{c}_{1-}$ & $\Ocal^{d}_{2}$ & $\Ocal^{f}_{1+}$ & $\Ocal^{f}_{1-}$  \\  
 A.viii.6  &  $R^{z'0}_{TT'}$  &  $\hspace*{1pt}^{s} R^{x'0}_{TT}$ & $R_{T}^{z' z}$  &  $R_{T}^{x' x}$ & $R_{L}^{z'x}$  &  $\hspace*{1pt}^{s} R^{0z}_{LT}$  &  $\hspace*{1pt}^{c} R^{0x}_{LT'}$  \\ 
 & $\Ocal^{b}_{2+}$ & $\Ocal^{b}_{2-}$ & $\Ocal^{c}_{2+}$  &  $\Ocal^{c}_{2-}$ & $\Ocal^{d}_{1}$ & $\Ocal^{f}_{1+}$ & $\Ocal^{f}_{1-}$  \\  
  A.viii.7  &  $\hspace*{1pt}^{s} R^{z'0}_{TT}$ & $R^{x'0}_{TT'}$  &  $R^{x'z}_{T}$ & $R^{z'x}_{T}$ & $R_{L}^{x'x}$  &  $\hspace*{1pt}^{s} R^{0z}_{LT}$  &  $\hspace*{1pt}^{c} R^{0x}_{LT'}$   \\ 
  & $\Ocal^{b}_{1+}$ & $\Ocal^{b}_{1-}$ & $\Ocal^{c}_{1+}$  &  $\Ocal^{c}_{1-}$ & $\Ocal^{d}_{2}$ & $\Ocal^{f}_{1+}$ & $\Ocal^{f}_{1-}$   \\  
  A.viii.8  &  $\hspace*{1pt}^{s} R^{z'0}_{TT}$ & $R^{x'0}_{TT'}$  &  $R_{T}^{z' z}$ & $R_{T}^{x' x}$ & $R_{L}^{z'x}$  &  $\hspace*{1pt}^{s} R^{0z}_{LT}$  &  $\hspace*{1pt}^{c} R^{0x}_{LT'}$  \\  
  & $\Ocal^{b}_{1+}$ & $\Ocal^{b}_{1-}$ & $\Ocal^{c}_{2+}$ & $\Ocal^{c}_{2-}$ & $\Ocal^{d}_{1}$ & $\Ocal^{f}_{1+}$ & $\Ocal^{f}_{1-}$  
\end{tabular}
\caption{Table~\ref{tab:MinimalMoravcsikSetsElectroIOneTwo} is continued here. The third subset of $24$ from a total of $64$ distinct possibilities is shown. The labelling-scheme for the sets is explained in Table~\ref{tab:MinimalMoravcsikSetsElectroI}.}
\label{tab:MinimalMoravcsikSetsElectroI2}
\end{table}

\begin{table}
\begin{small}
\begin{tabular}{l|cccccccc}
Set-Nr. & \multicolumn{8}{c}{Observables}  \\
\hline   
   B.i.1  &  $R^{x'z}_{T}$ & $R^{z'x}_{T}$ & $R_{T}^{z' z}$ & $R_{T}^{x' x}$ & $\hspace*{1pt}^{c} R^{z'0}_{LT'}$  &  $\hspace*{1pt}^{s} R^{x'0}_{LT}$ & $\hspace*{1pt}^{c} R^{x'x}_{LT}$  &  $\hspace*{1pt}^{s} R^{z'x}_{LT'}$   \\  
   & $\Ocal^{c}_{1+}$ & $\Ocal^{c}_{1-}$ & $\Ocal^{c}_{2+}$  &  $\Ocal^{c}_{2-}$ & $\Ocal^{g}_{2+}$ & $\Ocal^{g}_{2-}$ & $\Ocal^{h}_{2+}$ & $\Ocal^{h}_{2-}$   \\   
   B.i.2  &  $R_{T}^{z' z}$  &  $R_{T}^{x' x}$ & $\hspace*{1pt}^{s} R^{z'0}_{LT}$  &  $\hspace*{1pt}^{c} R^{x'0}_{LT'}$  &  $\hspace*{1pt}^{c} R^{z'0}_{LT'}$  &  $\hspace*{1pt}^{s} R^{x'0}_{LT}$  &  $\hspace*{1pt}^{c} R^{x'x}_{LT}$  &  $\hspace*{1pt}^{s} R^{z'x}_{LT'}$   \\  
   & $\Ocal^{c}_{2+}$ & $\Ocal^{c}_{2-}$ & $\Ocal^{g}_{1+}$  &  $\Ocal^{g}_{1-}$ & $\Ocal^{g}_{2+}$ & $\Ocal^{g}_{2-}$ & $\Ocal^{h}_{2+}$ & $\Ocal^{h}_{2-}$    \\  
   B.i.3  &  $R_{T}^{z' z}$  &  $R_{T}^{x' x}$ & $\hspace*{1pt}^{c} R^{z'0}_{LT'}$  &  $\hspace*{1pt}^{s} R^{x'0}_{LT}$  &  $\hspace*{1pt}^{s} R^{x'x}_{LT'}$  &  $\hspace*{1pt}^{c} R^{z'x}_{LT}$  &  $\hspace*{1pt}^{c} R^{x'x}_{LT}$  &  $\hspace*{1pt}^{s} R^{z'x}_{LT'}$   \\  
  & $\Ocal^{c}_{2+}$ & $\Ocal^{c}_{2-}$ & $\Ocal^{g}_{2+}$  &  $\Ocal^{g}_{2-}$ & $\Ocal^{h}_{1+}$ & $\Ocal^{h}_{1-}$ & $\Ocal^{h}_{2+}$ & $\Ocal^{h}_{2-}$    \\ 
  B.i.4  &  $R^{x'z}_{T}$  &  $R^{z'x}_{T}$ & $R_{T}^{z' z}$ & $R_{T}^{x' x}$  &  $\hspace*{1pt}^{s} R^{z'0}_{LT}$  &  $\hspace*{1pt}^{c} R^{x'0}_{LT'}$  &  $\hspace*{1pt}^{c} R^{x'x}_{LT}$  &  $\hspace*{1pt}^{s} R^{z'x}_{LT'}$   \\   
  & $\Ocal^{c}_{1+}$ & $\Ocal^{c}_{1-}$ & $\Ocal^{c}_{2+}$  &  $\Ocal^{c}_{2-}$ & $\Ocal^{g}_{1+}$ & $\Ocal^{g}_{1-}$ & $\Ocal^{h}_{2+}$ & $\Ocal^{h}_{2-}$    \\  
  B.i.5  &  $R^{x'z}_{T}$  &  $R^{z'x}_{T}$ & $\hspace*{1pt}^{s} R^{z'0}_{LT}$  &  $\hspace*{1pt}^{c} R^{x'0}_{LT'}$  &  $\hspace*{1pt}^{c} R^{z'0}_{LT'}$  &  $\hspace*{1pt}^{s} R^{x'0}_{LT}$  &  $\hspace*{1pt}^{c} R^{x'x}_{LT}$  &  $\hspace*{1pt}^{s} R^{z'x}_{LT'}$   \\    
  & $\Ocal^{c}_{1+}$ & $\Ocal^{c}_{1-}$ & $\Ocal^{g}_{1+}$  &  $\Ocal^{g}_{1-}$ & $\Ocal^{g}_{2+}$ & $\Ocal^{g}_{2-}$ & $\Ocal^{h}_{2+}$ & $\Ocal^{h}_{2-}$    \\ 
  B.i.6  &  $R^{x'z}_{T}$  &  $R^{z'x}_{T}$ & $\hspace*{1pt}^{c} R^{z'0}_{LT'}$  &  $\hspace*{1pt}^{s} R^{x'0}_{LT}$  &  $\hspace*{1pt}^{s} R^{x'x}_{LT'}$  &  $\hspace*{1pt}^{c} R^{z'x}_{LT}$  &  $\hspace*{1pt}^{c} R^{x'x}_{LT}$  &  $\hspace*{1pt}^{s} R^{z'x}_{LT'}$   \\  
  & $\Ocal^{c}_{1+}$ & $\Ocal^{c}_{1-}$ & $\Ocal^{g}_{2+}$  &  $\Ocal^{g}_{2-}$ & $\Ocal^{h}_{1+}$ & $\Ocal^{h}_{1-}$ & $\Ocal^{h}_{2+}$ & $\Ocal^{h}_{2-}$   \\
  B.i.7  &  $R^{x'z}_{T}$  &  $R^{z'x}_{T}$ & $R_{T}^{z' z}$ & $R_{T}^{x' x}$  &  $\hspace*{1pt}^{c} R^{z'0}_{LT'}$  &  $\hspace*{1pt}^{s} R^{x'0}_{LT}$  &  $\hspace*{1pt}^{s} R^{x'x}_{LT'}$  &  $\hspace*{1pt}^{c} R^{z'x}_{LT}$  \\ 
  & $\Ocal^{c}_{1+}$ & $\Ocal^{c}_{1-}$ & $\Ocal^{c}_{2+}$  &  $\Ocal^{c}_{2-}$ & $\Ocal^{g}_{2+}$ & $\Ocal^{g}_{2-}$ & $\Ocal^{h}_{1+}$ & $\Ocal^{h}_{1-}$   \\
  B.i.8  &  $R_{T}^{z' z}$  &  $R_{T}^{x' x}$ & $\hspace*{1pt}^{s} R^{z'0}_{LT}$  &  $\hspace*{1pt}^{c} R^{x'0}_{LT'}$  &  $\hspace*{1pt}^{s} R^{x'x}_{LT'}$  &  $\hspace*{1pt}^{c} R^{z'x}_{LT}$  &  $\hspace*{1pt}^{c} R^{x'x}_{LT}$  &  $\hspace*{1pt}^{s} R^{z'x}_{LT'}$    \\ 
  & $\Ocal^{c}_{2+}$ & $\Ocal^{c}_{2-}$ & $\Ocal^{g}_{1+}$  &  $\Ocal^{g}_{1-}$ & $\Ocal^{h}_{1+}$ & $\Ocal^{h}_{1-}$ & $\Ocal^{h}_{2+}$ & $\Ocal^{h}_{2-}$    \\
  B.i.9  &  $R_{T}^{z' z}$  &  $R_{T}^{x' x}$ & $\hspace*{1pt}^{s} R^{z'0}_{LT}$  &  $\hspace*{1pt}^{c} R^{x'0}_{LT'}$  &  $\hspace*{1pt}^{c} R^{z'0}_{LT'}$  &  $\hspace*{1pt}^{s} R^{x'0}_{LT}$  &  $\hspace*{1pt}^{s} R^{x'x}_{LT'}$  &  $\hspace*{1pt}^{c} R^{z'x}_{LT}$   \\    
  & $\Ocal^{c}_{2+}$ & $\Ocal^{c}_{2-}$ & $\Ocal^{g}_{1+}$  &  $\Ocal^{g}_{1-}$ & $\Ocal^{g}_{2+}$ & $\Ocal^{g}_{2-}$ & $\Ocal^{h}_{1+}$ & $\Ocal^{h}_{1-}$    \\  
  B.i.10  &  $R^{x'z}_{T}$  &  $R^{z'x}_{T}$ & $\hspace*{1pt}^{s} R^{z'0}_{LT}$  &  $\hspace*{1pt}^{c} R^{x'0}_{LT'}$  &  $\hspace*{1pt}^{s} R^{x'x}_{LT'}$  &  $\hspace*{1pt}^{c} R^{z'x}_{LT}$  &  $\hspace*{1pt}^{c} R^{x'x}_{LT}$  &  $\hspace*{1pt}^{s} R^{z'x}_{LT'}$    \\   
  & $\Ocal^{c}_{1+}$ & $\Ocal^{c}_{1-}$ & $\Ocal^{g}_{1+}$  &  $\Ocal^{g}_{1-}$ & $\Ocal^{h}_{1+}$ & $\Ocal^{h}_{1-}$ & $\Ocal^{h}_{2+}$ & $\Ocal^{h}_{2-}$    \\ 
  B.i.11  &  $R^{x'z}_{T}$  &  $R^{z'x}_{T}$ & $R_{T}^{z' z}$ & $R_{T}^{x' x}$  &  $\hspace*{1pt}^{s} R^{z'0}_{LT}$  &  $\hspace*{1pt}^{c} R^{x'0}_{LT'}$  &  $\hspace*{1pt}^{s} R^{x'x}_{LT'}$  &  $\hspace*{1pt}^{c} R^{z'x}_{LT}$  \\ 
  & $\Ocal^{c}_{1+}$ & $\Ocal^{c}_{1-}$ & $\Ocal^{c}_{2+}$  &  $\Ocal^{c}_{2-}$ & $\Ocal^{g}_{1+}$ & $\Ocal^{g}_{1-}$ & $\Ocal^{h}_{1+}$ & $\Ocal^{h}_{1-}$    \\ 
  B.i.12  &  $R^{x'z}_{T}$  &  $R^{z'x}_{T}$ & $\hspace*{1pt}^{s} R^{z'0}_{LT}$  &  $\hspace*{1pt}^{c} R^{x'0}_{LT'}$  &  $\hspace*{1pt}^{c} R^{z'0}_{LT'}$  &  $\hspace*{1pt}^{s} R^{x'0}_{LT}$  &  $\hspace*{1pt}^{s} R^{x'x}_{LT'}$  &  $\hspace*{1pt}^{c} R^{z'x}_{LT}$   \\
  & $\Ocal^{c}_{1+}$ & $\Ocal^{c}_{1-}$ & $\Ocal^{g}_{1+}$  &  $\Ocal^{g}_{1-}$ & $\Ocal^{g}_{2+}$ & $\Ocal^{g}_{2-}$ & $\Ocal^{h}_{1+}$ & $\Ocal^{h}_{1-}$   \\
  \hline
  B.ii.1  &  $R^{x'z}_{T}$  &  $R^{z'x}_{T}$ & $R_{T}^{z' z}$ & $R_{T}^{x' x}$  &  $\hspace*{1pt}^{c} R^{00}_{LT}$  &  $\hspace*{1pt}^{c} R^{0y}_{LT}$  &  $\hspace*{1pt}^{c} R^{z'0}_{LT'}$  &  $\hspace*{1pt}^{s} R^{x'0}_{LT}$   \\   
  & $\Ocal^{c}_{1+}$ & $\Ocal^{c}_{1-}$ & $\Ocal^{c}_{2+}$  &  $\Ocal^{c}_{2-}$ & $\Ocal^{e}_{2+}$ & $\Ocal^{e}_{2-}$ & $\Ocal^{g}_{2+}$ & $\Ocal^{g}_{2-}$    \\ 
  B.ii.2  &  $R_{T}^{z' z}$  &  $R_{T}^{x' x}$ & $\hspace*{1pt}^{c} R^{00}_{LT}$  &  $\hspace*{1pt}^{c} R^{0y}_{LT}$  &  $\hspace*{1pt}^{s} R^{z'0}_{LT}$  &  $\hspace*{1pt}^{c} R^{x'0}_{LT'}$  &  $\hspace*{1pt}^{c} R^{z'0}_{LT'}$  &  $\hspace*{1pt}^{s} R^{x'0}_{LT}$    \\   
  & $\Ocal^{c}_{2+}$ & $\Ocal^{c}_{2-}$ & $\Ocal^{e}_{2+}$  &  $\Ocal^{e}_{2-}$ & $\Ocal^{g}_{1+}$ & $\Ocal^{g}_{1-}$ & $\Ocal^{g}_{2+}$ & $\Ocal^{g}_{2-}$   \\  
  B.ii.3  &  $R_{T}^{z' z}$  &  $R_{T}^{x' x}$ & $\hspace*{1pt}^{s} R^{00}_{LT'}$  &  $\hspace*{1pt}^{s} R^{0y}_{LT'}$  &  $\hspace*{1pt}^{c} R^{00}_{LT}$  &  $\hspace*{1pt}^{c} R^{0y}_{LT}$  &  $\hspace*{1pt}^{c} R^{z'0}_{LT'}$  &  $\hspace*{1pt}^{s} R^{x'0}_{LT}$    \\  
  & $\Ocal^{c}_{2+}$ & $\Ocal^{c}_{2-}$ & $\Ocal^{e}_{1+}$  &  $\Ocal^{e}_{1-}$ & $\Ocal^{e}_{2+}$ & $\Ocal^{e}_{2-}$ & $\Ocal^{g}_{2+}$ & $\Ocal^{g}_{2-}$  \\  
  B.ii.4  &  $R^{x'z}_{T}$  &  $R^{z'x}_{T}$ & $R_{T}^{z' z}$ & $R_{T}^{x' x}$  &  $\hspace*{1pt}^{c} R^{00}_{LT}$  &  $\hspace*{1pt}^{c} R^{0y}_{LT}$  &  $\hspace*{1pt}^{s} R^{z'0}_{LT}$  &  $\hspace*{1pt}^{c} R^{x'0}_{LT'}$   \\  
  & $\Ocal^{c}_{1+}$ & $\Ocal^{c}_{1-}$ & $\Ocal^{c}_{2+}$  &  $\Ocal^{c}_{2-}$ & $\Ocal^{e}_{2+}$ & $\Ocal^{e}_{2-}$ & $\Ocal^{g}_{1+}$ & $\Ocal^{g}_{1-}$    \\  
  B.ii.5  &  $R^{x'z}_{T}$  &  $R^{z'x}_{T}$ & $\hspace*{1pt}^{c} R^{00}_{LT}$  &  $\hspace*{1pt}^{c} R^{0y}_{LT}$  &  $\hspace*{1pt}^{s} R^{z'0}_{LT}$  &  $\hspace*{1pt}^{c} R^{x'0}_{LT'}$  &  $\hspace*{1pt}^{c} R^{z'0}_{LT'}$  &  $\hspace*{1pt}^{s} R^{x'0}_{LT}$   \\   
  & $\Ocal^{c}_{1+}$ & $\Ocal^{c}_{1-}$ & $\Ocal^{e}_{2+}$  &  $\Ocal^{e}_{2-}$ & $\Ocal^{g}_{1+}$ & $\Ocal^{g}_{1-}$ & $\Ocal^{g}_{2+}$ & $\Ocal^{g}_{2-}$   \\    
  B.ii.6  &  $R^{x'z}_{T}$  &  $R^{z'x}_{T}$ & $\hspace*{1pt}^{s} R^{00}_{LT'}$  &  $\hspace*{1pt}^{s} R^{0y}_{LT'}$  &  $\hspace*{1pt}^{c} R^{00}_{LT}$  &  $\hspace*{1pt}^{c} R^{0y}_{LT}$  &  $\hspace*{1pt}^{c} R^{z'0}_{LT'}$  &  $\hspace*{1pt}^{s} R^{x'0}_{LT}$   \\    
  & $\Ocal^{c}_{1+}$ & $\Ocal^{c}_{1-}$ & $\Ocal^{e}_{1+}$  &  $\Ocal^{e}_{1-}$ & $\Ocal^{e}_{2+}$ & $\Ocal^{e}_{2-}$ & $\Ocal^{g}_{2+}$ & $\Ocal^{g}_{2-}$   \\  
  B.ii.7  &  $R^{x'z}_{T}$  &  $R^{z'x}_{T}$ & $R_{T}^{z' z}$ & $R_{T}^{x' x}$  &  $\hspace*{1pt}^{s} R^{00}_{LT'}$  &  $\hspace*{1pt}^{s} R^{0y}_{LT'}$  &  $\hspace*{1pt}^{c} R^{z'0}_{LT'}$  &  $\hspace*{1pt}^{s} R^{x'0}_{LT}$   \\   
  & $\Ocal^{c}_{1+}$ & $\Ocal^{c}_{1-}$ & $\Ocal^{c}_{2+}$  &  $\Ocal^{c}_{2-}$ & $\Ocal^{e}_{1+}$ & $\Ocal^{e}_{1-}$ & $\Ocal^{g}_{2+}$ & $\Ocal^{g}_{2-}$   \\
  B.ii.8  &  $R_{T}^{z' z}$  &  $R_{T}^{x' x}$ & $\hspace*{1pt}^{s} R^{00}_{LT'}$  &  $\hspace*{1pt}^{s} R^{0y}_{LT'}$  &  $\hspace*{1pt}^{c} R^{00}_{LT}$  &  $\hspace*{1pt}^{c} R^{0y}_{LT}$  &  $\hspace*{1pt}^{s} R^{z'0}_{LT}$  &  $\hspace*{1pt}^{c} R^{x'0}_{LT'}$   \\    
  & $\Ocal^{c}_{2+}$ & $\Ocal^{c}_{2-}$ & $\Ocal^{e}_{1+}$  &  $\Ocal^{e}_{1-}$ & $\Ocal^{e}_{2+}$ & $\Ocal^{e}_{2-}$ & $\Ocal^{g}_{1+}$ & $\Ocal^{g}_{1-}$   \\    
  B.ii.9  &  $R_{T}^{z' z}$  &  $R_{T}^{x' x}$ & $\hspace*{1pt}^{s} R^{00}_{LT'}$  &  $\hspace*{1pt}^{s} R^{0y}_{LT'}$  &  $\hspace*{1pt}^{s} R^{z'0}_{LT}$  &  $\hspace*{1pt}^{c} R^{x'0}_{LT'}$  &  $\hspace*{1pt}^{c} R^{z'0}_{LT'}$  &  $\hspace*{1pt}^{s} R^{x'0}_{LT}$   \\   
  & $\Ocal^{c}_{2+}$ & $\Ocal^{c}_{2-}$ & $\Ocal^{e}_{1+}$  &  $\Ocal^{e}_{1-}$ & $\Ocal^{g}_{1+}$ & $\Ocal^{g}_{1-}$ & $\Ocal^{g}_{2+}$ & $\Ocal^{g}_{2-}$   \\   
  B.ii.10  &  $R^{x'z}_{T}$  &  $R^{z'x}_{T}$ & $\hspace*{1pt}^{s} R^{00}_{LT'}$  &  $\hspace*{1pt}^{s} R^{0y}_{LT'}$  &  $\hspace*{1pt}^{c} R^{00}_{LT}$  &  $\hspace*{1pt}^{c} R^{0y}_{LT}$  &  $\hspace*{1pt}^{s} R^{z'0}_{LT}$  &  $\hspace*{1pt}^{c} R^{x'0}_{LT'}$  \\  
  & $\Ocal^{c}_{1+}$ & $\Ocal^{c}_{1-}$ & $\Ocal^{e}_{1+}$  &  $\Ocal^{e}_{1-}$ & $\Ocal^{e}_{2+}$ & $\Ocal^{e}_{2-}$ & $\Ocal^{g}_{1+}$ & $\Ocal^{g}_{1-}$   \\ 
  B.ii.11  &  $R^{x'z}_{T}$  &  $R^{z'x}_{T}$ & $R_{T}^{z' z}$ & $R_{T}^{x' x}$  &  $\hspace*{1pt}^{s} R^{00}_{LT'}$  &  $\hspace*{1pt}^{s} R^{0y}_{LT'}$  &  $\hspace*{1pt}^{s} R^{z'0}_{LT}$  &  $\hspace*{1pt}^{c} R^{x'0}_{LT'}$   \\  
  & $\Ocal^{c}_{1+}$ & $\Ocal^{c}_{1-}$ & $\Ocal^{c}_{2+}$  &  $\Ocal^{c}_{2-}$ & $\Ocal^{e}_{1+}$ & $\Ocal^{e}_{1-}$ & $\Ocal^{g}_{1+}$ & $\Ocal^{g}_{1-}$   \\
  B.ii.12  &  $R^{x'z}_{T}$  &  $R^{z'x}_{T}$ & $\hspace*{1pt}^{s} R^{00}_{LT'}$  &  $\hspace*{1pt}^{s} R^{0y}_{LT'}$  &  $\hspace*{1pt}^{s} R^{z'0}_{LT}$  &  $\hspace*{1pt}^{c} R^{x'0}_{LT'}$  &  $\hspace*{1pt}^{c} R^{z'0}_{LT'}$  &  $\hspace*{1pt}^{s} R^{x'0}_{LT}$   \\   
  & $\Ocal^{c}_{1+}$ & $\Ocal^{c}_{1-}$ & $\Ocal^{e}_{1+}$  &  $\Ocal^{e}_{1-}$ & $\Ocal^{g}_{1+}$ & $\Ocal^{g}_{1-}$ & $\Ocal^{g}_{2+}$ & $\Ocal^{g}_{2-}$  
\end{tabular}
\end{small}
\caption{The first $24$ cases from a total of $96$ distinct possibilities to form Moravcsik-complete sets composed of $14$ ob\-serva\-bles for electroproduction (set-numbers starting with a 'B'), are listed here (cf. section~\ref{sec:Electroproduction}).  In each case, the $8$ given ob\-serva\-bles have to be combined with the $6$ diagonal ob\-serva\-bles for electroproduction (cf. Table~\ref{tab:ElectroObservablesI}). $R^{\beta \alpha}_{i}$ is the physical notation, $\Ocal^{n}_{\nu \pm}$ the systematic mathematical notation.}
\label{tab:MinimalMoravcsikSetsElectroII}
\end{table}

\begin{table}
\begin{small}
\begin{tabular}{l|cccccccc}
Set-Nr. & \multicolumn{8}{c}{Observables}  \\
\hline   
     B.iii.1  &  $R^{x'z}_{T}$  &  $R^{z'x}_{T}$ & $R_{T}^{z' z}$ & $R_{T}^{x' x}$  &  $\hspace*{1pt}^{c} R^{0z}_{LT'}$  &  $\hspace*{1pt}^{s} R^{0x}_{LT}$  &  $\hspace*{1pt}^{c} R^{x'x}_{LT}$  &  $\hspace*{1pt}^{s} R^{z'x}_{LT'}$  \\  
     & $\Ocal^{c}_{1+}$ & $\Ocal^{c}_{1-}$ & $\Ocal^{c}_{2+}$  &  $\Ocal^{c}_{2-}$ & $\Ocal^{f}_{2+}$ & $\Ocal^{f}_{2-}$ & $\Ocal^{h}_{2+}$ & $\Ocal^{h}_{2-}$ \\   
     B.iii.2  &  $R_{T}^{z' z}$  &  $R_{T}^{x' x}$ & $\hspace*{1pt}^{s} R^{0z}_{LT}$  &  $\hspace*{1pt}^{c} R^{0x}_{LT'}$  &  $\hspace*{1pt}^{c} R^{0z}_{LT'}$  &  $\hspace*{1pt}^{s} R^{0x}_{LT}$  &  $\hspace*{1pt}^{c} R^{x'x}_{LT}$  &  $\hspace*{1pt}^{s} R^{z'x}_{LT'}$  \\  
     & $\Ocal^{c}_{2+}$ & $\Ocal^{c}_{2-}$ & $\Ocal^{f}_{1+}$  &  $\Ocal^{f}_{1-}$ & $\Ocal^{f}_{2+}$ & $\Ocal^{f}_{2-}$ & $\Ocal^{h}_{2+}$ & $\Ocal^{h}_{2-}$  \\  
     B.iii.3  &  $R_{T}^{z' z}$  &  $R_{T}^{x' x}$ & $\hspace*{1pt}^{c} R^{0z}_{LT'}$  &  $\hspace*{1pt}^{s} R^{0x}_{LT}$  &  $\hspace*{1pt}^{s} R^{x'x}_{LT'}$  &  $\hspace*{1pt}^{c} R^{z'x}_{LT}$  &  $\hspace*{1pt}^{c} R^{x'x}_{LT}$  &  $\hspace*{1pt}^{s} R^{z'x}_{LT'}$  \\  
    & $\Ocal^{c}_{2+}$ & $\Ocal^{c}_{2-}$ & $\Ocal^{f}_{2+}$  &  $\Ocal^{f}_{2-}$ & $\Ocal^{h}_{1+}$ & $\Ocal^{h}_{1-}$ & $\Ocal^{h}_{2+}$ & $\Ocal^{h}_{2-}$  \\ 
    B.iii.4  &  $R^{x'z}_{T}$  &  $R^{z'x}_{T}$ & $R_{T}^{z' z}$ & $R_{T}^{x' x}$  &  $\hspace*{1pt}^{s} R^{0z}_{LT}$  &  $\hspace*{1pt}^{c} R^{0x}_{LT'}$  &  $\hspace*{1pt}^{c} R^{x'x}_{LT}$  &  $\hspace*{1pt}^{s} R^{z'x}_{LT'}$  \\   
     & $\Ocal^{c}_{1+}$ & $\Ocal^{c}_{1-}$ & $\Ocal^{c}_{2+}$  &  $\Ocal^{c}_{2-}$ & $\Ocal^{f}_{1+}$ & $\Ocal^{f}_{1-}$ & $\Ocal^{h}_{2+}$ & $\Ocal^{h}_{2-}$  \\  
    B.iii.5  &  $R^{x'z}_{T}$  &  $R^{z'x}_{T}$ & $\hspace*{1pt}^{s} R^{0z}_{LT}$  &  $\hspace*{1pt}^{c} R^{0x}_{LT'}$  &  $\hspace*{1pt}^{c} R^{0z}_{LT'}$  &  $\hspace*{1pt}^{s} R^{0x}_{LT}$  &  $\hspace*{1pt}^{c} R^{x'x}_{LT}$  &  $\hspace*{1pt}^{s} R^{z'x}_{LT'}$  \\    
    & $\Ocal^{c}_{1+}$ & $\Ocal^{c}_{1-}$ & $\Ocal^{f}_{1+}$  &  $\Ocal^{f}_{1-}$ & $\Ocal^{f}_{2+}$ & $\Ocal^{f}_{2-}$ & $\Ocal^{h}_{2+}$ & $\Ocal^{h}_{2-}$  \\ 
   B.iii.6  &  $R^{x'z}_{T}$  &  $R^{z'x}_{T}$ & $\hspace*{1pt}^{c} R^{0z}_{LT'}$  &  $\hspace*{1pt}^{s} R^{0x}_{LT}$  &  $\hspace*{1pt}^{s} R^{x'x}_{LT'}$  &  $\hspace*{1pt}^{c} R^{z'x}_{LT}$  &  $\hspace*{1pt}^{c} R^{x'x}_{LT}$  &  $\hspace*{1pt}^{s} R^{z'x}_{LT'}$  \\  
    & $\Ocal^{c}_{1+}$ & $\Ocal^{c}_{1-}$ & $\Ocal^{f}_{2+}$  &  $\Ocal^{f}_{2-}$ & $\Ocal^{h}_{1+}$ & $\Ocal^{h}_{1-}$ & $\Ocal^{h}_{2+}$ & $\Ocal^{h}_{2-}$  \\
   B.iii.7  &  $R^{x'z}_{T}$  &  $R^{z'x}_{T}$ & $R_{T}^{z' z}$ & $R_{T}^{x' x}$  &  $\hspace*{1pt}^{c} R^{0z}_{LT'}$  &  $\hspace*{1pt}^{s} R^{0x}_{LT}$  &  $\hspace*{1pt}^{s} R^{x'x}_{LT'}$  &  $\hspace*{1pt}^{c} R^{z'x}_{LT}$  \\ 
   & $\Ocal^{c}_{1+}$ & $\Ocal^{c}_{1-}$ & $\Ocal^{c}_{2+}$  &  $\Ocal^{c}_{2-}$ & $\Ocal^{f}_{2+}$ & $\Ocal^{f}_{2-}$ & $\Ocal^{h}_{1+}$ & $\Ocal^{h}_{1-}$  \\
    B.iii.8  &  $R_{T}^{z' z}$  &  $R_{T}^{x' x}$ & $\hspace*{1pt}^{s} R^{0z}_{LT}$  &  $\hspace*{1pt}^{c} R^{0x}_{LT'}$  &  $\hspace*{1pt}^{s} R^{x'x}_{LT'}$  &  $\hspace*{1pt}^{c} R^{z'x}_{LT}$  &  $\hspace*{1pt}^{c} R^{x'x}_{LT}$  &  $\hspace*{1pt}^{s} R^{z'x}_{LT'}$  \\ 
  & $\Ocal^{c}_{2+}$ & $\Ocal^{c}_{2-}$ & $\Ocal^{f}_{1+}$  &  $\Ocal^{f}_{1-}$ & $\Ocal^{h}_{1+}$ & $\Ocal^{h}_{1-}$ & $\Ocal^{h}_{2+}$ & $\Ocal^{h}_{2-}$  \\
   B.iii.9  &  $R_{T}^{z' z}$  &  $R_{T}^{x' x}$ & $\hspace*{1pt}^{s} R^{0z}_{LT}$  &  $\hspace*{1pt}^{c} R^{0x}_{LT'}$  &  $\hspace*{1pt}^{c} R^{0z}_{LT'}$  &  $\hspace*{1pt}^{s} R^{0x}_{LT}$  &  $\hspace*{1pt}^{s} R^{x'x}_{LT'}$  &  $\hspace*{1pt}^{c} R^{z'x}_{LT}$  \\    
  & $\Ocal^{c}_{2+}$ & $\Ocal^{c}_{2-}$ & $\Ocal^{f}_{1+}$  &  $\Ocal^{f}_{1-}$ & $\Ocal^{f}_{2+}$ & $\Ocal^{f}_{2-}$ & $\Ocal^{h}_{1+}$ & $\Ocal^{h}_{1-}$  \\  
  B.iii.10  &  $R^{x'z}_{T}$  &  $R^{z'x}_{T}$ & $\hspace*{1pt}^{s} R^{0z}_{LT}$  &  $\hspace*{1pt}^{c} R^{0x}_{LT'}$  &  $\hspace*{1pt}^{s} R^{x'x}_{LT'}$  &  $\hspace*{1pt}^{c} R^{z'x}_{LT}$  &  $\hspace*{1pt}^{c} R^{x'x}_{LT}$  &  $\hspace*{1pt}^{s} R^{z'x}_{LT'}$  \\   
   & $\Ocal^{c}_{1+}$ & $\Ocal^{c}_{1-}$ & $\Ocal^{f}_{1+}$  &  $\Ocal^{f}_{1-}$ & $\Ocal^{h}_{1+}$ & $\Ocal^{h}_{1-}$ & $\Ocal^{h}_{2+}$ & $\Ocal^{h}_{2-}$  \\ 
  B.iii.11  &  $R^{x'z}_{T}$  &  $R^{z'x}_{T}$ & $R_{T}^{z' z}$ & $R_{T}^{x' x}$  &  $\hspace*{1pt}^{s} R^{0z}_{LT}$  &  $\hspace*{1pt}^{c} R^{0x}_{LT'}$  &  $\hspace*{1pt}^{s} R^{x'x}_{LT'}$  &  $\hspace*{1pt}^{c} R^{z'x}_{LT}$  \\ 
   & $\Ocal^{c}_{1+}$ & $\Ocal^{c}_{1-}$ & $\Ocal^{c}_{2+}$  &  $\Ocal^{c}_{2-}$ & $\Ocal^{f}_{1+}$ & $\Ocal^{f}_{1-}$ & $\Ocal^{h}_{1+}$ & $\Ocal^{h}_{1-}$  \\ 
  B.iii.12  &  $R^{x'z}_{T}$  &  $R^{z'x}_{T}$ & $\hspace*{1pt}^{s} R^{0z}_{LT}$  &  $\hspace*{1pt}^{c} R^{0x}_{LT'}$  &  $\hspace*{1pt}^{c} R^{0z}_{LT'}$  &  $\hspace*{1pt}^{s} R^{0x}_{LT}$  &  $\hspace*{1pt}^{s} R^{x'x}_{LT'}$  &  $\hspace*{1pt}^{c} R^{z'x}_{LT}$  \\
   & $\Ocal^{c}_{1+}$ & $\Ocal^{c}_{1-}$ & $\Ocal^{f}_{1+}$  &  $\Ocal^{f}_{1-}$ & $\Ocal^{f}_{2+}$ & $\Ocal^{f}_{2-}$ & $\Ocal^{h}_{1+}$ & $\Ocal^{h}_{1-}$  \\
  \hline
  B.iv.1  &  $R^{x'z}_{T}$  &  $R^{z'x}_{T}$ & $R_{T}^{z' z}$ & $R_{T}^{x' x}$  &  $\hspace*{1pt}^{c} R^{00}_{LT}$  &  $\hspace*{1pt}^{c} R^{0y}_{LT}$  &  $\hspace*{1pt}^{c} R^{0z}_{LT'}$  &  $\hspace*{1pt}^{s} R^{0x}_{LT}$  \\   
   & $\Ocal^{c}_{1+}$ & $\Ocal^{c}_{1-}$ & $\Ocal^{c}_{2+}$  &  $\Ocal^{c}_{2-}$ & $\Ocal^{e}_{2+}$ & $\Ocal^{e}_{2-}$ & $\Ocal^{f}_{2+}$ & $\Ocal^{f}_{2-}$  \\ 
  B.iv.2  &  $R_{T}^{z' z}$  &  $R_{T}^{x' x}$ & $\hspace*{1pt}^{c} R^{00}_{LT}$  &  $\hspace*{1pt}^{c} R^{0y}_{LT}$  &  $\hspace*{1pt}^{s} R^{0z}_{LT}$  &  $\hspace*{1pt}^{c} R^{0x}_{LT'}$  &  $\hspace*{1pt}^{c} R^{0z}_{LT'}$  &  $\hspace*{1pt}^{s} R^{0x}_{LT}$  \\   
  & $\Ocal^{c}_{2+}$ & $\Ocal^{c}_{2-}$ & $\Ocal^{e}_{2+}$  &  $\Ocal^{e}_{2-}$ & $\Ocal^{f}_{1+}$ & $\Ocal^{f}_{1-}$ & $\Ocal^{f}_{2+}$ & $\Ocal^{f}_{2-}$  \\  
  B.iv.3  &  $R_{T}^{z' z}$  &  $R_{T}^{x' x}$ & $\hspace*{1pt}^{s} R^{00}_{LT'}$  &  $\hspace*{1pt}^{s} R^{0y}_{LT'}$  &  $\hspace*{1pt}^{c} R^{00}_{LT}$  &  $\hspace*{1pt}^{c} R^{0y}_{LT}$  &  $\hspace*{1pt}^{c} R^{0z}_{LT'}$  &  $\hspace*{1pt}^{s} R^{0x}_{LT}$  \\  
   & $\Ocal^{c}_{2+}$ & $\Ocal^{c}_{2-}$ & $\Ocal^{e}_{1+}$  &  $\Ocal^{e}_{1-}$ & $\Ocal^{e}_{2+}$ & $\Ocal^{e}_{2-}$ & $\Ocal^{f}_{2+}$ & $\Ocal^{f}_{2-}$  \\  
  B.iv.4  &  $R^{x'z}_{T}$  &  $R^{z'x}_{T}$ & $R_{T}^{z' z}$ & $R_{T}^{x' x}$  &  $\hspace*{1pt}^{c} R^{00}_{LT}$  &  $\hspace*{1pt}^{c} R^{0y}_{LT}$  &  $\hspace*{1pt}^{s} R^{0z}_{LT}$  &  $\hspace*{1pt}^{c} R^{0x}_{LT'}$  \\  
  & $\Ocal^{c}_{1+}$ & $\Ocal^{c}_{1-}$ & $\Ocal^{c}_{2+}$  &  $\Ocal^{c}_{2-}$ & $\Ocal^{e}_{2+}$ & $\Ocal^{e}_{2-}$ & $\Ocal^{f}_{1+}$ & $\Ocal^{f}_{1-}$  \\  
  B.iv.5  &  $R^{x'z}_{T}$  &  $R^{z'x}_{T}$ & $\hspace*{1pt}^{c} R^{00}_{LT}$  &  $\hspace*{1pt}^{c} R^{0y}_{LT}$  &  $\hspace*{1pt}^{s} R^{0z}_{LT}$  &  $\hspace*{1pt}^{c} R^{0x}_{LT'}$  &  $\hspace*{1pt}^{c} R^{0z}_{LT'}$  &  $\hspace*{1pt}^{s} R^{0x}_{LT}$  \\   
  & $\Ocal^{c}_{1+}$ & $\Ocal^{c}_{1-}$ & $\Ocal^{e}_{2+}$  &  $\Ocal^{e}_{2-}$ & $\Ocal^{f}_{1+}$ & $\Ocal^{f}_{1-}$ & $\Ocal^{f}_{2+}$ & $\Ocal^{f}_{2-}$  \\    
  B.iv.6  &  $R^{x'z}_{T}$  &  $R^{z'x}_{T}$ & $\hspace*{1pt}^{s} R^{00}_{LT'}$  &  $\hspace*{1pt}^{s} R^{0y}_{LT'}$  &  $\hspace*{1pt}^{c} R^{00}_{LT}$  &  $\hspace*{1pt}^{c} R^{0y}_{LT}$  &  $\hspace*{1pt}^{c} R^{0z}_{LT'}$  &  $\hspace*{1pt}^{s} R^{0x}_{LT}$  \\    
  & $\Ocal^{c}_{1+}$ & $\Ocal^{c}_{1-}$ & $\Ocal^{e}_{1+}$  &  $\Ocal^{e}_{1-}$ & $\Ocal^{e}_{2+}$ & $\Ocal^{e}_{2-}$ & $\Ocal^{f}_{2+}$ & $\Ocal^{f}_{2-}$ \\  
  B.iv.7   &  $R^{x'z}_{T}$  &  $R^{z'x}_{T}$ & $R_{T}^{z' z}$ & $R_{T}^{x' x}$  &  $\hspace*{1pt}^{s} R^{00}_{LT'}$  &  $\hspace*{1pt}^{s} R^{0y}_{LT'}$  &  $\hspace*{1pt}^{c} R^{0z}_{LT'}$  &  $\hspace*{1pt}^{s} R^{0x}_{LT}$  \\   
    & $\Ocal^{c}_{1+}$ & $\Ocal^{c}_{1-}$ & $\Ocal^{c}_{2+}$  &  $\Ocal^{c}_{2-}$ & $\Ocal^{e}_{1+}$ & $\Ocal^{e}_{1-}$ & $\Ocal^{f}_{2+}$ & $\Ocal^{f}_{2-}$  \\
  B.iv.8  &  $R_{T}^{z' z}$  &  $R_{T}^{x' x}$ & $\hspace*{1pt}^{s} R^{00}_{LT'}$  &  $\hspace*{1pt}^{s} R^{0y}_{LT'}$  &  $\hspace*{1pt}^{c} R^{00}_{LT}$  &  $\hspace*{1pt}^{c} R^{0y}_{LT}$  &  $\hspace*{1pt}^{s} R^{0z}_{LT}$  &  $\hspace*{1pt}^{c} R^{0x}_{LT'}$  \\    
   & $\Ocal^{c}_{2+}$ & $\Ocal^{c}_{2-}$ & $\Ocal^{e}_{1+}$  &  $\Ocal^{e}_{1-}$ & $\Ocal^{e}_{2+}$ & $\Ocal^{e}_{2-}$ & $\Ocal^{f}_{1+}$ & $\Ocal^{f}_{1-}$  \\    
  B.iv.9  &  $R_{T}^{z' z}$  &  $R_{T}^{x' x}$ & $\hspace*{1pt}^{s} R^{00}_{LT'}$  &  $\hspace*{1pt}^{s} R^{0y}_{LT'}$  &  $\hspace*{1pt}^{s} R^{0z}_{LT}$  &  $\hspace*{1pt}^{c} R^{0x}_{LT'}$  &  $\hspace*{1pt}^{c} R^{0z}_{LT'}$  &  $\hspace*{1pt}^{s} R^{0x}_{LT}$  \\   
   & $\Ocal^{c}_{2+}$ & $\Ocal^{c}_{2-}$ & $\Ocal^{e}_{1+}$  &  $\Ocal^{e}_{1-}$ & $\Ocal^{f}_{1+}$ & $\Ocal^{f}_{1-}$ & $\Ocal^{f}_{2+}$ & $\Ocal^{f}_{2-}$  \\   
  B.iv.10  &  $R^{x'z}_{T}$  &  $R^{z'x}_{T}$ & $\hspace*{1pt}^{s} R^{00}_{LT'}$  &  $\hspace*{1pt}^{s} R^{0y}_{LT'}$  &  $\hspace*{1pt}^{c} R^{00}_{LT}$  &  $\hspace*{1pt}^{c} R^{0y}_{LT}$  &  $\hspace*{1pt}^{s} R^{0z}_{LT}$  &  $\hspace*{1pt}^{c} R^{0x}_{LT'}$  \\  
   & $\Ocal^{c}_{1+}$ & $\Ocal^{c}_{1-}$ & $\Ocal^{e}_{1+}$  &  $\Ocal^{e}_{1-}$ & $\Ocal^{e}_{2+}$ & $\Ocal^{e}_{2-}$ & $\Ocal^{f}_{1+}$ & $\Ocal^{f}_{1-}$  \\ 
   B.iv.11   &  $R^{x'z}_{T}$  &  $R^{z'x}_{T}$ & $R_{T}^{z' z}$ & $R_{T}^{x' x}$  &  $\hspace*{1pt}^{s} R^{00}_{LT'}$  &  $\hspace*{1pt}^{s} R^{0y}_{LT'}$  &  $\hspace*{1pt}^{s} R^{0z}_{LT}$  &  $\hspace*{1pt}^{c} R^{0x}_{LT'}$  \\  
   & $\Ocal^{c}_{1+}$ & $\Ocal^{c}_{1-}$ & $\Ocal^{c}_{2+}$  &  $\Ocal^{c}_{2-}$ & $\Ocal^{e}_{1+}$ & $\Ocal^{e}_{1-}$ & $\Ocal^{f}_{1+}$ & $\Ocal^{f}_{1-}$  \\
  B.iv.12   &  $R^{x'z}_{T}$  &  $R^{z'x}_{T}$ & $\hspace*{1pt}^{s} R^{00}_{LT'}$  &  $\hspace*{1pt}^{s} R^{0y}_{LT'}$  &  $\hspace*{1pt}^{s} R^{0z}_{LT}$  &  $\hspace*{1pt}^{c} R^{0x}_{LT'}$  &  $\hspace*{1pt}^{c} R^{0z}_{LT'}$  &  $\hspace*{1pt}^{s} R^{0x}_{LT}$  \\   
    & $\Ocal^{c}_{1+}$ & $\Ocal^{c}_{1-}$ & $\Ocal^{e}_{1+}$  &  $\Ocal^{e}_{1-}$ & $\Ocal^{f}_{1+}$ & $\Ocal^{f}_{1-}$ & $\Ocal^{f}_{2+}$ & $\Ocal^{f}_{2-}$  
\end{tabular}
\end{small}
\caption{Table~\ref{tab:MinimalMoravcsikSetsElectroII} is continued here. The second set of $24$ cases from a total of $96$ possibilities is listed.}
\label{tab:MinimalMoravcsikSetsElectroIIOneTwo}
\end{table}

\begin{table}
\begin{scriptsize}
\begin{tabular}{l|cccccccc}
Set-Nr. & \multicolumn{8}{c}{Observables}  \\
\hline   
  B.v.1  &  $\hspace*{1pt}^{s} R_{TT}^{0z}$ & $R_{TT'}^{0x}$  &  $R_{TT'}^{0z}$ & $\hspace*{1pt}^{s} R^{0x}_{TT}$  &  $\hspace*{1pt}^{c} R^{00}_{LT}$  &  $\hspace*{1pt}^{c} R^{0y}_{LT}$  &  $\hspace*{1pt}^{c} R^{z'0}_{LT'}$  &  $\hspace*{1pt}^{s} R^{x'0}_{LT}$   \\     
  & $\Ocal^{a}_{1+}$ & $\Ocal^{a}_{1-}$ & $\Ocal^{a}_{2+}$  &  $\Ocal^{a}_{2-}$ & $\Ocal^{e}_{2+}$ & $\Ocal^{e}_{2-}$ & $\Ocal^{g}_{2+}$ & $\Ocal^{g}_{2-}$  \\  
  B.v.2  &  $R_{TT'}^{0z}$  &  $\hspace*{1pt}^{s} R^{0x}_{TT}$  &  $\hspace*{1pt}^{c} R^{00}_{LT}$  &  $\hspace*{1pt}^{c} R^{0y}_{LT}$  &  $\hspace*{1pt}^{s} R^{z'0}_{LT}$  &  $\hspace*{1pt}^{c} R^{x'0}_{LT'}$  &  $\hspace*{1pt}^{c} R^{z'0}_{LT'}$  &  $\hspace*{1pt}^{s} R^{x'0}_{LT}$  \\     
  & $\Ocal^{a}_{2+}$ & $\Ocal^{a}_{2-}$ & $\Ocal^{e}_{2+}$  &  $\Ocal^{e}_{2-}$ & $\Ocal^{g}_{1+}$ & $\Ocal^{g}_{1-}$ & $\Ocal^{g}_{2+}$ & $\Ocal^{g}_{2-}$   \\  
  B.v.3  &  $R_{TT'}^{0z}$  &  $\hspace*{1pt}^{s} R^{0x}_{TT}$  &  $\hspace*{1pt}^{s} R^{00}_{LT'}$  &  $\hspace*{1pt}^{s} R^{0y}_{LT'}$  &  $\hspace*{1pt}^{c} R^{00}_{LT}$  &  $\hspace*{1pt}^{c} R^{0y}_{LT}$  &  $\hspace*{1pt}^{c} R^{z'0}_{LT'}$  &  $\hspace*{1pt}^{s} R^{x'0}_{LT}$   \\ 
  & $\Ocal^{a}_{2+}$ & $\Ocal^{a}_{2-}$ & $\Ocal^{e}_{1+}$  &  $\Ocal^{e}_{1-}$ & $\Ocal^{e}_{2+}$ & $\Ocal^{e}_{2-}$ & $\Ocal^{g}_{2+}$ & $\Ocal^{g}_{2-}$  \\   
  B.v.4  &  $\hspace*{1pt}^{s} R_{TT}^{0z}$ & $R_{TT'}^{0x}$  &  $\hspace*{1pt}^{c} R^{00}_{LT}$  &  $\hspace*{1pt}^{c} R^{0y}_{LT}$  &  $\hspace*{1pt}^{s} R^{z'0}_{LT}$  &  $\hspace*{1pt}^{c} R^{x'0}_{LT'}$  &  $\hspace*{1pt}^{c} R^{z'0}_{LT'}$  &  $\hspace*{1pt}^{s} R^{x'0}_{LT}$   \\    
  & $\Ocal^{a}_{1+}$ & $\Ocal^{a}_{1-}$ & $\Ocal^{e}_{2+}$  &  $\Ocal^{e}_{2-}$ & $\Ocal^{g}_{1+}$ & $\Ocal^{g}_{1-}$ & $\Ocal^{g}_{2+}$ & $\Ocal^{g}_{2-}$   \\  
  B.v.5  &  $\hspace*{1pt}^{s} R_{TT}^{0z}$ & $R_{TT'}^{0x}$  &  $R_{TT'}^{0z}$ & $\hspace*{1pt}^{s} R^{0x}_{TT}$  &  $\hspace*{1pt}^{c} R^{00}_{LT}$  &  $\hspace*{1pt}^{c} R^{0y}_{LT}$  &  $\hspace*{1pt}^{s} R^{z'0}_{LT}$  &  $\hspace*{1pt}^{c} R^{x'0}_{LT'}$   \\    
  & $\Ocal^{a}_{1+}$ & $\Ocal^{a}_{1-}$ & $\Ocal^{a}_{2+}$  &  $\Ocal^{a}_{2-}$ & $\Ocal^{e}_{2+}$ & $\Ocal^{e}_{2-}$ & $\Ocal^{g}_{1+}$ & $\Ocal^{g}_{1-}$   \\
  B.v.6  &  $\hspace*{1pt}^{s} R_{TT}^{0z}$ & $R_{TT'}^{0x}$  &  $\hspace*{1pt}^{s} R^{00}_{LT'}$  &  $\hspace*{1pt}^{s} R^{0y}_{LT'}$  &  $\hspace*{1pt}^{c} R^{00}_{LT}$  &  $\hspace*{1pt}^{c} R^{0y}_{LT}$  &  $\hspace*{1pt}^{c} R^{z'0}_{LT'}$  &  $\hspace*{1pt}^{s} R^{x'0}_{LT}$    \\   
  & $\Ocal^{a}_{1+}$ & $\Ocal^{a}_{1-}$ & $\Ocal^{e}_{1+}$  &  $\Ocal^{e}_{1-}$ & $\Ocal^{e}_{2+}$ & $\Ocal^{e}_{2-}$ & $\Ocal^{g}_{2+}$ & $\Ocal^{g}_{2-}$   \\  
  B.v.7  &  $\hspace*{1pt}^{s} R_{TT}^{0z}$ & $R_{TT'}^{0x}$  &  $R_{TT'}^{0z}$ & $\hspace*{1pt}^{s} R^{0x}_{TT}$  &  $\hspace*{1pt}^{s} R^{00}_{LT'}$  &  $\hspace*{1pt}^{s} R^{0y}_{LT'}$  &  $\hspace*{1pt}^{c} R^{z'0}_{LT'}$  &  $\hspace*{1pt}^{s} R^{x'0}_{LT}$  \\  
  & $\Ocal^{a}_{1+}$ & $\Ocal^{a}_{1-}$ & $\Ocal^{a}_{2+}$  &  $\Ocal^{a}_{2-}$ & $\Ocal^{e}_{1+}$ & $\Ocal^{e}_{1-}$ & $\Ocal^{g}_{2+}$ & $\Ocal^{g}_{2-}$  \\   
  B.v.8  &  $R_{TT'}^{0z}$  &  $\hspace*{1pt}^{s} R^{0x}_{TT}$  &  $\hspace*{1pt}^{s} R^{00}_{LT'}$  &  $\hspace*{1pt}^{s} R^{0y}_{LT'}$  &  $\hspace*{1pt}^{c} R^{00}_{LT}$  &  $\hspace*{1pt}^{c} R^{0y}_{LT}$  &  $\hspace*{1pt}^{s} R^{z'0}_{LT}$  &  $\hspace*{1pt}^{c} R^{x'0}_{LT'}$   \\   
  & $\Ocal^{a}_{2+}$ & $\Ocal^{a}_{2-}$ & $\Ocal^{e}_{1+}$  &  $\Ocal^{e}_{1-}$ & $\Ocal^{e}_{2+}$ & $\Ocal^{e}_{2-}$ & $\Ocal^{g}_{1+}$ & $\Ocal^{g}_{1-}$   \\ 
  B.v.9  &  $R_{TT'}^{0z}$  &  $\hspace*{1pt}^{s} R^{0x}_{TT}$  &  $\hspace*{1pt}^{s} R^{00}_{LT'}$  &  $\hspace*{1pt}^{s} R^{0y}_{LT'}$  &  $\hspace*{1pt}^{s} R^{z'0}_{LT}$  &  $\hspace*{1pt}^{c} R^{x'0}_{LT'}$  &  $\hspace*{1pt}^{c} R^{z'0}_{LT'}$  &  $\hspace*{1pt}^{s} R^{x'0}_{LT}$   \\   
  & $\Ocal^{a}_{2+}$ & $\Ocal^{a}_{2-}$ & $\Ocal^{e}_{1+}$  &  $\Ocal^{e}_{1-}$ & $\Ocal^{g}_{1+}$ & $\Ocal^{g}_{1-}$ & $\Ocal^{g}_{2+}$ & $\Ocal^{g}_{2-}$  \\   
  B.v.10  &  $\hspace*{1pt}^{s} R_{TT}^{0z}$ & $R_{TT'}^{0x}$  &  $\hspace*{1pt}^{s} R^{00}_{LT'}$  &  $\hspace*{1pt}^{s} R^{0y}_{LT'}$  &  $\hspace*{1pt}^{c} R^{00}_{LT}$  &  $\hspace*{1pt}^{c} R^{0y}_{LT}$  &  $\hspace*{1pt}^{s} R^{z'0}_{LT}$  &  $\hspace*{1pt}^{c} R^{x'0}_{LT'}$   \\    
  & $\Ocal^{a}_{1+}$ & $\Ocal^{a}_{1-}$ & $\Ocal^{e}_{1+}$  &  $\Ocal^{e}_{1-}$ & $\Ocal^{e}_{2+}$ & $\Ocal^{e}_{2-}$ & $\Ocal^{g}_{1+}$ & $\Ocal^{g}_{1-}$  \\ 
  B.v.11  &  $\hspace*{1pt}^{s} R_{TT}^{0z}$ & $R_{TT'}^{0x}$  &  $\hspace*{1pt}^{s} R^{00}_{LT'}$  &  $\hspace*{1pt}^{s} R^{0y}_{LT'}$  &  $\hspace*{1pt}^{s} R^{z'0}_{LT}$  &  $\hspace*{1pt}^{c} R^{x'0}_{LT'}$  &  $\hspace*{1pt}^{c} R^{z'0}_{LT'}$  &  $\hspace*{1pt}^{s} R^{x'0}_{LT}$   \\  
  & $\Ocal^{a}_{1+}$ & $\Ocal^{a}_{1-}$ & $\Ocal^{e}_{1+}$  &  $\Ocal^{e}_{1-}$ & $\Ocal^{g}_{1+}$ & $\Ocal^{g}_{1-}$ & $\Ocal^{g}_{2+}$ & $\Ocal^{g}_{2-}$   \\  
  B.v.12  &  $\hspace*{1pt}^{s} R_{TT}^{0z}$ & $R_{TT'}^{0x}$  &  $R_{TT'}^{0z}$ & $\hspace*{1pt}^{s} R^{0x}_{TT}$  &  $\hspace*{1pt}^{s} R^{00}_{LT'}$  &  $\hspace*{1pt}^{s} R^{0y}_{LT'}$  &  $\hspace*{1pt}^{s} R^{z'0}_{LT}$  &  $\hspace*{1pt}^{c} R^{x'0}_{LT'}$   \\ 
  & $\Ocal^{a}_{1+}$ & $\Ocal^{a}_{1-}$ & $\Ocal^{a}_{2+}$  &  $\Ocal^{a}_{2-}$ & $\Ocal^{e}_{1+}$ & $\Ocal^{e}_{1-}$ & $\Ocal^{g}_{1+}$ & $\Ocal^{g}_{1-}$   \\ 
  \hline
  B.vi.1  &  $\hspace*{1pt}^{s} R_{TT}^{0z}$ & $R_{TT'}^{0x}$  &  $R_{TT'}^{0z}$ & $\hspace*{1pt}^{s} R^{0x}_{TT}$  &  $\hspace*{1pt}^{c} R^{0z}_{LT'}$  &  $\hspace*{1pt}^{s} R^{0x}_{LT}$  &  $\hspace*{1pt}^{c} R^{x'x}_{LT}$  &  $\hspace*{1pt}^{s} R^{z'x}_{LT'}$   \\ 
  & $\Ocal^{a}_{1+}$ & $\Ocal^{a}_{1-}$ & $\Ocal^{a}_{2+}$  &  $\Ocal^{a}_{2-}$ & $\Ocal^{f}_{2+}$ & $\Ocal^{f}_{2-}$ & $\Ocal^{h}_{2+}$ & $\Ocal^{h}_{2-}$   \\ 
   B.vi.2  &  $R_{TT'}^{0z}$  &  $\hspace*{1pt}^{s} R^{0x}_{TT}$  &  $\hspace*{1pt}^{s} R^{0z}_{LT}$  &  $\hspace*{1pt}^{c} R^{0x}_{LT'}$  &  $\hspace*{1pt}^{c} R^{0z}_{LT'}$  &  $\hspace*{1pt}^{s} R^{0x}_{LT}$  &  $\hspace*{1pt}^{c} R^{x'x}_{LT}$  &  $\hspace*{1pt}^{s} R^{z'x}_{LT'}$   \\ 
  & $\Ocal^{a}_{2+}$ & $\Ocal^{a}_{2-}$ & $\Ocal^{f}_{1+}$  &  $\Ocal^{f}_{1-}$ & $\Ocal^{f}_{2+}$ & $\Ocal^{f}_{2-}$ & $\Ocal^{h}_{2+}$ & $\Ocal^{h}_{2-}$  \\ 
  B.vi.3  &  $R_{TT'}^{0z}$  &  $\hspace*{1pt}^{s} R^{0x}_{TT}$  &  $\hspace*{1pt}^{c} R^{0z}_{LT'}$  &  $\hspace*{1pt}^{s} R^{0x}_{LT}$  &  $\hspace*{1pt}^{s} R^{x'x}_{LT'}$  &  $\hspace*{1pt}^{c} R^{z'x}_{LT}$  &  $\hspace*{1pt}^{c} R^{x'x}_{LT}$  &  $\hspace*{1pt}^{s} R^{z'x}_{LT'}$   \\  
  & $\Ocal^{a}_{2+}$ & $\Ocal^{a}_{2-}$ & $\Ocal^{f}_{2+}$  &  $\Ocal^{f}_{2-}$ & $\Ocal^{h}_{1+}$ & $\Ocal^{h}_{1-}$ & $\Ocal^{h}_{2+}$ & $\Ocal^{h}_{2-}$   \\  
  B.vi.4  &  $\hspace*{1pt}^{s} R_{TT}^{0z}$ & $R_{TT'}^{0x}$  &  $\hspace*{1pt}^{s} R^{0z}_{LT}$  &  $\hspace*{1pt}^{c} R^{0x}_{LT'}$  &  $\hspace*{1pt}^{c} R^{0z}_{LT'}$  &  $\hspace*{1pt}^{s} R^{0x}_{LT}$  &  $\hspace*{1pt}^{c} R^{x'x}_{LT}$  &  $\hspace*{1pt}^{s} R^{z'x}_{LT'}$   \\  
  & $\Ocal^{a}_{1+}$ & $\Ocal^{a}_{1-}$ & $\Ocal^{f}_{1+}$  &  $\Ocal^{f}_{1-}$ & $\Ocal^{f}_{2+}$ & $\Ocal^{f}_{2-}$ & $\Ocal^{h}_{2+}$ & $\Ocal^{h}_{2-}$   \\ 
  B.vi.5  &  $\hspace*{1pt}^{s} R_{TT}^{0z}$ & $R_{TT'}^{0x}$  &  $R_{TT'}^{0z}$ & $\hspace*{1pt}^{s} R^{0x}_{TT}$  &  $\hspace*{1pt}^{s} R^{0z}_{LT}$  &  $\hspace*{1pt}^{c} R^{0x}_{LT'}$  &  $\hspace*{1pt}^{c} R^{x'x}_{LT}$  &  $\hspace*{1pt}^{s} R^{z'x}_{LT'}$   \\  
  & $\Ocal^{a}_{1+}$ & $\Ocal^{a}_{1-}$ & $\Ocal^{a}_{2+}$  &  $\Ocal^{a}_{2-}$ & $\Ocal^{f}_{1+}$ & $\Ocal^{f}_{1-}$ & $\Ocal^{h}_{2+}$ & $\Ocal^{h}_{2-}$   \\
  B.vi.6  &  $\hspace*{1pt}^{s} R_{TT}^{0z}$ & $R_{TT'}^{0x}$  &  $\hspace*{1pt}^{c} R^{0z}_{LT'}$  &  $\hspace*{1pt}^{s} R^{0x}_{LT}$  &  $\hspace*{1pt}^{s} R^{x'x}_{LT'}$  &  $\hspace*{1pt}^{c} R^{z'x}_{LT}$  &  $\hspace*{1pt}^{c} R^{x'x}_{LT}$  &  $\hspace*{1pt}^{s} R^{z'x}_{LT'}$   \\   
  & $\Ocal^{a}_{1+}$ & $\Ocal^{a}_{1-}$ & $\Ocal^{f}_{2+}$  &  $\Ocal^{f}_{2-}$ & $\Ocal^{h}_{1+}$ & $\Ocal^{h}_{1-}$ & $\Ocal^{h}_{2+}$ & $\Ocal^{h}_{2-}$   \\  
  B.vi.7  &  $\hspace*{1pt}^{s} R_{TT}^{0z}$ & $R_{TT'}^{0x}$  &  $R_{TT'}^{0z}$ & $\hspace*{1pt}^{s} R^{0x}_{TT}$  &  $\hspace*{1pt}^{c} R^{0z}_{LT'}$  &  $\hspace*{1pt}^{s} R^{0x}_{LT}$  &  $\hspace*{1pt}^{s} R^{x'x}_{LT'}$  &  $\hspace*{1pt}^{c} R^{z'x}_{LT}$   \\    
  & $\Ocal^{a}_{1+}$ & $\Ocal^{a}_{1-}$ & $\Ocal^{a}_{2+}$  &  $\Ocal^{a}_{2-}$ & $\Ocal^{f}_{2+}$ & $\Ocal^{f}_{2-}$ & $\Ocal^{h}_{1+}$ & $\Ocal^{h}_{1-}$   \\
  B.vi.8  &  $R_{TT'}^{0z}$  &  $\hspace*{1pt}^{s} R^{0x}_{TT}$  &  $\hspace*{1pt}^{s} R^{0z}_{LT}$  &  $\hspace*{1pt}^{c} R^{0x}_{LT'}$  &  $\hspace*{1pt}^{s} R^{x'x}_{LT'}$  &  $\hspace*{1pt}^{c} R^{z'x}_{LT}$  &  $\hspace*{1pt}^{c} R^{x'x}_{LT}$  &  $\hspace*{1pt}^{s} R^{z'x}_{LT'}$   \\ 
  & $\Ocal^{a}_{2+}$ & $\Ocal^{a}_{2-}$ & $\Ocal^{f}_{1+}$  &  $\Ocal^{f}_{1-}$ & $\Ocal^{h}_{1+}$ & $\Ocal^{h}_{1-}$ & $\Ocal^{h}_{2+}$ & $\Ocal^{h}_{2-}$   \\  
  B.vi.9  &  $R_{TT'}^{0z}$  &  $\hspace*{1pt}^{s} R^{0x}_{TT}$  &  $\hspace*{1pt}^{s} R^{0z}_{LT}$  &  $\hspace*{1pt}^{c} R^{0x}_{LT'}$  &  $\hspace*{1pt}^{c} R^{0z}_{LT'}$  &  $\hspace*{1pt}^{s} R^{0x}_{LT}$  &  $\hspace*{1pt}^{s} R^{x'x}_{LT'}$  &  $\hspace*{1pt}^{c} R^{z'x}_{LT}$  \\    
  & $\Ocal^{a}_{2+}$ & $\Ocal^{a}_{2-}$ & $\Ocal^{f}_{1+}$  &  $\Ocal^{f}_{1-}$ & $\Ocal^{f}_{2+}$ & $\Ocal^{f}_{2-}$ & $\Ocal^{h}_{1+}$ & $\Ocal^{h}_{1-}$    \\  
  B.vi.10  &  $\hspace*{1pt}^{s} R_{TT}^{0z}$ & $R_{TT'}^{0x}$  &  $\hspace*{1pt}^{s} R^{0z}_{LT}$  &  $\hspace*{1pt}^{c} R^{0x}_{LT'}$  &  $\hspace*{1pt}^{s} R^{x'x}_{LT'}$  &  $\hspace*{1pt}^{c} R^{z'x}_{LT}$  &  $\hspace*{1pt}^{c} R^{x'x}_{LT}$  &  $\hspace*{1pt}^{s} R^{z'x}_{LT'}$   \\    
  & $\Ocal^{a}_{1+}$ & $\Ocal^{a}_{1-}$ & $\Ocal^{f}_{1+}$  &  $\Ocal^{f}_{1-}$ & $\Ocal^{h}_{1+}$ & $\Ocal^{h}_{1-}$ & $\Ocal^{h}_{2+}$ & $\Ocal^{h}_{2-}$   \\ 
  B.vi.11  &  $\hspace*{1pt}^{s} R_{TT}^{0z}$ & $R_{TT'}^{0x}$  &  $\hspace*{1pt}^{s} R^{0z}_{LT}$  &  $\hspace*{1pt}^{c} R^{0x}_{LT'}$  &  $\hspace*{1pt}^{c} R^{0z}_{LT'}$  &  $\hspace*{1pt}^{s} R^{0x}_{LT}$  &  $\hspace*{1pt}^{s} R^{x'x}_{LT'}$  &  $\hspace*{1pt}^{c} R^{z'x}_{LT}$   \\   
  & $\Ocal^{a}_{1+}$ & $\Ocal^{a}_{1-}$ & $\Ocal^{f}_{1+}$  &  $\Ocal^{f}_{1-}$ & $\Ocal^{f}_{2+}$ & $\Ocal^{f}_{2-}$ & $\Ocal^{h}_{1+}$ & $\Ocal^{h}_{1-}$   \\  
  B.vi.12  &  $\hspace*{1pt}^{s} R_{TT}^{0z}$ & $R_{TT'}^{0x}$  &  $R_{TT'}^{0z}$ & $\hspace*{1pt}^{s} R^{0x}_{TT}$  &  $\hspace*{1pt}^{s} R^{0z}_{LT}$  &  $\hspace*{1pt}^{c} R^{0x}_{LT'}$  &  $\hspace*{1pt}^{s} R^{x'x}_{LT'}$  &  $\hspace*{1pt}^{c} R^{z'x}_{LT}$   \\ 
  & $\Ocal^{a}_{1+}$ & $\Ocal^{a}_{1-}$ & $\Ocal^{a}_{2+}$  &  $\Ocal^{a}_{2-}$ & $\Ocal^{f}_{1+}$ & $\Ocal^{f}_{1-}$ & $\Ocal^{h}_{1+}$ & $\Ocal^{h}_{1-}$    
\end{tabular}
\end{scriptsize}
\caption{Table~\ref{tab:MinimalMoravcsikSetsElectroIIOneTwo} is continued here. The third set of $24$ cases from a total of $96$ possibilities is listed.}
\label{tab:MinimalMoravcsikSetsElectroII2}
\end{table}

\begin{table}
\begin{scriptsize}
\begin{tabular}{l|cccccccc}
Set-Nr. & \multicolumn{8}{c}{Observables}  \\
\hline   
    B.vii.1   &  $\hspace*{1pt}^{s} R^{z'0}_{TT}$ & $R^{x'0}_{TT'}$  &  $R^{z'0}_{TT'}$ & $\hspace*{1pt}^{s} R^{x'0}_{TT}$  &  $\hspace*{1pt}^{c} R^{z'0}_{LT'}$  &  $\hspace*{1pt}^{s} R^{x'0}_{LT}$  &  $\hspace*{1pt}^{c} R^{x'x}_{LT}$  &  $\hspace*{1pt}^{s} R^{z'x}_{LT'}$  \\     
    & $\Ocal^{b}_{1+}$ & $\Ocal^{b}_{1-}$ & $\Ocal^{b}_{2+}$  &  $\Ocal^{b}_{2-}$ & $\Ocal^{g}_{2+}$ & $\Ocal^{g}_{2-}$ & $\Ocal^{h}_{2+}$ & $\Ocal^{h}_{2-}$  \\  
  B.vii.2   &  $R^{z'0}_{TT'}$  &  $\hspace*{1pt}^{s} R^{x'0}_{TT}$  &  $\hspace*{1pt}^{s} R^{z'0}_{LT}$  &  $\hspace*{1pt}^{c} R^{x'0}_{LT'}$  &  $\hspace*{1pt}^{c} R^{z'0}_{LT'}$  &  $\hspace*{1pt}^{s} R^{x'0}_{LT}$  &  $\hspace*{1pt}^{c} R^{x'x}_{LT}$  &  $\hspace*{1pt}^{s} R^{z'x}_{LT'}$  \\     
  & $\Ocal^{b}_{2+}$ & $\Ocal^{b}_{2-}$ & $\Ocal^{g}_{1+}$  &  $\Ocal^{g}_{1-}$ & $\Ocal^{g}_{2+}$ & $\Ocal^{g}_{2-}$ & $\Ocal^{h}_{2+}$ & $\Ocal^{h}_{2-}$  \\  
  B.vii.3   &  $R^{z'0}_{TT'}$  &  $\hspace*{1pt}^{s} R^{x'0}_{TT}$  &  $\hspace*{1pt}^{c} R^{z'0}_{LT'}$  &  $\hspace*{1pt}^{s} R^{x'0}_{LT}$  &  $\hspace*{1pt}^{s} R^{x'x}_{LT'}$  &  $\hspace*{1pt}^{c} R^{z'x}_{LT}$  &  $\hspace*{1pt}^{c} R^{x'x}_{LT}$  &  $\hspace*{1pt}^{s} R^{z'x}_{LT'}$  \\ 
  & $\Ocal^{b}_{2+}$ & $\Ocal^{b}_{2-}$ & $\Ocal^{g}_{2+}$  &  $\Ocal^{g}_{2-}$ & $\Ocal^{h}_{1+}$ & $\Ocal^{h}_{1-}$ & $\Ocal^{h}_{2+}$ & $\Ocal^{h}_{2-}$  \\   
  B.vii.4  &  $\hspace*{1pt}^{s} R^{z'0}_{TT}$ & $R^{x'0}_{TT'}$  &  $\hspace*{1pt}^{s} R^{z'0}_{LT}$  &  $\hspace*{1pt}^{c} R^{x'0}_{LT'}$  &  $\hspace*{1pt}^{c} R^{z'0}_{LT'}$  &  $\hspace*{1pt}^{s} R^{x'0}_{LT}$  &  $\hspace*{1pt}^{c} R^{x'x}_{LT}$  &  $\hspace*{1pt}^{s} R^{z'x}_{LT'}$  \\    
  & $\Ocal^{b}_{1+}$ & $\Ocal^{b}_{1-}$ & $\Ocal^{g}_{1+}$  &  $\Ocal^{g}_{1-}$ & $\Ocal^{g}_{2+}$ & $\Ocal^{g}_{2-}$ & $\Ocal^{h}_{2+}$ & $\Ocal^{h}_{2-}$  \\  
  B.vii.5   &  $\hspace*{1pt}^{s} R^{z'0}_{TT}$ & $R^{x'0}_{TT'}$  &  $R^{z'0}_{TT'}$ & $\hspace*{1pt}^{s} R^{x'0}_{TT}$  &  $\hspace*{1pt}^{s} R^{z'0}_{LT}$  &  $\hspace*{1pt}^{c} R^{x'0}_{LT'}$  &  $\hspace*{1pt}^{c} R^{x'x}_{LT}$  &  $\hspace*{1pt}^{s} R^{z'x}_{LT'}$  \\    
   & $\Ocal^{b}_{1+}$ & $\Ocal^{b}_{1-}$ & $\Ocal^{b}_{2+}$  &  $\Ocal^{b}_{2-}$ & $\Ocal^{g}_{1+}$ & $\Ocal^{g}_{1-}$ & $\Ocal^{h}_{2+}$ & $\Ocal^{h}_{2-}$  \\
  B.vii.6   &  $\hspace*{1pt}^{s} R^{z'0}_{TT}$ & $R^{x'0}_{TT'}$  &  $\hspace*{1pt}^{c} R^{z'0}_{LT'}$  &  $\hspace*{1pt}^{s} R^{x'0}_{LT}$  &  $\hspace*{1pt}^{s} R^{x'x}_{LT'}$  &  $\hspace*{1pt}^{c} R^{z'x}_{LT}$  &  $\hspace*{1pt}^{c} R^{x'x}_{LT}$  &  $\hspace*{1pt}^{s} R^{z'x}_{LT'}$  \\   
   & $\Ocal^{b}_{1+}$ & $\Ocal^{b}_{1-}$ & $\Ocal^{g}_{2+}$  &  $\Ocal^{g}_{2-}$ & $\Ocal^{h}_{1+}$ & $\Ocal^{h}_{1-}$ & $\Ocal^{h}_{2+}$ & $\Ocal^{h}_{2-}$  \\  
  B.vii.7   &  $\hspace*{1pt}^{s} R^{z'0}_{TT}$ & $R^{x'0}_{TT'}$  &  $R^{z'0}_{TT'}$ & $\hspace*{1pt}^{s} R^{x'0}_{TT}$  &  $\hspace*{1pt}^{c} R^{z'0}_{LT'}$  &  $\hspace*{1pt}^{s} R^{x'0}_{LT}$  &  $\hspace*{1pt}^{s} R^{x'x}_{LT'}$  &  $\hspace*{1pt}^{c} R^{z'x}_{LT}$  \\  
   & $\Ocal^{b}_{1+}$ & $\Ocal^{b}_{1-}$ & $\Ocal^{b}_{2+}$  &  $\Ocal^{b}_{2-}$ & $\Ocal^{g}_{2+}$ & $\Ocal^{g}_{2-}$ & $\Ocal^{h}_{1+}$ & $\Ocal^{h}_{1-}$  \\   
  B.vii.8   &  $R^{z'0}_{TT'}$  &  $\hspace*{1pt}^{s} R^{x'0}_{TT}$  &  $\hspace*{1pt}^{s} R^{z'0}_{LT}$  &  $\hspace*{1pt}^{c} R^{x'0}_{LT'}$  &  $\hspace*{1pt}^{s} R^{x'x}_{LT'}$  &  $\hspace*{1pt}^{c} R^{z'x}_{LT}$  &  $\hspace*{1pt}^{c} R^{x'x}_{LT}$  &  $\hspace*{1pt}^{s} R^{z'x}_{LT'}$  \\   
   & $\Ocal^{b}_{2+}$ & $\Ocal^{b}_{2-}$ & $\Ocal^{g}_{1+}$  &  $\Ocal^{g}_{1-}$ & $\Ocal^{h}_{1+}$ & $\Ocal^{h}_{1-}$ & $\Ocal^{h}_{2+}$ & $\Ocal^{h}_{2-}$  \\ 
  B.vii.9   &  $R^{z'0}_{TT'}$  &  $\hspace*{1pt}^{s} R^{x'0}_{TT}$  &  $\hspace*{1pt}^{s} R^{z'0}_{LT}$  &  $\hspace*{1pt}^{c} R^{x'0}_{LT'}$  &  $\hspace*{1pt}^{c} R^{z'0}_{LT'}$  &  $\hspace*{1pt}^{s} R^{x'0}_{LT}$  &  $\hspace*{1pt}^{s} R^{x'x}_{LT'}$  &  $\hspace*{1pt}^{c} R^{z'x}_{LT}$  \\   
   & $\Ocal^{b}_{2+}$ & $\Ocal^{b}_{2-}$ & $\Ocal^{g}_{1+}$  &  $\Ocal^{g}_{1-}$ & $\Ocal^{g}_{2+}$ & $\Ocal^{g}_{2-}$ & $\Ocal^{h}_{1+}$ & $\Ocal^{h}_{1-}$  \\   
  B.vii.10   &  $\hspace*{1pt}^{s} R^{z'0}_{TT}$ & $R^{x'0}_{TT'}$  &  $\hspace*{1pt}^{s} R^{z'0}_{LT}$  &  $\hspace*{1pt}^{c} R^{x'0}_{LT'}$  &  $\hspace*{1pt}^{s} R^{x'x}_{LT'}$  &  $\hspace*{1pt}^{c} R^{z'x}_{LT}$  &  $\hspace*{1pt}^{c} R^{x'x}_{LT}$  &  $\hspace*{1pt}^{s} R^{z'x}_{LT'}$  \\    
   & $\Ocal^{b}_{1+}$ & $\Ocal^{b}_{1-}$ & $\Ocal^{g}_{1+}$  &  $\Ocal^{g}_{1-}$ & $\Ocal^{h}_{1+}$ & $\Ocal^{h}_{1-}$ & $\Ocal^{h}_{2+}$ & $\Ocal^{h}_{2-}$  \\ 
  B.vii.11   &  $\hspace*{1pt}^{s} R^{z'0}_{TT}$ & $R^{x'0}_{TT'}$  &  $\hspace*{1pt}^{s} R^{z'0}_{LT}$  &  $\hspace*{1pt}^{c} R^{x'0}_{LT'}$  &  $\hspace*{1pt}^{c} R^{z'0}_{LT'}$  &  $\hspace*{1pt}^{s} R^{x'0}_{LT}$  &  $\hspace*{1pt}^{s} R^{x'x}_{LT'}$  &  $\hspace*{1pt}^{c} R^{z'x}_{LT}$  \\  
   & $\Ocal^{b}_{1+}$ & $\Ocal^{b}_{1-}$ & $\Ocal^{g}_{1+}$  &  $\Ocal^{g}_{1-}$ & $\Ocal^{g}_{2+}$ & $\Ocal^{g}_{2-}$ & $\Ocal^{h}_{1+}$ & $\Ocal^{h}_{1-}$  \\  
  B.vii.12   &  $\hspace*{1pt}^{s} R^{z'0}_{TT}$ & $R^{x'0}_{TT'}$  &  $R^{z'0}_{TT'}$ & $\hspace*{1pt}^{s} R^{x'0}_{TT}$  &  $\hspace*{1pt}^{s} R^{z'0}_{LT}$  &  $\hspace*{1pt}^{c} R^{x'0}_{LT'}$  &  $\hspace*{1pt}^{s} R^{x'x}_{LT'}$  &  $\hspace*{1pt}^{c} R^{z'x}_{LT}$  \\ 
   & $\Ocal^{b}_{1+}$ & $\Ocal^{b}_{1-}$ & $\Ocal^{b}_{2+}$  &  $\Ocal^{b}_{2-}$ & $\Ocal^{g}_{1+}$ & $\Ocal^{g}_{1-}$ & $\Ocal^{h}_{1+}$ & $\Ocal^{h}_{1-}$  \\ 
  \hline
  B.viii.1   &  $\hspace*{1pt}^{s} R^{z'0}_{TT}$ & $R^{x'0}_{TT'}$  &  $R^{z'0}_{TT'}$ & $\hspace*{1pt}^{s} R^{x'0}_{TT}$  &  $\hspace*{1pt}^{c} R^{00}_{LT}$  &  $\hspace*{1pt}^{c} R^{0y}_{LT}$  &  $\hspace*{1pt}^{c} R^{0z}_{LT'}$  &  $\hspace*{1pt}^{s} R^{0x}_{LT}$  \\ 
   & $\Ocal^{b}_{1+}$ & $\Ocal^{b}_{1-}$ & $\Ocal^{b}_{2+}$  &  $\Ocal^{b}_{2-}$ & $\Ocal^{e}_{2+}$ & $\Ocal^{e}_{2-}$ & $\Ocal^{f}_{2+}$ & $\Ocal^{f}_{2-}$  \\ 
   B.viii.2   &  $R^{z'0}_{TT'}$  &  $\hspace*{1pt}^{s} R^{x'0}_{TT}$  &  $\hspace*{1pt}^{c} R^{00}_{LT}$  &  $\hspace*{1pt}^{c} R^{0y}_{LT}$  &  $\hspace*{1pt}^{s} R^{0z}_{LT}$  &  $\hspace*{1pt}^{c} R^{0x}_{LT'}$  &  $\hspace*{1pt}^{c} R^{0z}_{LT'}$  &  $\hspace*{1pt}^{s} R^{0x}_{LT}$  \\ 
   & $\Ocal^{b}_{2+}$ & $\Ocal^{b}_{2-}$ & $\Ocal^{e}_{2+}$  &  $\Ocal^{e}_{2-}$ & $\Ocal^{f}_{1+}$ & $\Ocal^{f}_{1-}$ & $\Ocal^{f}_{2+}$ & $\Ocal^{f}_{2-}$  \\ 
  B.viii.3  &  $R^{z'0}_{TT'}$  &  $\hspace*{1pt}^{s} R^{x'0}_{TT}$  &  $\hspace*{1pt}^{s} R^{00}_{LT'}$  &  $\hspace*{1pt}^{s} R^{0y}_{LT'}$  &  $\hspace*{1pt}^{c} R^{00}_{LT}$  &  $\hspace*{1pt}^{c} R^{0y}_{LT}$  &  $\hspace*{1pt}^{c} R^{0z}_{LT'}$  &  $\hspace*{1pt}^{s} R^{0x}_{LT}$  \\  
   & $\Ocal^{b}_{2+}$ & $\Ocal^{b}_{2-}$ & $\Ocal^{e}_{1+}$  &  $\Ocal^{e}_{1-}$ & $\Ocal^{e}_{2+}$ & $\Ocal^{e}_{2-}$ & $\Ocal^{f}_{2+}$ & $\Ocal^{f}_{2-}$  \\  
  B.viii.4   &  $\hspace*{1pt}^{s} R^{z'0}_{TT}$ & $R^{x'0}_{TT'}$  &  $\hspace*{1pt}^{c} R^{00}_{LT}$  &  $\hspace*{1pt}^{c} R^{0y}_{LT}$  &  $\hspace*{1pt}^{s} R^{0z}_{LT}$  &  $\hspace*{1pt}^{c} R^{0x}_{LT'}$  &  $\hspace*{1pt}^{c} R^{0z}_{LT'}$  &  $\hspace*{1pt}^{s} R^{0x}_{LT}$  \\  
   & $\Ocal^{b}_{1+}$ & $\Ocal^{b}_{1-}$ & $\Ocal^{e}_{2+}$  &  $\Ocal^{e}_{2-}$ & $\Ocal^{f}_{1+}$ & $\Ocal^{f}_{1-}$ & $\Ocal^{f}_{2+}$ & $\Ocal^{f}_{2-}$  \\ 
   B.viii.5   &  $\hspace*{1pt}^{s} R^{z'0}_{TT}$ & $R^{x'0}_{TT'}$  &  $R^{z'0}_{TT'}$ & $\hspace*{1pt}^{s} R^{x'0}_{TT}$  &  $\hspace*{1pt}^{c} R^{00}_{LT}$  &  $\hspace*{1pt}^{c} R^{0y}_{LT}$  &  $\hspace*{1pt}^{s} R^{0z}_{LT}$  &  $\hspace*{1pt}^{c} R^{0x}_{LT'}$  \\  
   & $\Ocal^{b}_{1+}$ & $\Ocal^{b}_{1-}$ & $\Ocal^{b}_{2+}$  &  $\Ocal^{b}_{2-}$ & $\Ocal^{e}_{2+}$ & $\Ocal^{e}_{2-}$ & $\Ocal^{f}_{1+}$ & $\Ocal^{f}_{1-}$  \\
   B.viii.6   &  $\hspace*{1pt}^{s} R^{z'0}_{TT}$ & $R^{x'0}_{TT'}$  &  $\hspace*{1pt}^{s} R^{00}_{LT'}$  &  $\hspace*{1pt}^{s} R^{0y}_{LT'}$  &  $\hspace*{1pt}^{c} R^{00}_{LT}$  &  $\hspace*{1pt}^{c} R^{0y}_{LT}$  &  $\hspace*{1pt}^{c} R^{0z}_{LT'}$  &  $\hspace*{1pt}^{s} R^{0x}_{LT}$  \\   
   & $\Ocal^{b}_{1+}$ & $\Ocal^{b}_{1-}$ & $\Ocal^{e}_{1+}$  &  $\Ocal^{e}_{1-}$ & $\Ocal^{e}_{2+}$ & $\Ocal^{e}_{2-}$ & $\Ocal^{f}_{2+}$ & $\Ocal^{f}_{2-}$  \\  
  B.viii.7   &  $\hspace*{1pt}^{s} R^{z'0}_{TT}$ & $R^{x'0}_{TT'}$  &  $R^{z'0}_{TT'}$ & $\hspace*{1pt}^{s} R^{x'0}_{TT}$  &  $\hspace*{1pt}^{s} R^{00}_{LT'}$  &  $\hspace*{1pt}^{s} R^{0y}_{LT'}$  &  $\hspace*{1pt}^{c} R^{0z}_{LT'}$  &  $\hspace*{1pt}^{s} R^{0x}_{LT}$  \\    
    & $\Ocal^{b}_{1+}$ & $\Ocal^{b}_{1-}$ & $\Ocal^{b}_{2+}$  &  $\Ocal^{b}_{2-}$ & $\Ocal^{e}_{1+}$ & $\Ocal^{e}_{1-}$ & $\Ocal^{f}_{2+}$ & $\Ocal^{f}_{2-}$  \\
  B.viii.8   &  $R^{z'0}_{TT'}$  &  $\hspace*{1pt}^{s} R^{x'0}_{TT}$  &  $\hspace*{1pt}^{s} R^{00}_{LT'}$  &  $\hspace*{1pt}^{s} R^{0y}_{LT'}$  &  $\hspace*{1pt}^{c} R^{00}_{LT}$  &  $\hspace*{1pt}^{c} R^{0y}_{LT}$  &  $\hspace*{1pt}^{s} R^{0z}_{LT}$  &  $\hspace*{1pt}^{c} R^{0x}_{LT'}$  \\ 
   & $\Ocal^{b}_{2+}$ & $\Ocal^{b}_{2-}$ & $\Ocal^{e}_{1+}$  &  $\Ocal^{e}_{1-}$ & $\Ocal^{e}_{2+}$ & $\Ocal^{e}_{2-}$ & $\Ocal^{f}_{1+}$ & $\Ocal^{f}_{1-}$  \\  
  B.viii.9   &  $R^{z'0}_{TT'}$  &  $\hspace*{1pt}^{s} R^{x'0}_{TT}$  &  $\hspace*{1pt}^{s} R^{00}_{LT'}$  &  $\hspace*{1pt}^{s} R^{0y}_{LT'}$  &  $\hspace*{1pt}^{s} R^{0z}_{LT}$  &  $\hspace*{1pt}^{c} R^{0x}_{LT'}$  &  $\hspace*{1pt}^{c} R^{0z}_{LT'}$  &  $\hspace*{1pt}^{s} R^{0x}_{LT}$  \\    
   & $\Ocal^{b}_{2+}$ & $\Ocal^{b}_{2-}$ & $\Ocal^{e}_{1+}$  &  $\Ocal^{e}_{1-}$ & $\Ocal^{f}_{1+}$ & $\Ocal^{f}_{1-}$ & $\Ocal^{f}_{2+}$ & $\Ocal^{f}_{2-}$  \\  
  B.viii.10   &  $\hspace*{1pt}^{s} R^{z'0}_{TT}$ & $R^{x'0}_{TT'}$  &  $\hspace*{1pt}^{s} R^{00}_{LT'}$  &  $\hspace*{1pt}^{s} R^{0y}_{LT'}$  &  $\hspace*{1pt}^{c} R^{00}_{LT}$  &  $\hspace*{1pt}^{c} R^{0y}_{LT}$  &  $\hspace*{1pt}^{s} R^{0z}_{LT}$  &  $\hspace*{1pt}^{c} R^{0x}_{LT'}$  \\    
   & $\Ocal^{b}_{1+}$ & $\Ocal^{b}_{1-}$ & $\Ocal^{e}_{1+}$  &  $\Ocal^{e}_{1-}$ & $\Ocal^{e}_{2+}$ & $\Ocal^{e}_{2-}$ & $\Ocal^{f}_{1+}$ & $\Ocal^{f}_{1-}$  \\ 
  B.viii.11   &  $\hspace*{1pt}^{s} R^{z'0}_{TT}$ & $R^{x'0}_{TT'}$  &  $\hspace*{1pt}^{s} R^{00}_{LT'}$  &  $\hspace*{1pt}^{s} R^{0y}_{LT'}$  &  $\hspace*{1pt}^{s} R^{0z}_{LT}$  &  $\hspace*{1pt}^{c} R^{0x}_{LT'}$  &  $\hspace*{1pt}^{c} R^{0z}_{LT'}$  &  $\hspace*{1pt}^{s} R^{0x}_{LT}$  \\   
   & $\Ocal^{b}_{1+}$ & $\Ocal^{b}_{1-}$ & $\Ocal^{e}_{1+}$  &  $\Ocal^{e}_{1-}$ & $\Ocal^{f}_{1+}$ & $\Ocal^{f}_{1-}$ & $\Ocal^{f}_{2+}$ & $\Ocal^{f}_{2-}$  \\  
  B.viii.12   &  $\hspace*{1pt}^{s} R^{z'0}_{TT}$ & $R^{x'0}_{TT'}$  &  $R^{z'0}_{TT'}$ & $\hspace*{1pt}^{s} R^{x'0}_{TT}$  &  $\hspace*{1pt}^{s} R^{00}_{LT'}$  &  $\hspace*{1pt}^{s} R^{0y}_{LT'}$  &  $\hspace*{1pt}^{s} R^{0z}_{LT}$  &  $\hspace*{1pt}^{c} R^{0x}_{LT'}$  \\ 
   & $\Ocal^{b}_{1+}$ & $\Ocal^{b}_{1-}$ & $\Ocal^{b}_{2+}$  &  $\Ocal^{b}_{2-}$ & $\Ocal^{e}_{1+}$ & $\Ocal^{e}_{1-}$ & $\Ocal^{f}_{1+}$ & $\Ocal^{f}_{1-}$  
\end{tabular}
\end{scriptsize}
\caption{Table~\ref{tab:MinimalMoravcsikSetsElectroII2} is continued here. The fourth set of $24$ cases from a total of $96$ possibilities is listed.}
\label{tab:MinimalMoravcsikSetsElectroII2OneTwo}
\end{table}

\end{document}